\documentclass[%
 aip,
 amsmath,amssymb,
 preprint,%
]{revtex4-1}

\usepackage{graphicx}
\usepackage{dcolumn}
\usepackage{bm}
\usepackage[utf8]{inputenc}
\usepackage[T1]{fontenc}
\usepackage{mathptmx}
\usepackage{etoolbox}
\usepackage[colorlinks, linkcolor=blue,citecolor=blue, urlcolor=blue,bookmarks=true]{hyperref} 

\newcommand{\etal}{\textit{et al.}}

\usepackage{xr}
\makeatletter
\newcommand*{\addFileDependency}[1]{
\typeout{(#1)}
\@addtofilelist{#1}
\IfFileExists{#1}{}{\typeout{No file #1.}}
}\makeatother

\begin{document}

\preprint{AIP/123-QED}

\title[]{Systematic discrepancies between reference methods for non-covalent interactions within the S66 dataset}

\author{Benjamin X. Shi}
\thanks{These authors contributed equally to this work.}
\affiliation{Yusuf Hamied Department of Chemistry, University of Cambridge, Lensfield Road, Cambridge CB2 1EW, United Kingdom}%

\author{Flaviano Della Pia}
\thanks{These authors contributed equally to this work.}
\affiliation{Yusuf Hamied Department of Chemistry, University of Cambridge, Lensfield Road, Cambridge CB2 1EW, United Kingdom}%

\author{Yasmine S. Al-Hamdani}
\affiliation{Dipartimento di Fisica Ettore Pancini, Universita di Napoli Federico II, Monte Sant’Angelo, I-80126 Napoli, Italy }
\affiliation{Thomas Young Centre, University College London, London WC1E 6BT, United Kingdom }
\affiliation{Department of Earth Sciences, University College London, London WC1E 6BT, United Kingdom}

\author{Angelos Michaelides
}
\affiliation{Yusuf Hamied Department of Chemistry, University of Cambridge, Lensfield Road, Cambridge CB2 1EW, United Kingdom}%

\author{Dario Alfè}
\affiliation{Dipartimento di Fisica Ettore Pancini, Universita di Napoli Federico II, Monte Sant’Angelo, I-80126 Napoli, Italy }
\affiliation{London Centre for Nanotechnology, University College London, London WC1E 6BT,
United Kingdom}
\affiliation{Thomas Young Centre, University College London, London WC1E 6BT, United Kingdom }
\affiliation{Department of Earth Sciences, University College London, London WC1E 6BT, United Kingdom}

\author{Andrea Zen}
\email{andrea.zen@unina.it}
\affiliation{Dipartimento di Fisica Ettore Pancini, Universita di Napoli Federico II, Monte Sant’Angelo, I-80126 Napoli, Italy }
\affiliation{Department of Earth Sciences, University College London, London WC1E 6BT, United Kingdom}

\date{\today}

\begin{abstract}

The accurate treatment of non-covalent interactions is necessary to model a wide range of applications, from molecular crystals to surface catalysts to aqueous solutions and many more.
Quantum diffusion Monte Carlo (DMC) and coupled cluster theory with single, double and perturbative triple excitations [CCSD(T)] are considered two widely-trusted methods for treating non-covalent interactions.
However, while they have been well-validated for small molecules, recent work has indicated that these two methods can disagree by more than $7.5\,$kcal/mol for larger systems.
The origin of this discrepancy remains unknown.
Moreover, the lack of systematic comparisons, particularly for medium-sized complexes, has made it difficult to identify which systems may be prone to such disagreements and the potential scale of these differences.
In this work, we leverage the latest developments in DMC to compute interaction energies for the entire S66 dataset, containing 66 medium-sized complexes with a balanced representation of dispersion and electrostatic interactions.
Comparison to previous CCSD(T) references reveals systematic trends, with DMC predicting stronger binding than CCSD(T) for electrostatic-dominated systems, while the binding becomes weaker for dispersion-dominated systems.
We show that the relative strength of this discrepancy is correlated to the ratio of electrostatic and dispersion interactions, as obtained from energy decomposition analysis methods.
Finally, we have pinpointed model systems: the hydrogen-bonded acetic acid dimer (ID 20)
and dispersion-dominated uracil–cyclopentane dimer (ID 42), where these discrepancies are particularly prominent.
These systems offer cost-effective benchmarks to guide  future developments in DMC, CCSD(T) as well as the wider electronic structure theory community.
\end{abstract}

\maketitle

\section{Introduction}\label{sec:Introduction}

Non-covalent interactions play a crucial role in many areas of science.
These interactions govern the structure of molecular crystals~\cite{beranModelingPolymorphicMolecular2016} (e.g., in pharmaceutical drugs), biomolecules~\cite{rileyNoncovalentInteractionsBiochemistry2011} like DNA and proteins and are relevant to supramolecular~\cite{raynalSupramolecularCatalysisPart2014} science and nanotechnology.~\cite{planasMechanismCarbonDioxide2013,poloniLigandAssistedEnhancementCO22012}
They also underlie important processes across chemistry and biology, from protein-ligand binding~\cite{duInsightsProteinLigand2016}, to catalytic reactions, both on the surface~\cite{r.rehakIncludingDispersionDensity2020} and in solution.~\cite{raynalSupramolecularCatalysisPart2014}
Understanding and unlocking new processes for these applications will increasingly rely on accurate computational modeling tools that can treat non-covalent interactions.~\cite{muller-dethlefsNoncovalentInteractionsChallenge2000}

Two methods of choice for modeling non-covalent interactions are quantum diffusion Monte Carlo~\cite{foulkesQuantumMonteCarlo2001c} (DMC) and coupled cluster theory~\cite{bartlettCoupledclusterTheoryQuantum2007a} with single, double and perturbative triple excitations [CCSD(T)].
While these methods may not be as affordable as density functional theory~\cite{mattssonDesigningMeaningfulDensity2004} (DFT), the reference data they provide are pivotal for benchmarking and parametrizing the density functional approximations (DFAs) necessary for practical routine simulations.
For example, the local density approximation (LDA) and many extensions build upon a DMC-based parametrization of the correlation energy,~\cite{ceperleyGroundStateElectron1980a} while CCSD(T) interaction energy datasets have helped aid in the development of many modern dispersion corrections.~\cite{grimmeSemiempiricalGGAtypeDensity2006,grimmeConsistentAccurateInitio2010a,grimmeDispersionCorrectedMeanFieldElectronic2016,a.priceXDMcorrectedHybridDFT2023}
In particular, the applicability of these methods to larger systems have rapidly expanded in recent years, arising from computer hardware improvements and, more importantly, algorithmic/methodological developments to both DMC~\cite{zenBoostingAccuracySpeed2016c,krogelNexusModularWorkflow2016,bennettNewGenerationEffective2017,bennettNewGenerationEffective2018,annaberdiyevNewGenerationEffective2018,zenNewSchemeFixed2019,needsVariationalDiffusionQuantum2020,nakanoTurboRVBManybodyToolkit2020,kentQMCPACKAdvancesDevelopment2020,nakanoTurboGeniusPythonSuite2023} and CCSD(T).~\cite{riplingerEfficientLinearScaling2013b,riplingerSparseMapsSystematic2016,maExplicitlyCorrelatedLocal2018,nagyOptimizationLinearScalingLocal2018b,nagyApproachingBasisSet2019a,masiosAvertingInfraredCatastrophe2023c,jiangAccurateEfficientOpensource2024,yePeriodicLocalCoupledCluster2024a}

DMC and CCSD(T) solve the Schr\"odinger equation to model the systems with distinct approaches and corresponding approximations.
Despite these differences, there are many examples where DMC and CCSD(T) have come into alignment.
For example, besides small molecules,~\cite{dubeckyQuantumMonteCarlo2013,rezacExtensionsApplicationsA242015,raghavChemicalAccuracyUsing2023} agreement has been obtained for graphene bilayer binding energies~\cite{mostaaniQuantumMonteCarlo2015}, molecular crystal lattice energies,~\cite{zenFastAccurateQuantum2018a,dellapiaDMCICE13AmbientHigh2022b,dellapiaHowAccurateAre2024} molecule-surface interactions~\cite{karaltiAdsorptionWaterMolecule2012,al-hamdaniHowStronglyHydrogen2017,tsatsoulisComparisonQuantumChemistry2017a,al-hamdaniPropertiesWaterBoron2017a,brandenburgPhysisorptionWaterGraphene2019,shiManyBodyMethodsSurface2023a} and vacancy formation energies.~\cite{shiGeneralEmbeddedCluster2022a}
Recently, this agreement has been shown to start to falter~\cite{al-hamdaniInteractionsLargeMolecules2021,schaferUnderstandingDiscrepanciesWavefunction2024,ballesterosCoupledClusterBenchmarks2021,villotCoupledClusterBenchmarking2022a} for large dispersion-bound molecules, with differences as large as $7.5\,$kcal/mol for a buckyball-ring ($\mathrm{C}_{60}@[6]\mathrm{CPPA}$) complex.

The origin of the discrepancy between DMC and CCSD(T) for large dispersion-bound molecules is a topic of current debate,~\cite{schaferUnderstandingDiscrepanciesWavefunction2024,fishmanNewAngleBenchmarking2024,lambieApplicabilityCCSDTDispersion2024,laoCanonicalCoupledCluster2024} particularly on the validity of the perturbative triples (T) contribution in CCSD(T).
Sch{\"a}fer \etal{}~\cite{schaferUnderstandingDiscrepanciesWavefunction2024} have suggested that part of this discrepancy arises from missing contributions in (T) that can be accounted by the (cT) approach.
In addition, Semidalas~\etal{}~\cite{semidalasPostCCSDTCorrectionsS662025} have reported non-trivial discrepancies between CCSD(T) and post-CCSD(T) methods such as CCSDT(Q).
Conversely, Lambie~\etal{}~\cite{lambieApplicabilityCCSDTDispersion2024} have found that CCSD(T) does not differ significantly against CCSDT(Q) using the Pariser-Parr-Pople (PPP) model~\cite{pariserSemiEmpiricalTheoryElectronic1953,pariserSemiEmpiricalTheoryElectronic1953a,popleElectronInteractionUnsaturated1953} for large conjugated systems.
Similarly, Fishman~\etal{}~\cite{fishmanNewAngleBenchmarking2024} and Lao~\cite{laoCanonicalCoupledCluster2024} report only a slight overbinding of CCSD(T) against CCSDT(Q) that cannot explain the discrepancy against DMC.

Understanding these discrepancies between DMC and CCSD(T) for large molecules requires cross-validating these methods across systematic datasets, particularly those involving medium to large sized molecules~\cite{sureComprehensiveBenchmarkAssociation2015,ballesterosCoupledClusterBenchmarks2021,grimmeSupramolecularBindingThermodynamics2012} which sample a range of non-covalent interactions.
While DMC and CCSD(T) have both been compared (to great agreement) for the A24~\cite{rezacDescribingNoncovalentInteractions2013} and S22~\cite{jureckaBenchmarkDatabaseAccurate2006} datasets of small molecular complexes, DMC has not been frequently applied to study medium-sized datasets.
In particular, it has not been used to study the S66 dataset,~\cite{rezacS66WellbalancedDatabase2011} a compilation of 66 dimers that probes the two major types of non-covalent interactions: dispersion and hydrogen-bonding together with those of mixed character.
As well as covering a range of interactions, many of the molecules considered form the building blocks for larger biomolecules along different binding configurations.
Furthermore, the parallel-displaced benzene dimer~\cite{sinnokrotHighlyAccurateCoupled2004} is included in this set of complexes, making it an interesting modeling challenge.
Such a dataset has been pivotal towards benchmarking~\cite{goerigkBenchmarkingDensityFunctional2011,gaoMachineLearningCorrection2016,pengVersatileVanWaals2016a,yuDualhybridDirectRandom2020,grimmeR2SCAN3cSwissArmy2021,ehlertR2SCAND4DispersionCorrected2021,mullerOB97X3cCompositeRangeseparated2023,luSimpleEfficientUniversal2023,leeCorrectingDispersionCorrections2024} DFAs in DFT as well as lower-level approximations to wave-function methods~\cite{rezacExtensionsS66Data2011,rileyPerformanceMP25MP2X2012,altunExtrapolationLimitComplete2020a,sheeRegularizedSecondOrderMoller2021a,lupiJunChSJunChSF12Models2021,semidalasS66NoncovalentInteractions2022,beranImprovedDescriptionIntra2023a} and even machine-learning models.~\cite{christensenOrbNetDenaliMachine2021,villotInitioDispersionPotentials2024}

In this work, we leverage the latest developments in DMC to compute interaction energies for the entire S66 dataset.
When compared to CCSD(T) estimates (taken from the literature), we reveal a consistent weaker binding of dispersion interactions and consistent stronger binding of electrostatic interactions in DMC.
In particular, we show that their differences are correlated to the ratio of electrostatic and dispersive interactions within the system.
The discrepancies in dispersion-dominated systems are shown to be reduced when utilizing an (empirically fitted) CCSD(cT) formulation,~\cite{schaferUnderstandingDiscrepanciesWavefunction2024} although notable differences remain.
We identify specific systems with well-defined differences between DMC and CCSD(T) that can serve as model systems for testing future developments in both methods, setting the stage towards resolving their discrepancies.

\section{Methods}\label{sec:Methods}

\subsection{Diffusion Monte Carlo}
\noindent

The DMC interaction energies of the S66\cite{rezacS66WellbalancedDatabase2011} dataset are computed as:
\begin{equation}\label{equation_binding_energy}
   \Delta E_\textrm{int.} = E_\textrm{dimer} - E_\textrm{mon. 1} - E_\textrm{mon. 2}, 
\end{equation}
where $E_{\mathrm{dimer}}$ is the total energy of the dimer, and $E_\textrm{mon. 1}, E_\textrm{mon. 2}$ are the total energies of the constituent monomers.
In the S66 dataset, these monomers are kept fixed to their geometry in the dimer, which is in general different from their equilibrium geometry.
In this work, we first computed the energies of the monomers with DMC at a chosen reference geometry.
Subsequently, we added the deformation energy, i.e. the energy difference between the geometry of the monomer in the dimer and against this reference geometry using CCSD(T). 
We provide further details on these calculations in Sec.~S2.1 
of the supplementary material, and show for a subset of the S66 complexes that differences between DMC and CCSD(T) predictions of the deformation energies are  within $\sim 0.12\,$kcal/mol.

A detailed description of the DMC method can be found in Ref.~\citenum{foulkesQuantumMonteCarlo2001c}.
In this work, we compute fixed-node DMC interaction energies by using the CASINO code\cite{needsVariationalDiffusionQuantum2020}.
We use energy-consistent correlated electron pseudopotentials\cite{trailShapeEnergyConsistent2017} (eCEPP) with the determinant locality approximation (DLA)\cite{zenNewSchemeFixed2019}.
The trial wave-functions were of the Slater–Jastrow type with single Slater determinants, and the single-particle orbitals obtained from DFT local-density approximation (LDA) plane-wave calculations performed with PWscf\cite{giannozziQUANTUMESPRESSOModular2009a,giannozziQuantumESPRESSOExascale2020a} using an energy cut-off of 600 Ry and re-expanded in terms of B-splines\cite{alfeEfficientLocalizedBasis2004}.
The Jastrow factor included a two-body electron–electron (e–e) term, two-body electron–nucleus (e–n) terms, and three-body electron–electron–nucleus (e–e–n) terms.
The variational parameters of the Jastrow have been optimized by minimizing the variance of each system.
The final DMC estimates of $\Delta E_\text{int.}$ were extrapolated towards the zero time step limit ($\tau \to 0$) by making a cubic fit to a series of time step estimates from $0.1\,$au down to $0.003\,$au.
We estimate errors which capture both the stochastic errors in the fit as well as the errors in the cubic fit due to the changing behavior near the zero time step limit.
To do this, we make a linear fit on a subset of time steps below (and including) $0.02\,$au and calculate the difference of the extrapolated estimates from the linear fit against the original cubic fit. 
The final error estimate is taken to be the larger of the stochastic errors of the cubic fit or the difference between the linear fit and cubic fit, as discussed in Sec.~S2.4 
of the supplementary material.

The parameters chosen within the present work follow from previous DMC calculations for large molecules in Ref.~\citenum{al-hamdaniInteractionsLargeMolecules2021} as well as molecular crystals in Refs.~\citenum{dellapiaDMCICE13AmbientHigh2022b} and~\citenum{dellapiaHowAccurateAre2024}.
Within these studies of non-covalent interactions, the LDA trial wave-function was shown to be valid, either by comparison to experiments or when using trial wave-functions with other DFAs.
For the case of the AcOH dimer system (ID 20), we have performed our own validation tests on the choice of trial wave-function as well as localization approximation in Sec.~S7 
of the supplementary material.

\subsection{Coupled Cluster Theory}

Several CCSD(T) estimates~\cite{rezacS66WellbalancedDatabase2011,rezacExtensionsS66Data2011,brauerS66x8BenchmarkNoncovalent2016,kesharwaniS66NonCovalentInteractions2018,maAccurateIntermolecularInteraction2019,altunExtrapolationLimitComplete2020a,lupiJunChSJunChSF12Models2021,santraS66x8NoncovalentInteractions2022,semidalasS66NoncovalentInteractions2022,nagyPursuingBasisSet2023} of the S66 interaction energies are available in the literature. 
Here, we compare DMC to the average of three recent CCSD(T) calculations~\cite{rezacExtensionsS66Data2011,kesharwaniS66NonCovalentInteractions2018,nagyPursuingBasisSet2023}: the revised calculations from Řez{\'{a}}{\v{c}} \textit{et al.}\cite{rezacExtensionsS66Data2011}; the ``SILVER'' estimates from Kesharwani \textit{et al.}\cite{kesharwaniS66NonCovalentInteractions2018}; and the ``14k-GOLD'' estimates from Nagy \textit{et al.}\cite{nagyPursuingBasisSet2023}.
A brief description of the three different CCSD(T) calculations is reported in Sec.~S3 
of the supplementary material.\footnote{Note that the difficulty of reaching the complete basis set (CBS) limit is the principle reason for several CCSD(T) evaluations of S66. As reaching the CBS limit remains a challenge, particularly in dispersion bound complexes.}

Sch{\"a}fer \textit{et al.}~\cite{schaferUnderstandingDiscrepanciesWavefunction2024} have recently demonstrated that there exists an empirical relationship for dispersion-dominated complexes between the (cT) and the (T) correlation contributions to the total energy using the CCSD and MP2 correlation energies.
The resulting (cT)-fit is of the form:
\begin{equation}
    \dfrac{\textrm{(T)}}{\textrm{(cT)-fit}} = a + b \cdot \dfrac{\textrm{MP2 corr.}}{\textrm{CCSD corr.}},
\end{equation}
where $a$ and $b$ were parameters fitted from comparing CCSD(cT) to CCSD(T) calculations for a set of dispersion-bound complexes.
We have recomputed the CCSD(cT)-fit values from the original CCSD(T) ``SILVER'' estimates from Kesharwani \textit{et al.}, adding the difference between (cT) and (T) to the final (averaged) CCSD(T) estimates, as given in Sec.~S4 
of the supplementary material.

\section{Results and Discussion}\label{sec:Results}
\noindent

\begin{figure}[h!]
    \centering
    \includegraphics[width=1.\linewidth]{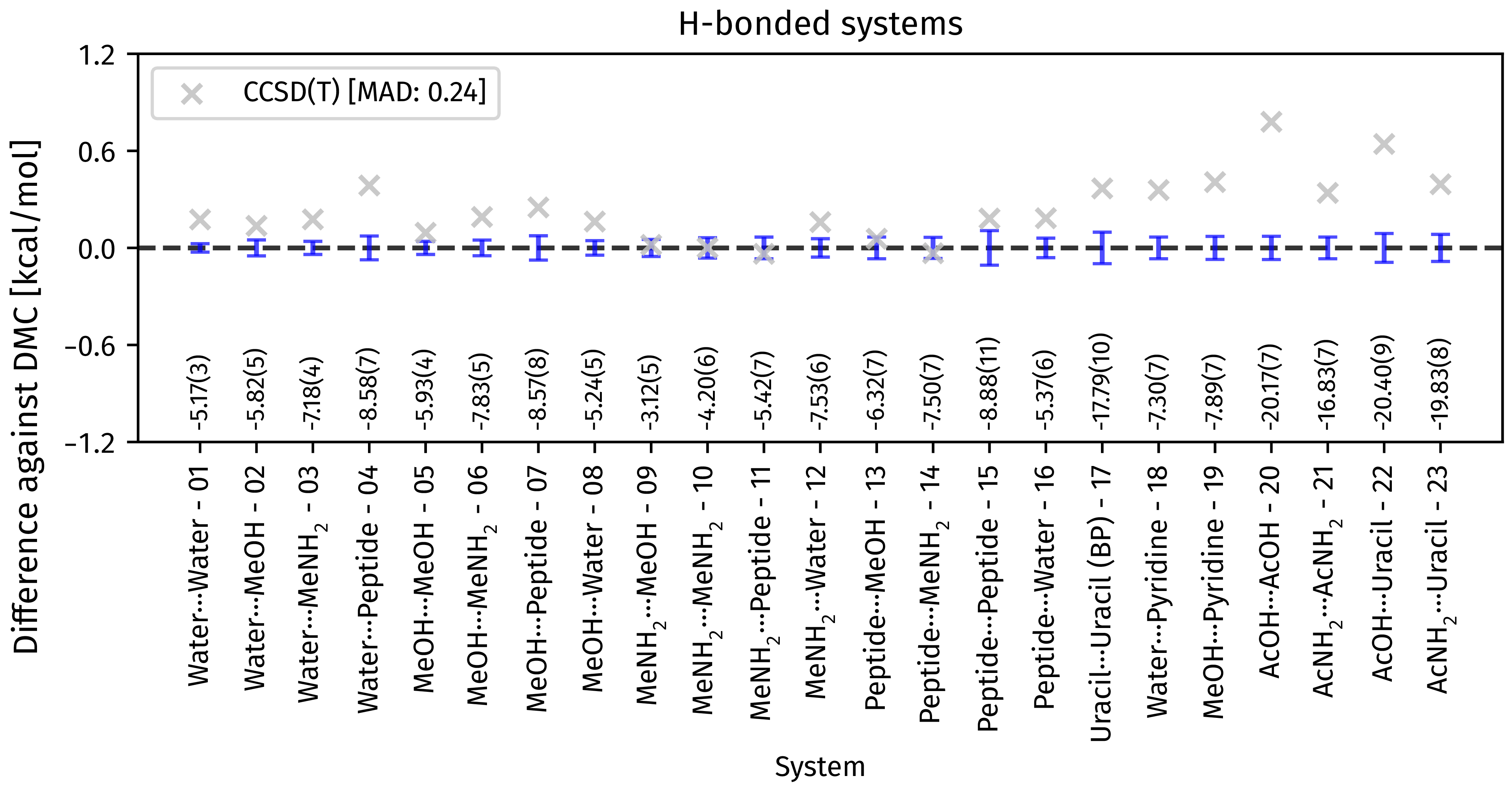}
    \caption{Comparison between DMC interaction energies $\Delta E_\textrm{int.}$ calculated in the present work against CCSD(T) for a subset of systems in the S66 dataset with hydrogen-bonds. The CCSD(T) estimate is taken as an average from three previous calculations~\cite{rezacExtensionsS66Data2011,kesharwaniS66NonCovalentInteractions2018,nagyPursuingBasisSet2023}, with corresponding standard deviation as error. The deviation of CCSD(T) from the DMC is plotted with grey crosses, with the statistical errors (corresponding to one standard deviation $\sigma$). The complex ID and label are provided below the x-axis, while the number above each x-axis tick represents the DMC $\Delta E_\textrm{int.}$ estimate, with the error on the last reported digit given in parentheses. The uracil dimer (ID 17) is in its base-pair (BP) configuration.}
    \label{fig:figure1}
\end{figure}
\begin{figure}[th!]
    \centering
    \includegraphics[width=1.\linewidth]{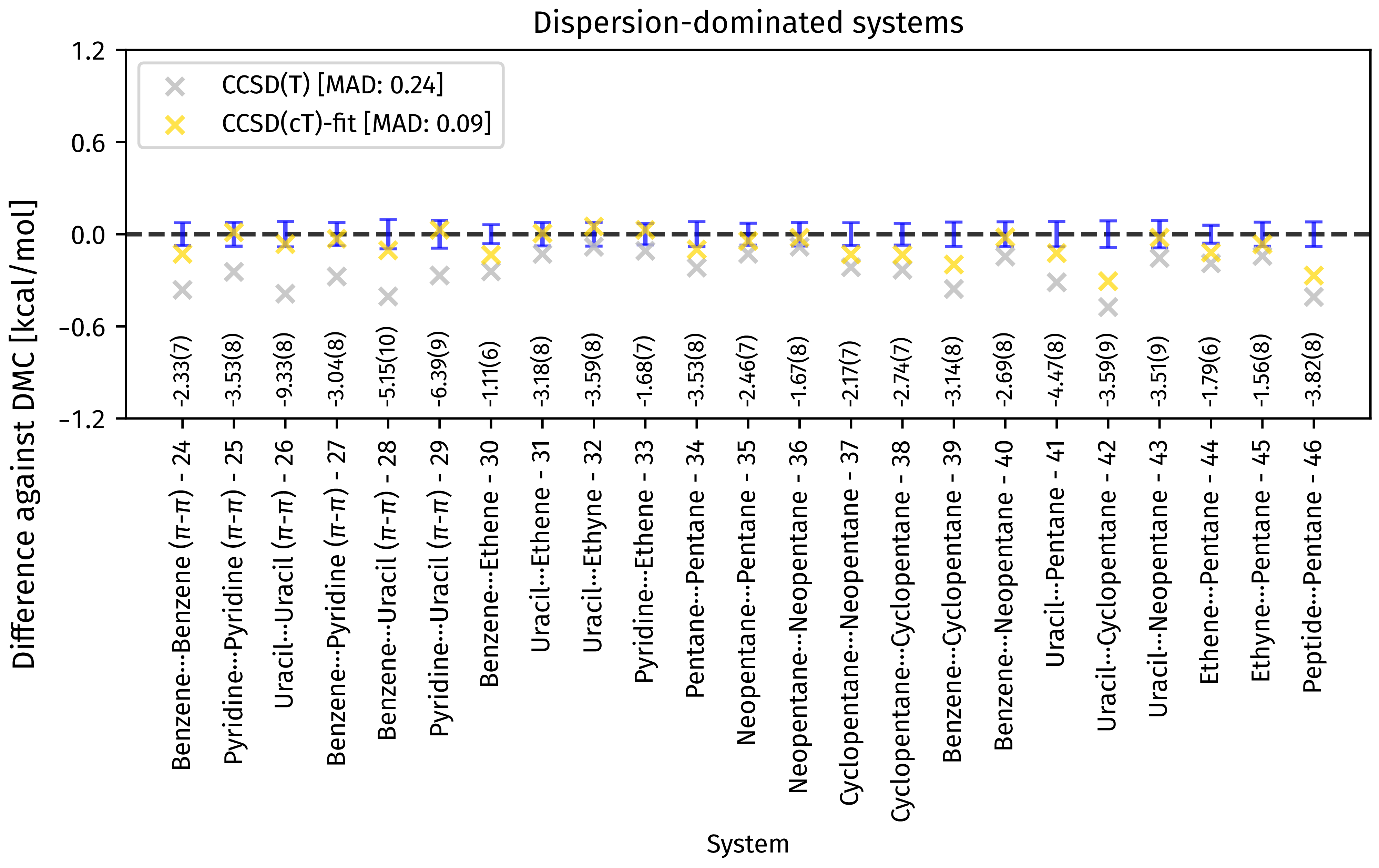}
    \caption{Comparison between DMC interaction energies $\Delta E_\textrm{int.}$ calculated in the present work against CCSD(T) for a subset of systems in the S66 dataset dominated by dispersion interactions. The first 6 dimers are $\pi {-} \pi$ stacked. Refer to the caption of Fig.~\ref{fig:figure1} for the plot details. The CCSD(T) estimate is taken as an average from three previous calculations~\cite{rezacExtensionsS66Data2011,kesharwaniS66NonCovalentInteractions2018,nagyPursuingBasisSet2023}, with corresponding standard deviation as error. Additional CCSD(cT)-fit estimates are reported with golden crosses. These are calculated by scaling the CCSD(T) estimates based on their MP2 and CCSD contributions with the approach described in Ref.~\citenum{schaferUnderstandingDiscrepanciesWavefunction2024}.}
    \label{fig:figure2}
\end{figure}
\begin{figure}[th!]
    \centering
    \includegraphics[width=1.\linewidth]{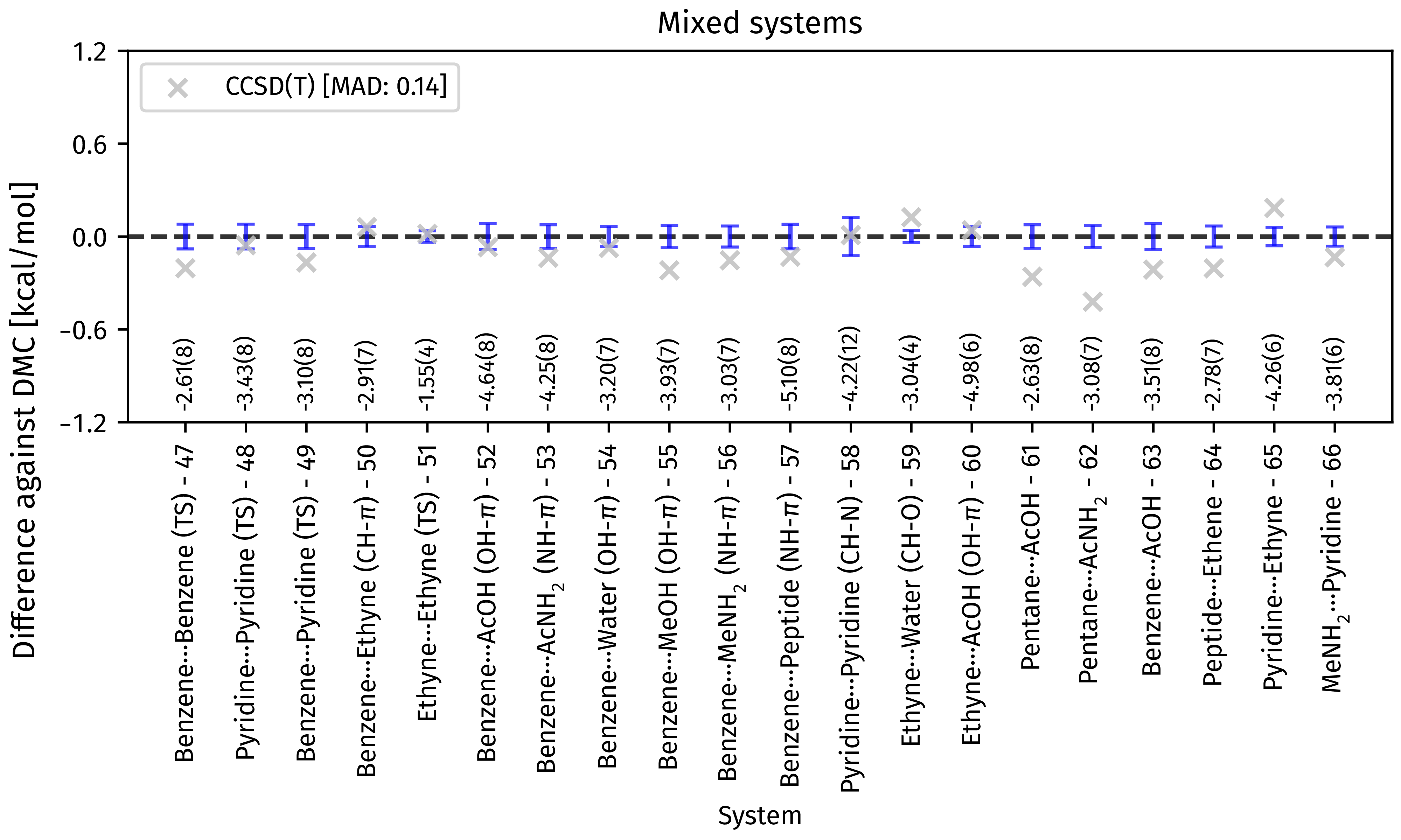}
    \caption{Comparison between DMC interaction energies $\Delta E_\textrm{int.}$ calculated in the present work against CCSD(T) for a subset of systems in the S66 dataset with mixed bonding character. The systems consist of T-shaped (TS) aromatic ring complexes as well as X-H${\cdots}\pi$ (X = C,O,N) interactions. Refer to the caption of Fig.~\ref{fig:figure1} for the plot details. The CCSD(T) estimate is taken as an average from three previous calculations~\cite{rezacExtensionsS66Data2011,kesharwaniS66NonCovalentInteractions2018,nagyPursuingBasisSet2023}, with corresponding standard deviation as error. }
    \label{fig:figure3}
\end{figure}

We report the final DMC estimates for the entire S66 dataset in Figs.~\ref{fig:figure1},~\ref{fig:figure2} and~\ref{fig:figure3}.
This dataset comprises of a diverse range of interactions and we have separated the systems according to the original S66 categories of hydrogen-bonded, dispersion-bonded and ``mixed''-character systems in Figs.~\ref{fig:figure1},~\ref{fig:figure2} and~\ref{fig:figure3}, respectively.
The corresponding dimer complexes are visualized in Fig.~S1 
of the supplementary material.
We report the DMC estimate of the interaction energy $\Delta E_\textrm{int.}$ above the label of each S66 complex.
In all cases, the errors on $\Delta E_\textrm{int.}$ estimates are below $0.12\,$kcal/mol, with the majority below $0.10\,$kcal/mol, facilitating reliable comparisons to CCSD(T).

The strength of $\Delta E_\textrm{int.}$ varies significantly across the systems, from as large as $-20.17 \pm 0.07\,$kcal/mol for complex 20 (acetic acid dimer) to as weak as $-1.11 \pm 0.06\,$kcal/mol for complex 30 (benzene-ethene dimer), being stronger in the H-bonded systems.
With gray crosses, we plot the difference between DMC and CCSD(T) estimates (as described in the Methods) for the three classes of interactions.
We use DMC as the reference (i.e., zero), and plot blue vertical bars along the horizontal zero-axis representing the errors on the DMC estimates.
There is overall excellent agreement, with a mean absolute deviation (MAD) of $0.21\,$kcal/mol across the entire S66 dataset.
We find systematic trends in the differences between CCSD(T) and DMC, with CCSD(T) predicting weaker binding compared to DMC for hydrogen-bonded systems in Fig.~\ref{fig:figure1}, with an MAD of $\sim 0.24\,$kcal/mol, while predicting a stronger binding for dispersion dominated systems in Fig.~\ref{fig:figure2}, with an MAD of $\sim 0.24\,$kcal/mol.
For the ``mixed'' character systems in Fig.~\ref{fig:figure3}, the MAD is lower at $0.14\,$kcal/mol.

The stronger binding of DMC over CCSD(T) has not been (systematically) reported before, with the acetic acid dimer (ID 20) giving the maximum deviation of $0.8\,$kcal/mol across all S66 systems.
Within Sec.~S7 
of the supplementary material, we have confirmed that the computed $\Delta E_\textrm{int.}$ estimate ($-20.17 \pm 0.07\,$kcal/mol) does not depend on the chosen pseudopotential (eCEPP), localization scheme (DLA) or trial wave-function (LDA).
For example, we have also performed all-electron calculations, giving an estimate of $-20.32 \pm 0.12\,$kcal/mol.
We have also performed tests using PBE and PBE0 trial wave-functions, showing that the DMC $\Delta E_\textrm{int.}$ has a negligible (${<}0.15\,$kcal/mol) dependence on the nodal surface for the DFAs considered.
Furthermore, we also computed estimates for two-other localization schemes: T-move and determinant localization T-move, both of which were within the statistical uncertainties of our original estimate.
It should be noted that while the absolute value of the difference can be significant for some hydrogen-bonded systems, the relative difference (normalized against the DMC $\Delta E_\textrm{int.}$) is significantly smaller, with a mean relative difference of 2.45\% compared to 8.21\% for the dispersion-dominated systems (see Sec.~S6 
of the supplementary material).

\begin{figure}[h!]
    \centering
    \includegraphics[width=.6\linewidth]{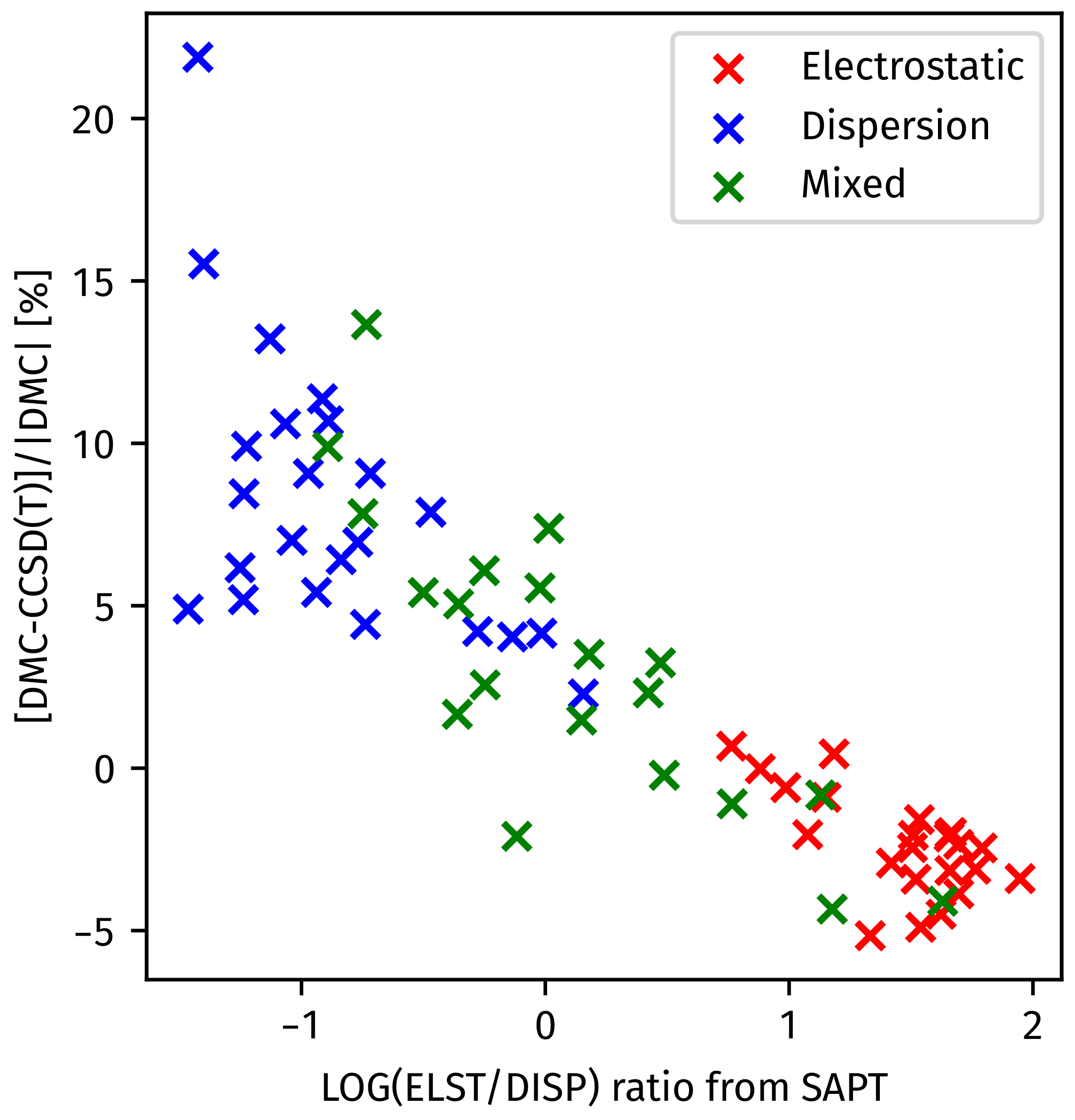}
    \caption{Error decomposition analysis. We report the difference between DMC and CCSD(T) relative to the DMC magnitude, i.e. $\left(E_{\mathrm{DMC}} - E_{\mathrm{CCSD(T)}}\right)/\left|E_{\mathrm{DMC}}\right|$, as a function of the natural logarithm of the electrostatic (ELST) to dispersion (DISP) ratio contribution to the binding energy. The ELST to DISP ratio is determined from the SAPT analysis from Ref.~\citenum{burnsBioFragmentDatabaseBFDb2017}. The color code is red for H-bonded systems (ID from 1 to 23), blue for dispersion dominated systems (ID from 24 to 46), and green for mixed systems (ID from 47 to 66).}
    \label{fig:figure4}
\end{figure}

The weaker binding of DMC over CCSD(T) for dispersion-dominated systems is now relatively well-documented~\cite{al-hamdaniInteractionsLargeMolecules2021,schaferUnderstandingDiscrepanciesWavefunction2024}, and there is evidence that it can be improved by replacing the perturbative triples (T) contribution with the recent (cT)~\cite{masiosAvertingInfraredCatastrophe2023c} contribution.
We plot the difference between an empirically CCSD(cT)-fit formulation as yellow crosses for the dispersion-dominated systems.
In all cases, CCSD(cT) has a weaker binding than CCSD(T), leaning closer towards DMC, leading to an MAD of $0.09\,$kcal/mol and mean relative difference of 3.44\%
However, this does not fully resolve the discrepancies across all of the dispersion systems, with significant discrepancy of ${\sim}0.5\,$kcal/mol remaining for the uracil-cyclopentane dimer (ID 42).
Such a significant discrepancy makes this a worthwhile system to investigate further and could give clues on remaining discrepancies between DMC and CCSD(T) observed in Ref.~\citenum{schaferUnderstandingDiscrepanciesWavefunction2024}.
The analysis reported above highlights an important outcome of this work: the identification of smaller, simpler systems that show notable discrepancies between DMC and coupled cluster methods. 
Specifically, we found a discrepancy of approximately ${\sim}0.8\,$kcal/mol for the hydrogen-bonded acetic acid dimer (ID 20) and ${\sim}0.5\,$kcal/mol for the dispersion-dominated uracil–cyclopentane dimer (ID 42), which contain 64 and 98 (total) electrons, respectively.
These medium-sized systems represent an almost tenfold reduction in electron count compared to the larger $\mathrm{C}_{60}@[6]\mathrm{CPPA}$ buckyball-ring system (672 electrons) studied in Ref.~\citenum{al-hamdaniInteractionsLargeMolecules2021}.
Thus, they might offer practical, cost-effective models for further exploring the discrepancy between DMC and coupled-cluster.

We now focus on the difference between DMC and CCSD(T) as a function of the dispersion and electrostatic contribution to the interaction energy.
In particular, we find that the relative differences between DMC and CCSD(T) for each system within the S66 dataset can be correlated to the relative strength of the dispersion and electrostatic interactions that make up its $\Delta E_\textrm{int.}$.
We used the Symmetry Adapted Perturbation Theory (SAPT) calculations (at the $s$SAPT0 level with the jun-cc-pVDZ basis set~\cite{parkerLevelsSymmetryAdapted2014}) from Burns \textit{et al.}\cite{burnsBioFragmentDatabaseBFDb2017}, which decomposes $\Delta E_\text{int.}$ into contributions from electrostatics (ELST), exchange, induction, and dispersion (DISP).
Notably, we show in Fig.~4 that there is a strong linear trend ($R^2{=}0.78$) between the natural logarithm of the ELST and DISP contributions, $\log\left(\frac{\text{ELST}}{\text{DISP}}\right)$, and the relative difference (in \%) between CCSD(T) and DMC.
In Sec.~S9 
of the supplementary material, we show that this strong linear trend remains at the more sophisticated SAPT2+(3)(CCD)/aug-cc-pVTZ level.~\cite{villotInitioDispersionPotentials2024}
This analysis confirms our prior observations on the trends between DMC and CCSD(T).
For example, the more dominant the DISP contribution to $\Delta E_\text{int.}$ (i.e., a more negative $\log\left(\frac{\text{ELST}}{\text{DISP}}\right)$), the more CCSD(T) is found to underbind with respect to DMC.
Similarly, the stronger the ELST contribution to $\Delta E_\textrm{int.}$ (i.e., a more positive $\log\left(\frac{\textrm{ELST}}{\textrm{DISP}}\right)$), the more CCSD(T) is found to overbind with respect to DMC.
We expect that this cheap descriptor can be used in the future to identify more challenging systems with larger discrepancies between DMC and CCSD(T).

Finally, we discuss briefly the potential origins of the observed discrepancies between DMC and CCSD(T) based upon the current literature.
For H-bonded systems, CCSDT(Q) estimates are available for the A24 dataset~\cite{rezacDescribingNoncovalentInteractions2013,rezacExtensionsApplicationsA242015} --  a set of $\Delta E_\text{int.}$ for small dimer complexes.
Nakano~\etal{}~\cite{nakanoBasisSetIncompleteness2024a} have performed DMC calculations for the entire A24 dataset, where there is a notable discrepancy of $0.26{\pm}0.07\,$kcal/mol and $0.34{\pm}0.07\,$kcal/mol for the water-ammonia and HCN dimer complexes, in line with our observation of stronger binding in DMC.
The CCSDT(Q) references, albeit at small basis sets, find negligible (${<}0.01\,$kcal/mol) changes relative to CCSD(T).
For dispersion-bound systems, there exist CCSDT(Q) estimates for the parallel-displaced (PD) benzene dimer by Semidalas \etal{}~\cite{semidalasPostCCSDTCorrectionsS662025} and by Karton and Martin~\cite{kartonPrototypicalPpDimers2021}, which report differences to CCSD(T) of $-0.085\,$kcal/mol and $-0.058\,$kcal/mol, respectively, using small truncated double-zeta quality basis sets.
The reported difference between DMC and CCSD(T) is $-0.37{\pm}0.08\,$kcal/mol for this system which indicates that the majority of this difference is not covered when going to CCSDT(Q).
For both H-bonded and dispersion-bound systems (as well as those of mixed-character), the observed differences could arise either from higher order excitations and larger basis sets needed from coupled cluster theory, or biases in the FN-DMC evaluations, likely coming from the fixed-node approximation.

\section{Conclusions}\label{sec:Conclusions}
\noindent
To summarize, we have computed highly accurate estimates for the S66 dateset -- one of the most widely used databases for non-covalent interactions in biological and organic molecules -- with fixed-node diffusion quantum Monte Carlo.
These estimates have provided new insights into recent discussions on its discrepancies with another widely-trusted method: coupled cluster theory with single, double and perturbative triple excitations [CCSD(T)].
Our data shows systematic trends, with DMC predicting stronger binding in hydrogen-bonded systems than CCSD(T), and weaker binding in dispersion dominated systems.
We show that there is a correlation between the relative strength of these discrepancies with the nature of the interaction, specifically the relative ratio of the electrostatic and dispersion contributions to the interaction energy as provided by previous Symmetry Adapted Perturbation Theory (SAPT) calculations.~\cite{burnsBioFragmentDatabaseBFDb2017}
In addition, we show that the discrepancy between DMC and CCSD(T) on dispersion-dominated systems can be reduced using a recently proposed CCSD(cT) formulation, albeit with still significant remaining differences.
While this work does not identify the origin of the disagreement between DMC and CCSD(T), it has identified the type of interactions where it is particularly prevalent and importantly, we have identified model systems within the S66 dataset where these errors are prominent.
These results have strong implications for the electronic structure theory community, addressing the knowledge gap on the trends of DMC interaction energies for non-covalent molecular complexes.
Furthermore, the accurate reference data produced within this work is expected to benefit the wider materials modeling community, being instrumental for benchmarking applications ranging from the development of machine learned interatomic potentials to crystal structure prediction, drug design, and renewable energy.

\section*{Supplementary Material}\label{sec:SI}
\noindent
See the supplementary material for details on the DMC calculations, comprising the convergence of the calculations with respect to the time step, the influence of the choice of the monomer geometry on the dimer interaction energy, as well as tests on the pseudopotential localization error and the Jastrow optimization.

\section*{Data Availability}
\noindent
The data that support the findings of this study are available within the article and its supplementary material. All analysis can be found on GitHub at \href{https://github.com/zenandrea/FNDMC-S66}{github.com/zenandrea/FNDMC-S66} and can be viewed interactively online through associated Jupyter notebooks (via Google Colab), with links provided in the corresponding GitHub repository.

\section*{Acknowledgments}\label{sec:acknowledgments}
\noindent
We acknowledge the computational resources from Cambridge Service for Data Driven Discovery (CSD3)  operated by the University of Cambridge Research Computing Service, provided by Dell EMC and Intel using Tier-2 funding from the Engineering and Physical Sciences Research Council (capital grant EP/T022159/1 and EP/P020259/1), and DiRAC funding from the Science and Technology Facilities Council (www.dirac.ac.uk). 
We are further grateful for computational support from the UK national high performance computing service, ARCHER2, for which access was obtained via the UKCP consortium and funded by EPSRC grant ref EP/X035891/1. This research also used resources of the Oak
Ridge Leadership Computing Facility at the Oak Ridge
National Laboratory, which is supported by the Office of
Science of the U.S. Department of Energy under Contract
No. DE-AC05-00OR22725.
A.M. and B.X.S acknowledge support from the European Union under the “n-AQUA” European Research
Council project (Grant No. 101071937). 
D.A. and A.Z. acknowledges support from Leverhulme grant no. RPG-2020-038, and from the European Union under the Next generation EU (projects 20222FXZ33 and P2022MC742).

\bibliography{refs}

\begin{thebibliography}{97}%
\makeatletter
\providecommand \@ifxundefined [1]{%
 \@ifx{#1\undefined}
}%
\providecommand \@ifnum [1]{%
 \ifnum #1\expandafter \@firstoftwo
 \else \expandafter \@secondoftwo
 \fi
}%
\providecommand \@ifx [1]{%
 \ifx #1\expandafter \@firstoftwo
 \else \expandafter \@secondoftwo
 \fi
}%
\providecommand \natexlab [1]{#1}%
\providecommand \enquote  [1]{``#1''}%
\providecommand \bibnamefont  [1]{#1}%
\providecommand \bibfnamefont [1]{#1}%
\providecommand \citenamefont [1]{#1}%
\providecommand \href@noop [0]{\@secondoftwo}%
\providecommand \href [0]{\begingroup \@sanitize@url \@href}%
\providecommand \@href[1]{\@@startlink{#1}\@@href}%
\providecommand \@@href[1]{\endgroup#1\@@endlink}%
\providecommand \@sanitize@url [0]{\catcode `\\12\catcode `\$12\catcode
  `\&12\catcode `\#12\catcode `\^12\catcode `\_12\catcode `\%12\relax}%
\providecommand \@@startlink[1]{}%
\providecommand \@@endlink[0]{}%
\providecommand \url  [0]{\begingroup\@sanitize@url \@url }%
\providecommand \@url [1]{\endgroup\@href {#1}{\urlprefix }}%
\providecommand \urlprefix  [0]{URL }%
\providecommand \Eprint [0]{\href }%
\providecommand \doibase [0]{http://dx.doi.org/}%
\providecommand \selectlanguage [0]{\@gobble}%
\providecommand \bibinfo  [0]{\@secondoftwo}%
\providecommand \bibfield  [0]{\@secondoftwo}%
\providecommand \translation [1]{[#1]}%
\providecommand \BibitemOpen [0]{}%
\providecommand \bibitemStop [0]{}%
\providecommand \bibitemNoStop [0]{.\EOS\space}%
\providecommand \EOS [0]{\spacefactor3000\relax}%
\providecommand \BibitemShut  [1]{\csname bibitem#1\endcsname}%
\let\auto@bib@innerbib\@empty
\bibitem [{\citenamefont
  {Beran}(2016)}]{beranModelingPolymorphicMolecular2016}%
  \BibitemOpen
  \bibfield  {author} {\bibinfo {author} {\bibfnamefont {G.~J.~O.}\
  \bibnamefont {Beran}},\ }\bibfield  {title} {\enquote {\bibinfo {title}
  {Modeling {{Polymorphic Molecular Crystals}} with {{Electronic Structure
  Theory}}},}\ }\href {\doibase 10.1021/acs.chemrev.5b00648} {\bibfield
  {journal} {\bibinfo  {journal} {Chem. Rev.}\ }\textbf {\bibinfo {volume}
  {116}},\ \bibinfo {pages} {5567--5613} (\bibinfo {year} {2016})}\BibitemShut
  {NoStop}%
\bibitem [{\citenamefont {Riley}\ and\ \citenamefont
  {Hobza}(2011)}]{rileyNoncovalentInteractionsBiochemistry2011}%
  \BibitemOpen
  \bibfield  {author} {\bibinfo {author} {\bibfnamefont {K.~E.}\ \bibnamefont
  {Riley}}\ and\ \bibinfo {author} {\bibfnamefont {P.}~\bibnamefont {Hobza}},\
  }\bibfield  {title} {\enquote {\bibinfo {title} {Noncovalent interactions in
  biochemistry},}\ }\href {\doibase 10.1002/wcms.8} {\bibfield  {journal}
  {\bibinfo  {journal} {WIREs Comput. Mol. Sci.}\ }\textbf {\bibinfo {volume}
  {1}},\ \bibinfo {pages} {3--17} (\bibinfo {year} {2011})}\BibitemShut
  {NoStop}%
\bibitem [{\citenamefont {Raynal}\ \emph {et~al.}(2014)\citenamefont {Raynal},
  \citenamefont {Ballester}, \citenamefont {{Vidal-Ferran}},\ and\
  \citenamefont {van Leeuwen}}]{raynalSupramolecularCatalysisPart2014}%
  \BibitemOpen
  \bibfield  {author} {\bibinfo {author} {\bibfnamefont {M.}~\bibnamefont
  {Raynal}}, \bibinfo {author} {\bibfnamefont {P.}~\bibnamefont {Ballester}},
  \bibinfo {author} {\bibfnamefont {A.}~\bibnamefont {{Vidal-Ferran}}}, \ and\
  \bibinfo {author} {\bibfnamefont {P.~W. N.~M.}\ \bibnamefont {van Leeuwen}},\
  }\bibfield  {title} {\enquote {\bibinfo {title} {Supramolecular catalysis.
  {{Part}} 1: Non-covalent interactions as a tool for building and modifying
  homogeneous catalysts},}\ }\href {\doibase 10.1039/C3CS60027K} {\bibfield
  {journal} {\bibinfo  {journal} {Chem. Soc. Rev.}\ }\textbf {\bibinfo {volume}
  {43}},\ \bibinfo {pages} {1660--1733} (\bibinfo {year} {2014})}\BibitemShut
  {NoStop}%
\bibitem [{\citenamefont {Planas}\ \emph {et~al.}(2013)\citenamefont {Planas},
  \citenamefont {Dzubak}, \citenamefont {Poloni}, \citenamefont {Lin},
  \citenamefont {McManus}, \citenamefont {McDonald}, \citenamefont {Neaton},
  \citenamefont {Long}, \citenamefont {Smit},\ and\ \citenamefont
  {Gagliardi}}]{planasMechanismCarbonDioxide2013}%
  \BibitemOpen
  \bibfield  {author} {\bibinfo {author} {\bibfnamefont {N.}~\bibnamefont
  {Planas}}, \bibinfo {author} {\bibfnamefont {A.~L.}\ \bibnamefont {Dzubak}},
  \bibinfo {author} {\bibfnamefont {R.}~\bibnamefont {Poloni}}, \bibinfo
  {author} {\bibfnamefont {L.-C.}\ \bibnamefont {Lin}}, \bibinfo {author}
  {\bibfnamefont {A.}~\bibnamefont {McManus}}, \bibinfo {author} {\bibfnamefont
  {T.~M.}\ \bibnamefont {McDonald}}, \bibinfo {author} {\bibfnamefont {J.~B.}\
  \bibnamefont {Neaton}}, \bibinfo {author} {\bibfnamefont {J.~R.}\
  \bibnamefont {Long}}, \bibinfo {author} {\bibfnamefont {B.}~\bibnamefont
  {Smit}}, \ and\ \bibinfo {author} {\bibfnamefont {L.}~\bibnamefont
  {Gagliardi}},\ }\bibfield  {title} {\enquote {\bibinfo {title} {The
  {{Mechanism}} of {{Carbon Dioxide Adsorption}} in an
  {{Alkylamine-Functionalized Metal}}--{{Organic Framework}}},}\ }\href
  {\doibase 10.1021/ja4004766} {\bibfield  {journal} {\bibinfo  {journal} {J.
  Am. Chem. Soc.}\ }\textbf {\bibinfo {volume} {135}},\ \bibinfo {pages}
  {7402--7405} (\bibinfo {year} {2013})}\BibitemShut {NoStop}%
\bibitem [{\citenamefont {Poloni}, \citenamefont {Smit},\ and\ \citenamefont
  {Neaton}(2012)}]{poloniLigandAssistedEnhancementCO22012}%
  \BibitemOpen
  \bibfield  {author} {\bibinfo {author} {\bibfnamefont {R.}~\bibnamefont
  {Poloni}}, \bibinfo {author} {\bibfnamefont {B.}~\bibnamefont {Smit}}, \ and\
  \bibinfo {author} {\bibfnamefont {J.~B.}\ \bibnamefont {Neaton}},\ }\bibfield
   {title} {\enquote {\bibinfo {title} {Ligand-{{Assisted Enhancement}} of
  {{CO2 Capture}} in {{Metal}}--{{Organic Frameworks}}},}\ }\href {\doibase
  10.1021/ja2118943} {\bibfield  {journal} {\bibinfo  {journal} {J. Am. Chem.
  Soc.}\ }\textbf {\bibinfo {volume} {134}},\ \bibinfo {pages} {6714--6719}
  (\bibinfo {year} {2012})}\BibitemShut {NoStop}%
\bibitem [{\citenamefont {Du}\ \emph {et~al.}(2016)\citenamefont {Du},
  \citenamefont {Li}, \citenamefont {Xia}, \citenamefont {Ai}, \citenamefont
  {Liang}, \citenamefont {Sang}, \citenamefont {Ji},\ and\ \citenamefont
  {Liu}}]{duInsightsProteinLigand2016}%
  \BibitemOpen
  \bibfield  {author} {\bibinfo {author} {\bibfnamefont {X.}~\bibnamefont
  {Du}}, \bibinfo {author} {\bibfnamefont {Y.}~\bibnamefont {Li}}, \bibinfo
  {author} {\bibfnamefont {Y.-L.}\ \bibnamefont {Xia}}, \bibinfo {author}
  {\bibfnamefont {S.-M.}\ \bibnamefont {Ai}}, \bibinfo {author} {\bibfnamefont
  {J.}~\bibnamefont {Liang}}, \bibinfo {author} {\bibfnamefont
  {P.}~\bibnamefont {Sang}}, \bibinfo {author} {\bibfnamefont {X.-L.}\
  \bibnamefont {Ji}}, \ and\ \bibinfo {author} {\bibfnamefont {S.-Q.}\
  \bibnamefont {Liu}},\ }\bibfield  {title} {\enquote {\bibinfo {title}
  {Insights into {{Protein}}--{{Ligand Interactions}}: {{Mechanisms}},
  {{Models}}, and {{Methods}}},}\ }\href {\doibase 10.3390/ijms17020144}
  {\bibfield  {journal} {\bibinfo  {journal} {Int. J. Mol. Sci.}\ }\textbf
  {\bibinfo {volume} {17}},\ \bibinfo {pages} {144} (\bibinfo {year}
  {2016})}\BibitemShut {NoStop}%
\bibitem [{\citenamefont {R.~Rehak}\ \emph {et~al.}(2020)\citenamefont
  {R.~Rehak}, \citenamefont {Piccini}, \citenamefont {Alessio},\ and\
  \citenamefont {Sauer}}]{r.rehakIncludingDispersionDensity2020}%
  \BibitemOpen
  \bibfield  {author} {\bibinfo {author} {\bibfnamefont {F.}~\bibnamefont
  {R.~Rehak}}, \bibinfo {author} {\bibfnamefont {G.}~\bibnamefont {Piccini}},
  \bibinfo {author} {\bibfnamefont {M.}~\bibnamefont {Alessio}}, \ and\
  \bibinfo {author} {\bibfnamefont {J.}~\bibnamefont {Sauer}},\ }\bibfield
  {title} {\enquote {\bibinfo {title} {Including dispersion in density
  functional theory for adsorption on flat oxide surfaces, in metal--organic
  frameworks and in acidic zeolites},}\ }\href {\doibase 10.1039/D0CP00394H}
  {\bibfield  {journal} {\bibinfo  {journal} {Phys. Chem. Chem. Phys.}\
  }\textbf {\bibinfo {volume} {22}},\ \bibinfo {pages} {7577--7585} (\bibinfo
  {year} {2020})}\BibitemShut {NoStop}%
\bibitem [{\citenamefont {{M{\"u}ller-Dethlefs}}\ and\ \citenamefont
  {Hobza}(2000)}]{muller-dethlefsNoncovalentInteractionsChallenge2000}%
  \BibitemOpen
  \bibfield  {author} {\bibinfo {author} {\bibfnamefont {K.}~\bibnamefont
  {{M{\"u}ller-Dethlefs}}}\ and\ \bibinfo {author} {\bibfnamefont
  {P.}~\bibnamefont {Hobza}},\ }\bibfield  {title} {\enquote {\bibinfo {title}
  {Noncovalent {{Interactions}}:\, {{A Challenge}} for {{Experiment}} and
  {{Theory}}},}\ }\href {\doibase 10.1021/cr9900331} {\bibfield  {journal}
  {\bibinfo  {journal} {Chem. Rev.}\ }\textbf {\bibinfo {volume} {100}},\
  \bibinfo {pages} {143--168} (\bibinfo {year} {2000})}\BibitemShut {NoStop}%
\bibitem [{\citenamefont {Foulkes}\ \emph {et~al.}(2001)\citenamefont
  {Foulkes}, \citenamefont {Mitas}, \citenamefont {Needs},\ and\ \citenamefont
  {Rajagopal}}]{foulkesQuantumMonteCarlo2001c}%
  \BibitemOpen
  \bibfield  {author} {\bibinfo {author} {\bibfnamefont {W.~M.~C.}\
  \bibnamefont {Foulkes}}, \bibinfo {author} {\bibfnamefont {L.}~\bibnamefont
  {Mitas}}, \bibinfo {author} {\bibfnamefont {R.~J.}\ \bibnamefont {Needs}}, \
  and\ \bibinfo {author} {\bibfnamefont {G.}~\bibnamefont {Rajagopal}},\
  }\bibfield  {title} {\enquote {\bibinfo {title} {Quantum {{Monte Carlo}}
  simulations of solids},}\ }\href {\doibase 10.1103/RevModPhys.73.33}
  {\bibfield  {journal} {\bibinfo  {journal} {Rev. Mod. Phys.}\ }\textbf
  {\bibinfo {volume} {73}},\ \bibinfo {pages} {33--83} (\bibinfo {year}
  {2001})}\BibitemShut {NoStop}%
\bibitem [{\citenamefont {Bartlett}\ and\ \citenamefont
  {Musia{\l}}(2007)}]{bartlettCoupledclusterTheoryQuantum2007a}%
  \BibitemOpen
  \bibfield  {author} {\bibinfo {author} {\bibfnamefont {R.~J.}\ \bibnamefont
  {Bartlett}}\ and\ \bibinfo {author} {\bibfnamefont {M.}~\bibnamefont
  {Musia{\l}}},\ }\bibfield  {title} {\enquote {\bibinfo {title}
  {Coupled-cluster theory in quantum chemistry},}\ }\href {\doibase
  10.1103/RevModPhys.79.291} {\bibfield  {journal} {\bibinfo  {journal} {Rev.
  Mod. Phys.}\ }\textbf {\bibinfo {volume} {79}},\ \bibinfo {pages} {291--352}
  (\bibinfo {year} {2007})}\BibitemShut {NoStop}%
\bibitem [{\citenamefont {Mattsson}\ \emph {et~al.}(2004)\citenamefont
  {Mattsson}, \citenamefont {Schultz}, \citenamefont {Desjarlais},
  \citenamefont {Mattsson},\ and\ \citenamefont
  {Leung}}]{mattssonDesigningMeaningfulDensity2004}%
  \BibitemOpen
  \bibfield  {author} {\bibinfo {author} {\bibfnamefont {A.~E.}\ \bibnamefont
  {Mattsson}}, \bibinfo {author} {\bibfnamefont {P.~A.}\ \bibnamefont
  {Schultz}}, \bibinfo {author} {\bibfnamefont {M.~P.}\ \bibnamefont
  {Desjarlais}}, \bibinfo {author} {\bibfnamefont {T.~R.}\ \bibnamefont
  {Mattsson}}, \ and\ \bibinfo {author} {\bibfnamefont {K.}~\bibnamefont
  {Leung}},\ }\bibfield  {title} {\enquote {\bibinfo {title} {Designing
  meaningful density functional theory calculations in materials science---a
  primer},}\ }\href {\doibase 10.1088/0965-0393/13/1/R01} {\bibfield  {journal}
  {\bibinfo  {journal} {Modelling Simul. Mater. Sci. Eng.}\ }\textbf {\bibinfo
  {volume} {13}},\ \bibinfo {pages} {R1--R31} (\bibinfo {year}
  {2004})}\BibitemShut {NoStop}%
\bibitem [{\citenamefont {Ceperley}\ and\ \citenamefont
  {Alder}(1980)}]{ceperleyGroundStateElectron1980a}%
  \BibitemOpen
  \bibfield  {author} {\bibinfo {author} {\bibfnamefont {D.~M.}\ \bibnamefont
  {Ceperley}}\ and\ \bibinfo {author} {\bibfnamefont {B.~J.}\ \bibnamefont
  {Alder}},\ }\bibfield  {title} {\enquote {\bibinfo {title} {Ground {{State}}
  of the {{Electron Gas}} by a {{Stochastic Method}}},}\ }\href {\doibase
  10.1103/PhysRevLett.45.566} {\bibfield  {journal} {\bibinfo  {journal} {Phys.
  Rev. Lett.}\ }\textbf {\bibinfo {volume} {45}},\ \bibinfo {pages} {566--569}
  (\bibinfo {year} {1980})}\BibitemShut {NoStop}%
\bibitem [{\citenamefont
  {Grimme}(2006)}]{grimmeSemiempiricalGGAtypeDensity2006}%
  \BibitemOpen
  \bibfield  {author} {\bibinfo {author} {\bibfnamefont {S.}~\bibnamefont
  {Grimme}},\ }\bibfield  {title} {\enquote {\bibinfo {title} {{Semiempirical
  GGA-type density functional constructed with a long-range dispersion
  correction}},}\ }\href {\doibase 10.1002/jcc.20495} {\bibfield  {journal}
  {\bibinfo  {journal} {J. Comp. Chem.}\ }\textbf {\bibinfo {volume} {27}},\
  \bibinfo {pages} {1787--1799} (\bibinfo {year} {2006})}\BibitemShut {NoStop}%
\bibitem [{\citenamefont {Grimme}\ \emph {et~al.}(2010)\citenamefont {Grimme},
  \citenamefont {Antony}, \citenamefont {Ehrlich},\ and\ \citenamefont
  {Krieg}}]{grimmeConsistentAccurateInitio2010a}%
  \BibitemOpen
  \bibfield  {author} {\bibinfo {author} {\bibfnamefont {S.}~\bibnamefont
  {Grimme}}, \bibinfo {author} {\bibfnamefont {J.}~\bibnamefont {Antony}},
  \bibinfo {author} {\bibfnamefont {S.}~\bibnamefont {Ehrlich}}, \ and\
  \bibinfo {author} {\bibfnamefont {H.}~\bibnamefont {Krieg}},\ }\bibfield
  {title} {\enquote {\bibinfo {title} {A consistent and accurate ab initio
  parametrization of density functional dispersion correction ({{DFT-D}}) for
  the 94 elements {{H-Pu}}},}\ }\href {\doibase 10.1063/1.3382344} {\bibfield
  {journal} {\bibinfo  {journal} {J. Chem. Phys.}\ }\textbf {\bibinfo {volume}
  {132}},\ \bibinfo {pages} {154104} (\bibinfo {year} {2010})}\BibitemShut
  {NoStop}%
\bibitem [{\citenamefont {Grimme}\ \emph {et~al.}(2016)\citenamefont {Grimme},
  \citenamefont {Hansen}, \citenamefont {Brandenburg},\ and\ \citenamefont
  {Bannwarth}}]{grimmeDispersionCorrectedMeanFieldElectronic2016}%
  \BibitemOpen
  \bibfield  {author} {\bibinfo {author} {\bibfnamefont {S.}~\bibnamefont
  {Grimme}}, \bibinfo {author} {\bibfnamefont {A.}~\bibnamefont {Hansen}},
  \bibinfo {author} {\bibfnamefont {J.~G.}\ \bibnamefont {Brandenburg}}, \ and\
  \bibinfo {author} {\bibfnamefont {C.}~\bibnamefont {Bannwarth}},\ }\bibfield
  {title} {\enquote {\bibinfo {title} {Dispersion-{{Corrected Mean-Field
  Electronic Structure Methods}}},}\ }\href {\doibase
  10.1021/acs.chemrev.5b00533} {\bibfield  {journal} {\bibinfo  {journal}
  {Chem. Rev.}\ }\textbf {\bibinfo {volume} {116}},\ \bibinfo {pages}
  {5105--5154} (\bibinfo {year} {2016})}\BibitemShut {NoStop}%
\bibitem [{\citenamefont {A.~Price}, \citenamefont {{Otero-de-la-Roza}},\ and\
  \citenamefont {R.~Johnson}(2023)}]{a.priceXDMcorrectedHybridDFT2023}%
  \BibitemOpen
  \bibfield  {author} {\bibinfo {author} {\bibfnamefont {A.~J.}\ \bibnamefont
  {A.~Price}}, \bibinfo {author} {\bibfnamefont {A.}~\bibnamefont
  {{Otero-de-la-Roza}}}, \ and\ \bibinfo {author} {\bibfnamefont
  {E.}~\bibnamefont {R.~Johnson}},\ }\bibfield  {title} {\enquote {\bibinfo
  {title} {{{XDM-corrected}} hybrid {{DFT}} with numerical atomic orbitals
  predicts molecular crystal lattice energies with unprecedented accuracy},}\
  }\href {\doibase 10.1039/D2SC05997E} {\bibfield  {journal} {\bibinfo
  {journal} {Chem. Sci.}\ }\textbf {\bibinfo {volume} {14}},\ \bibinfo {pages}
  {1252--1262} (\bibinfo {year} {2023})}\BibitemShut {NoStop}%
\bibitem [{\citenamefont {Zen}\ \emph {et~al.}(2016)\citenamefont {Zen},
  \citenamefont {Sorella}, \citenamefont {Gillan}, \citenamefont
  {Michaelides},\ and\ \citenamefont
  {Alf{\`e}}}]{zenBoostingAccuracySpeed2016c}%
  \BibitemOpen
  \bibfield  {author} {\bibinfo {author} {\bibfnamefont {A.}~\bibnamefont
  {Zen}}, \bibinfo {author} {\bibfnamefont {S.}~\bibnamefont {Sorella}},
  \bibinfo {author} {\bibfnamefont {M.~J.}\ \bibnamefont {Gillan}}, \bibinfo
  {author} {\bibfnamefont {A.}~\bibnamefont {Michaelides}}, \ and\ \bibinfo
  {author} {\bibfnamefont {D.}~\bibnamefont {Alf{\`e}}},\ }\bibfield  {title}
  {\enquote {\bibinfo {title} {Boosting the accuracy and speed of quantum
  {{Monte Carlo}}: {{Size}} consistency and time step},}\ }\href {\doibase
  10.1103/PhysRevB.93.241118} {\bibfield  {journal} {\bibinfo  {journal} {Phys.
  Rev. B}\ }\textbf {\bibinfo {volume} {93}},\ \bibinfo {pages} {241118}
  (\bibinfo {year} {2016})}\BibitemShut {NoStop}%
\bibitem [{\citenamefont {Krogel}(2016)}]{krogelNexusModularWorkflow2016}%
  \BibitemOpen
  \bibfield  {author} {\bibinfo {author} {\bibfnamefont {J.~T.}\ \bibnamefont
  {Krogel}},\ }\bibfield  {title} {\enquote {\bibinfo {title} {Nexus: {{A}}
  modular workflow management system for quantum simulation codes},}\ }\href
  {\doibase 10.1016/j.cpc.2015.08.012} {\bibfield  {journal} {\bibinfo
  {journal} {Comput. Phys. Commun.}\ }\textbf {\bibinfo {volume} {198}},\
  \bibinfo {pages} {154--168} (\bibinfo {year} {2016})}\BibitemShut {NoStop}%
\bibitem [{\citenamefont {Bennett}\ \emph {et~al.}(2017)\citenamefont
  {Bennett}, \citenamefont {Melton}, \citenamefont {Annaberdiyev},
  \citenamefont {Wang}, \citenamefont {Shulenburger},\ and\ \citenamefont
  {Mitas}}]{bennettNewGenerationEffective2017}%
  \BibitemOpen
  \bibfield  {author} {\bibinfo {author} {\bibfnamefont {M.~C.}\ \bibnamefont
  {Bennett}}, \bibinfo {author} {\bibfnamefont {C.~A.}\ \bibnamefont {Melton}},
  \bibinfo {author} {\bibfnamefont {A.}~\bibnamefont {Annaberdiyev}}, \bibinfo
  {author} {\bibfnamefont {G.}~\bibnamefont {Wang}}, \bibinfo {author}
  {\bibfnamefont {L.}~\bibnamefont {Shulenburger}}, \ and\ \bibinfo {author}
  {\bibfnamefont {L.}~\bibnamefont {Mitas}},\ }\bibfield  {title} {\enquote
  {\bibinfo {title} {A new generation of effective core potentials for
  correlated calculations},}\ }\href {\doibase 10.1063/1.4995643} {\bibfield
  {journal} {\bibinfo  {journal} {J. Chem. Phys.}\ }\textbf {\bibinfo {volume}
  {147}},\ \bibinfo {pages} {224106} (\bibinfo {year} {2017})}\BibitemShut
  {NoStop}%
\bibitem [{\citenamefont {Bennett}\ \emph {et~al.}(2018)\citenamefont
  {Bennett}, \citenamefont {Wang}, \citenamefont {Annaberdiyev}, \citenamefont
  {Melton}, \citenamefont {Shulenburger},\ and\ \citenamefont
  {Mitas}}]{bennettNewGenerationEffective2018}%
  \BibitemOpen
  \bibfield  {author} {\bibinfo {author} {\bibfnamefont {M.~C.}\ \bibnamefont
  {Bennett}}, \bibinfo {author} {\bibfnamefont {G.}~\bibnamefont {Wang}},
  \bibinfo {author} {\bibfnamefont {A.}~\bibnamefont {Annaberdiyev}}, \bibinfo
  {author} {\bibfnamefont {C.~A.}\ \bibnamefont {Melton}}, \bibinfo {author}
  {\bibfnamefont {L.}~\bibnamefont {Shulenburger}}, \ and\ \bibinfo {author}
  {\bibfnamefont {L.}~\bibnamefont {Mitas}},\ }\bibfield  {title} {\enquote
  {\bibinfo {title} {A new generation of effective core potentials from
  correlated calculations: 2nd row elements},}\ }\href {\doibase
  10.1063/1.5038135} {\bibfield  {journal} {\bibinfo  {journal} {J. Chem.
  Phys.}\ }\textbf {\bibinfo {volume} {149}},\ \bibinfo {pages} {104108}
  (\bibinfo {year} {2018})}\BibitemShut {NoStop}%
\bibitem [{\citenamefont {Annaberdiyev}\ \emph {et~al.}(2018)\citenamefont
  {Annaberdiyev}, \citenamefont {Wang}, \citenamefont {Melton}, \citenamefont
  {Bennett}, \citenamefont {Shulenburger},\ and\ \citenamefont
  {Mitas}}]{annaberdiyevNewGenerationEffective2018}%
  \BibitemOpen
  \bibfield  {author} {\bibinfo {author} {\bibfnamefont {A.}~\bibnamefont
  {Annaberdiyev}}, \bibinfo {author} {\bibfnamefont {G.}~\bibnamefont {Wang}},
  \bibinfo {author} {\bibfnamefont {C.~A.}\ \bibnamefont {Melton}}, \bibinfo
  {author} {\bibfnamefont {M.~C.}\ \bibnamefont {Bennett}}, \bibinfo {author}
  {\bibfnamefont {L.}~\bibnamefont {Shulenburger}}, \ and\ \bibinfo {author}
  {\bibfnamefont {L.}~\bibnamefont {Mitas}},\ }\bibfield  {title} {\enquote
  {\bibinfo {title} {A new generation of effective core potentials from
  correlated calculations: 3d transition metal series},}\ }\href {\doibase
  10.1063/1.5040472} {\bibfield  {journal} {\bibinfo  {journal} {J. Chem.
  Phys.}\ }\textbf {\bibinfo {volume} {149}},\ \bibinfo {pages} {134108}
  (\bibinfo {year} {2018})}\BibitemShut {NoStop}%
\bibitem [{\citenamefont {Zen}\ \emph {et~al.}(2019)\citenamefont {Zen},
  \citenamefont {Brandenburg}, \citenamefont {Michaelides},\ and\ \citenamefont
  {Alf{\`e}}}]{zenNewSchemeFixed2019}%
  \BibitemOpen
  \bibfield  {author} {\bibinfo {author} {\bibfnamefont {A.}~\bibnamefont
  {Zen}}, \bibinfo {author} {\bibfnamefont {J.~G.}\ \bibnamefont
  {Brandenburg}}, \bibinfo {author} {\bibfnamefont {A.}~\bibnamefont
  {Michaelides}}, \ and\ \bibinfo {author} {\bibfnamefont {D.}~\bibnamefont
  {Alf{\`e}}},\ }\bibfield  {title} {\enquote {\bibinfo {title} {A new scheme
  for fixed node diffusion quantum {{Monte Carlo}} with pseudopotentials:
  {{Improving}} reproducibility and reducing the trial-wave-function bias},}\
  }\href {\doibase 10.1063/1.5119729} {\bibfield  {journal} {\bibinfo
  {journal} {J. Chem. Phys.}\ }\textbf {\bibinfo {volume} {151}},\ \bibinfo
  {pages} {134105} (\bibinfo {year} {2019})}\BibitemShut {NoStop}%
\bibitem [{\citenamefont {Needs}\ \emph {et~al.}(2020)\citenamefont {Needs},
  \citenamefont {Towler}, \citenamefont {Drummond}, \citenamefont
  {L{\'o}pez~R{\'i}os},\ and\ \citenamefont
  {Trail}}]{needsVariationalDiffusionQuantum2020}%
  \BibitemOpen
  \bibfield  {author} {\bibinfo {author} {\bibfnamefont {R.~J.}\ \bibnamefont
  {Needs}}, \bibinfo {author} {\bibfnamefont {M.~D.}\ \bibnamefont {Towler}},
  \bibinfo {author} {\bibfnamefont {N.~D.}\ \bibnamefont {Drummond}}, \bibinfo
  {author} {\bibfnamefont {P.}~\bibnamefont {L{\'o}pez~R{\'i}os}}, \ and\
  \bibinfo {author} {\bibfnamefont {J.~R.}\ \bibnamefont {Trail}},\ }\bibfield
  {title} {\enquote {\bibinfo {title} {Variational and diffusion quantum
  {{Monte Carlo}} calculations with the {{CASINO}} code},}\ }\href {\doibase
  10.1063/1.5144288} {\bibfield  {journal} {\bibinfo  {journal} {J. Chem.
  Phys.}\ }\textbf {\bibinfo {volume} {152}},\ \bibinfo {pages} {154106}
  (\bibinfo {year} {2020})}\BibitemShut {NoStop}%
\bibitem [{\citenamefont {Nakano}\ \emph {et~al.}(2020)\citenamefont {Nakano},
  \citenamefont {Attaccalite}, \citenamefont {Barborini}, \citenamefont
  {Capriotti}, \citenamefont {Casula}, \citenamefont {Coccia}, \citenamefont
  {Dagrada}, \citenamefont {Genovese}, \citenamefont {Luo}, \citenamefont
  {Mazzola}, \citenamefont {Zen},\ and\ \citenamefont
  {Sorella}}]{nakanoTurboRVBManybodyToolkit2020}%
  \BibitemOpen
  \bibfield  {author} {\bibinfo {author} {\bibfnamefont {K.}~\bibnamefont
  {Nakano}}, \bibinfo {author} {\bibfnamefont {C.}~\bibnamefont {Attaccalite}},
  \bibinfo {author} {\bibfnamefont {M.}~\bibnamefont {Barborini}}, \bibinfo
  {author} {\bibfnamefont {L.}~\bibnamefont {Capriotti}}, \bibinfo {author}
  {\bibfnamefont {M.}~\bibnamefont {Casula}}, \bibinfo {author} {\bibfnamefont
  {E.}~\bibnamefont {Coccia}}, \bibinfo {author} {\bibfnamefont
  {M.}~\bibnamefont {Dagrada}}, \bibinfo {author} {\bibfnamefont
  {C.}~\bibnamefont {Genovese}}, \bibinfo {author} {\bibfnamefont
  {Y.}~\bibnamefont {Luo}}, \bibinfo {author} {\bibfnamefont {G.}~\bibnamefont
  {Mazzola}}, \bibinfo {author} {\bibfnamefont {A.}~\bibnamefont {Zen}}, \ and\
  \bibinfo {author} {\bibfnamefont {S.}~\bibnamefont {Sorella}},\ }\bibfield
  {title} {\enquote {\bibinfo {title} {{{TurboRVB}}: {{A}} many-body toolkit
  for ab initio electronic simulations by quantum {{Monte Carlo}}},}\ }\href
  {\doibase 10.1063/5.0005037} {\bibfield  {journal} {\bibinfo  {journal} {J.
  Chem. Phys.}\ }\textbf {\bibinfo {volume} {152}},\ \bibinfo {pages} {204121}
  (\bibinfo {year} {2020})}\BibitemShut {NoStop}%
\bibitem [{\citenamefont {Kent}\ \emph {et~al.}(2020)\citenamefont {Kent},
  \citenamefont {Annaberdiyev}, \citenamefont {Benali}, \citenamefont
  {Bennett}, \citenamefont {Landinez~Borda}, \citenamefont {Doak},
  \citenamefont {Hao}, \citenamefont {Jordan}, \citenamefont {Krogel},
  \citenamefont {Kyl{\"a}np{\"a}{\"a}}, \citenamefont {Lee}, \citenamefont
  {Luo}, \citenamefont {Malone}, \citenamefont {Melton}, \citenamefont {Mitas},
  \citenamefont {Morales}, \citenamefont {Neuscamman}, \citenamefont
  {Reboredo}, \citenamefont {Rubenstein}, \citenamefont {Saritas},
  \citenamefont {Upadhyay}, \citenamefont {Wang}, \citenamefont {Zhang},\ and\
  \citenamefont {Zhao}}]{kentQMCPACKAdvancesDevelopment2020}%
  \BibitemOpen
  \bibfield  {author} {\bibinfo {author} {\bibfnamefont {P.~R.~C.}\
  \bibnamefont {Kent}}, \bibinfo {author} {\bibfnamefont {A.}~\bibnamefont
  {Annaberdiyev}}, \bibinfo {author} {\bibfnamefont {A.}~\bibnamefont
  {Benali}}, \bibinfo {author} {\bibfnamefont {M.~C.}\ \bibnamefont {Bennett}},
  \bibinfo {author} {\bibfnamefont {E.~J.}\ \bibnamefont {Landinez~Borda}},
  \bibinfo {author} {\bibfnamefont {P.}~\bibnamefont {Doak}}, \bibinfo {author}
  {\bibfnamefont {H.}~\bibnamefont {Hao}}, \bibinfo {author} {\bibfnamefont
  {K.~D.}\ \bibnamefont {Jordan}}, \bibinfo {author} {\bibfnamefont {J.~T.}\
  \bibnamefont {Krogel}}, \bibinfo {author} {\bibfnamefont {I.}~\bibnamefont
  {Kyl{\"a}np{\"a}{\"a}}}, \bibinfo {author} {\bibfnamefont {J.}~\bibnamefont
  {Lee}}, \bibinfo {author} {\bibfnamefont {Y.}~\bibnamefont {Luo}}, \bibinfo
  {author} {\bibfnamefont {F.~D.}\ \bibnamefont {Malone}}, \bibinfo {author}
  {\bibfnamefont {C.~A.}\ \bibnamefont {Melton}}, \bibinfo {author}
  {\bibfnamefont {L.}~\bibnamefont {Mitas}}, \bibinfo {author} {\bibfnamefont
  {M.~A.}\ \bibnamefont {Morales}}, \bibinfo {author} {\bibfnamefont
  {E.}~\bibnamefont {Neuscamman}}, \bibinfo {author} {\bibfnamefont {F.~A.}\
  \bibnamefont {Reboredo}}, \bibinfo {author} {\bibfnamefont {B.}~\bibnamefont
  {Rubenstein}}, \bibinfo {author} {\bibfnamefont {K.}~\bibnamefont {Saritas}},
  \bibinfo {author} {\bibfnamefont {S.}~\bibnamefont {Upadhyay}}, \bibinfo
  {author} {\bibfnamefont {G.}~\bibnamefont {Wang}}, \bibinfo {author}
  {\bibfnamefont {S.}~\bibnamefont {Zhang}}, \ and\ \bibinfo {author}
  {\bibfnamefont {L.}~\bibnamefont {Zhao}},\ }\bibfield  {title} {\enquote
  {\bibinfo {title} {{{QMCPACK}}: {{Advances}} in the development, efficiency,
  and application of auxiliary field and real-space variational and diffusion
  quantum {{Monte Carlo}}},}\ }\href {\doibase 10.1063/5.0004860} {\bibfield
  {journal} {\bibinfo  {journal} {J. Chem. Phys.}\ }\textbf {\bibinfo {volume}
  {152}},\ \bibinfo {pages} {174105} (\bibinfo {year} {2020})}\BibitemShut
  {NoStop}%
\bibitem [{\citenamefont {Nakano}\ \emph {et~al.}(2023)\citenamefont {Nakano},
  \citenamefont {Kohul{\'a}k}, \citenamefont {Raghav}, \citenamefont {Casula},\
  and\ \citenamefont {Sorella}}]{nakanoTurboGeniusPythonSuite2023}%
  \BibitemOpen
  \bibfield  {author} {\bibinfo {author} {\bibfnamefont {K.}~\bibnamefont
  {Nakano}}, \bibinfo {author} {\bibfnamefont {O.}~\bibnamefont {Kohul{\'a}k}},
  \bibinfo {author} {\bibfnamefont {A.}~\bibnamefont {Raghav}}, \bibinfo
  {author} {\bibfnamefont {M.}~\bibnamefont {Casula}}, \ and\ \bibinfo {author}
  {\bibfnamefont {S.}~\bibnamefont {Sorella}},\ }\bibfield  {title} {\enquote
  {\bibinfo {title} {{{TurboGenius}}: {{Python}} suite for high-throughput
  calculations of ab~initio quantum {{Monte Carlo}} methods},}\ }\href
  {\doibase 10.1063/5.0179003} {\bibfield  {journal} {\bibinfo  {journal} {J.
  Chem. Phys.}\ }\textbf {\bibinfo {volume} {159}},\ \bibinfo {pages} {224801}
  (\bibinfo {year} {2023})}\BibitemShut {NoStop}%
\bibitem [{\citenamefont {Riplinger}\ and\ \citenamefont
  {Neese}(2013)}]{riplingerEfficientLinearScaling2013b}%
  \BibitemOpen
  \bibfield  {author} {\bibinfo {author} {\bibfnamefont {C.}~\bibnamefont
  {Riplinger}}\ and\ \bibinfo {author} {\bibfnamefont {F.}~\bibnamefont
  {Neese}},\ }\bibfield  {title} {\enquote {\bibinfo {title} {An efficient and
  near linear scaling pair natural orbital based local coupled cluster
  method},}\ }\href {\doibase 10.1063/1.4773581} {\bibfield  {journal}
  {\bibinfo  {journal} {J. Chem. Phys.}\ }\textbf {\bibinfo {volume} {138}},\
  \bibinfo {pages} {034106} (\bibinfo {year} {2013})}\BibitemShut {NoStop}%
\bibitem [{\citenamefont {Riplinger}\ \emph {et~al.}(2016)\citenamefont
  {Riplinger}, \citenamefont {Pinski}, \citenamefont {Becker}, \citenamefont
  {Valeev},\ and\ \citenamefont {Neese}}]{riplingerSparseMapsSystematic2016}%
  \BibitemOpen
  \bibfield  {author} {\bibinfo {author} {\bibfnamefont {C.}~\bibnamefont
  {Riplinger}}, \bibinfo {author} {\bibfnamefont {P.}~\bibnamefont {Pinski}},
  \bibinfo {author} {\bibfnamefont {U.}~\bibnamefont {Becker}}, \bibinfo
  {author} {\bibfnamefont {E.~F.}\ \bibnamefont {Valeev}}, \ and\ \bibinfo
  {author} {\bibfnamefont {F.}~\bibnamefont {Neese}},\ }\bibfield  {title}
  {\enquote {\bibinfo {title} {Sparse maps---{{A}} systematic infrastructure
  for reduced-scaling electronic structure methods. {{II}}. {{Linear}} scaling
  domain based pair natural orbital coupled cluster theory},}\ }\href {\doibase
  10.1063/1.4939030} {\bibfield  {journal} {\bibinfo  {journal} {J. Chem.
  Phys.}\ }\textbf {\bibinfo {volume} {144}},\ \bibinfo {pages} {024109}
  (\bibinfo {year} {2016})}\BibitemShut {NoStop}%
\bibitem [{\citenamefont {Ma}\ and\ \citenamefont
  {Werner}(2018)}]{maExplicitlyCorrelatedLocal2018}%
  \BibitemOpen
  \bibfield  {author} {\bibinfo {author} {\bibfnamefont {Q.}~\bibnamefont
  {Ma}}\ and\ \bibinfo {author} {\bibfnamefont {H.-J.}\ \bibnamefont
  {Werner}},\ }\bibfield  {title} {\enquote {\bibinfo {title} {Explicitly
  correlated local coupled-cluster methods using pair natural orbitals},}\
  }\href {\doibase 10.1002/wcms.1371} {\bibfield  {journal} {\bibinfo
  {journal} {Wiley Interdiscip. Rev.: Comput. Mol. Sci.}\ }\textbf {\bibinfo
  {volume} {8}},\ \bibinfo {pages} {e1371} (\bibinfo {year}
  {2018})}\BibitemShut {NoStop}%
\bibitem [{\citenamefont {Nagy}, \citenamefont {Samu},\ and\ \citenamefont
  {K{\'a}llay}(2018)}]{nagyOptimizationLinearScalingLocal2018b}%
  \BibitemOpen
  \bibfield  {author} {\bibinfo {author} {\bibfnamefont {P.~R.}\ \bibnamefont
  {Nagy}}, \bibinfo {author} {\bibfnamefont {G.}~\bibnamefont {Samu}}, \ and\
  \bibinfo {author} {\bibfnamefont {M.}~\bibnamefont {K{\'a}llay}},\ }\bibfield
   {title} {\enquote {\bibinfo {title} {Optimization of the {{Linear-Scaling
  Local Natural Orbital CCSD}}({{T}}) {{Method}}: {{Improved Algorithm}} and
  {{Benchmark Applications}}},}\ }\href {\doibase 10.1021/acs.jctc.8b00442}
  {\bibfield  {journal} {\bibinfo  {journal} {J. Chem. Theory Comput.}\
  }\textbf {\bibinfo {volume} {14}},\ \bibinfo {pages} {4193--4215} (\bibinfo
  {year} {2018})}\BibitemShut {NoStop}%
\bibitem [{\citenamefont {Nagy}\ and\ \citenamefont
  {K{\'a}llay}(2019)}]{nagyApproachingBasisSet2019a}%
  \BibitemOpen
  \bibfield  {author} {\bibinfo {author} {\bibfnamefont {P.~R.}\ \bibnamefont
  {Nagy}}\ and\ \bibinfo {author} {\bibfnamefont {M.}~\bibnamefont
  {K{\'a}llay}},\ }\bibfield  {title} {\enquote {\bibinfo {title} {Approaching
  the {{Basis Set Limit}} of {{CCSD}}({{T}}) {{Energies}} for {{Large
  Molecules}} with {{Local Natural Orbital Coupled-Cluster Methods}}},}\ }\href
  {\doibase 10.1021/acs.jctc.9b00511} {\bibfield  {journal} {\bibinfo
  {journal} {J. Chem. Theory Comput.}\ }\textbf {\bibinfo {volume} {15}},\
  \bibinfo {pages} {5275--5298} (\bibinfo {year} {2019})}\BibitemShut {NoStop}%
\bibitem [{\citenamefont {Masios}\ \emph {et~al.}(2023)\citenamefont {Masios},
  \citenamefont {Irmler}, \citenamefont {Sch{\"a}fer},\ and\ \citenamefont
  {Gr{\"u}neis}}]{masiosAvertingInfraredCatastrophe2023c}%
  \BibitemOpen
  \bibfield  {author} {\bibinfo {author} {\bibfnamefont {N.}~\bibnamefont
  {Masios}}, \bibinfo {author} {\bibfnamefont {A.}~\bibnamefont {Irmler}},
  \bibinfo {author} {\bibfnamefont {T.}~\bibnamefont {Sch{\"a}fer}}, \ and\
  \bibinfo {author} {\bibfnamefont {A.}~\bibnamefont {Gr{\"u}neis}},\
  }\bibfield  {title} {\enquote {\bibinfo {title} {Averting the {{Infrared
  Catastrophe}} in the {{Gold Standard}} of {{Quantum Chemistry}}},}\ }\href
  {\doibase 10.1103/PhysRevLett.131.186401} {\bibfield  {journal} {\bibinfo
  {journal} {Phys. Rev. Lett.}\ }\textbf {\bibinfo {volume} {131}},\ \bibinfo
  {pages} {186401} (\bibinfo {year} {2023})}\BibitemShut {NoStop}%
\bibitem [{\citenamefont {Jiang}\ \emph {et~al.}(2024)\citenamefont {Jiang},
  \citenamefont {Glick}, \citenamefont {Poole}, \citenamefont {Turney},
  \citenamefont {Sherrill},\ and\ \citenamefont
  {Schaefer}}]{jiangAccurateEfficientOpensource2024}%
  \BibitemOpen
  \bibfield  {author} {\bibinfo {author} {\bibfnamefont {A.}~\bibnamefont
  {Jiang}}, \bibinfo {author} {\bibfnamefont {Z.~L.}\ \bibnamefont {Glick}},
  \bibinfo {author} {\bibfnamefont {D.}~\bibnamefont {Poole}}, \bibinfo
  {author} {\bibfnamefont {J.~M.}\ \bibnamefont {Turney}}, \bibinfo {author}
  {\bibfnamefont {C.~D.}\ \bibnamefont {Sherrill}}, \ and\ \bibinfo {author}
  {\bibfnamefont {H.~F.}\ \bibnamefont {Schaefer}, \bibfnamefont {III}},\
  }\bibfield  {title} {\enquote {\bibinfo {title} {Accurate and efficient
  open-source implementation of domain-based local pair natural orbital
  ({{DLPNO}}) coupled-cluster theory using a t1-transformed {{Hamiltonian}}},}\
  }\href {\doibase 10.1063/5.0219963} {\bibfield  {journal} {\bibinfo
  {journal} {J. Chem. Phys.}\ }\textbf {\bibinfo {volume} {161}},\ \bibinfo
  {pages} {082502} (\bibinfo {year} {2024})}\BibitemShut {NoStop}%
\bibitem [{\citenamefont {Ye}\ and\ \citenamefont
  {Berkelbach}(2024)}]{yePeriodicLocalCoupledCluster2024a}%
  \BibitemOpen
  \bibfield  {author} {\bibinfo {author} {\bibfnamefont {H.-Z.}\ \bibnamefont
  {Ye}}\ and\ \bibinfo {author} {\bibfnamefont {T.~C.}\ \bibnamefont
  {Berkelbach}},\ }\bibfield  {title} {\enquote {\bibinfo {title} {Periodic
  {{Local Coupled-Cluster Theory}} for {{Insulators}} and {{Metals}}},}\ }\href
  {\doibase 10.1021/acs.jctc.4c00936} {\bibfield  {journal} {\bibinfo
  {journal} {J. Chem. Theory Comput.}\ }\textbf {\bibinfo {volume} {20}},\
  \bibinfo {pages} {8948--8959} (\bibinfo {year} {2024})}\BibitemShut {NoStop}%
\bibitem [{\citenamefont {Dubeck{\'y}}\ \emph {et~al.}(2013)\citenamefont
  {Dubeck{\'y}}, \citenamefont {Jure{\v c}ka}, \citenamefont {Derian},
  \citenamefont {Hobza}, \citenamefont {Otyepka},\ and\ \citenamefont
  {Mitas}}]{dubeckyQuantumMonteCarlo2013}%
  \BibitemOpen
  \bibfield  {author} {\bibinfo {author} {\bibfnamefont {M.}~\bibnamefont
  {Dubeck{\'y}}}, \bibinfo {author} {\bibfnamefont {P.}~\bibnamefont {Jure{\v
  c}ka}}, \bibinfo {author} {\bibfnamefont {R.}~\bibnamefont {Derian}},
  \bibinfo {author} {\bibfnamefont {P.}~\bibnamefont {Hobza}}, \bibinfo
  {author} {\bibfnamefont {M.}~\bibnamefont {Otyepka}}, \ and\ \bibinfo
  {author} {\bibfnamefont {L.}~\bibnamefont {Mitas}},\ }\bibfield  {title}
  {\enquote {\bibinfo {title} {Quantum {{Monte Carlo Methods Describe
  Noncovalent Interactions}} with {{Subchemical Accuracy}}},}\ }\href {\doibase
  10.1021/ct4006739} {\bibfield  {journal} {\bibinfo  {journal} {J. Chem.
  Theory Comput.}\ }\textbf {\bibinfo {volume} {9}},\ \bibinfo {pages}
  {4287--4292} (\bibinfo {year} {2013})}\BibitemShut {NoStop}%
\bibitem [{\citenamefont {{\v R}ez{\'a}{\v c}}\ \emph
  {et~al.}(2015)\citenamefont {{\v R}ez{\'a}{\v c}}, \citenamefont
  {Dubeck{\'y}}, \citenamefont {Jure{\v c}ka},\ and\ \citenamefont
  {Hobza}}]{rezacExtensionsApplicationsA242015}%
  \BibitemOpen
  \bibfield  {author} {\bibinfo {author} {\bibfnamefont {J.}~\bibnamefont {{\v
  R}ez{\'a}{\v c}}}, \bibinfo {author} {\bibfnamefont {M.}~\bibnamefont
  {Dubeck{\'y}}}, \bibinfo {author} {\bibfnamefont {P.}~\bibnamefont {Jure{\v
  c}ka}}, \ and\ \bibinfo {author} {\bibfnamefont {P.}~\bibnamefont {Hobza}},\
  }\bibfield  {title} {\enquote {\bibinfo {title} {Extensions and applications
  of the {{A24}} data set of accurate interaction energies},}\ }\href {\doibase
  10.1039/C5CP03151F} {\bibfield  {journal} {\bibinfo  {journal} {Phys. Chem.
  Chem. Phys.}\ }\textbf {\bibinfo {volume} {17}},\ \bibinfo {pages}
  {19268--19277} (\bibinfo {year} {2015})}\BibitemShut {NoStop}%
\bibitem [{\citenamefont {Raghav}\ \emph {et~al.}(2023)\citenamefont {Raghav},
  \citenamefont {Maezono}, \citenamefont {Hongo}, \citenamefont {Sorella},\
  and\ \citenamefont {Nakano}}]{raghavChemicalAccuracyUsing2023}%
  \BibitemOpen
  \bibfield  {author} {\bibinfo {author} {\bibfnamefont {A.}~\bibnamefont
  {Raghav}}, \bibinfo {author} {\bibfnamefont {R.}~\bibnamefont {Maezono}},
  \bibinfo {author} {\bibfnamefont {K.}~\bibnamefont {Hongo}}, \bibinfo
  {author} {\bibfnamefont {S.}~\bibnamefont {Sorella}}, \ and\ \bibinfo
  {author} {\bibfnamefont {K.}~\bibnamefont {Nakano}},\ }\bibfield  {title}
  {\enquote {\bibinfo {title} {Toward {{Chemical Accuracy Using}} the {{Jastrow
  Correlated Antisymmetrized Geminal Power Ansatz}}},}\ }\href {\doibase
  10.1021/acs.jctc.2c01141} {\bibfield  {journal} {\bibinfo  {journal} {J.
  Chem. Theory Comput.}\ }\textbf {\bibinfo {volume} {19}},\ \bibinfo {pages}
  {2222--2229} (\bibinfo {year} {2023})}\BibitemShut {NoStop}%
\bibitem [{\citenamefont {Mostaani}, \citenamefont {Drummond},\ and\
  \citenamefont {Fal'ko}(2015)}]{mostaaniQuantumMonteCarlo2015}%
  \BibitemOpen
  \bibfield  {author} {\bibinfo {author} {\bibfnamefont {E.}~\bibnamefont
  {Mostaani}}, \bibinfo {author} {\bibfnamefont {N.~D.}\ \bibnamefont
  {Drummond}}, \ and\ \bibinfo {author} {\bibfnamefont {V.~I.}\ \bibnamefont
  {Fal'ko}},\ }\bibfield  {title} {\enquote {\bibinfo {title} {Quantum {{Monte
  Carlo Calculation}} of the {{Binding Energy}} of {{Bilayer Graphene}}},}\
  }\href {\doibase 10.1103/PhysRevLett.115.115501} {\bibfield  {journal}
  {\bibinfo  {journal} {Phys. Rev. Lett.}\ }\textbf {\bibinfo {volume} {115}},\
  \bibinfo {pages} {115501} (\bibinfo {year} {2015})}\BibitemShut {NoStop}%
\bibitem [{\citenamefont {Zen}\ \emph {et~al.}(2018)\citenamefont {Zen},
  \citenamefont {Brandenburg}, \citenamefont {Klime{\v s}}, \citenamefont
  {Tkatchenko}, \citenamefont {Alf{\`e}},\ and\ \citenamefont
  {Michaelides}}]{zenFastAccurateQuantum2018a}%
  \BibitemOpen
  \bibfield  {author} {\bibinfo {author} {\bibfnamefont {A.}~\bibnamefont
  {Zen}}, \bibinfo {author} {\bibfnamefont {J.~G.}\ \bibnamefont
  {Brandenburg}}, \bibinfo {author} {\bibfnamefont {J.}~\bibnamefont {Klime{\v
  s}}}, \bibinfo {author} {\bibfnamefont {A.}~\bibnamefont {Tkatchenko}},
  \bibinfo {author} {\bibfnamefont {D.}~\bibnamefont {Alf{\`e}}}, \ and\
  \bibinfo {author} {\bibfnamefont {A.}~\bibnamefont {Michaelides}},\
  }\bibfield  {title} {\enquote {\bibinfo {title} {Fast and accurate quantum
  {{Monte Carlo}} for molecular crystals},}\ }\href {\doibase
  10.1073/pnas.1715434115} {\bibfield  {journal} {\bibinfo  {journal} {Proc.
  Natl. Acad. Sci. U. S. A.}\ }\textbf {\bibinfo {volume} {115}},\ \bibinfo
  {pages} {1724--1729} (\bibinfo {year} {2018})}\BibitemShut {NoStop}%
\bibitem [{\citenamefont {Della~Pia}\ \emph {et~al.}(2022)\citenamefont
  {Della~Pia}, \citenamefont {Zen}, \citenamefont {Alf{\`e}},\ and\
  \citenamefont {Michaelides}}]{dellapiaDMCICE13AmbientHigh2022b}%
  \BibitemOpen
  \bibfield  {author} {\bibinfo {author} {\bibfnamefont {F.}~\bibnamefont
  {Della~Pia}}, \bibinfo {author} {\bibfnamefont {A.}~\bibnamefont {Zen}},
  \bibinfo {author} {\bibfnamefont {D.}~\bibnamefont {Alf{\`e}}}, \ and\
  \bibinfo {author} {\bibfnamefont {A.}~\bibnamefont {Michaelides}},\
  }\bibfield  {title} {\enquote {\bibinfo {title} {{{DMC-ICE13}}: {{Ambient}}
  and high pressure polymorphs of ice from diffusion {{Monte Carlo}} and
  density functional theory},}\ }\href {\doibase 10.1063/5.0102645} {\bibfield
  {journal} {\bibinfo  {journal} {J. Chem. Phys.}\ }\textbf {\bibinfo {volume}
  {157}},\ \bibinfo {pages} {134701} (\bibinfo {year} {2022})}\BibitemShut
  {NoStop}%
\bibitem [{\citenamefont {Della~Pia}\ \emph {et~al.}(2024)\citenamefont
  {Della~Pia}, \citenamefont {Zen}, \citenamefont {Alf{\`e}},\ and\
  \citenamefont {Michaelides}}]{dellapiaHowAccurateAre2024}%
  \BibitemOpen
  \bibfield  {author} {\bibinfo {author} {\bibfnamefont {F.}~\bibnamefont
  {Della~Pia}}, \bibinfo {author} {\bibfnamefont {A.}~\bibnamefont {Zen}},
  \bibinfo {author} {\bibfnamefont {D.}~\bibnamefont {Alf{\`e}}}, \ and\
  \bibinfo {author} {\bibfnamefont {A.}~\bibnamefont {Michaelides}},\
  }\bibfield  {title} {\enquote {\bibinfo {title} {How {{Accurate Are
  Simulations}} and {{Experiments}} for the {{Lattice Energies}} of {{Molecular
  Crystals}}?}}\ }\href {\doibase 10.1103/PhysRevLett.133.046401} {\bibfield
  {journal} {\bibinfo  {journal} {Phys. Rev. Lett.}\ }\textbf {\bibinfo
  {volume} {133}},\ \bibinfo {pages} {046401} (\bibinfo {year}
  {2024})}\BibitemShut {NoStop}%
\bibitem [{\citenamefont {Karalti}\ \emph {et~al.}(2012)\citenamefont
  {Karalti}, \citenamefont {Alf{\`e}}, \citenamefont {Gillan},\ and\
  \citenamefont {Jordan}}]{karaltiAdsorptionWaterMolecule2012}%
  \BibitemOpen
  \bibfield  {author} {\bibinfo {author} {\bibfnamefont {O.}~\bibnamefont
  {Karalti}}, \bibinfo {author} {\bibfnamefont {D.}~\bibnamefont {Alf{\`e}}},
  \bibinfo {author} {\bibfnamefont {M.~J.}\ \bibnamefont {Gillan}}, \ and\
  \bibinfo {author} {\bibfnamefont {K.~D.}\ \bibnamefont {Jordan}},\ }\bibfield
   {title} {\enquote {\bibinfo {title} {Adsorption of a water molecule on the
  {{MgO}}(100) surface as described by cluster and slab models},}\ }\href
  {\doibase 10.1039/C2CP00015F} {\bibfield  {journal} {\bibinfo  {journal}
  {Phys. Chem. Chem. Phys.}\ }\textbf {\bibinfo {volume} {14}},\ \bibinfo
  {pages} {7846--7853} (\bibinfo {year} {2012})}\BibitemShut {NoStop}%
\bibitem [{\citenamefont {{Al-Hamdani}}, \citenamefont {Alf{\`e}},\ and\
  \citenamefont {Michaelides}(2017)}]{al-hamdaniHowStronglyHydrogen2017}%
  \BibitemOpen
  \bibfield  {author} {\bibinfo {author} {\bibfnamefont {Y.~S.}\ \bibnamefont
  {{Al-Hamdani}}}, \bibinfo {author} {\bibfnamefont {D.}~\bibnamefont
  {Alf{\`e}}}, \ and\ \bibinfo {author} {\bibfnamefont {A.}~\bibnamefont
  {Michaelides}},\ }\bibfield  {title} {\enquote {\bibinfo {title} {How
  strongly do hydrogen and water molecules stick to carbon nanomaterials?}}\
  }\href {\doibase 10.1063/1.4977180} {\bibfield  {journal} {\bibinfo
  {journal} {J. Chem. Phys.}\ }\textbf {\bibinfo {volume} {146}},\ \bibinfo
  {pages} {094701} (\bibinfo {year} {2017})}\BibitemShut {NoStop}%
\bibitem [{\citenamefont {Tsatsoulis}\ \emph {et~al.}(2017)\citenamefont
  {Tsatsoulis}, \citenamefont {Hummel}, \citenamefont {Usvyat}, \citenamefont
  {Sch{\"u}tz}, \citenamefont {Booth}, \citenamefont {Binnie}, \citenamefont
  {Gillan}, \citenamefont {Alf{\`e}}, \citenamefont {Michaelides},\ and\
  \citenamefont {Gr{\"u}neis}}]{tsatsoulisComparisonQuantumChemistry2017a}%
  \BibitemOpen
  \bibfield  {author} {\bibinfo {author} {\bibfnamefont {T.}~\bibnamefont
  {Tsatsoulis}}, \bibinfo {author} {\bibfnamefont {F.}~\bibnamefont {Hummel}},
  \bibinfo {author} {\bibfnamefont {D.}~\bibnamefont {Usvyat}}, \bibinfo
  {author} {\bibfnamefont {M.}~\bibnamefont {Sch{\"u}tz}}, \bibinfo {author}
  {\bibfnamefont {G.~H.}\ \bibnamefont {Booth}}, \bibinfo {author}
  {\bibfnamefont {S.~S.}\ \bibnamefont {Binnie}}, \bibinfo {author}
  {\bibfnamefont {M.~J.}\ \bibnamefont {Gillan}}, \bibinfo {author}
  {\bibfnamefont {D.}~\bibnamefont {Alf{\`e}}}, \bibinfo {author}
  {\bibfnamefont {A.}~\bibnamefont {Michaelides}}, \ and\ \bibinfo {author}
  {\bibfnamefont {A.}~\bibnamefont {Gr{\"u}neis}},\ }\bibfield  {title}
  {\enquote {\bibinfo {title} {A comparison between quantum chemistry and
  quantum {{Monte Carlo}} techniques for the adsorption of water on the (001)
  {{LiH}} surface},}\ }\href {\doibase 10.1063/1.4984048} {\bibfield  {journal}
  {\bibinfo  {journal} {J. Chem. Phys.}\ }\textbf {\bibinfo {volume} {146}},\
  \bibinfo {pages} {204108} (\bibinfo {year} {2017})}\BibitemShut {NoStop}%
\bibitem [{\citenamefont {{Al-Hamdani}}\ \emph {et~al.}(2017)\citenamefont
  {{Al-Hamdani}}, \citenamefont {Rossi}, \citenamefont {Alf{\`e}},
  \citenamefont {Tsatsoulis}, \citenamefont {Ramberger}, \citenamefont
  {Brandenburg}, \citenamefont {Zen}, \citenamefont {Kresse}, \citenamefont
  {Gr{\"u}neis}, \citenamefont {Tkatchenko},\ and\ \citenamefont
  {Michaelides}}]{al-hamdaniPropertiesWaterBoron2017a}%
  \BibitemOpen
  \bibfield  {author} {\bibinfo {author} {\bibfnamefont {Y.~S.}\ \bibnamefont
  {{Al-Hamdani}}}, \bibinfo {author} {\bibfnamefont {M.}~\bibnamefont {Rossi}},
  \bibinfo {author} {\bibfnamefont {D.}~\bibnamefont {Alf{\`e}}}, \bibinfo
  {author} {\bibfnamefont {T.}~\bibnamefont {Tsatsoulis}}, \bibinfo {author}
  {\bibfnamefont {B.}~\bibnamefont {Ramberger}}, \bibinfo {author}
  {\bibfnamefont {J.~G.}\ \bibnamefont {Brandenburg}}, \bibinfo {author}
  {\bibfnamefont {A.}~\bibnamefont {Zen}}, \bibinfo {author} {\bibfnamefont
  {G.}~\bibnamefont {Kresse}}, \bibinfo {author} {\bibfnamefont
  {A.}~\bibnamefont {Gr{\"u}neis}}, \bibinfo {author} {\bibfnamefont
  {A.}~\bibnamefont {Tkatchenko}}, \ and\ \bibinfo {author} {\bibfnamefont
  {A.}~\bibnamefont {Michaelides}},\ }\bibfield  {title} {\enquote {\bibinfo
  {title} {Properties of the water to boron nitride interaction: {{From}} zero
  to two dimensions with benchmark accuracy},}\ }\href {\doibase
  10.1063/1.4985878} {\bibfield  {journal} {\bibinfo  {journal} {J. Chem.
  Phys.}\ }\textbf {\bibinfo {volume} {147}},\ \bibinfo {pages} {044710}
  (\bibinfo {year} {2017})}\BibitemShut {NoStop}%
\bibitem [{\citenamefont {Brandenburg}\ \emph {et~al.}(2019)\citenamefont
  {Brandenburg}, \citenamefont {Zen}, \citenamefont {Fitzner}, \citenamefont
  {Ramberger}, \citenamefont {Kresse}, \citenamefont {Tsatsoulis},
  \citenamefont {Gr{\"u}neis}, \citenamefont {Michaelides},\ and\ \citenamefont
  {Alf{\`e}}}]{brandenburgPhysisorptionWaterGraphene2019}%
  \BibitemOpen
  \bibfield  {author} {\bibinfo {author} {\bibfnamefont {J.~G.}\ \bibnamefont
  {Brandenburg}}, \bibinfo {author} {\bibfnamefont {A.}~\bibnamefont {Zen}},
  \bibinfo {author} {\bibfnamefont {M.}~\bibnamefont {Fitzner}}, \bibinfo
  {author} {\bibfnamefont {B.}~\bibnamefont {Ramberger}}, \bibinfo {author}
  {\bibfnamefont {G.}~\bibnamefont {Kresse}}, \bibinfo {author} {\bibfnamefont
  {T.}~\bibnamefont {Tsatsoulis}}, \bibinfo {author} {\bibfnamefont
  {A.}~\bibnamefont {Gr{\"u}neis}}, \bibinfo {author} {\bibfnamefont
  {A.}~\bibnamefont {Michaelides}}, \ and\ \bibinfo {author} {\bibfnamefont
  {D.}~\bibnamefont {Alf{\`e}}},\ }\bibfield  {title} {\enquote {\bibinfo
  {title} {Physisorption of {{Water}} on {{Graphene}}: {{Subchemical Accuracy}}
  from {{Many-Body Electronic Structure Methods}}},}\ }\href {\doibase
  10.1021/acs.jpclett.8b03679} {\bibfield  {journal} {\bibinfo  {journal} {J.
  Phys. Chem. Lett.}\ }\textbf {\bibinfo {volume} {10}},\ \bibinfo {pages}
  {358--368} (\bibinfo {year} {2019})}\BibitemShut {NoStop}%
\bibitem [{\citenamefont {Shi}\ \emph {et~al.}(2023)\citenamefont {Shi},
  \citenamefont {Zen}, \citenamefont {Kapil}, \citenamefont {Nagy},
  \citenamefont {Gr{\"u}neis},\ and\ \citenamefont
  {Michaelides}}]{shiManyBodyMethodsSurface2023a}%
  \BibitemOpen
  \bibfield  {author} {\bibinfo {author} {\bibfnamefont {B.~X.}\ \bibnamefont
  {Shi}}, \bibinfo {author} {\bibfnamefont {A.}~\bibnamefont {Zen}}, \bibinfo
  {author} {\bibfnamefont {V.}~\bibnamefont {Kapil}}, \bibinfo {author}
  {\bibfnamefont {P.~R.}\ \bibnamefont {Nagy}}, \bibinfo {author}
  {\bibfnamefont {A.}~\bibnamefont {Gr{\"u}neis}}, \ and\ \bibinfo {author}
  {\bibfnamefont {A.}~\bibnamefont {Michaelides}},\ }\bibfield  {title}
  {\enquote {\bibinfo {title} {Many-{{Body Methods}} for {{Surface Chemistry
  Come}} of {{Age}}: {{Achieving Consensus}} with {{Experiments}}},}\ }\href
  {\doibase 10.1021/jacs.3c09616} {\bibfield  {journal} {\bibinfo  {journal}
  {J. Am. Chem. Soc.}\ }\textbf {\bibinfo {volume} {145}},\ \bibinfo {pages}
  {25372--25381} (\bibinfo {year} {2023})}\BibitemShut {NoStop}%
\bibitem [{\citenamefont {Shi}\ \emph {et~al.}(2022)\citenamefont {Shi},
  \citenamefont {Kapil}, \citenamefont {Zen}, \citenamefont {Chen},
  \citenamefont {Alavi},\ and\ \citenamefont
  {Michaelides}}]{shiGeneralEmbeddedCluster2022a}%
  \BibitemOpen
  \bibfield  {author} {\bibinfo {author} {\bibfnamefont {B.~X.}\ \bibnamefont
  {Shi}}, \bibinfo {author} {\bibfnamefont {V.}~\bibnamefont {Kapil}}, \bibinfo
  {author} {\bibfnamefont {A.}~\bibnamefont {Zen}}, \bibinfo {author}
  {\bibfnamefont {J.}~\bibnamefont {Chen}}, \bibinfo {author} {\bibfnamefont
  {A.}~\bibnamefont {Alavi}}, \ and\ \bibinfo {author} {\bibfnamefont
  {A.}~\bibnamefont {Michaelides}},\ }\bibfield  {title} {\enquote {\bibinfo
  {title} {General embedded cluster protocol for accurate modeling of oxygen
  vacancies in metal-oxides},}\ }\href {\doibase 10.1063/5.0087031} {\bibfield
  {journal} {\bibinfo  {journal} {J. Chem. Phys.}\ }\textbf {\bibinfo {volume}
  {156}},\ \bibinfo {pages} {124704} (\bibinfo {year} {2022})}\BibitemShut
  {NoStop}%
\bibitem [{\citenamefont {{Al-Hamdani}}\ \emph {et~al.}(2021)\citenamefont
  {{Al-Hamdani}}, \citenamefont {Nagy}, \citenamefont {Zen}, \citenamefont
  {Barton}, \citenamefont {K{\'a}llay}, \citenamefont {Brandenburg},\ and\
  \citenamefont {Tkatchenko}}]{al-hamdaniInteractionsLargeMolecules2021}%
  \BibitemOpen
  \bibfield  {author} {\bibinfo {author} {\bibfnamefont {Y.~S.}\ \bibnamefont
  {{Al-Hamdani}}}, \bibinfo {author} {\bibfnamefont {P.~R.}\ \bibnamefont
  {Nagy}}, \bibinfo {author} {\bibfnamefont {A.}~\bibnamefont {Zen}}, \bibinfo
  {author} {\bibfnamefont {D.}~\bibnamefont {Barton}}, \bibinfo {author}
  {\bibfnamefont {M.}~\bibnamefont {K{\'a}llay}}, \bibinfo {author}
  {\bibfnamefont {J.~G.}\ \bibnamefont {Brandenburg}}, \ and\ \bibinfo {author}
  {\bibfnamefont {A.}~\bibnamefont {Tkatchenko}},\ }\bibfield  {title}
  {\enquote {\bibinfo {title} {Interactions between large molecules pose a
  puzzle for reference quantum mechanical methods},}\ }\href {\doibase
  10.1038/s41467-021-24119-3} {\bibfield  {journal} {\bibinfo  {journal} {Nat.
  Commun.}\ }\textbf {\bibinfo {volume} {12}},\ \bibinfo {pages} {3927}
  (\bibinfo {year} {2021})}\BibitemShut {NoStop}%
\bibitem [{\citenamefont {Sch{\"a}fer}\ \emph {et~al.}(2024)\citenamefont
  {Sch{\"a}fer}, \citenamefont {Irmler}, \citenamefont {Gallo},\ and\
  \citenamefont
  {Gr{\"u}neis}}]{schaferUnderstandingDiscrepanciesWavefunction2024}%
  \BibitemOpen
  \bibfield  {author} {\bibinfo {author} {\bibfnamefont {T.}~\bibnamefont
  {Sch{\"a}fer}}, \bibinfo {author} {\bibfnamefont {A.}~\bibnamefont {Irmler}},
  \bibinfo {author} {\bibfnamefont {A.}~\bibnamefont {Gallo}}, \ and\ \bibinfo
  {author} {\bibfnamefont {A.}~\bibnamefont {Gr{\"u}neis}},\ }\href {\doibase
  10.48550/arXiv.2407.01442} {\enquote {\bibinfo {title} {Understanding
  {{Discrepancies}} of {{Wavefunction Theories}} for {{Large Molecules}}},}\ }
  (\bibinfo {year} {2024}),\ \Eprint {http://arxiv.org/abs/2407.01442}
  {arXiv:2407.01442} \BibitemShut {NoStop}%
\bibitem [{\citenamefont {Ballesteros}, \citenamefont {Dunivan},\ and\
  \citenamefont {Lao}(2021)}]{ballesterosCoupledClusterBenchmarks2021}%
  \BibitemOpen
  \bibfield  {author} {\bibinfo {author} {\bibfnamefont {F.}~\bibnamefont
  {Ballesteros}}, \bibinfo {author} {\bibfnamefont {S.}~\bibnamefont
  {Dunivan}}, \ and\ \bibinfo {author} {\bibfnamefont {K.~U.}\ \bibnamefont
  {Lao}},\ }\bibfield  {title} {\enquote {\bibinfo {title} {Coupled cluster
  benchmarks of large noncovalent complexes: {{The L7}} dataset as well as
  {{DNA}}--ellipticine and buckycatcher--fullerene},}\ }\href {\doibase
  10.1063/5.0042906} {\bibfield  {journal} {\bibinfo  {journal} {J. Chem.
  Phys.}\ }\textbf {\bibinfo {volume} {154}},\ \bibinfo {pages} {154104}
  (\bibinfo {year} {2021})}\BibitemShut {NoStop}%
\bibitem [{\citenamefont {Villot}\ \emph {et~al.}(2022)\citenamefont {Villot},
  \citenamefont {Ballesteros}, \citenamefont {Wang},\ and\ \citenamefont
  {Lao}}]{villotCoupledClusterBenchmarking2022a}%
  \BibitemOpen
  \bibfield  {author} {\bibinfo {author} {\bibfnamefont {C.}~\bibnamefont
  {Villot}}, \bibinfo {author} {\bibfnamefont {F.}~\bibnamefont {Ballesteros}},
  \bibinfo {author} {\bibfnamefont {D.}~\bibnamefont {Wang}}, \ and\ \bibinfo
  {author} {\bibfnamefont {K.~U.}\ \bibnamefont {Lao}},\ }\bibfield  {title}
  {\enquote {\bibinfo {title} {Coupled {{Cluster Benchmarking}} of {{Large
  Noncovalent Complexes}} in {{L7}} and {{S12L}} as {{Well}} as the {{C60
  Dimer}}, {{DNA}}--{{Ellipticine}}, and {{HIV}}--{{Indinavir}}},}\ }\href
  {\doibase 10.1021/acs.jpca.2c01421} {\bibfield  {journal} {\bibinfo
  {journal} {J. Phys. Chem. A}\ }\textbf {\bibinfo {volume} {126}},\ \bibinfo
  {pages} {4326--4341} (\bibinfo {year} {2022})}\BibitemShut {NoStop}%
\bibitem [{\citenamefont {Fishman}\ \emph {et~al.}(2024)\citenamefont
  {Fishman}, \citenamefont {Lesiuk}, \citenamefont {Martin},\ and\
  \citenamefont {Boese}}]{fishmanNewAngleBenchmarking2024}%
  \BibitemOpen
  \bibfield  {author} {\bibinfo {author} {\bibfnamefont {V.}~\bibnamefont
  {Fishman}}, \bibinfo {author} {\bibfnamefont {M.}~\bibnamefont {Lesiuk}},
  \bibinfo {author} {\bibfnamefont {J.~M.~L.}\ \bibnamefont {Martin}}, \ and\
  \bibinfo {author} {\bibfnamefont {A.~D.}\ \bibnamefont {Boese}},\ }\href
  {\doibase 10.48550/arXiv.2410.12603} {\enquote {\bibinfo {title} {A {{New
  Angle}} on {{Benchmarking Noncovalent Interactions}}},}\ } (\bibinfo {year}
  {2024}),\ \Eprint {http://arxiv.org/abs/2410.12603} {arXiv:2410.12603}
  \BibitemShut {NoStop}%
\bibitem [{\citenamefont {Lambie}\ \emph {et~al.}(2024)\citenamefont {Lambie},
  \citenamefont {Kats}, \citenamefont {Usyvat},\ and\ \citenamefont
  {Alavi}}]{lambieApplicabilityCCSDTDispersion2024}%
  \BibitemOpen
  \bibfield  {author} {\bibinfo {author} {\bibfnamefont {S.}~\bibnamefont
  {Lambie}}, \bibinfo {author} {\bibfnamefont {D.}~\bibnamefont {Kats}},
  \bibinfo {author} {\bibfnamefont {D.}~\bibnamefont {Usyvat}}, \ and\ \bibinfo
  {author} {\bibfnamefont {A.}~\bibnamefont {Alavi}},\ }\href {\doibase
  10.48550/arXiv.2411.13986} {\enquote {\bibinfo {title} {On the applicability
  of {{CCSD}}({{T}}) for dispersion interactions in large conjugated
  systems},}\ } (\bibinfo {year} {2024}),\ \Eprint
  {http://arxiv.org/abs/2411.13986} {arXiv:2411.13986 [physics]} \BibitemShut
  {NoStop}%
\bibitem [{\citenamefont {Lao}(2024)}]{laoCanonicalCoupledCluster2024}%
  \BibitemOpen
  \bibfield  {author} {\bibinfo {author} {\bibfnamefont {K.~U.}\ \bibnamefont
  {Lao}},\ }\bibfield  {title} {\enquote {\bibinfo {title} {Canonical coupled
  cluster binding benchmark for nanoscale noncovalent complexes at the
  hundred-atom scale},}\ }\href {\doibase 10.1063/5.0242359} {\bibfield
  {journal} {\bibinfo  {journal} {J. Chem. Phys.}\ }\textbf {\bibinfo {volume}
  {161}},\ \bibinfo {pages} {234103} (\bibinfo {year} {2024})}\BibitemShut
  {NoStop}%
\bibitem [{\citenamefont {Semidalas}, \citenamefont {Boese},\ and\
  \citenamefont {Martin}(2025)}]{semidalasPostCCSDTCorrectionsS662025}%
  \BibitemOpen
  \bibfield  {author} {\bibinfo {author} {\bibfnamefont {E.}~\bibnamefont
  {Semidalas}}, \bibinfo {author} {\bibfnamefont {A.~D.}\ \bibnamefont
  {Boese}}, \ and\ \bibinfo {author} {\bibfnamefont {J.~M.~L.}\ \bibnamefont
  {Martin}},\ }\bibfield  {title} {\enquote {\bibinfo {title}
  {Post-{{CCSD}}({{T}}) corrections in the {{S66}} noncovalent interactions
  benchmark},}\ }\href {\doibase 10.1016/j.cplett.2025.141874} {\bibfield
  {journal} {\bibinfo  {journal} {Chem. Phys. Lett.}\ }\textbf {\bibinfo
  {volume} {863}},\ \bibinfo {pages} {141874} (\bibinfo {year}
  {2025})}\BibitemShut {NoStop}%
\bibitem [{\citenamefont {Pariser}\ and\ \citenamefont
  {Parr}(1953{\natexlab{a}})}]{pariserSemiEmpiricalTheoryElectronic1953}%
  \BibitemOpen
  \bibfield  {author} {\bibinfo {author} {\bibfnamefont {R.}~\bibnamefont
  {Pariser}}\ and\ \bibinfo {author} {\bibfnamefont {R.~G.}\ \bibnamefont
  {Parr}},\ }\bibfield  {title} {\enquote {\bibinfo {title} {A
  {{Semi}}-{{Empirical Theory}} of the {{Electronic Spectra}} and {{Electronic
  Structure}} of {{Complex Unsaturated Molecules}}. {{I}}.}}\ }\href {\doibase
  10.1063/1.1698929} {\bibfield  {journal} {\bibinfo  {journal} {J. Chem.
  Phys.}\ }\textbf {\bibinfo {volume} {21}},\ \bibinfo {pages} {466--471}
  (\bibinfo {year} {1953}{\natexlab{a}})}\BibitemShut {NoStop}%
\bibitem [{\citenamefont {Pariser}\ and\ \citenamefont
  {Parr}(1953{\natexlab{b}})}]{pariserSemiEmpiricalTheoryElectronic1953a}%
  \BibitemOpen
  \bibfield  {author} {\bibinfo {author} {\bibfnamefont {R.}~\bibnamefont
  {Pariser}}\ and\ \bibinfo {author} {\bibfnamefont {R.~G.}\ \bibnamefont
  {Parr}},\ }\bibfield  {title} {\enquote {\bibinfo {title} {A
  {{Semi}}-{{Empirical Theory}} of the {{Electronic Spectra}} and {{Electronic
  Structure}} of {{Complex Unsaturated Molecules}}. {{II}}},}\ }\href {\doibase
  10.1063/1.1699030} {\bibfield  {journal} {\bibinfo  {journal} {J. Chem.
  Phys.}\ }\textbf {\bibinfo {volume} {21}},\ \bibinfo {pages} {767--776}
  (\bibinfo {year} {1953}{\natexlab{b}})}\BibitemShut {NoStop}%
\bibitem [{\citenamefont
  {Pople}(1953)}]{popleElectronInteractionUnsaturated1953}%
  \BibitemOpen
  \bibfield  {author} {\bibinfo {author} {\bibfnamefont {J.~A.}\ \bibnamefont
  {Pople}},\ }\bibfield  {title} {\enquote {\bibinfo {title} {Electron
  interaction in unsaturated hydrocarbons},}\ }\href {\doibase
  10.1039/TF9534901375} {\bibfield  {journal} {\bibinfo  {journal} {Trans.
  Faraday Soc.}\ }\textbf {\bibinfo {volume} {49}},\ \bibinfo {pages}
  {1375--1385} (\bibinfo {year} {1953})}\BibitemShut {NoStop}%
\bibitem [{\citenamefont {Sure}\ and\ \citenamefont
  {Grimme}(2015)}]{sureComprehensiveBenchmarkAssociation2015}%
  \BibitemOpen
  \bibfield  {author} {\bibinfo {author} {\bibfnamefont {R.}~\bibnamefont
  {Sure}}\ and\ \bibinfo {author} {\bibfnamefont {S.}~\bibnamefont {Grimme}},\
  }\bibfield  {title} {\enquote {\bibinfo {title} {Comprehensive {{Benchmark}}
  of {{Association}} ({{Free}}) {{Energies}} of {{Realistic Host}}--{{Guest
  Complexes}}},}\ }\href {\doibase 10.1021/acs.jctc.5b00296} {\bibfield
  {journal} {\bibinfo  {journal} {J. Chem. Theory Comput.}\ }\textbf {\bibinfo
  {volume} {11}},\ \bibinfo {pages} {3785--3801} (\bibinfo {year}
  {2015})}\BibitemShut {NoStop}%
\bibitem [{\citenamefont
  {Grimme}(2012)}]{grimmeSupramolecularBindingThermodynamics2012}%
  \BibitemOpen
  \bibfield  {author} {\bibinfo {author} {\bibfnamefont {S.}~\bibnamefont
  {Grimme}},\ }\bibfield  {title} {\enquote {\bibinfo {title} {Supramolecular
  {{Binding Thermodynamics}} by {{Dispersion-Corrected Density Functional
  Theory}}},}\ }\href {\doibase 10.1002/chem.201200497} {\bibfield  {journal}
  {\bibinfo  {journal} {Chem. - Eur. J.}\ }\textbf {\bibinfo {volume} {18}},\
  \bibinfo {pages} {9955--9964} (\bibinfo {year} {2012})}\BibitemShut {NoStop}%
\bibitem [{\citenamefont {{\v R}ez{\'a}{\v c}}\ and\ \citenamefont
  {Hobza}(2013)}]{rezacDescribingNoncovalentInteractions2013}%
  \BibitemOpen
  \bibfield  {author} {\bibinfo {author} {\bibfnamefont {J.}~\bibnamefont {{\v
  R}ez{\'a}{\v c}}}\ and\ \bibinfo {author} {\bibfnamefont {P.}~\bibnamefont
  {Hobza}},\ }\bibfield  {title} {\enquote {\bibinfo {title} {Describing
  {{Noncovalent Interactions}} beyond the {{Common Approximations}}: {{How
  Accurate Is}} the ``{{Gold Standard}},'' {{CCSD}}({{T}}) at the {{Complete
  Basis Set Limit}}?}}\ }\href {\doibase 10.1021/ct400057w} {\bibfield
  {journal} {\bibinfo  {journal} {J. Chem. Theory Comput.}\ }\textbf {\bibinfo
  {volume} {9}},\ \bibinfo {pages} {2151--2155} (\bibinfo {year}
  {2013})}\BibitemShut {NoStop}%
\bibitem [{\citenamefont {Jure{\v c}ka}\ \emph {et~al.}(2006)\citenamefont
  {Jure{\v c}ka}, \citenamefont {{\v S}poner}, \citenamefont {{\v C}ern{\'y}},\
  and\ \citenamefont {Hobza}}]{jureckaBenchmarkDatabaseAccurate2006}%
  \BibitemOpen
  \bibfield  {author} {\bibinfo {author} {\bibfnamefont {P.}~\bibnamefont
  {Jure{\v c}ka}}, \bibinfo {author} {\bibfnamefont {J.}~\bibnamefont {{\v
  S}poner}}, \bibinfo {author} {\bibfnamefont {J.}~\bibnamefont {{\v
  C}ern{\'y}}}, \ and\ \bibinfo {author} {\bibfnamefont {P.}~\bibnamefont
  {Hobza}},\ }\bibfield  {title} {\enquote {\bibinfo {title} {Benchmark
  database of accurate ({{MP2}} and {{CCSD}}({{T}}) complete basis set limit)
  interaction energies of small model complexes, {{DNA}} base pairs, and amino
  acid pairs},}\ }\href {\doibase 10.1039/B600027D} {\bibfield  {journal}
  {\bibinfo  {journal} {Phys. Chem. Chem. Phys.}\ }\textbf {\bibinfo {volume}
  {8}},\ \bibinfo {pages} {1985--1993} (\bibinfo {year} {2006})}\BibitemShut
  {NoStop}%
\bibitem [{\citenamefont {{\v R}ez{\'a}{\v c}}, \citenamefont {Riley},\ and\
  \citenamefont
  {Hobza}(2011{\natexlab{a}})}]{rezacS66WellbalancedDatabase2011}%
  \BibitemOpen
  \bibfield  {author} {\bibinfo {author} {\bibfnamefont {J.}~\bibnamefont {{\v
  R}ez{\'a}{\v c}}}, \bibinfo {author} {\bibfnamefont {K.~E.}\ \bibnamefont
  {Riley}}, \ and\ \bibinfo {author} {\bibfnamefont {P.}~\bibnamefont
  {Hobza}},\ }\bibfield  {title} {\enquote {\bibinfo {title} {S66: {{A
  Well-balanced Database}} of {{Benchmark Interaction Energies Relevant}} to
  {{Biomolecular Structures}}},}\ }\href {\doibase 10.1021/ct2002946}
  {\bibfield  {journal} {\bibinfo  {journal} {J. Chem. Theory Comput.}\
  }\textbf {\bibinfo {volume} {7}},\ \bibinfo {pages} {2427--2438} (\bibinfo
  {year} {2011}{\natexlab{a}})}\BibitemShut {NoStop}%
\bibitem [{\citenamefont {Sinnokrot}\ and\ \citenamefont
  {Sherrill}(2004)}]{sinnokrotHighlyAccurateCoupled2004}%
  \BibitemOpen
  \bibfield  {author} {\bibinfo {author} {\bibfnamefont {M.~O.}\ \bibnamefont
  {Sinnokrot}}\ and\ \bibinfo {author} {\bibfnamefont {C.~D.}\ \bibnamefont
  {Sherrill}},\ }\bibfield  {title} {\enquote {\bibinfo {title} {Highly
  {{Accurate Coupled Cluster Potential Energy Curves}} for the {{Benzene
  Dimer}}:\, {{Sandwich}}, {{T-Shaped}}, and {{Parallel-Displaced
  Configurations}}},}\ }\href {\doibase 10.1021/jp0469517} {\bibfield
  {journal} {\bibinfo  {journal} {J. Phys. Chem. A}\ }\textbf {\bibinfo
  {volume} {108}},\ \bibinfo {pages} {10200--10207} (\bibinfo {year}
  {2004})}\BibitemShut {NoStop}%
\bibitem [{\citenamefont {Goerigk}, \citenamefont {Kruse},\ and\ \citenamefont
  {Grimme}(2011)}]{goerigkBenchmarkingDensityFunctional2011}%
  \BibitemOpen
  \bibfield  {author} {\bibinfo {author} {\bibfnamefont {L.}~\bibnamefont
  {Goerigk}}, \bibinfo {author} {\bibfnamefont {H.}~\bibnamefont {Kruse}}, \
  and\ \bibinfo {author} {\bibfnamefont {S.}~\bibnamefont {Grimme}},\
  }\bibfield  {title} {\enquote {\bibinfo {title} {Benchmarking {{Density
  Functional Methods}} against the {{S66}} and {{S66x8 Datasets}} for
  {{Non-Covalent Interactions}}},}\ }\href {\doibase 10.1002/cphc.201100826}
  {\bibfield  {journal} {\bibinfo  {journal} {ChemPhysChem}\ }\textbf {\bibinfo
  {volume} {12}},\ \bibinfo {pages} {3421--3433} (\bibinfo {year}
  {2011})}\BibitemShut {NoStop}%
\bibitem [{\citenamefont {Gao}\ \emph {et~al.}(2016)\citenamefont {Gao},
  \citenamefont {Li}, \citenamefont {Li}, \citenamefont {Li}, \citenamefont
  {Fang}, \citenamefont {Li}, \citenamefont {Hu}, \citenamefont {Lu},\ and\
  \citenamefont {Su}}]{gaoMachineLearningCorrection2016}%
  \BibitemOpen
  \bibfield  {author} {\bibinfo {author} {\bibfnamefont {T.}~\bibnamefont
  {Gao}}, \bibinfo {author} {\bibfnamefont {H.}~\bibnamefont {Li}}, \bibinfo
  {author} {\bibfnamefont {W.}~\bibnamefont {Li}}, \bibinfo {author}
  {\bibfnamefont {L.}~\bibnamefont {Li}}, \bibinfo {author} {\bibfnamefont
  {C.}~\bibnamefont {Fang}}, \bibinfo {author} {\bibfnamefont {H.}~\bibnamefont
  {Li}}, \bibinfo {author} {\bibfnamefont {L.}~\bibnamefont {Hu}}, \bibinfo
  {author} {\bibfnamefont {Y.}~\bibnamefont {Lu}}, \ and\ \bibinfo {author}
  {\bibfnamefont {Z.-M.}\ \bibnamefont {Su}},\ }\bibfield  {title} {\enquote
  {\bibinfo {title} {A machine learning correction for {{DFT}} non-covalent
  interactions based on the {{S22}}, {{S66}} and {{X40}} benchmark
  databases},}\ }\href {\doibase 10.1186/s13321-016-0133-7} {\bibfield
  {journal} {\bibinfo  {journal} {J. Cheminform.}\ }\textbf {\bibinfo {volume}
  {8}},\ \bibinfo {pages} {24} (\bibinfo {year} {2016})}\BibitemShut {NoStop}%
\bibitem [{\citenamefont {Peng}\ \emph {et~al.}(2016)\citenamefont {Peng},
  \citenamefont {Yang}, \citenamefont {Perdew},\ and\ \citenamefont
  {Sun}}]{pengVersatileVanWaals2016a}%
  \BibitemOpen
  \bibfield  {author} {\bibinfo {author} {\bibfnamefont {H.}~\bibnamefont
  {Peng}}, \bibinfo {author} {\bibfnamefont {Z.-H.}\ \bibnamefont {Yang}},
  \bibinfo {author} {\bibfnamefont {J.~P.}\ \bibnamefont {Perdew}}, \ and\
  \bibinfo {author} {\bibfnamefont {J.}~\bibnamefont {Sun}},\ }\bibfield
  {title} {\enquote {\bibinfo {title} {Versatile van der {{Waals Density
  Functional Based}} on a {{Meta-Generalized Gradient Approximation}}},}\
  }\href {\doibase 10.1103/PhysRevX.6.041005} {\bibfield  {journal} {\bibinfo
  {journal} {Phys. Rev. X}\ }\textbf {\bibinfo {volume} {6}},\ \bibinfo {pages}
  {041005} (\bibinfo {year} {2016})}\BibitemShut {NoStop}%
\bibitem [{\citenamefont {Yu}\ and\ \citenamefont
  {Wang}(2020)}]{yuDualhybridDirectRandom2020}%
  \BibitemOpen
  \bibfield  {author} {\bibinfo {author} {\bibfnamefont {F.}~\bibnamefont
  {Yu}}\ and\ \bibinfo {author} {\bibfnamefont {Y.}~\bibnamefont {Wang}},\
  }\bibfield  {title} {\enquote {\bibinfo {title} {Dual-hybrid direct random
  phase approximation and second-order screened exchange with nonlocal van der
  {{Waals}} correlations for noncovalent interactions},}\ }\href {\doibase
  10.1002/jcc.26149} {\bibfield  {journal} {\bibinfo  {journal} {J. Comp.
  Chem.}\ }\textbf {\bibinfo {volume} {41}},\ \bibinfo {pages} {1018--1025}
  (\bibinfo {year} {2020})}\BibitemShut {NoStop}%
\bibitem [{\citenamefont {Grimme}\ \emph {et~al.}(2021)\citenamefont {Grimme},
  \citenamefont {Hansen}, \citenamefont {Ehlert},\ and\ \citenamefont
  {Mewes}}]{grimmeR2SCAN3cSwissArmy2021}%
  \BibitemOpen
  \bibfield  {author} {\bibinfo {author} {\bibfnamefont {S.}~\bibnamefont
  {Grimme}}, \bibinfo {author} {\bibfnamefont {A.}~\bibnamefont {Hansen}},
  \bibinfo {author} {\bibfnamefont {S.}~\bibnamefont {Ehlert}}, \ and\ \bibinfo
  {author} {\bibfnamefont {J.-M.}\ \bibnamefont {Mewes}},\ }\bibfield  {title}
  {\enquote {\bibinfo {title} {{{r2SCAN-3c}}: {{A}} ``{{Swiss}} army knife''
  composite electronic-structure method},}\ }\href {\doibase 10.1063/5.0040021}
  {\bibfield  {journal} {\bibinfo  {journal} {J. Chem. Phys.}\ }\textbf
  {\bibinfo {volume} {154}},\ \bibinfo {pages} {064103} (\bibinfo {year}
  {2021})}\BibitemShut {NoStop}%
\bibitem [{\citenamefont {Ehlert}\ \emph {et~al.}(2021)\citenamefont {Ehlert},
  \citenamefont {Huniar}, \citenamefont {Ning}, \citenamefont {Furness},
  \citenamefont {Sun}, \citenamefont {Kaplan}, \citenamefont {Perdew},\ and\
  \citenamefont {Brandenburg}}]{ehlertR2SCAND4DispersionCorrected2021}%
  \BibitemOpen
  \bibfield  {author} {\bibinfo {author} {\bibfnamefont {S.}~\bibnamefont
  {Ehlert}}, \bibinfo {author} {\bibfnamefont {U.}~\bibnamefont {Huniar}},
  \bibinfo {author} {\bibfnamefont {J.}~\bibnamefont {Ning}}, \bibinfo {author}
  {\bibfnamefont {J.~W.}\ \bibnamefont {Furness}}, \bibinfo {author}
  {\bibfnamefont {J.}~\bibnamefont {Sun}}, \bibinfo {author} {\bibfnamefont
  {A.~D.}\ \bibnamefont {Kaplan}}, \bibinfo {author} {\bibfnamefont {J.~P.}\
  \bibnamefont {Perdew}}, \ and\ \bibinfo {author} {\bibfnamefont {J.~G.}\
  \bibnamefont {Brandenburg}},\ }\bibfield  {title} {\enquote {\bibinfo {title}
  {{{r2SCAN-D4}}: {{Dispersion}} corrected meta-generalized gradient
  approximation for general chemical applications},}\ }\href {\doibase
  10.1063/5.0041008} {\bibfield  {journal} {\bibinfo  {journal} {J. Chem.
  Phys.}\ }\textbf {\bibinfo {volume} {154}},\ \bibinfo {pages} {061101}
  (\bibinfo {year} {2021})}\BibitemShut {NoStop}%
\bibitem [{\citenamefont {M{\"u}ller}, \citenamefont {Hansen},\ and\
  \citenamefont {Grimme}(2023)}]{mullerOB97X3cCompositeRangeseparated2023}%
  \BibitemOpen
  \bibfield  {author} {\bibinfo {author} {\bibfnamefont {M.}~\bibnamefont
  {M{\"u}ller}}, \bibinfo {author} {\bibfnamefont {A.}~\bibnamefont {Hansen}},
  \ and\ \bibinfo {author} {\bibfnamefont {S.}~\bibnamefont {Grimme}},\
  }\bibfield  {title} {\enquote {\bibinfo {title} {{{$\omega$B97X-3c}}: {{A}}
  composite range-separated hybrid {{DFT}} method with a molecule-optimized
  polarized valence double-{$\zeta$} basis set},}\ }\href {\doibase
  10.1063/5.0133026} {\bibfield  {journal} {\bibinfo  {journal} {J. Chem.
  Phys.}\ }\textbf {\bibinfo {volume} {158}},\ \bibinfo {pages} {014103}
  (\bibinfo {year} {2023})}\BibitemShut {NoStop}%
\bibitem [{\citenamefont {Lu}\ and\ \citenamefont
  {Chen}(2023)}]{luSimpleEfficientUniversal2023}%
  \BibitemOpen
  \bibfield  {author} {\bibinfo {author} {\bibfnamefont {T.}~\bibnamefont
  {Lu}}\ and\ \bibinfo {author} {\bibfnamefont {Q.}~\bibnamefont {Chen}},\
  }\bibfield  {title} {\enquote {\bibinfo {title} {Simple, {{Efficient}}, and
  {{Universal Energy Decomposition Analysis Method Based}} on
  {{Dispersion-Corrected Density Functional Theory}}},}\ }\href {\doibase
  10.1021/acs.jpca.3c04374} {\bibfield  {journal} {\bibinfo  {journal} {J.
  Phys. Chem. A}\ }\textbf {\bibinfo {volume} {127}},\ \bibinfo {pages}
  {7023--7035} (\bibinfo {year} {2023})}\BibitemShut {NoStop}%
\bibitem [{\citenamefont {Lee}\ \emph {et~al.}(2024)\citenamefont {Lee},
  \citenamefont {Kim}, \citenamefont {Sim}, \citenamefont {Sogal},
  \citenamefont {Kim}, \citenamefont {Yu}, \citenamefont {Burke},\ and\
  \citenamefont {Sim}}]{leeCorrectingDispersionCorrections2024}%
  \BibitemOpen
  \bibfield  {author} {\bibinfo {author} {\bibfnamefont {M.}~\bibnamefont
  {Lee}}, \bibinfo {author} {\bibfnamefont {B.}~\bibnamefont {Kim}}, \bibinfo
  {author} {\bibfnamefont {M.}~\bibnamefont {Sim}}, \bibinfo {author}
  {\bibfnamefont {M.}~\bibnamefont {Sogal}}, \bibinfo {author} {\bibfnamefont
  {Y.}~\bibnamefont {Kim}}, \bibinfo {author} {\bibfnamefont {H.}~\bibnamefont
  {Yu}}, \bibinfo {author} {\bibfnamefont {K.}~\bibnamefont {Burke}}, \ and\
  \bibinfo {author} {\bibfnamefont {E.}~\bibnamefont {Sim}},\ }\bibfield
  {title} {\enquote {\bibinfo {title} {Correcting {{Dispersion Corrections}}
  with {{Density-Corrected DFT}}},}\ }\href {\doibase 10.1021/acs.jctc.4c00689}
  {\bibfield  {journal} {\bibinfo  {journal} {J. Chem. Theory Comput.}\
  }\textbf {\bibinfo {volume} {20}},\ \bibinfo {pages} {7155--7167} (\bibinfo
  {year} {2024})}\BibitemShut {NoStop}%
\bibitem [{\citenamefont {{\v R}ez{\'a}{\v c}}, \citenamefont {Riley},\ and\
  \citenamefont {Hobza}(2011{\natexlab{b}})}]{rezacExtensionsS66Data2011}%
  \BibitemOpen
  \bibfield  {author} {\bibinfo {author} {\bibfnamefont {J.}~\bibnamefont {{\v
  R}ez{\'a}{\v c}}}, \bibinfo {author} {\bibfnamefont {K.~E.}\ \bibnamefont
  {Riley}}, \ and\ \bibinfo {author} {\bibfnamefont {P.}~\bibnamefont
  {Hobza}},\ }\bibfield  {title} {\enquote {\bibinfo {title} {Extensions of the
  {{S66 Data Set}}: {{More Accurate Interaction Energies}} and
  {{Angular-Displaced Nonequilibrium Geometries}}},}\ }\href {\doibase
  10.1021/ct200523a} {\bibfield  {journal} {\bibinfo  {journal} {J. Chem.
  Theory Comput.}\ }\textbf {\bibinfo {volume} {7}},\ \bibinfo {pages}
  {3466--3470} (\bibinfo {year} {2011}{\natexlab{b}})}\BibitemShut {NoStop}%
\bibitem [{\citenamefont {Riley}, \citenamefont {{\v R}ez{\'a}{\v c}},\ and\
  \citenamefont {Hobza}(2012)}]{rileyPerformanceMP25MP2X2012}%
  \BibitemOpen
  \bibfield  {author} {\bibinfo {author} {\bibfnamefont {K.~E.}\ \bibnamefont
  {Riley}}, \bibinfo {author} {\bibfnamefont {J.}~\bibnamefont {{\v
  R}ez{\'a}{\v c}}}, \ and\ \bibinfo {author} {\bibfnamefont {P.}~\bibnamefont
  {Hobza}},\ }\bibfield  {title} {\enquote {\bibinfo {title} {The performance
  of {{MP2}}.5 and {{MP2}}.{{X}} methods for nonequilibrium geometries of
  molecular complexes},}\ }\href {\doibase 10.1039/C2CP41874F} {\bibfield
  {journal} {\bibinfo  {journal} {Phys. Chem. Chem. Phys.}\ }\textbf {\bibinfo
  {volume} {14}},\ \bibinfo {pages} {13187--13193} (\bibinfo {year}
  {2012})}\BibitemShut {NoStop}%
\bibitem [{\citenamefont {Altun}, \citenamefont {Neese},\ and\ \citenamefont
  {Bistoni}(2020)}]{altunExtrapolationLimitComplete2020a}%
  \BibitemOpen
  \bibfield  {author} {\bibinfo {author} {\bibfnamefont {A.}~\bibnamefont
  {Altun}}, \bibinfo {author} {\bibfnamefont {F.}~\bibnamefont {Neese}}, \ and\
  \bibinfo {author} {\bibfnamefont {G.}~\bibnamefont {Bistoni}},\ }\bibfield
  {title} {\enquote {\bibinfo {title} {Extrapolation to the {{Limit}} of a
  {{Complete Pair Natural Orbital Space}} in {{Local Coupled-Cluster
  Calculations}}},}\ }\href {\doibase 10.1021/acs.jctc.0c00344} {\bibfield
  {journal} {\bibinfo  {journal} {J. Chem. Theory Comput.}\ }\textbf {\bibinfo
  {volume} {16}},\ \bibinfo {pages} {6142--6149} (\bibinfo {year}
  {2020})}\BibitemShut {NoStop}%
\bibitem [{\citenamefont {Shee}\ \emph {et~al.}(2021)\citenamefont {Shee},
  \citenamefont {Loipersberger}, \citenamefont {Rettig}, \citenamefont {Lee},\
  and\ \citenamefont {{Head-Gordon}}}]{sheeRegularizedSecondOrderMoller2021a}%
  \BibitemOpen
  \bibfield  {author} {\bibinfo {author} {\bibfnamefont {J.}~\bibnamefont
  {Shee}}, \bibinfo {author} {\bibfnamefont {M.}~\bibnamefont {Loipersberger}},
  \bibinfo {author} {\bibfnamefont {A.}~\bibnamefont {Rettig}}, \bibinfo
  {author} {\bibfnamefont {J.}~\bibnamefont {Lee}}, \ and\ \bibinfo {author}
  {\bibfnamefont {M.}~\bibnamefont {{Head-Gordon}}},\ }\bibfield  {title}
  {\enquote {\bibinfo {title} {Regularized {{Second-Order
  M{\o}ller}}--{{Plesset Theory}}: {{A More Accurate Alternative}} to
  {{Conventional MP2}} for {{Noncovalent Interactions}} and {{Transition Metal
  Thermochemistry}} for the {{Same Computational Cost}}},}\ }\href {\doibase
  10.1021/acs.jpclett.1c03468} {\bibfield  {journal} {\bibinfo  {journal} {J.
  Phys. Chem. Lett.}\ }\textbf {\bibinfo {volume} {12}},\ \bibinfo {pages}
  {12084--12097} (\bibinfo {year} {2021})}\BibitemShut {NoStop}%
\bibitem [{\citenamefont {Lupi}\ \emph {et~al.}(2021)\citenamefont {Lupi},
  \citenamefont {Alessandrini}, \citenamefont {Puzzarini},\ and\ \citenamefont
  {Barone}}]{lupiJunChSJunChSF12Models2021}%
  \BibitemOpen
  \bibfield  {author} {\bibinfo {author} {\bibfnamefont {J.}~\bibnamefont
  {Lupi}}, \bibinfo {author} {\bibfnamefont {S.}~\bibnamefont {Alessandrini}},
  \bibinfo {author} {\bibfnamefont {C.}~\bibnamefont {Puzzarini}}, \ and\
  \bibinfo {author} {\bibfnamefont {V.}~\bibnamefont {Barone}},\ }\bibfield
  {title} {\enquote {\bibinfo {title} {{{junChS}} and {{junChS-F12 Models}}:
  {{Parameter-free Efficient}} yet {{Accurate Composite Schemes}} for
  {{Energies}} and {{Structures}} of {{Noncovalent Complexes}}},}\ }\href
  {\doibase 10.1021/acs.jctc.1c00869} {\bibfield  {journal} {\bibinfo
  {journal} {J. Chem. Theory Comput.}\ }\textbf {\bibinfo {volume} {17}},\
  \bibinfo {pages} {6974--6992} (\bibinfo {year} {2021})}\BibitemShut {NoStop}%
\bibitem [{\citenamefont {Semidalas}\ \emph {et~al.}(2022)\citenamefont
  {Semidalas}, \citenamefont {Santra}, \citenamefont {Mehta},\ and\
  \citenamefont {Martin}}]{semidalasS66NoncovalentInteractions2022}%
  \BibitemOpen
  \bibfield  {author} {\bibinfo {author} {\bibfnamefont {E.}~\bibnamefont
  {Semidalas}}, \bibinfo {author} {\bibfnamefont {G.}~\bibnamefont {Santra}},
  \bibinfo {author} {\bibfnamefont {N.}~\bibnamefont {Mehta}}, \ and\ \bibinfo
  {author} {\bibfnamefont {J.~M.~L.}\ \bibnamefont {Martin}},\ }\bibfield
  {title} {\enquote {\bibinfo {title} {S66 noncovalent interactions benchmark
  re-examined: {{Composite}} localized coupled cluster approaches},}\ }\href
  {\doibase 10.1063/5.0119282} {\bibfield  {journal} {\bibinfo  {journal} {AIP
  Conf. Proc.}\ }\textbf {\bibinfo {volume} {2611}},\ \bibinfo {pages} {020016}
  (\bibinfo {year} {2022})}\BibitemShut {NoStop}%
\bibitem [{\citenamefont {Beran}\ \emph {et~al.}(2023)\citenamefont {Beran},
  \citenamefont {Greenwell}, \citenamefont {Cook},\ and\ \citenamefont {{\v
  R}ez{\'a}{\v c}}}]{beranImprovedDescriptionIntra2023a}%
  \BibitemOpen
  \bibfield  {author} {\bibinfo {author} {\bibfnamefont {G.~J.~O.}\
  \bibnamefont {Beran}}, \bibinfo {author} {\bibfnamefont {C.}~\bibnamefont
  {Greenwell}}, \bibinfo {author} {\bibfnamefont {C.}~\bibnamefont {Cook}}, \
  and\ \bibinfo {author} {\bibfnamefont {J.}~\bibnamefont {{\v R}ez{\'a}{\v
  c}}},\ }\bibfield  {title} {\enquote {\bibinfo {title} {Improved
  {{Description}} of {{Intra-}} and {{Intermolecular Interactions}} through
  {{Dispersion-Corrected Second-Order M{\o}ller}}--{{Plesset Perturbation
  Theory}}},}\ }\href {\doibase 10.1021/acs.accounts.3c00578} {\bibfield
  {journal} {\bibinfo  {journal} {Acc. Chem. Res.}\ }\textbf {\bibinfo {volume}
  {56}},\ \bibinfo {pages} {3525--3534} (\bibinfo {year} {2023})}\BibitemShut
  {NoStop}%
\bibitem [{\citenamefont {Christensen}\ \emph {et~al.}(2021)\citenamefont
  {Christensen}, \citenamefont {Sirumalla}, \citenamefont {Qiao}, \citenamefont
  {O'Connor}, \citenamefont {Smith}, \citenamefont {Ding}, \citenamefont
  {Bygrave}, \citenamefont {Anandkumar}, \citenamefont {Welborn}, \citenamefont
  {Manby},\ and\ \citenamefont {Miller}}]{christensenOrbNetDenaliMachine2021}%
  \BibitemOpen
  \bibfield  {author} {\bibinfo {author} {\bibfnamefont {A.~S.}\ \bibnamefont
  {Christensen}}, \bibinfo {author} {\bibfnamefont {S.~K.}\ \bibnamefont
  {Sirumalla}}, \bibinfo {author} {\bibfnamefont {Z.}~\bibnamefont {Qiao}},
  \bibinfo {author} {\bibfnamefont {M.~B.}\ \bibnamefont {O'Connor}}, \bibinfo
  {author} {\bibfnamefont {D.~G.~A.}\ \bibnamefont {Smith}}, \bibinfo {author}
  {\bibfnamefont {F.}~\bibnamefont {Ding}}, \bibinfo {author} {\bibfnamefont
  {P.~J.}\ \bibnamefont {Bygrave}}, \bibinfo {author} {\bibfnamefont
  {A.}~\bibnamefont {Anandkumar}}, \bibinfo {author} {\bibfnamefont
  {M.}~\bibnamefont {Welborn}}, \bibinfo {author} {\bibfnamefont {F.~R.}\
  \bibnamefont {Manby}}, \ and\ \bibinfo {author} {\bibfnamefont {T.~F.}\
  \bibnamefont {Miller}, \bibfnamefont {III}},\ }\bibfield  {title} {\enquote
  {\bibinfo {title} {{{OrbNet Denali}}: {{A}} machine learning potential for
  biological and organic chemistry with semi-empirical cost and {{DFT}}
  accuracy},}\ }\href {\doibase 10.1063/5.0061990} {\bibfield  {journal}
  {\bibinfo  {journal} {J. Chem. Phys.}\ }\textbf {\bibinfo {volume} {155}},\
  \bibinfo {pages} {204103} (\bibinfo {year} {2021})}\BibitemShut {NoStop}%
\bibitem [{\citenamefont {Villot}\ and\ \citenamefont
  {Lao}(2024)}]{villotInitioDispersionPotentials2024}%
  \BibitemOpen
  \bibfield  {author} {\bibinfo {author} {\bibfnamefont {C.}~\bibnamefont
  {Villot}}\ and\ \bibinfo {author} {\bibfnamefont {K.~U.}\ \bibnamefont
  {Lao}},\ }\bibfield  {title} {\enquote {\bibinfo {title} {Ab initio
  dispersion potentials based on physics-based functional forms with machine
  learning},}\ }\href {\doibase 10.1063/5.0204064} {\bibfield  {journal}
  {\bibinfo  {journal} {J. Chem. Phys.}\ }\textbf {\bibinfo {volume} {160}},\
  \bibinfo {pages} {184103} (\bibinfo {year} {2024})}\BibitemShut {NoStop}%
\bibitem [{\citenamefont {Trail}\ and\ \citenamefont
  {Needs}(2017)}]{trailShapeEnergyConsistent2017}%
  \BibitemOpen
  \bibfield  {author} {\bibinfo {author} {\bibfnamefont {J.~R.}\ \bibnamefont
  {Trail}}\ and\ \bibinfo {author} {\bibfnamefont {R.~J.}\ \bibnamefont
  {Needs}},\ }\bibfield  {title} {\enquote {\bibinfo {title} {Shape and energy
  consistent pseudopotentials for correlated electron systems},}\ }\href
  {\doibase 10.1063/1.4984046} {\bibfield  {journal} {\bibinfo  {journal} {J.
  Chem. Phys.}\ }\textbf {\bibinfo {volume} {146}},\ \bibinfo {pages} {204107}
  (\bibinfo {year} {2017})}\BibitemShut {NoStop}%
\bibitem [{\citenamefont {Giannozzi}\ \emph {et~al.}(2009)\citenamefont
  {Giannozzi}, \citenamefont {Baroni}, \citenamefont {Bonini}, \citenamefont
  {Calandra}, \citenamefont {Car}, \citenamefont {Cavazzoni}, \citenamefont
  {Ceresoli}, \citenamefont {Chiarotti}, \citenamefont {Cococcioni},
  \citenamefont {Dabo}, \citenamefont {Corso}, \citenamefont {de~Gironcoli},
  \citenamefont {Fabris}, \citenamefont {Fratesi}, \citenamefont {Gebauer},
  \citenamefont {Gerstmann}, \citenamefont {Gougoussis}, \citenamefont
  {Kokalj}, \citenamefont {Lazzeri}, \citenamefont {{Martin-Samos}},
  \citenamefont {Marzari}, \citenamefont {Mauri}, \citenamefont {Mazzarello},
  \citenamefont {Paolini}, \citenamefont {Pasquarello}, \citenamefont
  {Paulatto}, \citenamefont {Sbraccia}, \citenamefont {Scandolo}, \citenamefont
  {Sclauzero}, \citenamefont {Seitsonen}, \citenamefont {Smogunov},
  \citenamefont {Umari},\ and\ \citenamefont
  {Wentzcovitch}}]{giannozziQUANTUMESPRESSOModular2009a}%
  \BibitemOpen
  \bibfield  {author} {\bibinfo {author} {\bibfnamefont {P.}~\bibnamefont
  {Giannozzi}}, \bibinfo {author} {\bibfnamefont {S.}~\bibnamefont {Baroni}},
  \bibinfo {author} {\bibfnamefont {N.}~\bibnamefont {Bonini}}, \bibinfo
  {author} {\bibfnamefont {M.}~\bibnamefont {Calandra}}, \bibinfo {author}
  {\bibfnamefont {R.}~\bibnamefont {Car}}, \bibinfo {author} {\bibfnamefont
  {C.}~\bibnamefont {Cavazzoni}}, \bibinfo {author} {\bibfnamefont
  {D.}~\bibnamefont {Ceresoli}}, \bibinfo {author} {\bibfnamefont {G.~L.}\
  \bibnamefont {Chiarotti}}, \bibinfo {author} {\bibfnamefont {M.}~\bibnamefont
  {Cococcioni}}, \bibinfo {author} {\bibfnamefont {I.}~\bibnamefont {Dabo}},
  \bibinfo {author} {\bibfnamefont {A.~D.}\ \bibnamefont {Corso}}, \bibinfo
  {author} {\bibfnamefont {S.}~\bibnamefont {de~Gironcoli}}, \bibinfo {author}
  {\bibfnamefont {S.}~\bibnamefont {Fabris}}, \bibinfo {author} {\bibfnamefont
  {G.}~\bibnamefont {Fratesi}}, \bibinfo {author} {\bibfnamefont
  {R.}~\bibnamefont {Gebauer}}, \bibinfo {author} {\bibfnamefont
  {U.}~\bibnamefont {Gerstmann}}, \bibinfo {author} {\bibfnamefont
  {C.}~\bibnamefont {Gougoussis}}, \bibinfo {author} {\bibfnamefont
  {A.}~\bibnamefont {Kokalj}}, \bibinfo {author} {\bibfnamefont
  {M.}~\bibnamefont {Lazzeri}}, \bibinfo {author} {\bibfnamefont
  {L.}~\bibnamefont {{Martin-Samos}}}, \bibinfo {author} {\bibfnamefont
  {N.}~\bibnamefont {Marzari}}, \bibinfo {author} {\bibfnamefont
  {F.}~\bibnamefont {Mauri}}, \bibinfo {author} {\bibfnamefont
  {R.}~\bibnamefont {Mazzarello}}, \bibinfo {author} {\bibfnamefont
  {S.}~\bibnamefont {Paolini}}, \bibinfo {author} {\bibfnamefont
  {A.}~\bibnamefont {Pasquarello}}, \bibinfo {author} {\bibfnamefont
  {L.}~\bibnamefont {Paulatto}}, \bibinfo {author} {\bibfnamefont
  {C.}~\bibnamefont {Sbraccia}}, \bibinfo {author} {\bibfnamefont
  {S.}~\bibnamefont {Scandolo}}, \bibinfo {author} {\bibfnamefont
  {G.}~\bibnamefont {Sclauzero}}, \bibinfo {author} {\bibfnamefont {A.~P.}\
  \bibnamefont {Seitsonen}}, \bibinfo {author} {\bibfnamefont {A.}~\bibnamefont
  {Smogunov}}, \bibinfo {author} {\bibfnamefont {P.}~\bibnamefont {Umari}}, \
  and\ \bibinfo {author} {\bibfnamefont {R.~M.}\ \bibnamefont {Wentzcovitch}},\
  }\bibfield  {title} {\enquote {\bibinfo {title} {{{QUANTUM ESPRESSO}}: A
  modular and open-source software project for quantum simulations of
  materials},}\ }\href {\doibase 10.1088/0953-8984/21/39/395502} {\bibfield
  {journal} {\bibinfo  {journal} {J. Phys.: Condens. Matter}\ }\textbf
  {\bibinfo {volume} {21}},\ \bibinfo {pages} {395502} (\bibinfo {year}
  {2009})}\BibitemShut {NoStop}%
\bibitem [{\citenamefont {Giannozzi}\ \emph {et~al.}(2020)\citenamefont
  {Giannozzi}, \citenamefont {Baseggio}, \citenamefont {Bonf{\`a}},
  \citenamefont {Brunato}, \citenamefont {Car}, \citenamefont {Carnimeo},
  \citenamefont {Cavazzoni}, \citenamefont {{de Gironcoli}}, \citenamefont
  {Delugas}, \citenamefont {Ferrari~Ruffino}, \citenamefont {Ferretti},
  \citenamefont {Marzari}, \citenamefont {Timrov}, \citenamefont {Urru},\ and\
  \citenamefont {Baroni}}]{giannozziQuantumESPRESSOExascale2020a}%
  \BibitemOpen
  \bibfield  {author} {\bibinfo {author} {\bibfnamefont {P.}~\bibnamefont
  {Giannozzi}}, \bibinfo {author} {\bibfnamefont {O.}~\bibnamefont {Baseggio}},
  \bibinfo {author} {\bibfnamefont {P.}~\bibnamefont {Bonf{\`a}}}, \bibinfo
  {author} {\bibfnamefont {D.}~\bibnamefont {Brunato}}, \bibinfo {author}
  {\bibfnamefont {R.}~\bibnamefont {Car}}, \bibinfo {author} {\bibfnamefont
  {I.}~\bibnamefont {Carnimeo}}, \bibinfo {author} {\bibfnamefont
  {C.}~\bibnamefont {Cavazzoni}}, \bibinfo {author} {\bibfnamefont
  {S.}~\bibnamefont {{de Gironcoli}}}, \bibinfo {author} {\bibfnamefont
  {P.}~\bibnamefont {Delugas}}, \bibinfo {author} {\bibfnamefont
  {F.}~\bibnamefont {Ferrari~Ruffino}}, \bibinfo {author} {\bibfnamefont
  {A.}~\bibnamefont {Ferretti}}, \bibinfo {author} {\bibfnamefont
  {N.}~\bibnamefont {Marzari}}, \bibinfo {author} {\bibfnamefont
  {I.}~\bibnamefont {Timrov}}, \bibinfo {author} {\bibfnamefont
  {A.}~\bibnamefont {Urru}}, \ and\ \bibinfo {author} {\bibfnamefont
  {S.}~\bibnamefont {Baroni}},\ }\bibfield  {title} {\enquote {\bibinfo {title}
  {Quantum {{ESPRESSO}} toward the exascale},}\ }\href {\doibase
  10.1063/5.0005082} {\bibfield  {journal} {\bibinfo  {journal} {J. Chem.
  Phys.}\ }\textbf {\bibinfo {volume} {152}},\ \bibinfo {pages} {154105}
  (\bibinfo {year} {2020})}\BibitemShut {NoStop}%
\bibitem [{\citenamefont {Alf{\`e}}\ and\ \citenamefont
  {Gillan}(2004)}]{alfeEfficientLocalizedBasis2004}%
  \BibitemOpen
  \bibfield  {author} {\bibinfo {author} {\bibfnamefont {D.}~\bibnamefont
  {Alf{\`e}}}\ and\ \bibinfo {author} {\bibfnamefont {M.~J.}\ \bibnamefont
  {Gillan}},\ }\bibfield  {title} {\enquote {\bibinfo {title} {Efficient
  localized basis set for quantum {{Monte Carlo}} calculations on condensed
  matter},}\ }\href {\doibase 10.1103/PhysRevB.70.161101} {\bibfield  {journal}
  {\bibinfo  {journal} {Phys. Rev. B}\ }\textbf {\bibinfo {volume} {70}},\
  \bibinfo {pages} {161101} (\bibinfo {year} {2004})}\BibitemShut {NoStop}%
\bibitem [{\citenamefont {Brauer}\ \emph {et~al.}(2016)\citenamefont {Brauer},
  \citenamefont {Kesharwani}, \citenamefont {Kozuch},\ and\ \citenamefont
  {Martin}}]{brauerS66x8BenchmarkNoncovalent2016}%
  \BibitemOpen
  \bibfield  {author} {\bibinfo {author} {\bibfnamefont {B.}~\bibnamefont
  {Brauer}}, \bibinfo {author} {\bibfnamefont {M.~K.}\ \bibnamefont
  {Kesharwani}}, \bibinfo {author} {\bibfnamefont {S.}~\bibnamefont {Kozuch}},
  \ and\ \bibinfo {author} {\bibfnamefont {J.~M.~L.}\ \bibnamefont {Martin}},\
  }\bibfield  {title} {\enquote {\bibinfo {title} {The {{S66x8}} benchmark for
  noncovalent interactions revisited: Explicitly correlated ab initio methods
  and density functional theory},}\ }\href {\doibase 10.1039/C6CP00688D}
  {\bibfield  {journal} {\bibinfo  {journal} {Phys. Chem. Chem. Phys.}\
  }\textbf {\bibinfo {volume} {18}},\ \bibinfo {pages} {20905--20925} (\bibinfo
  {year} {2016})}\BibitemShut {NoStop}%
\bibitem [{\citenamefont {Kesharwani}\ \emph {et~al.}(2018)\citenamefont
  {Kesharwani}, \citenamefont {Karton}, \citenamefont {Sylvetsky},\ and\
  \citenamefont {Martin}}]{kesharwaniS66NonCovalentInteractions2018}%
  \BibitemOpen
  \bibfield  {author} {\bibinfo {author} {\bibfnamefont {M.~K.}\ \bibnamefont
  {Kesharwani}}, \bibinfo {author} {\bibfnamefont {A.}~\bibnamefont {Karton}},
  \bibinfo {author} {\bibfnamefont {N.}~\bibnamefont {Sylvetsky}}, \ and\
  \bibinfo {author} {\bibfnamefont {J.~M.~L.}\ \bibnamefont {Martin}},\
  }\bibfield  {title} {\enquote {\bibinfo {title} {The {{S66 Non-Covalent
  Interactions Benchmark Reconsidered Using Explicitly Correlated Methods
  Near}} the {{Basis Set Limit}}*},}\ }\href {\doibase 10.1071/CH17588}
  {\bibfield  {journal} {\bibinfo  {journal} {Aust. J. Chem.}\ }\textbf
  {\bibinfo {volume} {71}},\ \bibinfo {pages} {238--248} (\bibinfo {year}
  {2018})}\BibitemShut {NoStop}%
\bibitem [{\citenamefont {Ma}\ and\ \citenamefont
  {Werner}(2019)}]{maAccurateIntermolecularInteraction2019}%
  \BibitemOpen
  \bibfield  {author} {\bibinfo {author} {\bibfnamefont {Q.}~\bibnamefont
  {Ma}}\ and\ \bibinfo {author} {\bibfnamefont {H.-J.}\ \bibnamefont
  {Werner}},\ }\bibfield  {title} {\enquote {\bibinfo {title} {Accurate
  {{Intermolecular Interaction Energies Using Explicitly Correlated Local
  Coupled Cluster Methods}} [{{PNO-LCCSD}}({{T}})-{{F12}}]},}\ }\href {\doibase
  10.1021/acs.jctc.8b01098} {\bibfield  {journal} {\bibinfo  {journal} {J.
  Chem. Theory Comput.}\ }\textbf {\bibinfo {volume} {15}},\ \bibinfo {pages}
  {1044--1052} (\bibinfo {year} {2019})}\BibitemShut {NoStop}%
\bibitem [{\citenamefont {Santra}\ \emph {et~al.}(2022)\citenamefont {Santra},
  \citenamefont {Semidalas}, \citenamefont {Mehta}, \citenamefont {Karton},\
  and\ \citenamefont {L.~Martin}}]{santraS66x8NoncovalentInteractions2022}%
  \BibitemOpen
  \bibfield  {author} {\bibinfo {author} {\bibfnamefont {G.}~\bibnamefont
  {Santra}}, \bibinfo {author} {\bibfnamefont {E.}~\bibnamefont {Semidalas}},
  \bibinfo {author} {\bibfnamefont {N.}~\bibnamefont {Mehta}}, \bibinfo
  {author} {\bibfnamefont {A.}~\bibnamefont {Karton}}, \ and\ \bibinfo {author}
  {\bibfnamefont {J.~M.}\ \bibnamefont {L.~Martin}},\ }\bibfield  {title}
  {\enquote {\bibinfo {title} {S66x8 noncovalent interactions revisited: New
  benchmark and performance of composite localized coupled-cluster methods},}\
  }\href {\doibase 10.1039/D2CP03938A} {\bibfield  {journal} {\bibinfo
  {journal} {Phys. Chem. Chem. Phys.}\ }\textbf {\bibinfo {volume} {24}},\
  \bibinfo {pages} {25555--25570} (\bibinfo {year} {2022})}\BibitemShut
  {NoStop}%
\bibitem [{\citenamefont {Nagy}\ \emph {et~al.}(2023)\citenamefont {Nagy},
  \citenamefont {{Gyevi-Nagy}}, \citenamefont {L{\H o}rincz},\ and\
  \citenamefont {K{\'a}llay}}]{nagyPursuingBasisSet2023}%
  \BibitemOpen
  \bibfield  {author} {\bibinfo {author} {\bibfnamefont {P.~R.}\ \bibnamefont
  {Nagy}}, \bibinfo {author} {\bibfnamefont {L.}~\bibnamefont {{Gyevi-Nagy}}},
  \bibinfo {author} {\bibfnamefont {B.~D.}\ \bibnamefont {L{\H o}rincz}}, \
  and\ \bibinfo {author} {\bibfnamefont {M.}~\bibnamefont {K{\'a}llay}},\
  }\bibfield  {title} {\enquote {\bibinfo {title} {Pursuing the basis set limit
  of {{CCSD}}({{T}}) non-covalent interaction energies for medium-sized
  complexes: Case study on the {{S66}} compilation},}\ }\href {\doibase
  10.1080/00268976.2022.2109526} {\bibfield  {journal} {\bibinfo  {journal}
  {Mol. Phys.}\ }\textbf {\bibinfo {volume} {121}},\ \bibinfo {pages}
  {e2109526} (\bibinfo {year} {2023})}\BibitemShut {NoStop}%
\bibitem [{Note1()}]{Note1}%
  \BibitemOpen
  \bibinfo {note} {Note that the difficulty of reaching the complete basis set
  (CBS) limit is the principle reason for several CCSD(T) evaluations of S66.
  As reaching the CBS limit remains a challenge, particularly in dispersion
  bound complexes.}\BibitemShut {Stop}%
\bibitem [{\citenamefont {Burns}\ \emph {et~al.}(2017)\citenamefont {Burns},
  \citenamefont {Faver}, \citenamefont {Zheng}, \citenamefont {Marshall},
  \citenamefont {Smith}, \citenamefont {Vanommeslaeghe}, \citenamefont
  {MacKerell}, \citenamefont {Merz},\ and\ \citenamefont
  {Sherrill}}]{burnsBioFragmentDatabaseBFDb2017}%
  \BibitemOpen
  \bibfield  {author} {\bibinfo {author} {\bibfnamefont {L.~A.}\ \bibnamefont
  {Burns}}, \bibinfo {author} {\bibfnamefont {J.~C.}\ \bibnamefont {Faver}},
  \bibinfo {author} {\bibfnamefont {Z.}~\bibnamefont {Zheng}}, \bibinfo
  {author} {\bibfnamefont {M.~S.}\ \bibnamefont {Marshall}}, \bibinfo {author}
  {\bibfnamefont {D.~G.~A.}\ \bibnamefont {Smith}}, \bibinfo {author}
  {\bibfnamefont {K.}~\bibnamefont {Vanommeslaeghe}}, \bibinfo {author}
  {\bibfnamefont {A.~D.}\ \bibnamefont {MacKerell}, \bibfnamefont {Jr.}},
  \bibinfo {author} {\bibfnamefont {K.~M.}\ \bibnamefont {Merz}, \bibfnamefont
  {Jr.}}, \ and\ \bibinfo {author} {\bibfnamefont {C.~D.}\ \bibnamefont
  {Sherrill}},\ }\bibfield  {title} {\enquote {\bibinfo {title} {The
  {{BioFragment Database}} ({{BFDb}}): {{An}} open-data platform for
  computational chemistry analysis of noncovalent interactions},}\ }\href
  {\doibase 10.1063/1.5001028} {\bibfield  {journal} {\bibinfo  {journal} {J.
  Chem. Phys.}\ }\textbf {\bibinfo {volume} {147}},\ \bibinfo {pages} {161727}
  (\bibinfo {year} {2017})}\BibitemShut {NoStop}%
\bibitem [{\citenamefont {Parker}\ \emph {et~al.}(2014)\citenamefont {Parker},
  \citenamefont {Burns}, \citenamefont {Parrish}, \citenamefont {Ryno},\ and\
  \citenamefont {Sherrill}}]{parkerLevelsSymmetryAdapted2014}%
  \BibitemOpen
  \bibfield  {author} {\bibinfo {author} {\bibfnamefont {T.~M.}\ \bibnamefont
  {Parker}}, \bibinfo {author} {\bibfnamefont {L.~A.}\ \bibnamefont {Burns}},
  \bibinfo {author} {\bibfnamefont {R.~M.}\ \bibnamefont {Parrish}}, \bibinfo
  {author} {\bibfnamefont {A.~G.}\ \bibnamefont {Ryno}}, \ and\ \bibinfo
  {author} {\bibfnamefont {C.~D.}\ \bibnamefont {Sherrill}},\ }\bibfield
  {title} {\enquote {\bibinfo {title} {Levels of symmetry adapted perturbation
  theory ({{SAPT}}). {{I}}. {{Efficiency}} and performance for interaction
  energies},}\ }\href {\doibase 10.1063/1.4867135} {\bibfield  {journal}
  {\bibinfo  {journal} {J. Chem. Phys.}\ }\textbf {\bibinfo {volume} {140}},\
  \bibinfo {pages} {094106} (\bibinfo {year} {2014})}\BibitemShut {NoStop}%
\bibitem [{\citenamefont {Nakano}\ \emph {et~al.}(2024)\citenamefont {Nakano},
  \citenamefont {Shi}, \citenamefont {Alf{\`e}},\ and\ \citenamefont
  {Zen}}]{nakanoBasisSetIncompleteness2024a}%
  \BibitemOpen
  \bibfield  {author} {\bibinfo {author} {\bibfnamefont {K.}~\bibnamefont
  {Nakano}}, \bibinfo {author} {\bibfnamefont {B.~X.}\ \bibnamefont {Shi}},
  \bibinfo {author} {\bibfnamefont {D.}~\bibnamefont {Alf{\`e}}}, \ and\
  \bibinfo {author} {\bibfnamefont {A.}~\bibnamefont {Zen}},\ }\href {\doibase
  10.48550/arXiv.2412.00368} {\enquote {\bibinfo {title} {Basis set
  incompleteness errors in fixed-node diffusion {{Monte Carlo}} calculations on
  non-covalent interactions},}\ } (\bibinfo {year} {2024}),\ \Eprint
  {http://arxiv.org/abs/2412.00368} {arXiv:2412.00368 [physics]} \BibitemShut
  {NoStop}%
\bibitem [{\citenamefont {Karton}\ and\ \citenamefont
  {Martin}(2021)}]{kartonPrototypicalPpDimers2021}%
  \BibitemOpen
  \bibfield  {author} {\bibinfo {author} {\bibfnamefont {A.}~\bibnamefont
  {Karton}}\ and\ \bibinfo {author} {\bibfnamefont {J.~M.~L.}\ \bibnamefont
  {Martin}},\ }\bibfield  {title} {\enquote {\bibinfo {title} {Prototypical
  {$\pi$}--{$\pi$} dimers re-examined by means of high-level {{CCSDT}}({{Q}})
  composite ab initio methods},}\ }\href {\doibase 10.1063/5.0043046}
  {\bibfield  {journal} {\bibinfo  {journal} {J. Chem. Phys.}\ }\textbf
  {\bibinfo {volume} {154}},\ \bibinfo {pages} {124117} (\bibinfo {year}
  {2021})}\BibitemShut {NoStop}%
\end{thebibliography}%


\providecommand{\latin}[1]{#1}
\makeatletter
\providecommand{\doi}
  {\begingroup\let\do\@makeother\dospecials
  \catcode`\{=1 \catcode`\}=2 \doi@aux}
\providecommand{\doi@aux}[1]{\endgroup\texttt{#1}}
\makeatother
\providecommand*\mcitethebibliography{\thebibliography}
\csname @ifundefined\endcsname{endmcitethebibliography}
  {\let\endmcitethebibliography\endthebibliography}{}
\begin{mcitethebibliography}{25}
\providecommand*\natexlab[1]{#1}
\providecommand*\mciteSetBstSublistMode[1]{}
\providecommand*\mciteSetBstMaxWidthForm[2]{}
\providecommand*\mciteBstWouldAddEndPuncttrue
  {\def\EndOfBibitem{\unskip.}}
\providecommand*\mciteBstWouldAddEndPunctfalse
  {\let\EndOfBibitem\relax}
\providecommand*\mciteSetBstMidEndSepPunct[3]{}
\providecommand*\mciteSetBstSublistLabelBeginEnd[3]{}
\providecommand*\EndOfBibitem{}
\mciteSetBstSublistMode{f}
\mciteSetBstMaxWidthForm{subitem}{(\alph{mcitesubitemcount})}
\mciteSetBstSublistLabelBeginEnd
  {\mcitemaxwidthsubitemform\space}
  {\relax}
  {\relax}

\bibitem[Řez{\'{a}}{\v{c}} \latin{et~al.}(2011)Řez{\'{a}}{\v{c}}, Riley, and
  Hobza]{S66}
Řez{\'{a}}{\v{c}},~J.; Riley,~K.~E.; Hobza,~P. {S66: A well-balanced database
  of benchmark interaction energies relevant to biomolecular structures}.
  \emph{Journal of Chemical Theory and Computation} \textbf{2011}, \emph{7},
  2427--2438\relax
\mciteBstWouldAddEndPuncttrue
\mciteSetBstMidEndSepPunct{\mcitedefaultmidpunct}
{\mcitedefaultendpunct}{\mcitedefaultseppunct}\relax
\EndOfBibitem
\bibitem[Neese(2018)]{Orca_v4}
Neese,~F. Software update: the ORCA program system, version 4.0. \emph{WIREs
  Computational Molecular Science} \textbf{2018}, \emph{8}, e1327\relax
\mciteBstWouldAddEndPuncttrue
\mciteSetBstMidEndSepPunct{\mcitedefaultmidpunct}
{\mcitedefaultendpunct}{\mcitedefaultseppunct}\relax
\EndOfBibitem
\bibitem[Riplinger and Neese(2013)Riplinger, and Neese]{DLPNO-CCSD}
Riplinger,~C.; Neese,~F. An efficient and near linear scaling pair natural
  orbital based local coupled cluster method. \emph{The Journal of Chemical
  Physics} \textbf{2013}, \emph{138}, 034106\relax
\mciteBstWouldAddEndPuncttrue
\mciteSetBstMidEndSepPunct{\mcitedefaultmidpunct}
{\mcitedefaultendpunct}{\mcitedefaultseppunct}\relax
\EndOfBibitem
\bibitem[Riplinger \latin{et~al.}(2013)Riplinger, Sandhoefer, Hansen, and
  Neese]{DLPNO-CCSD(T)}
Riplinger,~C.; Sandhoefer,~B.; Hansen,~A.; Neese,~F. Natural triple excitations
  in local coupled cluster calculations with pair natural orbitals. \emph{The
  Journal of Chemical Physics} \textbf{2013}, \emph{139}, 134101\relax
\mciteBstWouldAddEndPuncttrue
\mciteSetBstMidEndSepPunct{\mcitedefaultmidpunct}
{\mcitedefaultendpunct}{\mcitedefaultseppunct}\relax
\EndOfBibitem
\bibitem[Zhong \latin{et~al.}(2008)Zhong, Barnes, and Petersson]{CBS-SCF}
Zhong,~S.; Barnes,~E.~C.; Petersson,~G.~A. Uniformly convergent n-tuple-zeta
  augmented polarized (nZaP) basis sets for complete basis set extrapolations.
  I. Self-consistent field energies. \emph{The Journal of Chemical Physics}
  \textbf{2008}, \emph{129}, 184116\relax
\mciteBstWouldAddEndPuncttrue
\mciteSetBstMidEndSepPunct{\mcitedefaultmidpunct}
{\mcitedefaultendpunct}{\mcitedefaultseppunct}\relax
\EndOfBibitem
\bibitem[Helgaker \latin{et~al.}(1997)Helgaker, Klopper, Koch, and
  Noga]{CBS-corr}
Helgaker,~T.; Klopper,~W.; Koch,~H.; Noga,~J. Basis-set convergence of
  correlated calculations on water. \emph{The Journal of Chemical Physics}
  \textbf{1997}, \emph{106}, 9639--9646\relax
\mciteBstWouldAddEndPuncttrue
\mciteSetBstMidEndSepPunct{\mcitedefaultmidpunct}
{\mcitedefaultendpunct}{\mcitedefaultseppunct}\relax
\EndOfBibitem
\bibitem[Neese and Valeev(2011)Neese, and Valeev]{CBS-exponents}
Neese,~F.; Valeev,~E.~F. Revisiting the Atomic Natural Orbital Approach for
  Basis Sets: Robust Systematic Basis Sets for Explicitly Correlated and
  Conventional Correlated ab initio Methods? \emph{Journal of Chemical Theory
  and Computation} \textbf{2011}, \emph{7}, 33--43\relax
\mciteBstWouldAddEndPuncttrue
\mciteSetBstMidEndSepPunct{\mcitedefaultmidpunct}
{\mcitedefaultendpunct}{\mcitedefaultseppunct}\relax
\EndOfBibitem
\bibitem[Foulkes \latin{et~al.}(2001)Foulkes, Mitas, Needs, and
  Rajagopal]{foulkes_qmc}
Foulkes,~W. M.~C.; Mitas,~L.; Needs,~R.~J.; Rajagopal,~G. Quantum Monte Carlo
  simulations of solids. \emph{Reviews of Modern Physics} \textbf{2001},
  \emph{73}, 33--83\relax
\mciteBstWouldAddEndPuncttrue
\mciteSetBstMidEndSepPunct{\mcitedefaultmidpunct}
{\mcitedefaultendpunct}{\mcitedefaultseppunct}\relax
\EndOfBibitem
\bibitem[Mitáš \latin{et~al.}(1991)Mitáš, Shirley, and Ceperley]{la_mitas}
Mitáš,~L.; Shirley,~E.~L.; Ceperley,~D.~M. {Nonlocal pseudopotentials and
  diffusion Monte Carlo}. \emph{The Journal of Chemical Physics} \textbf{1991},
  \emph{95}, 3467--3475\relax
\mciteBstWouldAddEndPuncttrue
\mciteSetBstMidEndSepPunct{\mcitedefaultmidpunct}
{\mcitedefaultendpunct}{\mcitedefaultseppunct}\relax
\EndOfBibitem
\bibitem[Hammond \latin{et~al.}(1987)Hammond, Reynolds, and
  Lester]{hammond1987}
Hammond,~B.; Reynolds,~P.; Lester,~W. Valence quantum Monte Carlo with ab
  initio effective core potentials. \emph{The Journal of chemical physics}
  \textbf{1987}, \emph{87}, 1130--1136\relax
\mciteBstWouldAddEndPuncttrue
\mciteSetBstMidEndSepPunct{\mcitedefaultmidpunct}
{\mcitedefaultendpunct}{\mcitedefaultseppunct}\relax
\EndOfBibitem
\bibitem[Casula(2006)]{tm_casula_1}
Casula,~M. Beyond the locality approximation in the standard diffusion Monte
  Carlo method. \emph{Physical Review B} \textbf{2006}, \emph{74}, 161102\relax
\mciteBstWouldAddEndPuncttrue
\mciteSetBstMidEndSepPunct{\mcitedefaultmidpunct}
{\mcitedefaultendpunct}{\mcitedefaultseppunct}\relax
\EndOfBibitem
\bibitem[Casula \latin{et~al.}(2010)Casula, Moroni, Sorella, and
  Filippi]{tm_casula_2}
Casula,~M.; Moroni,~S.; Sorella,~S.; Filippi,~C. {Size-consistent variational
  approaches to nonlocal pseudopotentials: Standard and lattice regularized
  diffusion Monte Carlo methods revisited}. \emph{The Journal of Chemical
  Physics} \textbf{2010}, \emph{132}, 154113\relax
\mciteBstWouldAddEndPuncttrue
\mciteSetBstMidEndSepPunct{\mcitedefaultmidpunct}
{\mcitedefaultendpunct}{\mcitedefaultseppunct}\relax
\EndOfBibitem
\bibitem[Zen \latin{et~al.}(2019)Zen, Brandenburg, Michaelides, and
  Alfè]{dla_zen}
Zen,~A.; Brandenburg,~J.~G.; Michaelides,~A.; Alfè,~D. {A new scheme for fixed
  node diffusion quantum Monte Carlo with pseudopotentials: Improving
  reproducibility and reducing the trial-wave-function bias}. \emph{The Journal
  of Chemical Physics} \textbf{2019}, \emph{151}, 134105\relax
\mciteBstWouldAddEndPuncttrue
\mciteSetBstMidEndSepPunct{\mcitedefaultmidpunct}
{\mcitedefaultendpunct}{\mcitedefaultseppunct}\relax
\EndOfBibitem
\bibitem[Zen \latin{et~al.}(2016)Zen, Sorella, Gillan, Michaelides, and
  Alf\`e]{ZSGMA}
Zen,~A.; Sorella,~S.; Gillan,~M.~J.; Michaelides,~A.; Alf\`e,~D. Boosting the
  accuracy and speed of quantum Monte Carlo: Size consistency and time step.
  \emph{Physical Review B} \textbf{2016}, \emph{93}, 241118\relax
\mciteBstWouldAddEndPuncttrue
\mciteSetBstMidEndSepPunct{\mcitedefaultmidpunct}
{\mcitedefaultendpunct}{\mcitedefaultseppunct}\relax
\EndOfBibitem
\bibitem[Zen \latin{et~al.}(2018)Zen, Brandenburg, Klimeš, Tkatchenko, Alfè,
  and Michaelides]{ZenPNAS2018}
Zen,~A.; Brandenburg,~J.~G.; Klimeš,~J.; Tkatchenko,~A.; Alfè,~D.;
  Michaelides,~A. Fast and accurate quantum Monte Carlo for molecular crystals.
  \emph{Proceedings of the National Academy of Sciences} \textbf{2018},
  \emph{115}, 1724--1729\relax
\mciteBstWouldAddEndPuncttrue
\mciteSetBstMidEndSepPunct{\mcitedefaultmidpunct}
{\mcitedefaultendpunct}{\mcitedefaultseppunct}\relax
\EndOfBibitem
\bibitem[Virtanen \latin{et~al.}(2020)Virtanen, Gommers, Oliphant, Haberland,
  Reddy, Cournapeau, Burovski, Peterson, Weckesser, Bright, {van der Walt},
  Brett, Wilson, Millman, Mayorov, Nelson, Jones, Kern, Larson, Carey, Polat,
  Feng, Moore, VanderPlas, Laxalde, Perktold, Cimrman, Henriksen, Quintero,
  Harris, Archibald, Ribeiro, Pedregosa, and {van
  Mulbregt}]{si-virtanenSciPy10Fundamental2020}
Virtanen,~P.; Gommers,~R.; Oliphant,~T.~E.; Haberland,~M.; Reddy,~T.;
  Cournapeau,~D.; Burovski,~E.; Peterson,~P.; Weckesser,~W.; Bright,~J.; {van
  der Walt},~S.~J.; Brett,~M.; Wilson,~J.; Millman,~K.~J.; Mayorov,~N.;
  Nelson,~A. R.~J.; Jones,~E.; Kern,~R.; Larson,~E.; Carey,~C.~J.;
  Polat,~{\.I}.; Feng,~Y.; Moore,~E.~W.; VanderPlas,~J.; Laxalde,~D.;
  Perktold,~J.; Cimrman,~R.; Henriksen,~I.; Quintero,~E.~A.; Harris,~C.~R.;
  Archibald,~A.~M.; Ribeiro,~A.~H.; Pedregosa,~F.; {van Mulbregt},~P. {{SciPy}}
  1.0: Fundamental Algorithms for Scientific Computing in {{Python}}.
  \emph{Nat. Methods} \textbf{2020}, \emph{17}, 261--272\relax
\mciteBstWouldAddEndPuncttrue
\mciteSetBstMidEndSepPunct{\mcitedefaultmidpunct}
{\mcitedefaultendpunct}{\mcitedefaultseppunct}\relax
\EndOfBibitem
\bibitem[{\v R}ez{\'a}{\v c} \latin{et~al.}(2011){\v R}ez{\'a}{\v c}, Riley,
  and Hobza]{si_rezacExtensionsS66Data2011}
{\v R}ez{\'a}{\v c},~J.; Riley,~K.~E.; Hobza,~P. Extensions of the {{S66 Data
  Set}}: {{More Accurate Interaction Energies}} and {{Angular-Displaced
  Nonequilibrium Geometries}}. \emph{J. Chem. Theory Comput.} \textbf{2011},
  \emph{7}, 3466--3470\relax
\mciteBstWouldAddEndPuncttrue
\mciteSetBstMidEndSepPunct{\mcitedefaultmidpunct}
{\mcitedefaultendpunct}{\mcitedefaultseppunct}\relax
\EndOfBibitem
\bibitem[Kesharwani \latin{et~al.}(2018)Kesharwani, Karton, Sylvetsky, and
  Martin]{S66_revJM_2018}
Kesharwani,~M.~K.; Karton,~A.; Sylvetsky,~N.; Martin,~J. M.~L. {The S66
  Non-Covalent Interactions Benchmark Reconsidered Using Explicitly Correlated
  Methods Near the Basis Set Limit}. \emph{Australian Journal of Chemistry}
  \textbf{2018}, \emph{71}, 238\relax
\mciteBstWouldAddEndPuncttrue
\mciteSetBstMidEndSepPunct{\mcitedefaultmidpunct}
{\mcitedefaultendpunct}{\mcitedefaultseppunct}\relax
\EndOfBibitem
\bibitem[Péter R.~Nagy and Kállay(2023)Péter R.~Nagy, and
  Kállay]{nagy_gold}
Péter R.~Nagy,~B. D.~L.,~László Gyevi-Nagy; Kállay,~M. Pursuing the basis
  set limit of CCSD(T) non-covalent interaction energies for medium-sized
  complexes: case study on the S66 compilation. \emph{Molecular Physics}
  \textbf{2023}, \emph{121}, e2109526\relax
\mciteBstWouldAddEndPuncttrue
\mciteSetBstMidEndSepPunct{\mcitedefaultmidpunct}
{\mcitedefaultendpunct}{\mcitedefaultseppunct}\relax
\EndOfBibitem
\bibitem[Řez{\'{a}}{\v{c}} \latin{et~al.}(2011)Řez{\'{a}}{\v{c}}, Riley, and
  Hobza]{S66x8}
Řez{\'{a}}{\v{c}},~J.; Riley,~K.~E.; Hobza,~P. {Extensions of the S66 Data
  Set: More Accurate Interaction Energies and Angular-Displaced Nonequilibrium
  Geometries}. \emph{Journal of Chemical Theory and Computation} \textbf{2011},
  \emph{7}, 3466--3470\relax
\mciteBstWouldAddEndPuncttrue
\mciteSetBstMidEndSepPunct{\mcitedefaultmidpunct}
{\mcitedefaultendpunct}{\mcitedefaultseppunct}\relax
\EndOfBibitem
\bibitem[Schäfer \latin{et~al.}(2024)Schäfer, Irmler, Gallo, and
  Grüneis]{CCSDcT_Gruneis_2}
Schäfer,~T.; Irmler,~A.; Gallo,~A.; Grüneis,~A. Understanding Discrepancies
  of Wavefunction Theories for Large Molecules. 2024;
  \url{https://arxiv.org/abs/2407.01442}\relax
\mciteBstWouldAddEndPuncttrue
\mciteSetBstMidEndSepPunct{\mcitedefaultmidpunct}
{\mcitedefaultendpunct}{\mcitedefaultseppunct}\relax
\EndOfBibitem
\bibitem[Trail and Needs(2017)Trail, and Needs]{CASINO_PSEUDO_eCEPP}
Trail,~J.~R.; Needs,~R.~J. Shape and energy consistent pseudopotentials for
  correlated electron systems. \emph{The Journal of Chemical Physics}
  \textbf{2017}, \emph{146}, 204107\relax
\mciteBstWouldAddEndPuncttrue
\mciteSetBstMidEndSepPunct{\mcitedefaultmidpunct}
{\mcitedefaultendpunct}{\mcitedefaultseppunct}\relax
\EndOfBibitem
\bibitem[Burns \latin{et~al.}(2017)Burns, Faver, Zheng, Marshall, Smith,
  Vanommeslaeghe, MacKerell, Merz, and Sherrill]{SAPT_Sherrill}
Burns,~L.~A.; Faver,~J.~C.; Zheng,~Z.; Marshall,~M.~S.; Smith,~D. G.~A.;
  Vanommeslaeghe,~K.; MacKerell,~J.,~Alexander~D.; Merz,~J.,~Kenneth~M.;
  Sherrill,~C.~D. {The BioFragment Database (BFDb): An open-data platform for
  computational chemistry analysis of noncovalent interactions}. \emph{The
  Journal of Chemical Physics} \textbf{2017}, \emph{147}, 161727\relax
\mciteBstWouldAddEndPuncttrue
\mciteSetBstMidEndSepPunct{\mcitedefaultmidpunct}
{\mcitedefaultendpunct}{\mcitedefaultseppunct}\relax
\EndOfBibitem
\bibitem[Villot and Lao(2024)Villot, and
  Lao]{si-villotInitioDispersionPotentials2024}
Villot,~C.; Lao,~K.~U. Ab Initio Dispersion Potentials Based on Physics-Based
  Functional Forms with Machine Learning. \emph{J. Chem. Phys.} \textbf{2024},
  \emph{160}, 184103\relax
\mciteBstWouldAddEndPuncttrue
\mciteSetBstMidEndSepPunct{\mcitedefaultmidpunct}
{\mcitedefaultendpunct}{\mcitedefaultseppunct}\relax
\EndOfBibitem
\end{mcitethebibliography}
\end{document}


\newpage
\tableofcontents
\newpage

We provide here additional supporting data as well as contextual information for the manuscript ``On the systematic discrepancies between
reference methods on noncovalent interaction
energies within the S66 dataset''. 
All output files are provided on \href{}{GitHub}, which contains a Jupyter Notebook file that analyzes the data. 
This data can also be viewed and analyzed on the browser with \href{}{Colab}.

In particular, in this supplemental material we provide:
\begin{itemize}
    \item a brief description of the three previous CCSD(T) estimates of the binding energy of the S66 dataset in Sec.~\ref{sec:CCSDT}, with the final estimates given in \ref{sec:final_cc_estimates}.
    \item the total energy of each dimer and the corresponding monomers used to compute the binding energies reported in the main manuscript in Sec.~\ref{sec:table};
    \item an analysis on the mean relative differences between DMC and CCSD(T) on the S66 dataset in Sec.~\ref{sec:relative_difference};
    \item a quantitative analysis of the localization error on the binding energy for the case of acetic acid in Sec.~\ref{sec:localization_acetic};
    \item the convergence of the DMC estimates with respect to the simulation time step for all the dimers in Sec.~\ref{sec:timestep}.
    \item an energy decomposition analysis (into electrostatic, dispersive, induction and exchange contributions) of the S66 dataset with the SAPT method in Sec.~\ref{sec:int_ene_eda}.
\end{itemize}
\clearpage

\section{The S66 dataset}
\begin{figure}
    \includegraphics[width=0.8\textwidth]{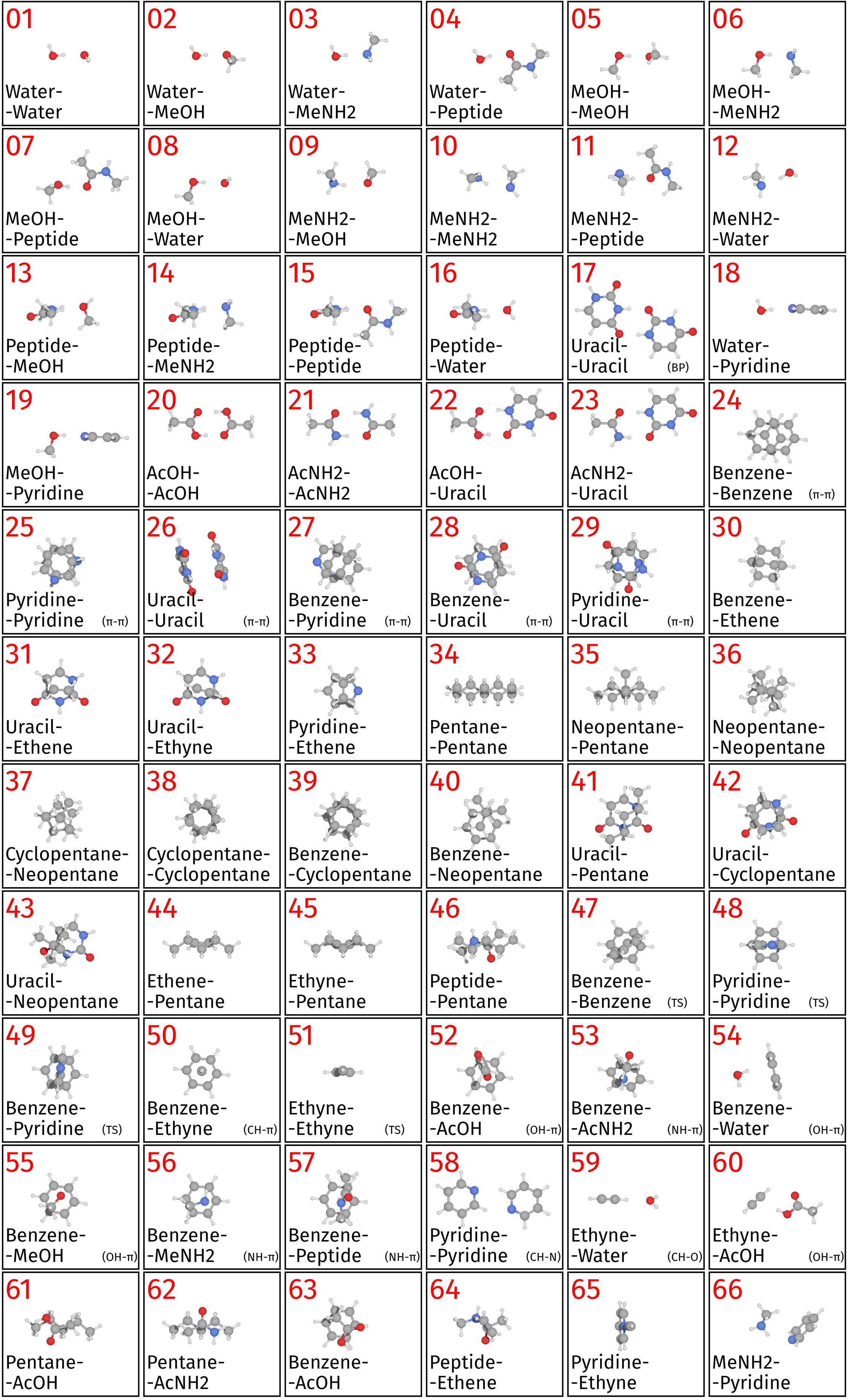}
    \caption{\label{fig:s66_viz_dataset}A visualization of the 66 dimer complexes within the S66 dataset. The IDs for each system is provided on the top right with additional description of their geometry given in the bottom right.}
\end{figure}

The entire S66 dataset is visualised in Fig.~\ref{fig:s66_viz_dataset}.
It consists of 66 dimer complexes, composed from combinations of 14 monomer molecules.
These monomers consist of only carbon, oxygen, nitrogen, and hydrogen -- the most commonly encountered elements in biochemistry.
Within the dimers, the monomers are combined and placed at different geometries, for example through parallel $\pi$-$\pi$ stacking or in a T-shape (TS) or with NH, CH or OH groups pointing perpendicular to the plane of an aromatic $\pi$ ring, among others.
The geometries of these dimers were obtained from second-order M{\o}ller-Plesset perturbation theory (MP2) performed with the Dunning cc-pVTZ basis set.

The dimers of the S66 dataset were chosen to sample a balanced range of noncovalent interactions, consisting of 23 electrostatic-dominated systems [IDs 1-23], 23 dispersion-dominated systems [IDs 24-46], and 20 systems [IDs 47-66] with mixed (electrostatic/dispersion) interactions.
It should be noted all these classifications are rather arbitrary and can differ based on the choice of energy decomposition analysis schemes.
Regardless, they have been chosen to sample some important types of interactions within each category.
For examples, the electrostatic-dominated systems covers all possible combinations of hydrogen bonding donors and acceptors of the water molecule, hydroxyl group, amine group, and carbonyl group, alongside the type of hydrogen bonding expected in nucleic acid base pairs.
The dispersion-dominated systems consists of combinations between planar aromatic molecules and aliphatic hydrocarbons, leading to three types of interactions: $\pi$–$\pi$ stacking (10 systems), aliphatic–aliphatic (5 systems), and $\pi$–aliphatic (8 systems) interactions.

\clearpage

\section{Validating diffusion Monte Carlo}

\subsection{Computing the interaction energy} \label{sec:deformation_ene}

The DMC interaction energies of the S66\cite{S66} dataset are computed as:
\begin{equation}\label{si_equation_binding_energy}
    \Delta E_\textrm{int.}^\textrm{DMC} = E_\textrm{dimer}^\textrm{DMC} - E_\textrm{mon. 1}^\textrm{DMC} - E_\textrm{mon. 2}^\textrm{DMC}, 
\end{equation}
where $E_{\mathrm{dimer}}$ is the total energy of the dimer, and $E_\textrm{mon. 1}, E_\textrm{mon. 2}$ are the total energies of the constituent monomers.
Here, the constituent monomers take on the geometry they adopt in the dimer -- away from their equilibrium position.

\begin{table}
\caption{\label{tab:monomer_tot_ene}Total energy of the 14 monomers which make up the S66 dataset. These geometries are taken from specific dimer complexes within the S66 dataset that are identified in the table and the order in which the monomer appears (important for dimers consisting of the same molecule) is reported. The method (see Sec.~\ref{sec:timestep_error}) used to estimate the error on the DMC estimate is also provided.}
\begin{adjustbox}{center}
\begin{tabular}{lrrrr}
\toprule
Monomer & Dimer Geometry & Order & Total Energy [kcal/mol] & Error Type \\ 
\midrule
AcNH$_2$ & AcNH$_2$$\cdots$AcNH$_2$ (ID 21) & 1 & -25290.30$\pm$0.03 & $\sigma_\text{cubic fit}$ \\
AcOH & AcOH$\cdots$AcOH (ID 20) & 1 & -28725.26$\pm$0.04 & $\sigma_\text{cubic fit}$ \\
Benzene & Benzene$\cdots$Benzene ($\pi$-$\pi$) (ID 24) & 1 & -23624.42$\pm$0.04 & $\sigma_\text{cubic fit}$ \\
Cyclopentane & Cyclopentane$\cdots$Neopentane (ID 37) & 1 & -21586.06$\pm$0.03 & $\sigma_\text{cubic fit}$ \\
Ethene & Benzene$\cdots$Ethene (ID 30) & 2 & -8610.43$\pm$0.02 & $\sigma_\text{cubic fit}$ \\
Ethyne & Uracil$\cdots$Ethyne (ID 32) & 2 & -7823.07$\pm$0.02 & $\sigma_\text{cubic fit}$ \\
MeNH$_2$ & Benzene$\cdots$MeNH$_2$ (NH-$\pi$) (ID 56) & 2 & -11671.48$\pm$0.02 & $\sigma_\text{cubic fit}$ \\
MeOH & Benzene$\cdots$MeOH (OH-$\pi$) (ID 55) & 2 & -15103.83$\pm$0.02 & $\sigma_\text{cubic fit}$ \\
Neopentane & Neopentane$\cdots$Neopentane (ID 36) & 1 & -22350.00$\pm$0.04 & $\sigma_\text{cubic fit}$ \\
Pentane & Pentane$\cdots$Pentane (ID 34) & 1 & -22346.69$\pm$0.03 & $\sigma_\text{cubic fit}$ \\
Peptide & Benzene$\cdots$Peptide (NH-$\pi$) (ID 57) & 2 & -29604.32$\pm$0.04 & $\sigma_\text{cubic fit}$ \\
Pyridine & Pyridine$\cdots$Pyridine ($\pi$-$\pi$) (ID 25) & 1 & -25905.64$\pm$0.04 & $\sigma_\text{cubic fit}$ \\
Uracil & Uracil$\cdots$Uracil ($\pi$-$\pi$) (ID 26) & 1 & -48309.04$\pm$0.04 & $\sigma_\text{cubic fit}$ \\
Water & Water$\cdots$Water (ID 1) & 2 & -10799.44$\pm$0.01 & $\sigma_\text{cubic fit}$ \\
\bottomrule
\end{tabular}
\end{adjustbox}
\end{table}

There are a total of only 14 different monomer species (listed in Table~\ref{tab:monomer_tot_ene} that are combined to make up the 66 dimers.
Importantly, the above definition requires the calculation of the total energy of 132 monomers, which can add significant manual expense and cost to compute with DMC.
We reach an estimate of the energy of each monomer by computing the DMC total energy at a reference geometry (chosen from a dimer in the S66 dataset) $E_\textrm{monomer 1/2, ref.}^\textrm{DMC}$, combined with a deformation energy (to reach its geometry in the dimer) computed at the CCSD(T) level (details given in section~\ref{sec:deformation_energy_setup_ccsdt}):
\begin{equation} \label{eq:deformation_energy}
     E_\textrm{mon. 1}^\textrm{DMC} = E_\textrm{mon. 1, ref.}^\textrm{DMC} + \Delta E_\textrm{mon. 1, def.}^\textrm{CCSD(T)}.
\end{equation}
Thus, this requires DMC estimates on the total energy of only 14 monomers.
We give the final estimate to the DMC total energy for each of the 14 monomers in Table~\ref{tab:monomer_tot_ene}, with the corresponding dimer geometry where this monomer was taken from identified and the type of extrapolation used to reach the zero time step limit.

\begin{figure}[!h]
    \includegraphics[width=3.365in]{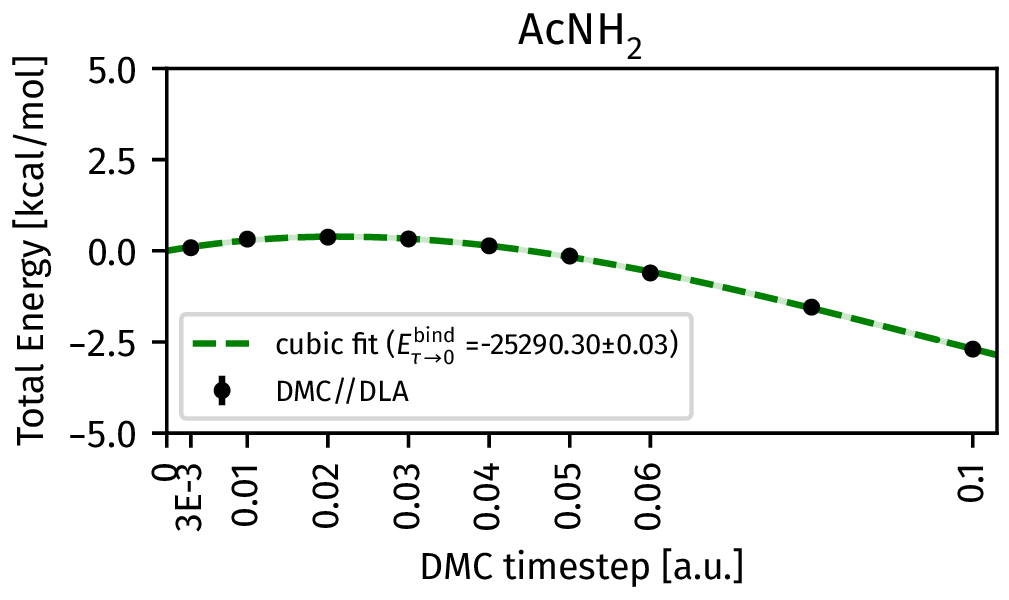}
    \caption{\label{fig:monomer_01} The time step dependence of the AcNH$_2$ monomer in the AcNH$_2$$\cdots$AcNH$_2$ dimer (ID 21) geometry.}
\end{figure}

\begin{figure}[!h]
    \includegraphics[width=3.365in]{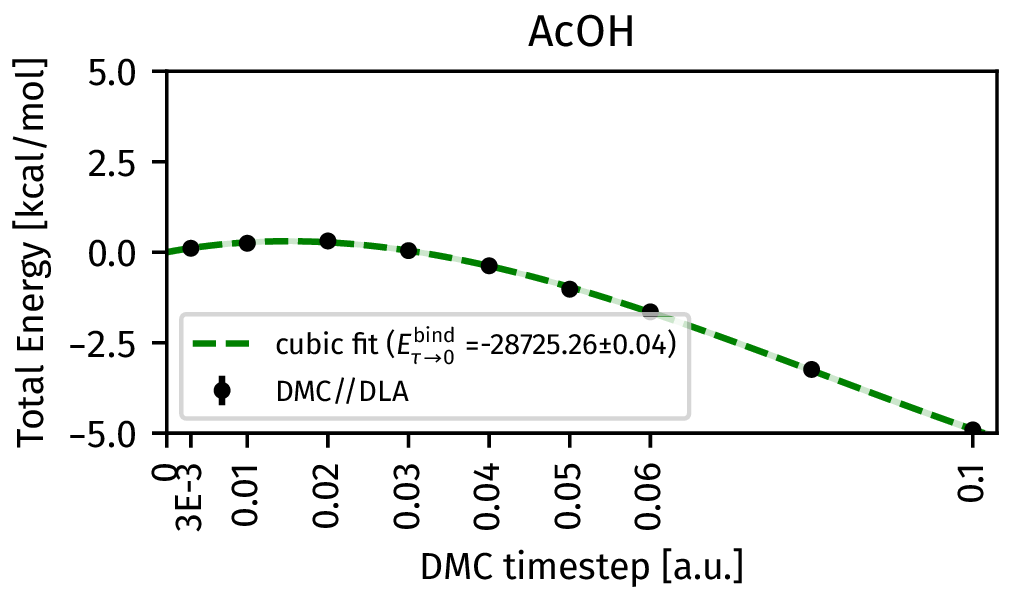}
    \caption{\label{fig:monomer_02} The time step dependence of the AcOH monomer in the AcOH$\cdots$AcOH dimer (ID 20) geometry.}
\end{figure}

\begin{figure}[!h]
    \includegraphics[width=3.365in]{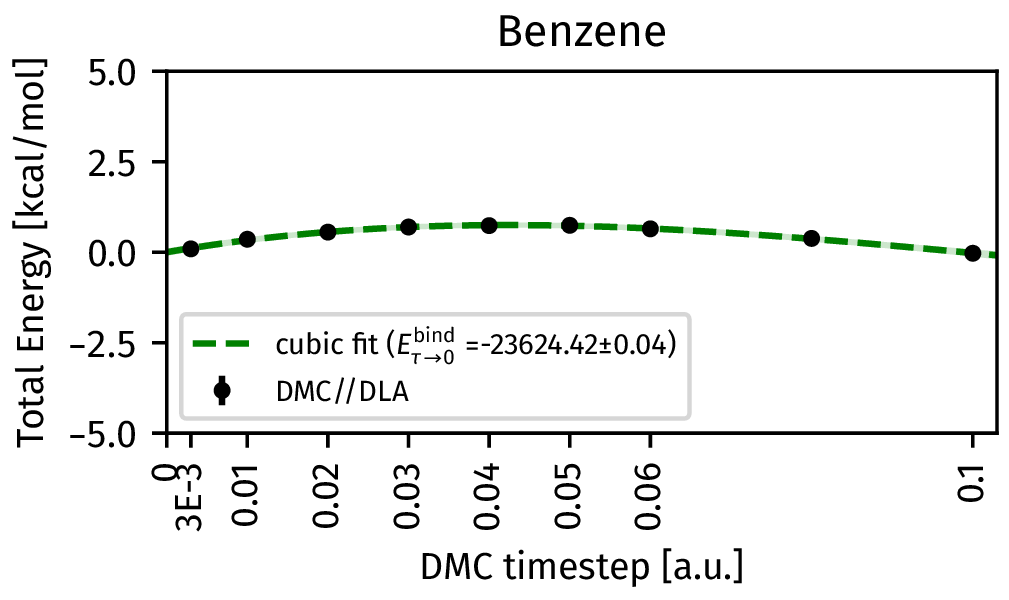}
    \caption{\label{fig:monomer_03} The time step dependence of the Benzene monomer in the Benzene$\cdots$Benzene ($\pi$-$\pi$) dimer (ID 24) geometry.}
\end{figure}

\begin{figure}[!h]
    \includegraphics[width=3.365in]{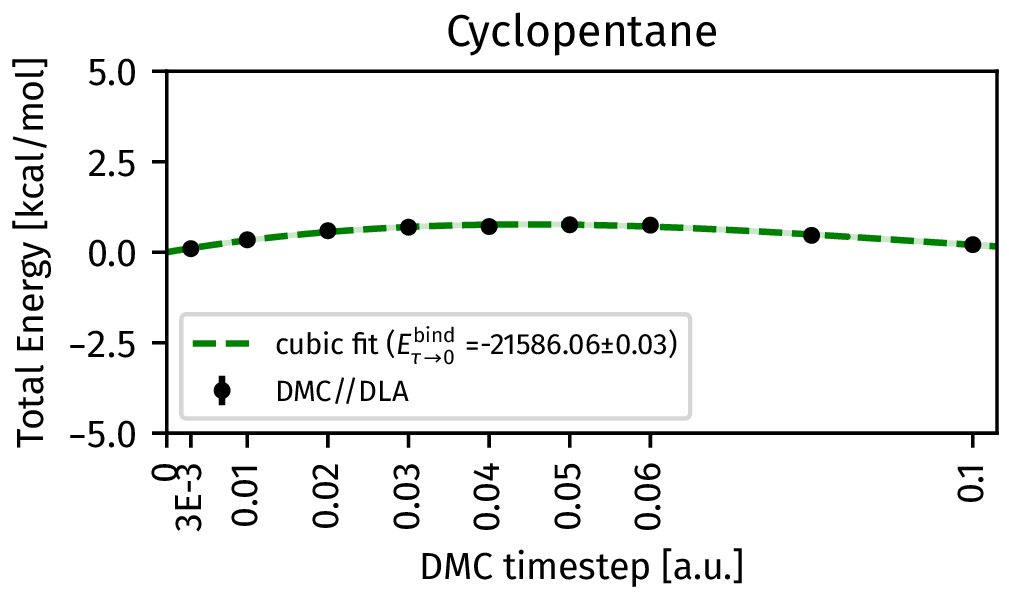}
    \caption{\label{fig:monomer_04} The time step dependence of the Cyclopentane monomer in the Cyclopentane$\cdots$Neopentane dimer (ID 37) geometry.}
\end{figure}

\begin{figure}[!h]
    \includegraphics[width=3.365in]{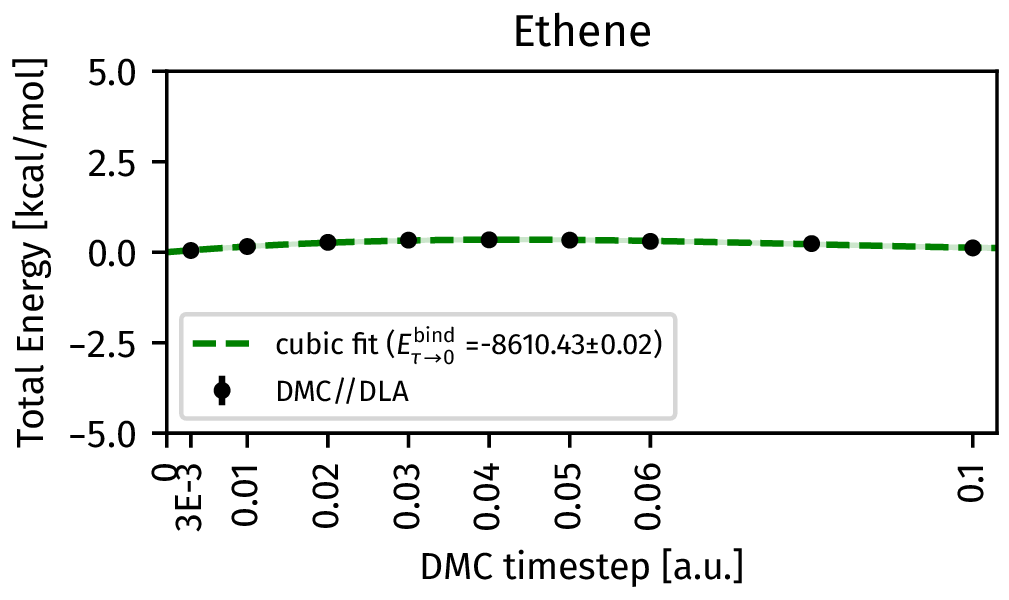}
    \caption{\label{fig:monomer_05} The time step dependence of the Ethene monomer in the Benzene$\cdots$Ethene dimer (ID 30) geometry.}
\end{figure}

\begin{figure}[!h]
    \includegraphics[width=3.365in]{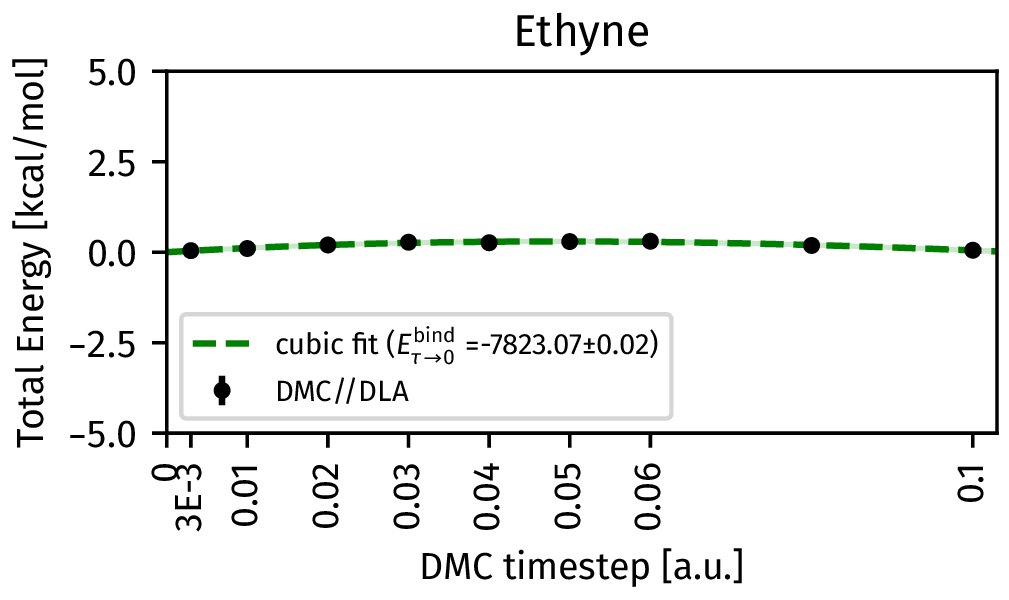}
    \caption{\label{fig:monomer_06} The time step dependence of the Ethyne monomer in the Uracil$\cdots$Ethyne dimer (ID 32) geometry.}
\end{figure}

\begin{figure}[!h]
    \includegraphics[width=3.365in]{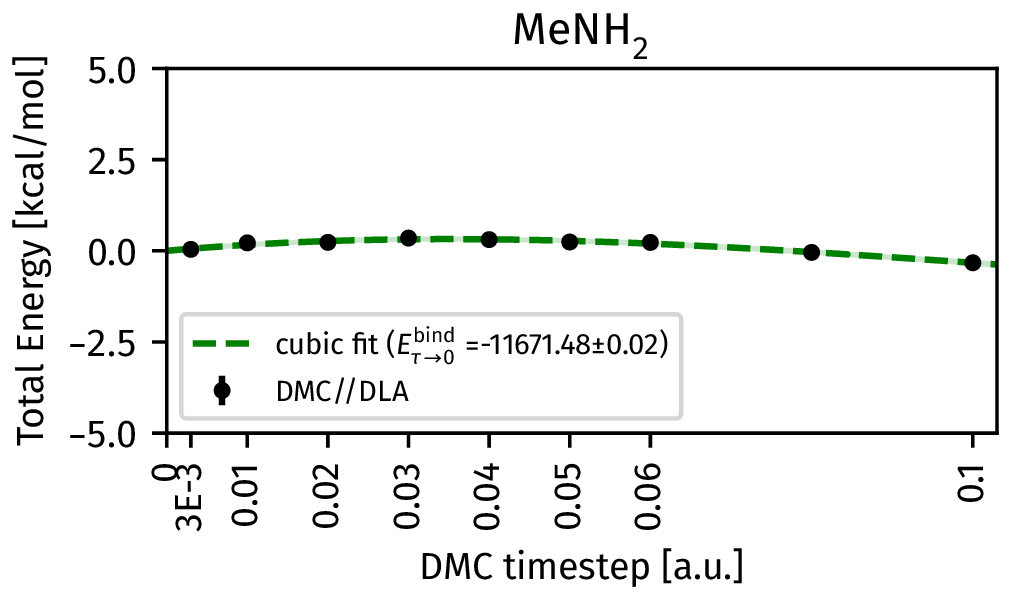}
    \caption{\label{fig:monomer_07} The time step dependence of the MeNH$_2$ monomer in the Benzene$\cdots$MeNH$_2$ (NH-$\pi$) dimer (ID 56) geometry.}
\end{figure}

\begin{figure}[!h]
    \includegraphics[width=3.365in]{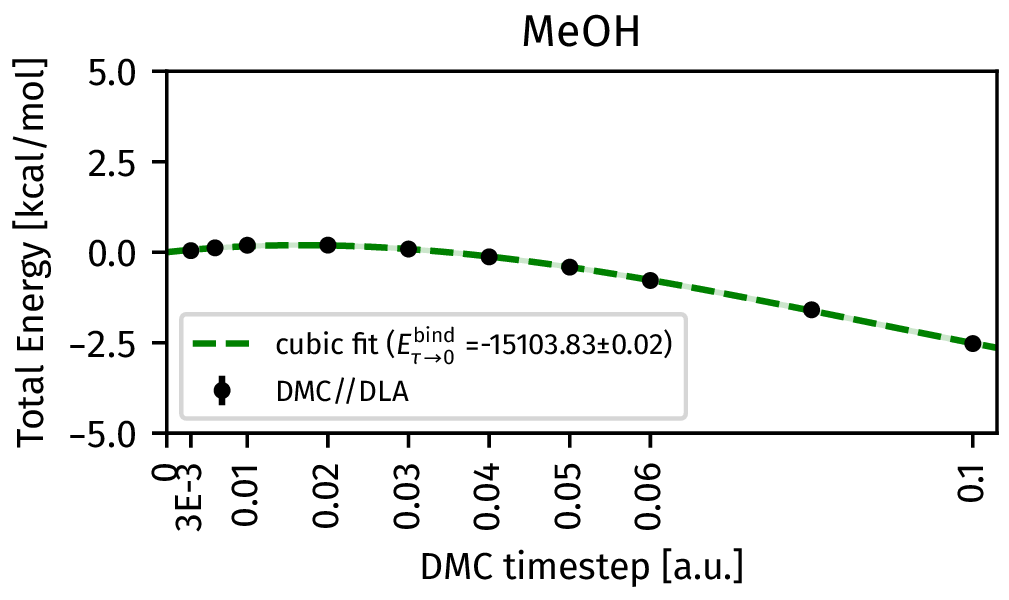}
    \caption{\label{fig:monomer_08} The time step dependence of the MeOH monomer in the Benzene$\cdots$MeOH (OH-$\pi$) dimer (ID 55) geometry.}
\end{figure}

\begin{figure}[!h]
    \includegraphics[width=3.365in]{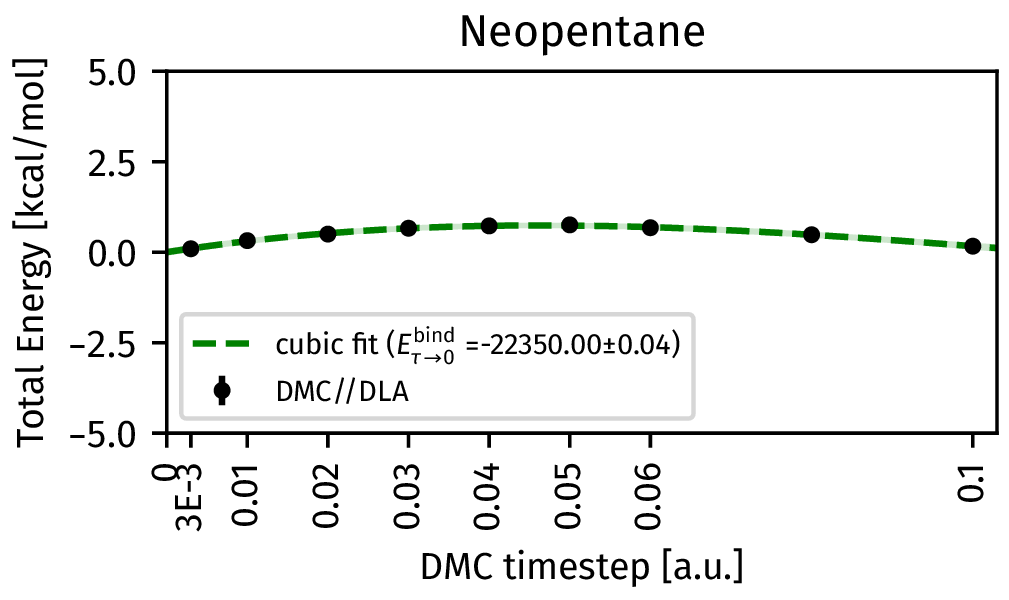}
    \caption{\label{fig:monomer_09} The time step dependence of the Neopentane monomer in the Neopentane$\cdots$Neopentane dimer (ID 36) geometry.}
\end{figure}

\begin{figure}[!h]
    \includegraphics[width=3.365in]{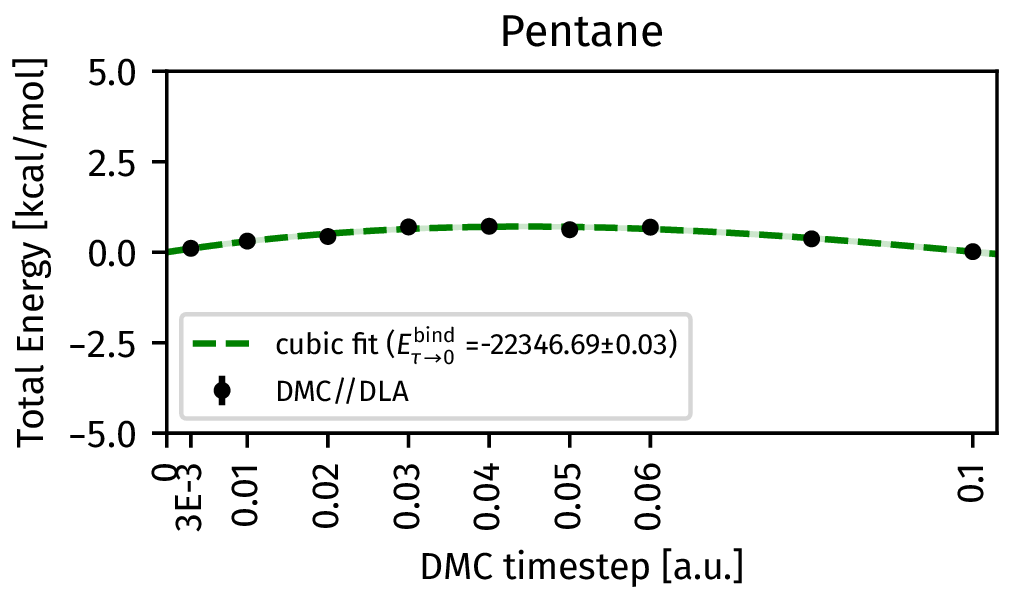}
    \caption{\label{fig:monomer_10} The time step dependence of the Pentane monomer in the Pentane$\cdots$Pentane dimer (ID 34) geometry.}
\end{figure}

\begin{figure}[!h]
    \includegraphics[width=3.365in]{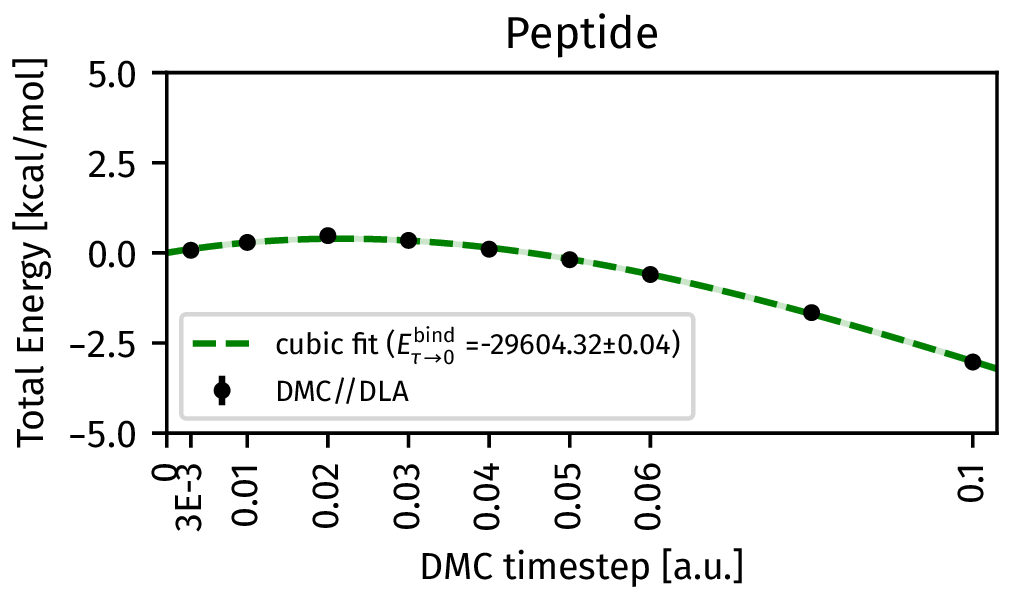}
    \caption{\label{fig:monomer_11} The time step dependence of the Peptide monomer in the Benzene$\cdots$Peptide (NH-$\pi$) dimer (ID 57) geometry.}
\end{figure}

\begin{figure}[!h]
    \includegraphics[width=3.365in]{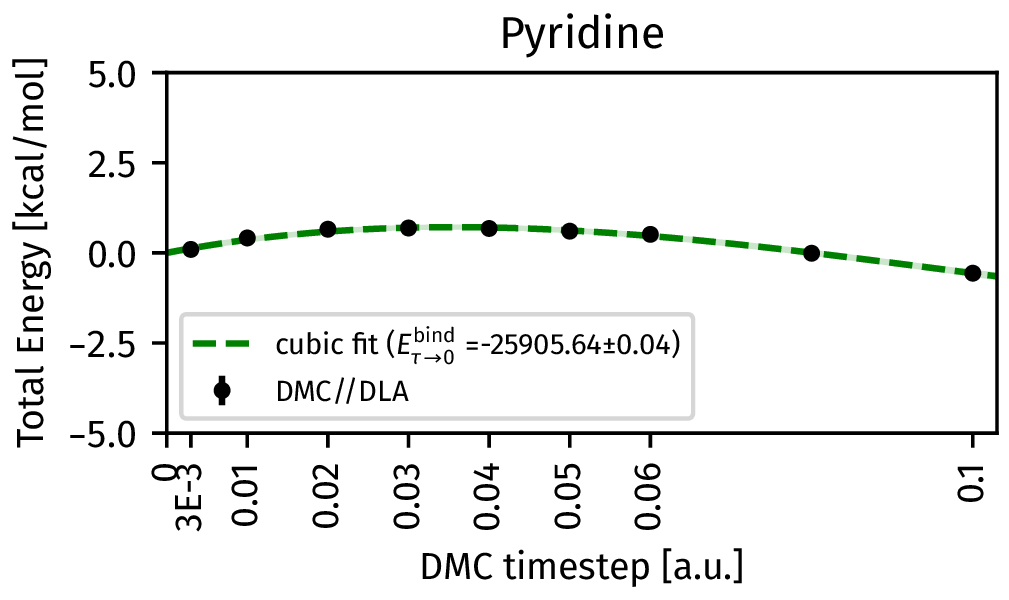}
    \caption{\label{fig:monomer_12} The time step dependence of the Pyridine monomer in the Pyridine$\cdots$Pyridine ($\pi$-$\pi$) dimer (ID 25) geometry.}
\end{figure}

\begin{figure}[!h]
    \includegraphics[width=3.365in]{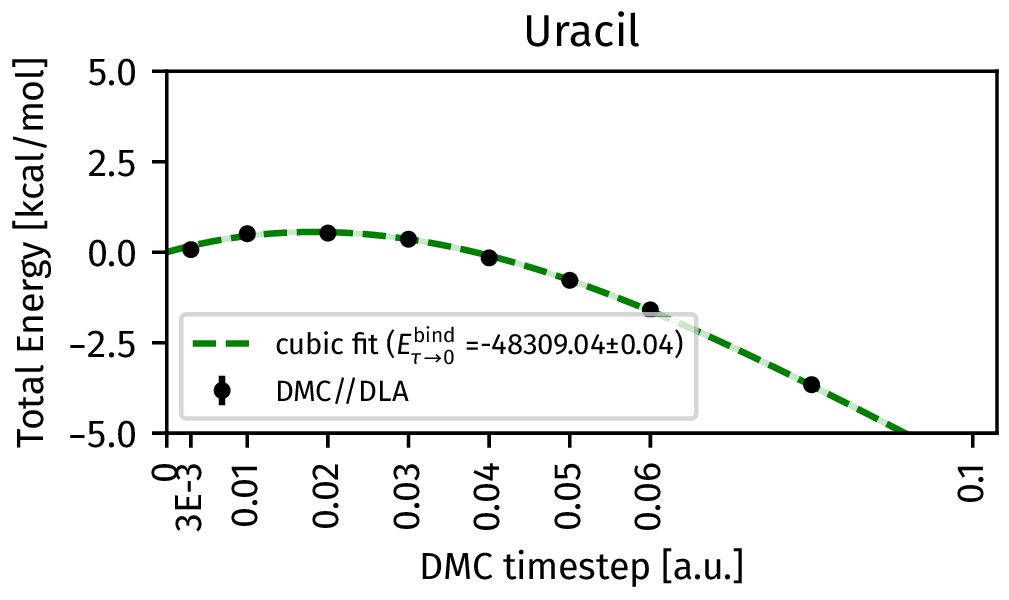}
    \caption{\label{fig:monomer_13} The time step dependence of the Uracil monomer in the Uracil$\cdots$Uracil ($\pi$-$\pi$) dimer (ID 26) geometry.}
\end{figure}

\begin{figure}[!h]
    \includegraphics[width=3.365in]{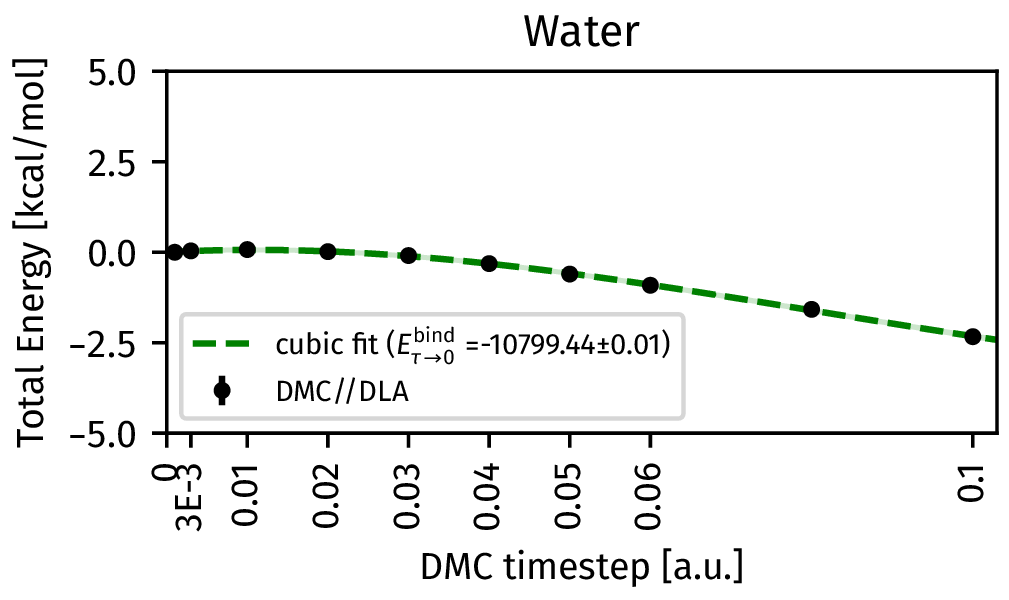}
    \caption{\label{fig:monomer_14} The time step dependence of the Water monomer in the Water$\cdots$Water dimer (ID 1) geometry.}
\end{figure}

Figs.~\ref{fig:monomer_01}--~\ref{fig:monomer_14} illustrate the time step dependence of the total energy for each individual monomer.
Table~\ref{tab:monomer_deformation_ene} illustrates the CCSD(T) deformation energy calculated for each of the two monomers of the dimers of the S66 dataset with respect to the corresponding geometries used with DMC.
We show in Table~\ref{tab:monomer_deformation_ene_validation} that the CCSD(T) deformation energy matches DMC estimates to within $0.12\,$kcal/mol for a subset of systems.
The DMC estimates were reported for the $0.01\,$au time step.

\LTcapwidth=\textwidth
    
\begin{longtable}{llrr}
\caption{\label{tab:monomer_deformation_ene}Deformation energy for the two monomers within each of the dimers of the S66 dataset. This energy is with respect to the geometry used in Table~\ref{tab:monomer_tot_ene}.} \\

\toprule
ID & Dimer Name & $\Delta E_\text{mon. 1, def.}^\text{CCSD(T)}$ [kcal/mol] & $\Delta E_\text{mon. 2, def.}^\text{CCSD(T)}$ [kcal/mol] \\
\midrule
\endfirsthead

\caption[]{(continued)} \\
\endhead

\multicolumn{4}{r}{{Continued on next page}} \\
\endfoot

\bottomrule
\endlastfoot

1 & Water$\cdots$Water & 0.031 & 0.000 \\
2 & Water$\cdots$MeOH & 0.042 & -0.016 \\
3 & Water$\cdots$MeNH$_2$ & 0.109 & -0.026 \\
4 & Water$\cdots$Peptide & 0.087 & 0.067 \\
5 & MeOH$\cdots$MeOH & 0.056 & -0.022 \\
6 & MeOH$\cdots$MeNH$_2$ & 0.222 & -0.026 \\
7 & MeOH$\cdots$Peptide & 0.147 & -0.006 \\
8 & MeOH$\cdots$Water & 0.038 & -0.001 \\
9 & MeNH$_2$$\cdots$MeOH & -0.003 & -0.033 \\
10 & MeNH$_2$$\cdots$MeNH$_2$ & 0.005 & -0.015 \\
11 & MeNH$_2$$\cdots$Peptide & 0.018 & -0.102 \\
12 & MeNH$_2$$\cdots$Water & -0.018 & 0.116 \\
13 & Peptide$\cdots$MeOH & -0.048 & -0.034 \\
14 & Peptide$\cdots$MeNH$_2$ & 0.076 & -0.016 \\
15 & Peptide$\cdots$Peptide & 0.160 & 0.078 \\
16 & Peptide$\cdots$Water & 0.050 & -0.003 \\
17 & Uracil$\cdots$Uracil (BP) & 0.348 & 0.230 \\
18 & Water$\cdots$Pyridine & 0.101 & 0.004 \\
19 & MeOH$\cdots$Pyridine & 0.208 & 0.008 \\
20 & AcOH$\cdots$AcOH & 0.000 & -0.002 \\
21 & AcNH$_2$$\cdots$AcNH$_2$ & 0.000 & -0.002 \\
22 & AcOH$\cdots$Uracil & 0.070 & 0.390 \\
23 & AcNH$_2$$\cdots$Uracil & 0.056 & 0.500 \\
24 & Benzene$\cdots$Benzene ($\pi$-$\pi$) & 0.000 & 0.000 \\
25 & Pyridine$\cdots$Pyridine ($\pi$-$\pi$) & 0.000 & -0.003 \\
26 & Uracil$\cdots$Uracil ($\pi$-$\pi$) & 0.000 & 0.000 \\
27 & Benzene$\cdots$Pyridine ($\pi$-$\pi$) & -0.002 & -0.005 \\
28 & Benzene$\cdots$Uracil ($\pi$-$\pi$) & 0.010 & -0.305 \\
29 & Pyridine$\cdots$Uracil ($\pi$-$\pi$) & 0.014 & -0.239 \\
30 & Benzene$\cdots$Ethene & -0.006 & 0.000 \\
31 & Uracil$\cdots$Ethene & -0.318 & -0.000 \\
32 & Uracil$\cdots$Ethyne & -0.246 & 0.000 \\
33 & Pyridine$\cdots$Ethene & -0.016 & -0.000 \\
34 & Pentane$\cdots$Pentane & 0.000 & -0.000 \\
35 & Neopentane$\cdots$Pentane & 0.000 & -0.002 \\
36 & Neopentane$\cdots$Neopentane & 0.000 & 0.000 \\
37 & Cyclopentane$\cdots$Neopentane & 0.000 & 0.000 \\
38 & Cyclopentane$\cdots$Cyclopentane & 0.007 & 0.007 \\
39 & Benzene$\cdots$Cyclopentane & -0.006 & 0.011 \\
40 & Benzene$\cdots$Neopentane & -0.004 & 0.006 \\
41 & Uracil$\cdots$Pentane & -0.338 & 0.051 \\
42 & Uracil$\cdots$Cyclopentane & -0.349 & 0.024 \\
43 & Uracil$\cdots$Neopentane & -0.311 & 0.012 \\
44 & Ethene$\cdots$Pentane & -0.003 & 0.003 \\
45 & Ethyne$\cdots$Pentane & -0.027 & 0.035 \\
46 & Peptide$\cdots$Pentane & -0.013 & 0.029 \\
47 & Benzene$\cdots$Benzene (TS) & -0.003 & 0.004 \\
48 & Pyridine$\cdots$Pyridine (TS) & -0.003 & 0.002 \\
49 & Benzene$\cdots$Pyridine (TS) & -0.001 & 0.004 \\
50 & Benzene$\cdots$Ethyne (CH-$\pi$) & 0.001 & -0.017 \\
51 & Ethyne$\cdots$Ethyne (TS) & -0.030 & -0.026 \\
52 & Benzene$\cdots$AcOH (OH-$\pi$) & 0.018 & -1.276 \\
53 & Benzene$\cdots$AcNH$_2$ (NH-$\pi$) & 0.042 & -0.646 \\
54 & Benzene$\cdots$Water (OH-$\pi$) & 0.000 & 0.044 \\
55 & Benzene$\cdots$MeOH (OH-$\pi$) & 0.004 & 0.000 \\
56 & Benzene$\cdots$MeNH$_2$ (NH-$\pi$) & -0.001 & 0.000 \\
57 & Benzene$\cdots$Peptide (NH-$\pi$) & 0.003 & 0.000 \\
58 & Pyridine$\cdots$Pyridine (CH-N) & 0.018 & 0.018 \\
59 & Ethyne$\cdots$Water (CH-O) & -0.010 & -0.001 \\
60 & Ethyne$\cdots$AcOH (OH-$\pi$) & 0.024 & -1.253 \\
61 & Pentane$\cdots$AcOH & 0.033 & -1.330 \\
62 & Pentane$\cdots$AcNH$_2$ & 0.031 & -0.705 \\
63 & Benzene$\cdots$AcOH & 0.002 & -1.295 \\
64 & Peptide$\cdots$Ethene & -0.027 & 0.018 \\
65 & Pyridine$\cdots$Ethyne & -0.006 & 0.029 \\
66 & MeNH$_2$$\cdots$Pyridine & 0.002 & -0.001 \\
\end{longtable}

\begin{table}
\caption{\label{tab:monomer_deformation_ene_validation}Comparison between DMC ($0.01\,$au time step) and CCSD(T) for the deformation energy $E_\text{def.}$ of a subset of AcNH$_2$, AcOH, cyclopentane, peptide and urcail monomers found in the S66 dataset. The order in which the monomer appears in the dimer (in the provided .xyz geometry) is given. The reference monomer configuration to calculate $E_\text{def.}$ is given in Table~\ref{tab:monomer_deformation_ene}.}
\begin{adjustbox}{center}
\begin{tabular}{lrrrrr}
\toprule
Monomer & Dimer Geometry & Order & $\Delta E_\text{def.}^\text{DMC}$ & $\Delta E_\text{def.}^\text{CCSD(T)}$ & Deviation \\ 
\midrule
AcNH$_2$ & AcNH$_2$$\cdots$AcNH$_2$ & 2 & -0.07 $\pm$ 0.05 & -0.00 & 0.07 $\pm$ 0.05 \\
AcNH$_2$ & AcNH$_2$$\cdots$AcNH$_2$ & 1 & 0.00 $\pm$ 0.00 & 0.00 & 0.00 $\pm$ 0.00 \\
AcNH$_2$ & Benzene$\cdots$AcNH$_2$ (NH-$\pi$) & 2 & -0.68 $\pm$ 0.04 & -0.65 & 0.03 $\pm$ 0.04 \\
AcNH$_2$ & Pentane$\cdots$AcNH$_2$ & 2 & -0.64 $\pm$ 0.05 & -0.70 & -0.07 $\pm$ 0.05 \\
AcNH$_2$ & AcNH$_2$$\cdots$Uracil & 1 & 0.10 $\pm$ 0.05 & 0.06 & -0.04 $\pm$ 0.05 \\
AcOH & AcOH$\cdots$Uracil & 1 & 0.15 $\pm$ 0.06 & 0.07 & -0.08 $\pm$ 0.06 \\
AcOH & AcOH$\cdots$AcOH & 2 & 0.07 $\pm$ 0.05 & -0.00 & -0.07 $\pm$ 0.05 \\
AcOH & Pentane$\cdots$AcOH & 2 & -1.27 $\pm$ 0.05 & -1.33 & -0.06 $\pm$ 0.05 \\
AcOH & Ethyne$\cdots$AcOH (OH-$\pi$) & 2 & -1.18 $\pm$ 0.06 & -1.25 & -0.07 $\pm$ 0.06 \\
AcOH & Benzene$\cdots$AcOH & 2 & -1.18 $\pm$ 0.04 & -1.29 & -0.12 $\pm$ 0.04 \\
AcOH & AcOH$\cdots$AcOH & 1 & 0.00 $\pm$ 0.00 & 0.00 & 0.00 $\pm$ 0.00 \\
AcOH & Benzene$\cdots$AcOH (OH-$\pi$) & 2 & -1.18 $\pm$ 0.05 & -1.28 & -0.10 $\pm$ 0.05 \\
Cyclopentane & Benzene$\cdots$Cyclopentane & 2 & 0.08 $\pm$ 0.05 & 0.01 & -0.07 $\pm$ 0.05 \\
Cyclopentane & Cyclopentane$\cdots$Neopentane & 1 & 0.00 $\pm$ 0.00 & 0.00 & 0.00 $\pm$ 0.00 \\
Cyclopentane & Uracil$\cdots$Cyclopentane & 2 & 0.08 $\pm$ 0.05 & 0.02 & -0.06 $\pm$ 0.05 \\
Cyclopentane & Cyclopentane$\cdots$Cyclopentane & 1 & 0.06 $\pm$ 0.05 & 0.01 & -0.05 $\pm$ 0.05 \\
Cyclopentane & Cyclopentane$\cdots$Cyclopentane & 2 & 0.07 $\pm$ 0.05 & 0.01 & -0.06 $\pm$ 0.05 \\
Peptide & Peptide$\cdots$MeOH & 1 & -0.05 $\pm$ 0.06 & -0.05 & -0.00 $\pm$ 0.06 \\
Peptide & Peptide$\cdots$MeNH$_2$ & 1 & 0.14 $\pm$ 0.05 & 0.08 & -0.06 $\pm$ 0.05 \\
Peptide & Peptide$\cdots$Peptide & 1 & 0.18 $\pm$ 0.05 & 0.16 & -0.02 $\pm$ 0.05 \\
Peptide & MeNH$_2$$\cdots$Peptide & 2 & -0.03 $\pm$ 0.05 & -0.10 & -0.07 $\pm$ 0.05 \\
Peptide & Peptide$\cdots$Ethene & 1 & 0.00 $\pm$ 0.05 & -0.03 & -0.03 $\pm$ 0.05 \\
Peptide & MeOH$\cdots$Peptide & 2 & 0.11 $\pm$ 0.05 & -0.01 & -0.11 $\pm$ 0.05 \\
Peptide & Peptide$\cdots$Peptide & 2 & 0.11 $\pm$ 0.05 & 0.08 & -0.03 $\pm$ 0.05 \\
Peptide & Peptide$\cdots$Pentane & 1 & 0.06 $\pm$ 0.05 & -0.01 & -0.07 $\pm$ 0.05 \\
Peptide & Benzene$\cdots$Peptide (NH-$\pi$) & 2 & 0.00 $\pm$ 0.00 & 0.00 & 0.00 $\pm$ 0.00 \\
Peptide & Peptide$\cdots$Water & 1 & 0.07 $\pm$ 0.05 & 0.05 & -0.02 $\pm$ 0.05 \\
Peptide & Water$\cdots$Peptide & 2 & 0.08 $\pm$ 0.05 & 0.07 & -0.02 $\pm$ 0.05 \\
Uracil & Benzene$\cdots$Uracil ($\pi$-$\pi$) & 2 & -0.41 $\pm$ 0.05 & -0.31 & 0.10 $\pm$ 0.05 \\
Uracil & Uracil$\cdots$Uracil (BP) & 1 & 0.25 $\pm$ 0.06 & 0.35 & 0.10 $\pm$ 0.06 \\
Uracil & Uracil$\cdots$Ethene & 1 & -0.43 $\pm$ 0.05 & -0.32 & 0.12 $\pm$ 0.05 \\
Uracil & Uracil$\cdots$Uracil ($\pi$-$\pi$) & 1 & 0.00 $\pm$ 0.00 & 0.00 & 0.00 $\pm$ 0.00 \\
Uracil & Pyridine$\cdots$Uracil ($\pi$-$\pi$) & 2 & -0.34 $\pm$ 0.06 & -0.24 & 0.10 $\pm$ 0.06 \\
Uracil & AcOH$\cdots$Uracil & 2 & 0.33 $\pm$ 0.06 & 0.39 & 0.06 $\pm$ 0.06 \\
Uracil & Uracil$\cdots$Cyclopentane & 1 & -0.42 $\pm$ 0.06 & -0.35 & 0.07 $\pm$ 0.06 \\
Uracil & Uracil$\cdots$Ethyne & 1 & -0.30 $\pm$ 0.06 & -0.25 & 0.06 $\pm$ 0.06 \\
Uracil & AcNH$_2$$\cdots$Uracil & 2 & 0.49 $\pm$ 0.06 & 0.50 & 0.02 $\pm$ 0.06 \\
Uracil & Uracil$\cdots$Pentane & 1 & -0.34 $\pm$ 0.05 & -0.34 & 0.00 $\pm$ 0.05 \\
Uracil & Uracil$\cdots$Neopentane & 1 & -0.36 $\pm$ 0.05 & -0.31 & 0.05 $\pm$ 0.05 \\
Uracil & Uracil$\cdots$Uracil ($\pi$-$\pi$) & 2 & -0.02 $\pm$ 0.06 & 0.00 & 0.02 $\pm$ 0.06 \\
Uracil & Uracil$\cdots$Uracil (BP) & 2 & 0.18 $\pm$ 0.06 & 0.23 & 0.05 $\pm$ 0.06 \\
\bottomrule
\end{tabular}
\end{adjustbox}
\end{table}

\clearpage

\subsection{Setup for the CCSD(T) calculations used to evaluate the deformation energy}\label{sec:deformation_energy_setup_ccsdt} 

The deformation energy, appearing in Eq.~\ref{eq:deformation_energy}, had been estimated using 
the Orca program system~\cite{Orca_v4} version 4.2.1. 
In particular, we performed 
Domain-Based Local Pair Natural Orbital Coupled Cluster with Single, Double, and Perturbative Triple excitations,~\cite{DLPNO-CCSD, DLPNO-CCSD(T)} or DLPNO-CCSD(T), calculations.
We used Dunning's correlation consistent polarized valence triple-zeta (cc-pVTZ) and quadruple-zeta (cc-pVQZ) basis sets, and we extrapolated the complete basis set limit independently for the self-consistent field energy, with the scheme defined in Ref.~\citenum{CBS-SCF}, and for the correlation energy, with the scheme defined in Ref.~\citenum{CBS-corr}, using the exponents given in Ref.~\citenum{CBS-exponents}.

\clearpage

\clearpage

\subsection{Brief summary}
In this section, we briefly describe the fixed node DMC algorithm and the main factors affecting its accuracy within practical calculations.

Fixed-node DMC, is a stochastic projector method for solving the imaginary-time many-body Schrodinger equation, where a trial many-electron wave-function $\Psi_\mathrm{T} (\mathbf{R})$, where $\mathbf{R}$ is the electronic configuration, is chosen and used to define a trial many-electron nodal surface (the hyper-surface where $\Psi_\mathrm{T} (\mathbf{R})=0$). 
With the given nodal surface, fixed-node DMC will project out the lowest-energy many-electron state.
The trial wave-function has a critical role in determining the accuracy of fixed-node DMC. The trial wave-function is the product $\Psi_\mathrm{T} (\mathbf{R}) = \mathcal{D}(\mathbf{R}) * \exp{\mathcal{J}(\mathbf{R})} $ of an antisymmetric function $\mathcal{D}(\mathbf{R}) $ and a symmetric (bosonic) function $\exp{\mathcal{J}(\mathbf{R})}$, called the Jastrow factor, describing the dynamical correlation between the electrons by including explicit functions of the electron-electron distances and electron-nucleus distances.
The common practice for the Jastrow factor is to decide a functional form for $\mathcal{J}$ and optimize its parameters by  minimizing either the energy or the variance, using the variational Monte Carlo (VMC)\cite{foulkes_qmc} scheme.
The stochastic optimization of the Jastrow factor implies an optimization uncertainty on its parameters. 

When dealing with large systems ($>100$ atoms), fixed-node DMC employs pseudopotentials to substantially improve its efficiency.
To deal with nonlocal terms of pseudopotentials, the fixed-node DMC algorithm must use an additional approximation, leading to the so-called localization error.
The first approximation to solve the localization error 
consisted of ``localizing'' nonlocal pseudopotential operators using the trial wave function,~\cite{la_mitas} or part of the wave function.~\cite{hammond1987}
Currently there are four schemes based on this approach: the locality approximation (LA)\cite{la_mitas}, the T-move (TM)\cite{tm_casula_1,tm_casula_2}, the determinant locality approximation (DLA)\cite{dla_zen}, and the determinant locality T-move (DTM)\cite{dla_zen}.
Here, we use the DLA scheme, and report additional tests against TM and DTM in Table~\ref{tab:loc_scheme_test}.

\begin{table}
\caption{\label{tab:loc_scheme_test}Comparison of the extrapolated interaction energy $\Delta E_\text{int.}$ for the TM and DLA localization schemes for the H$_2$O$\cdots$H$_2$O (ID 1) and AcOH$\cdots$AcOH dimers (ID 20).}
\begin{adjustbox}{center}
\begin{tabular}{llr}
\toprule
 &  & $\Delta E_\text{int.}$ [kcal/mol] \\ 
\midrule
\multirow[t]{2}{*}{H2O$\cdots$H2O} & TM & -5.06$\pm$0.03 \\
 & DLA & -5.17$\pm$0.03 \\
\cline{1-3}
\multirow[t]{3}{*}{AcOH$\cdots$AcOH} & TM & -20.06$\pm$0.08 \\
 & DLA & -20.17$\pm$0.07 \\
 & DTM & -20.30$\pm$0.08 \\
\cline{1-3}
\bottomrule
\end{tabular}
\end{adjustbox}
\end{table}

\subsection{\label{sec:timestep_error}Reaching the time step limit and estimating errors}

A key aspect affecting the accuracy of DMC is the simulation time step.
In fact, as mentioned above, in DMC a propagation according to the imaginary time Schrödinger equation is performed to project out the exact ground state from a trial wave-function.
A time step $\tau$ must be chosen, but the projection is exact only in the continuous limit $\tau \to 0$.
The bias due to the finite time step is usually called the time step error.
We note that the time step dependence can be affected by the chosen Jastrow, the trial wave-function as well as the algorithm used to perform the walker propagation.
Many such algorithms exist and we use the ZSGMA\cite{ZSGMA,ZenPNAS2018} DMC algorithm.

We extrapolate to the zero time step limit ($\tau \to 0$) using a set of time steps (0.1, 0.08, 0.06, 0.05, 0.04, 0.03, 0.02, 0.01 and 0.003 au).
In calculating any of the energy terms, for all the time steps up until and including $0.1\,$au, we fit a curve to a cubic polynomial of the form:
\begin{equation}
    E(\tau) = A + B\tau + C\tau^2 + D\tau^3,
\end{equation}
where A, B, C, and D are fit parameters, with A being the value in the limit of $\tau \to 0$.
Here $E$ can be either a total energy (i.e., $E_\text{dimer}^\text{DMC}$, $E_\text{mon. 1}^\text{DMC}$ and $E_\text{mon. 2}^\text{DMC}$) or an interaction energy $\Delta E_\text{int}$.
For the time steps below and including $0.02\,$au, we have also fitted a (linear) line.
We use the SciPy~\cite{si-virtanenSciPy10Fundamental2020} \texttt{curve\_fit} function to fit the data-points, which weights the contribution to the residual according to their 1$\sigma$ stochastic error bars.
%

We use the zero time step estimate from the cubic fit for all of our estimates.
For the majority of systems, we set the error on this estimate to the predicted standard deviation of the cubic fit at zero time step: $\sigma_\textrm{cubic fit}$.
However, for some systems, the predicted zero time step prediction with the linear fit can be outside the standard deviation of the cubic fit, indicating that there is a significant change in behavior at smaller time steps.
Here, we instead estimate the error as the difference in the zero time step prediction between the linear and cubic fits: $\Delta_\textrm{cubic fit}^\textrm{linear fit}$



\clearpage

\section{CCSD(T) estimates from the literature}\label{sec:CCSDT}

Reference values at the CCSD(T) level for the S66 dataset has been computed within several studies.
The largest dimer of S66 dataset -- the uracil dimer -- can reach up to 34 atoms and 116 electrons (of which 84 are valence).
This can pose considerable difficulty with performing CCSD(T) on the calculations, particularly when aiming to reach the complete basis set (CBS) limit.
For example, the uracil dimer with a saug-ano-pVQZ basis set calculation `took eight days wall clock time running in parallel on 96 CPUs with a total of 1.5 TB RAM and 18 TB of solid state scratch disk'.
In fact, reaching accurate estimates at the CBS limit requires going beyond the above quadruple-$\zeta$ (QZ) basis set.
As such, the previous studies take on various composite schemes to approximate the CBS limit, performing larger basis set or extrapolated calculations for the MP2 or CCSD contribution to the binding energy, with the remaining contributions to CCSD(T) performed at a smaller basis set.
In particular, they exploit the decomposition of the CCSD(T) binding energy into a Hartree-Fock (HF) component $\Delta E_\textrm{HF}$, a MP2 correlation component $\Delta E_\textrm{MP2}$, a CCSD correction to MP2 $\Delta E_\textrm{CCSD-MP2}$ alongside a final (T) contribution $\Delta E_\textrm{(T)}$:
\begin{equation}
    \Delta E^\textrm{CCSD(T)}_\textrm{int} = \Delta E_\textrm{int}^\textrm{HF} + \Delta E^\textrm{MP2}_\textrm{int} + \Delta E^\textrm{CCSD-MP2}_\textrm{int} + \Delta E^\textrm{(T)}_\textrm{int}.
\end{equation}
Each of these components has a differing dependence on the basis set, with higher order correlations [i.e., (T)] typically requiring smaller basis sets.

The most common class of basis sets used to treat noncovalent interactions are the Dunning cc-pV$X$Z and aug-cc-pV$X$Z (augmented with diffuse functions) basis, which we will refer to as $X$Z and a$X$Z respectively.
In many studies, The $X$Z basis sets are used on the H atoms, with a$X$Z on the remaining elements, leading to the heavy-aug-cc-pV$X$Z basis set, shortened to ha$X$Z.
The cardinal number $X$ can be either a double-$\zeta$ (DZ), triple-$\zeta$ (TZ), quadruple-$\zeta$ (QZ) or quintuple-$\zeta$ (5Z) basis set, in order of increasing basis set size.
Adjacent pairs  of basis sets can be combined within a two-point extrapolation scheme to approximate the CBS limit, which we shall indicate with CBS(DZ/TZ) for the DZ and TZ pair.
Additionally, there is the choice of employing a counterpoise correction, where within the calculation of $\Delta_\textrm{int}$, the energies of the individual monomers are computed together with `ghost' basis functions from the other monomer, removing some errors arising from basis-set superposition error.
Calculations can either employ no counterpoise correction (no-CP), counterpoise correction (CP) or the average of the two (half-CP).
Regardless all these estimates should reach the same value in the CBS limit.

In this work, we reach a final CCSD(T) estimate which takes the average of CCSD(T) estimates from three separate studies, all of which approximate the CCSD(T) CBS limit with differing treatments.
These differences are summarized below:
\begin{itemize}
    \item \textbf{Řez{\'{a}}{\v{c}} \textit{et al.}~\cite{si_rezacExtensionsS66Data2011}} --- The S66 dataset and its first CCSD(T) estimates were introduced by Řez{\'{a}}{\v{c}} \textit{et al.} in Ref.~\citenum{S66}, and these CCSD(T) estimates were subsequently revised and improved in Ref.~\citenum{si_rezacExtensionsS66Data2011}.
    Here, both $\Delta E^\textrm{HF}_\textrm{int}$ and $\Delta E^\textrm{MP2}_\textrm{int}$ were computed with a two-point extrapolation using the aTZ and aQZ basis sets [i.e., CBS(aTZ/aQZ)].
    The remaining $\Delta E^\textrm{CCSD-MP2}_\textrm{int}$ and $\Delta E^\textrm{(T)}_\textrm{int}$ contributions were computed with a CBS(haDZ/haTZ) treatment.
    All contributions utilised CP corrections.
    \item \textbf{Kesharwani \textit{et al.}~\cite{S66_revJM_2018}} --- Kesharwani \textit{et al.} re-evaluated the S66 dataset using explicitly correlated F12-based methods.
    They came up with four different tiers which trade accuracy for cost, namely the `GOLD', `SILVER', `BRONZE' and `STERLING' levels.
    As GOLD was only feasible for a subset of 18 systems, we focus here on the SILVER estimates, which were computed for the entire dataset.
    The $\Delta E^\textrm{HF}_\textrm{int}$ and $\Delta E^\textrm{MP2}_\textrm{int}$ components were computed using MP2-F12 with the cc-pV5Z-F12 basis set, with further (minor) corrections to the HF treatment using a complementary auxiliary basis set (CABS) treatment.
    The $\Delta E^\textrm{CCSD-MP2}_\textrm{int}$ contribution was computed with CCSD(F12*) employing the aug-cc-pVTZ-F12 basis set.
    The final $\Delta E^\textrm{(T)}_\textrm{int}$ contribution did not employ any F12 treatment and was reached using a CBS(haDZ/haTZ) extrapolation.
    All contributions utilised a half-CP correction.
    \item \textbf{Nagy \textit{et al.}~\cite{nagy_gold}} --- Nagy \textit{et al.} reported an improvement upon the previous S66 CCSD(T) references -- termed `14k-GOLD'.
    Here, $\Delta E^\textrm{HF}_\textrm{int}$ was treated with using the aQZ-F12 basis set together with a CABS treatment, while $\Delta E^\textrm{MP2}_\textrm{int}$ was treated using a CBS(aTZ-F12/aQZ-F12) two-point extrapolation.
    The $\Delta E^\textrm{CCSD-MP2}_\textrm{int}$ contribution was computed with CCSD(F12*) and MP2-F12 using a CBS(haDZ-F12/haTZ-F12) extrapolation.
    The final $\Delta E^\textrm{(T)}_\textrm{int}$ contribution did not employ any F12 treatment and was reached using a CBS(haTZ/haQZ) extrapolation.
    The $\Delta E^\textrm{HF}_\textrm{int}$ was performed with half-CP while all other contributions were performed with (full) CP correction.
\end{itemize}

\LTcapwidth=\textwidth
    
\begin{longtable}{llrrrrrr}
\caption{\label{tab:cc_references}CCSD(T) references for the S66 dataset. The final CCSD(T) and CCSD(cT)-fit values are computed as the average of the values from the three references. The error is taken to be the standard deviation from the three references.} \\

\toprule
 & \rotatebox{90}{System} & \rotatebox{90}{\v{R}ez\'a\v{c} \textit{et al.} (2006)} & \rotatebox{90}{Kesharwani \textit{et al.} (2018)} & \rotatebox{90}{Nagy \textit{et al.} (2023)} & \rotatebox{90}{Final CCSD(T)} & \rotatebox{90}{Final CCSD(cT)-fit} \\ 
\midrule
\endfirsthead

\caption[]{(continued)} \\
\endhead

\multicolumn{8}{r}{{Continued on next page}} \\
\endfoot

\bottomrule
\endlastfoot

1 & Water$\cdots$Water & -5.01 & -4.98 & -4.99 & -4.99$\pm$0.01 & -4.96$\pm$0.01 \\
2 & Water$\cdots$MeOH & -5.70 & -5.67 & -5.67 & -5.68$\pm$0.01 & -5.63$\pm$0.01 \\
3 & Water$\cdots$MeNH$_2$ & -7.04 & -6.99 & -7.00 & -7.01$\pm$0.02 & -6.94$\pm$0.02 \\
4 & Water$\cdots$Peptide & -8.22 & -8.18 & -8.19 & -8.20$\pm$0.02 & -8.15$\pm$0.02 \\
5 & MeOH$\cdots$MeOH & -5.85 & -5.82 & -5.83 & -5.83$\pm$0.01 & -5.78$\pm$0.01 \\
6 & MeOH$\cdots$MeNH$_2$ & -7.67 & -7.62 & -7.62 & -7.64$\pm$0.02 & -7.55$\pm$0.02 \\
7 & MeOH$\cdots$Peptide & -8.34 & -8.31 & -8.31 & -8.32$\pm$0.01 & -8.25$\pm$0.01 \\
8 & MeOH$\cdots$Water & -5.09 & -5.06 & -5.07 & -5.08$\pm$0.01 & -5.03$\pm$0.01 \\
9 & MeNH$_2$$\cdots$MeOH & -3.11 & -3.09 & -3.09 & -3.10$\pm$0.01 & -3.05$\pm$0.01 \\
10 & MeNH$_2$$\cdots$MeNH$_2$ & -4.22 & -4.18 & -4.19 & -4.20$\pm$0.01 & -4.13$\pm$0.01 \\
11 & MeNH$_2$$\cdots$Peptide & -5.48 & -5.44 & -5.44 & -5.45$\pm$0.02 & -5.37$\pm$0.02 \\
12 & MeNH$_2$$\cdots$Water & -7.40 & -7.35 & -7.36 & -7.37$\pm$0.02 & -7.29$\pm$0.02 \\
13 & Peptide$\cdots$MeOH & -6.28 & -6.25 & -6.25 & -6.26$\pm$0.01 & -6.20$\pm$0.01 \\
14 & Peptide$\cdots$MeNH$_2$ & -7.56 & -7.52 & -7.52 & -7.53$\pm$0.02 & -7.44$\pm$0.02 \\
15 & Peptide$\cdots$Peptide & -8.72 & -8.69 & -8.69 & -8.70$\pm$0.01 & -8.61$\pm$0.01 \\
16 & Peptide$\cdots$Water & -5.20 & -5.18 & -5.18 & -5.19$\pm$0.01 & -5.15$\pm$0.01 \\
17 & Uracil$\cdots$Uracil (BP) & -17.45 & -17.41 & -17.40 & -17.42$\pm$0.02 & -17.29$\pm$0.02 \\
18 & Water$\cdots$Pyridine & -6.97 & -6.93 & -6.93 & -6.94$\pm$0.02 & -6.87$\pm$0.02 \\
19 & MeOH$\cdots$Pyridine & -7.51 & -7.47 & -7.46 & -7.48$\pm$0.02 & -7.39$\pm$0.02 \\
20 & AcOH$\cdots$AcOH & -19.41 & -19.36 & -19.38 & -19.39$\pm$0.02 & -19.27$\pm$0.02 \\
21 & AcNH$_2$$\cdots$AcNH$_2$ & -16.52 & -16.47 & -16.48 & -16.49$\pm$0.02 & -16.40$\pm$0.02 \\
22 & AcOH$\cdots$Uracil & -19.78 & -19.74 & -19.75 & -19.75$\pm$0.02 & -19.64$\pm$0.02 \\
23 & AcNH$_2$$\cdots$Uracil & -19.47 & -19.42 & -19.42 & -19.44$\pm$0.02 & -19.33$\pm$0.02 \\
24 & Benzene$\cdots$Benzene ($\pi$-$\pi$) & -2.72 & -2.68 & -2.69 & -2.70$\pm$0.02 & -2.46$\pm$0.02 \\
25 & Pyridine$\cdots$Pyridine ($\pi$-$\pi$) & -3.80 & -3.75 & -3.76 & -3.77$\pm$0.02 & -3.51$\pm$0.02 \\
26 & Uracil$\cdots$Uracil ($\pi$-$\pi$) & -9.75 & -9.67 & -9.72 & -9.71$\pm$0.03 & -9.39$\pm$0.03 \\
27 & Benzene$\cdots$Pyridine ($\pi$-$\pi$) & -3.34 & -3.30 & -3.30 & -3.31$\pm$0.02 & -3.07$\pm$0.02 \\
28 & Benzene$\cdots$Uracil ($\pi$-$\pi$) & -5.59 & -5.52 & -5.54 & -5.55$\pm$0.03 & -5.25$\pm$0.03 \\
29 & Pyridine$\cdots$Uracil ($\pi$-$\pi$) & -6.70 & -6.63 & -6.66 & -6.66$\pm$0.03 & -6.37$\pm$0.03 \\
30 & Benzene$\cdots$Ethene & -1.36 & -1.36 & -1.34 & -1.35$\pm$0.01 & -1.24$\pm$0.01 \\
31 & Uracil$\cdots$Ethene & -3.33 & -3.29 & -3.31 & -3.31$\pm$0.02 & -3.17$\pm$0.02 \\
32 & Uracil$\cdots$Ethyne & -3.69 & -3.65 & -3.68 & -3.67$\pm$0.02 & -3.54$\pm$0.02 \\
33 & Pyridine$\cdots$Ethene & -1.80 & -1.78 & -1.78 & -1.79$\pm$0.01 & -1.66$\pm$0.01 \\
34 & Pentane$\cdots$Pentane & -3.76 & -3.74 & -3.73 & -3.74$\pm$0.01 & -3.63$\pm$0.01 \\
35 & Neopentane$\cdots$Pentane & -2.60 & -2.58 & -2.58 & -2.59$\pm$0.01 & -2.50$\pm$0.01 \\
36 & Neopentane$\cdots$Neopentane & -1.76 & -1.74 & -1.75 & -1.75$\pm$0.01 & -1.69$\pm$0.01 \\
37 & Cyclopentane$\cdots$Neopentane & -2.40 & -2.38 & -2.38 & -2.38$\pm$0.01 & -2.30$\pm$0.01 \\
38 & Cyclopentane$\cdots$Cyclopentane & -2.99 & -2.97 & -2.96 & -2.97$\pm$0.01 & -2.87$\pm$0.01 \\
39 & Benzene$\cdots$Cyclopentane & -3.51 & -3.49 & -3.48 & -3.49$\pm$0.01 & -3.33$\pm$0.01 \\
40 & Benzene$\cdots$Neopentane & -2.85 & -2.82 & -2.82 & -2.83$\pm$0.01 & -2.71$\pm$0.01 \\
41 & Uracil$\cdots$Pentane & -4.81 & -4.76 & -4.77 & -4.78$\pm$0.02 & -4.59$\pm$0.02 \\
42 & Uracil$\cdots$Cyclopentane & -4.09 & -4.05 & -4.06 & -4.07$\pm$0.02 & -3.90$\pm$0.02 \\
43 & Uracil$\cdots$Neopentane & -3.69 & -3.65 & -3.66 & -3.67$\pm$0.02 & -3.53$\pm$0.02 \\
44 & Ethene$\cdots$Pentane & -1.99 & -1.97 & -1.98 & -1.98$\pm$0.01 & -1.91$\pm$0.01 \\
45 & Ethyne$\cdots$Pentane & -1.72 & -1.70 & -1.70 & -1.71$\pm$0.01 & -1.63$\pm$0.01 \\
46 & Peptide$\cdots$Pentane & -4.26 & -4.22 & -4.22 & -4.23$\pm$0.02 & -4.09$\pm$0.02 \\
47 & Benzene$\cdots$Benzene (TS) & -2.83 & -2.80 & -2.81 & -2.81$\pm$0.01 & -2.68$\pm$0.01 \\
48 & Pyridine$\cdots$Pyridine (TS) & -3.51 & -3.47 & -3.48 & -3.49$\pm$0.02 & -3.35$\pm$0.02 \\
49 & Benzene$\cdots$Pyridine (TS) & -3.29 & -3.26 & -3.27 & -3.27$\pm$0.01 & -3.15$\pm$0.01 \\
50 & Benzene$\cdots$Ethyne (CH-$\pi$) & -2.86 & -2.83 & -2.84 & -2.84$\pm$0.01 & -2.76$\pm$0.01 \\
51 & Ethyne$\cdots$Ethyne (TS) & -1.54 & -1.52 & -1.53 & -1.53$\pm$0.01 & -1.49$\pm$0.01 \\
52 & Benzene$\cdots$AcOH (OH-$\pi$) & -4.73 & -4.69 & -4.69 & -4.70$\pm$0.02 & -4.59$\pm$0.02 \\
53 & Benzene$\cdots$AcNH$_2$ (NH-$\pi$) & -4.40 & -4.38 & -4.38 & -4.38$\pm$0.01 & -4.29$\pm$0.01 \\
54 & Benzene$\cdots$Water (OH-$\pi$) & -3.29 & -3.27 & -3.26 & -3.27$\pm$0.01 & -3.21$\pm$0.01 \\
55 & Benzene$\cdots$MeOH (OH-$\pi$) & -4.17 & -4.14 & -4.14 & -4.15$\pm$0.02 & -4.04$\pm$0.02 \\
56 & Benzene$\cdots$MeNH$_2$ (NH-$\pi$) & -3.20 & -3.17 & -3.17 & -3.18$\pm$0.01 & -3.07$\pm$0.01 \\
57 & Benzene$\cdots$Peptide (NH-$\pi$) & -5.26 & -5.22 & -5.22 & -5.23$\pm$0.02 & -5.08$\pm$0.02 \\
58 & Pyridine$\cdots$Pyridine (CH-N) & -4.24 & -4.19 & -4.19 & -4.21$\pm$0.02 & -4.12$\pm$0.02 \\
59 & Ethyne$\cdots$Water (CH-O) & -2.93 & -2.90 & -2.91 & -2.92$\pm$0.01 & -2.89$\pm$0.01 \\
60 & Ethyne$\cdots$AcOH (OH-$\pi$) & -4.97 & -4.92 & -4.93 & -4.94$\pm$0.02 & -4.86$\pm$0.02 \\
61 & Pentane$\cdots$AcOH & -2.91 & -2.88 & -2.88 & -2.89$\pm$0.02 & -2.79$\pm$0.02 \\
62 & Pentane$\cdots$AcNH$_2$ & -3.53 & -3.49 & -3.50 & -3.51$\pm$0.02 & -3.39$\pm$0.02 \\
63 & Benzene$\cdots$AcOH & -3.75 & -3.71 & -3.72 & -3.72$\pm$0.02 & -3.59$\pm$0.02 \\
64 & Peptide$\cdots$Ethene & -3.00 & -2.97 & -2.98 & -2.98$\pm$0.01 & -2.90$\pm$0.01 \\
65 & Pyridine$\cdots$Ethyne & -4.10 & -4.06 & -4.07 & -4.08$\pm$0.02 & -4.02$\pm$0.02 \\
66 & MeNH$_2$$\cdots$Pyridine & -3.97 & -3.93 & -3.93 & -3.94$\pm$0.02 & -3.83$\pm$0.02 \\
\end{longtable}

\clearpage

\section{\label{sec:final_cc_estimates}Final CCSD(T) and CCSD(cT)-fit estimates} 

As discussed above, in this work, we arrive at a final estimate of CCSD (T) taking the average of the CCSD(T) estimates from Kesharwani \etal{}, \v{R}ez\'a\v{c} \textit{et al.} and Nagy \textit{et al.}.
The resulting error bars are the standard deviation of the three references.
The final CCSD(T) estimates are reported in Table~\ref{tab:cc_references}.
In addition, we also give an estimate at the CCSD(cT) for the dispersion-dominated systems.
CCSD(cT) incorporates additional higher-order terms to the triples excitation amplitudes compared CCSD(T), crucial for studying systems with large polarizability.
In particular, it has been shown that CCSD(cT) can be approximated from the $\Delta_\textrm{int}^\textrm{MP2}$, $\Delta_\textrm{int}^\textrm{CCSD-MP2}$ and $\Delta_\textrm{int}^\textrm{(T)}$ values.
Specifically, it is given by the following expression:
\begin{equation}
    \dfrac{\Delta_\textrm{int}^\textrm{(T)}}{\Delta_\textrm{int}^\textrm{(cT)}} = a + b \cdot \dfrac{\Delta_\textrm{int}^\textrm{MP2}}{\Delta_\textrm{int}^\textrm{CCSD}},
\end{equation}
where $a = 0.7764$ and $b = 0.2780$ were fitted to CCSD(cT) data.

\clearpage
\section{Final DMC estimates}\label{sec:table}
We report the final DMC estimates in Table~\ref{tab:dmc-final-energies}.
This calculates the interaction energy as given in Eq.~\ref{si_equation_binding_energy} with deformation energies computed as in Eq.~\ref{eq:deformation_energy} and Table~\ref{tab:monomer_deformation_ene}.
We note that the extrapolation towards the zero time step limit is performed directly on $\Delta E_\textrm{int}$ rather than the individual total energy components.
In Table~\ref{tab:dmc-cc-comparison}, we have computed the deviation of these final estimates to the CCSD(T) estimates in Table~\ref{tab:cc_references}.

\LTcapwidth=\textwidth
    
\begin{longtable}{llrr}
\caption{\label{tab:dmc-final-energies}Final DMC $\Delta E_\text{int}$ estimates for the S66 dataset. The method (see Sec.~\ref{sec:timestep_error}) used to estimate the error on the DMC estimate is also provided.} \\

\toprule
 & System & $\Delta E_\text{int}$ [kcal/mol] & Fit type \\
\midrule
\endfirsthead

\caption[]{(continued)} \\
\endhead

\multicolumn{4}{r}{{Continued on next page}} \\
\endfoot

\bottomrule
\endlastfoot

1 & Water$\cdots$Water & -5.17$\pm$0.03 & $\sigma_\text{cubic fit}$ \\
2 & Water$\cdots$MeOH & -5.82$\pm$0.05 & $\Delta_\text{cubic fit}^\text{linear fit}$ \\
3 & Water$\cdots$MeNH$_2$ & -7.18$\pm$0.04 & $\sigma_\text{cubic fit}$ \\
4 & Water$\cdots$Peptide & -8.58$\pm$0.07 & $\Delta_\text{cubic fit}^\text{linear fit}$ \\
5 & MeOH$\cdots$MeOH & -5.93$\pm$0.04 & $\sigma_\text{cubic fit}$ \\
6 & MeOH$\cdots$MeNH$_2$ & -7.83$\pm$0.05 & $\sigma_\text{cubic fit}$ \\
7 & MeOH$\cdots$Peptide & -8.57$\pm$0.08 & $\Delta_\text{cubic fit}^\text{linear fit}$ \\
8 & MeOH$\cdots$Water & -5.24$\pm$0.05 & $\sigma_\text{cubic fit}$ \\
9 & MeNH$_2$$\cdots$MeOH & -3.12$\pm$0.05 & $\sigma_\text{cubic fit}$ \\
10 & MeNH$_2$$\cdots$MeNH$_2$ & -4.20$\pm$0.06 & $\Delta_\text{cubic fit}^\text{linear fit}$ \\
11 & MeNH$_2$$\cdots$Peptide & -5.42$\pm$0.07 & $\sigma_\text{cubic fit}$ \\
12 & MeNH$_2$$\cdots$Water & -7.53$\pm$0.06 & $\Delta_\text{cubic fit}^\text{linear fit}$ \\
13 & Peptide$\cdots$MeOH & -6.32$\pm$0.07 & $\sigma_\text{cubic fit}$ \\
14 & Peptide$\cdots$MeNH$_2$ & -7.50$\pm$0.07 & $\sigma_\text{cubic fit}$ \\
15 & Peptide$\cdots$Peptide & -8.88$\pm$0.11 & $\Delta_\text{cubic fit}^\text{linear fit}$ \\
16 & Peptide$\cdots$Water & -5.37$\pm$0.06 & $\sigma_\text{cubic fit}$ \\
17 & Uracil$\cdots$Uracil (BP) & -17.79$\pm$0.10 & $\sigma_\text{cubic fit}$ \\
18 & Water$\cdots$Pyridine & -7.30$\pm$0.07 & $\Delta_\text{cubic fit}^\text{linear fit}$ \\
19 & MeOH$\cdots$Pyridine & -7.89$\pm$0.07 & $\sigma_\text{cubic fit}$ \\
20 & AcOH$\cdots$AcOH & -20.17$\pm$0.07 & $\sigma_\text{cubic fit}$ \\
21 & AcNH$_2$$\cdots$AcNH$_2$ & -16.83$\pm$0.07 & $\sigma_\text{cubic fit}$ \\
22 & AcOH$\cdots$Uracil & -20.40$\pm$0.09 & $\sigma_\text{cubic fit}$ \\
23 & AcNH$_2$$\cdots$Uracil & -19.83$\pm$0.08 & $\sigma_\text{cubic fit}$ \\
24 & Benzene$\cdots$Benzene ($\pi$-$\pi$) & -2.33$\pm$0.07 & $\sigma_\text{cubic fit}$ \\
25 & Pyridine$\cdots$Pyridine ($\pi$-$\pi$) & -3.53$\pm$0.08 & $\sigma_\text{cubic fit}$ \\
26 & Uracil$\cdots$Uracil ($\pi$-$\pi$) & -9.33$\pm$0.08 & $\sigma_\text{cubic fit}$ \\
27 & Benzene$\cdots$Pyridine ($\pi$-$\pi$) & -3.04$\pm$0.08 & $\sigma_\text{cubic fit}$ \\
28 & Benzene$\cdots$Uracil ($\pi$-$\pi$) & -5.15$\pm$0.10 & $\sigma_\text{cubic fit}$ \\
29 & Pyridine$\cdots$Uracil ($\pi$-$\pi$) & -6.39$\pm$0.09 & $\sigma_\text{cubic fit}$ \\
30 & Benzene$\cdots$Ethene & -1.11$\pm$0.06 & $\sigma_\text{cubic fit}$ \\
31 & Uracil$\cdots$Ethene & -3.18$\pm$0.08 & $\sigma_\text{cubic fit}$ \\
32 & Uracil$\cdots$Ethyne & -3.59$\pm$0.08 & $\sigma_\text{cubic fit}$ \\
33 & Pyridine$\cdots$Ethene & -1.68$\pm$0.07 & $\sigma_\text{cubic fit}$ \\
34 & Pentane$\cdots$Pentane & -3.53$\pm$0.08 & $\Delta_\text{cubic fit}^\text{linear fit}$ \\
35 & Neopentane$\cdots$Pentane & -2.46$\pm$0.07 & $\sigma_\text{cubic fit}$ \\
36 & Neopentane$\cdots$Neopentane & -1.67$\pm$0.08 & $\sigma_\text{cubic fit}$ \\
37 & Cyclopentane$\cdots$Neopentane & -2.17$\pm$0.07 & $\sigma_\text{cubic fit}$ \\
38 & Cyclopentane$\cdots$Cyclopentane & -2.74$\pm$0.07 & $\sigma_\text{cubic fit}$ \\
39 & Benzene$\cdots$Cyclopentane & -3.14$\pm$0.08 & $\sigma_\text{cubic fit}$ \\
40 & Benzene$\cdots$Neopentane & -2.69$\pm$0.08 & $\sigma_\text{cubic fit}$ \\
41 & Uracil$\cdots$Pentane & -4.47$\pm$0.08 & $\sigma_\text{cubic fit}$ \\
42 & Uracil$\cdots$Cyclopentane & -3.59$\pm$0.09 & $\sigma_\text{cubic fit}$ \\
43 & Uracil$\cdots$Neopentane & -3.51$\pm$0.09 & $\sigma_\text{cubic fit}$ \\
44 & Ethene$\cdots$Pentane & -1.79$\pm$0.06 & $\sigma_\text{cubic fit}$ \\
45 & Ethyne$\cdots$Pentane & -1.56$\pm$0.08 & $\Delta_\text{cubic fit}^\text{linear fit}$ \\
46 & Peptide$\cdots$Pentane & -3.82$\pm$0.08 & $\sigma_\text{cubic fit}$ \\
47 & Benzene$\cdots$Benzene (TS) & -2.61$\pm$0.08 & $\sigma_\text{cubic fit}$ \\
48 & Pyridine$\cdots$Pyridine (TS) & -3.43$\pm$0.08 & $\sigma_\text{cubic fit}$ \\
49 & Benzene$\cdots$Pyridine (TS) & -3.10$\pm$0.08 & $\sigma_\text{cubic fit}$ \\
50 & Benzene$\cdots$Ethyne (CH-$\pi$) & -2.91$\pm$0.07 & $\sigma_\text{cubic fit}$ \\
51 & Ethyne$\cdots$Ethyne (TS) & -1.55$\pm$0.04 & $\sigma_\text{cubic fit}$ \\
52 & Benzene$\cdots$AcOH (OH-$\pi$) & -4.64$\pm$0.08 & $\sigma_\text{cubic fit}$ \\
53 & Benzene$\cdots$AcNH$_2$ (NH-$\pi$) & -4.25$\pm$0.08 & $\sigma_\text{cubic fit}$ \\
54 & Benzene$\cdots$Water (OH-$\pi$) & -3.20$\pm$0.07 & $\sigma_\text{cubic fit}$ \\
55 & Benzene$\cdots$MeOH (OH-$\pi$) & -3.93$\pm$0.07 & $\sigma_\text{cubic fit}$ \\
56 & Benzene$\cdots$MeNH$_2$ (NH-$\pi$) & -3.03$\pm$0.07 & $\sigma_\text{cubic fit}$ \\
57 & Benzene$\cdots$Peptide (NH-$\pi$) & -5.10$\pm$0.08 & $\sigma_\text{cubic fit}$ \\
58 & Pyridine$\cdots$Pyridine (CH-N) & -4.22$\pm$0.12 & $\Delta_\text{cubic fit}^\text{linear fit}$ \\
59 & Ethyne$\cdots$Water (CH-O) & -3.04$\pm$0.04 & $\sigma_\text{cubic fit}$ \\
60 & Ethyne$\cdots$AcOH (OH-$\pi$) & -4.98$\pm$0.06 & $\sigma_\text{cubic fit}$ \\
61 & Pentane$\cdots$AcOH & -2.63$\pm$0.08 & $\sigma_\text{cubic fit}$ \\
62 & Pentane$\cdots$AcNH$_2$ & -3.08$\pm$0.07 & $\sigma_\text{cubic fit}$ \\
63 & Benzene$\cdots$AcOH & -3.51$\pm$0.08 & $\sigma_\text{cubic fit}$ \\
64 & Peptide$\cdots$Ethene & -2.78$\pm$0.07 & $\sigma_\text{cubic fit}$ \\
65 & Pyridine$\cdots$Ethyne & -4.26$\pm$0.06 & $\sigma_\text{cubic fit}$ \\
66 & MeNH$_2$$\cdots$Pyridine & -3.81$\pm$0.06 & $\sigma_\text{cubic fit}$ \\
\end{longtable}
\LTcapwidth=\textwidth
\small
\begin{longtable}{llrrr}
\caption{\label{tab:dmc-cc-comparison}Final DMC and CCSD(T) $\Delta E_\text{int.}$ estimates for the S66 dataset in kcal/mol, with their deviation of CCSD(T) from DMC given.} \\

\toprule
 & System & $\Delta E_\text{int.}^\text{DMC}$ [kcal/mol] & $\Delta E_\text{int.}^\text{CCSD(T)}$ [kcal/mol] & Deviation [kcal/mol] \\
\midrule
\endfirsthead

\caption[]{(continued)} \\
\endhead

\multicolumn{4}{r}{{Continued on next page}} \\
\endfoot

\bottomrule
\endlastfoot

1 & Water$\cdots$Water & -5.17$\pm$0.03 & -4.99$\pm$0.01 & 0.18$\pm$0.03  \\
2 & Water$\cdots$MeOH & -5.82$\pm$0.05 & -5.68$\pm$0.01 & 0.14$\pm$0.05  \\
3 & Water$\cdots$MeNH$_2$ & -7.18$\pm$0.04 & -7.01$\pm$0.02 & 0.18$\pm$0.05  \\
4 & Water$\cdots$Peptide & -8.58$\pm$0.07 & -8.20$\pm$0.02 & 0.39$\pm$0.08  \\
5 & MeOH$\cdots$MeOH & -5.93$\pm$0.04 & -5.83$\pm$0.01 & 0.09$\pm$0.04  \\
6 & MeOH$\cdots$MeNH$_2$ & -7.83$\pm$0.05 & -7.64$\pm$0.02 & 0.19$\pm$0.05  \\
7 & MeOH$\cdots$Peptide & -8.57$\pm$0.08 & -8.32$\pm$0.01 & 0.25$\pm$0.08  \\
8 & MeOH$\cdots$Water & -5.24$\pm$0.05 & -5.08$\pm$0.01 & 0.16$\pm$0.05  \\
9 & MeNH$_2$$\cdots$MeOH & -3.12$\pm$0.05 & -3.10$\pm$0.01 & 0.02$\pm$0.05  \\
10 & MeNH$_2$$\cdots$MeNH$_2$ & -4.20$\pm$0.06 & -4.20$\pm$0.01 & 0.00$\pm$0.06  \\
11 & MeNH$_2$$\cdots$Peptide & -5.42$\pm$0.07 & -5.45$\pm$0.02 & -0.04$\pm$0.07  \\
12 & MeNH$_2$$\cdots$Water & -7.53$\pm$0.06 & -7.37$\pm$0.02 & 0.16$\pm$0.06  \\
13 & Peptide$\cdots$MeOH & -6.32$\pm$0.07 & -6.26$\pm$0.01 & 0.06$\pm$0.07  \\
14 & Peptide$\cdots$MeNH$_2$ & -7.50$\pm$0.07 & -7.53$\pm$0.02 & -0.03$\pm$0.07  \\
15 & Peptide$\cdots$Peptide & -8.88$\pm$0.11 & -8.70$\pm$0.01 & 0.18$\pm$0.11  \\
16 & Peptide$\cdots$Water & -5.37$\pm$0.06 & -5.19$\pm$0.01 & 0.18$\pm$0.06  \\
17 & Uracil$\cdots$Uracil (BP) & -17.79$\pm$0.10 & -17.42$\pm$0.02 & 0.37$\pm$0.10  \\
18 & Water$\cdots$Pyridine & -7.30$\pm$0.07 & -6.94$\pm$0.02 & 0.36$\pm$0.07  \\
19 & MeOH$\cdots$Pyridine & -7.89$\pm$0.07 & -7.48$\pm$0.02 & 0.41$\pm$0.07  \\
20 & AcOH$\cdots$AcOH & -20.17$\pm$0.07 & -19.39$\pm$0.02 & 0.78$\pm$0.07  \\
21 & AcNH$_2$$\cdots$AcNH$_2$ & -16.83$\pm$0.07 & -16.49$\pm$0.02 & 0.34$\pm$0.07  \\
22 & AcOH$\cdots$Uracil & -20.40$\pm$0.09 & -19.75$\pm$0.02 & 0.64$\pm$0.09  \\
23 & AcNH$_2$$\cdots$Uracil & -19.83$\pm$0.08 & -19.44$\pm$0.02 & 0.39$\pm$0.09  \\
24 & Benzene$\cdots$Benzene ($\pi$-$\pi$) & -2.33$\pm$0.07 & -2.70$\pm$0.02 & -0.36$\pm$0.08  \\
25 & Pyridine$\cdots$Pyridine ($\pi$-$\pi$) & -3.53$\pm$0.08 & -3.77$\pm$0.02 & -0.25$\pm$0.08  \\
26 & Uracil$\cdots$Uracil ($\pi$-$\pi$) & -9.33$\pm$0.08 & -9.71$\pm$0.03 & -0.39$\pm$0.09  \\
27 & Benzene$\cdots$Pyridine ($\pi$-$\pi$) & -3.04$\pm$0.08 & -3.31$\pm$0.02 & -0.28$\pm$0.08  \\
28 & Benzene$\cdots$Uracil ($\pi$-$\pi$) & -5.15$\pm$0.10 & -5.55$\pm$0.03 & -0.41$\pm$0.10  \\
29 & Pyridine$\cdots$Uracil ($\pi$-$\pi$) & -6.39$\pm$0.09 & -6.66$\pm$0.03 & -0.27$\pm$0.10  \\
30 & Benzene$\cdots$Ethene & -1.11$\pm$0.06 & -1.35$\pm$0.01 & -0.24$\pm$0.06  \\
31 & Uracil$\cdots$Ethene & -3.18$\pm$0.08 & -3.31$\pm$0.02 & -0.13$\pm$0.08  \\
32 & Uracil$\cdots$Ethyne & -3.59$\pm$0.08 & -3.67$\pm$0.02 & -0.08$\pm$0.08  \\
33 & Pyridine$\cdots$Ethene & -1.68$\pm$0.07 & -1.79$\pm$0.01 & -0.11$\pm$0.07  \\
34 & Pentane$\cdots$Pentane & -3.53$\pm$0.08 & -3.74$\pm$0.01 & -0.22$\pm$0.08  \\
35 & Neopentane$\cdots$Pentane & -2.46$\pm$0.07 & -2.59$\pm$0.01 & -0.13$\pm$0.07  \\
36 & Neopentane$\cdots$Neopentane & -1.67$\pm$0.08 & -1.75$\pm$0.01 & -0.08$\pm$0.08  \\
37 & Cyclopentane$\cdots$Neopentane & -2.17$\pm$0.07 & -2.38$\pm$0.01 & -0.21$\pm$0.08  \\
38 & Cyclopentane$\cdots$Cyclopentane & -2.74$\pm$0.07 & -2.97$\pm$0.01 & -0.23$\pm$0.07  \\
39 & Benzene$\cdots$Cyclopentane & -3.14$\pm$0.08 & -3.49$\pm$0.01 & -0.36$\pm$0.08  \\
40 & Benzene$\cdots$Neopentane & -2.69$\pm$0.08 & -2.83$\pm$0.01 & -0.15$\pm$0.08  \\
41 & Uracil$\cdots$Pentane & -4.47$\pm$0.08 & -4.78$\pm$0.02 & -0.31$\pm$0.09  \\
42 & Uracil$\cdots$Cyclopentane & -3.59$\pm$0.09 & -4.07$\pm$0.02 & -0.47$\pm$0.09  \\
43 & Uracil$\cdots$Neopentane & -3.51$\pm$0.09 & -3.67$\pm$0.02 & -0.16$\pm$0.09  \\
44 & Ethene$\cdots$Pentane & -1.79$\pm$0.06 & -1.98$\pm$0.01 & -0.19$\pm$0.06  \\
45 & Ethyne$\cdots$Pentane & -1.56$\pm$0.08 & -1.71$\pm$0.01 & -0.14$\pm$0.08  \\
46 & Peptide$\cdots$Pentane & -3.82$\pm$0.08 & -4.23$\pm$0.02 & -0.41$\pm$0.08  \\
47 & Benzene$\cdots$Benzene (TS) & -2.61$\pm$0.08 & -2.81$\pm$0.01 & -0.20$\pm$0.08  \\
48 & Pyridine$\cdots$Pyridine (TS) & -3.43$\pm$0.08 & -3.49$\pm$0.02 & -0.06$\pm$0.08  \\
49 & Benzene$\cdots$Pyridine (TS) & -3.10$\pm$0.08 & -3.27$\pm$0.01 & -0.17$\pm$0.08  \\
50 & Benzene$\cdots$Ethyne (CH-$\pi$) & -2.91$\pm$0.07 & -2.84$\pm$0.01 & 0.06$\pm$0.07  \\
51 & Ethyne$\cdots$Ethyne (TS) & -1.55$\pm$0.04 & -1.53$\pm$0.01 & 0.02$\pm$0.04  \\
52 & Benzene$\cdots$AcOH (OH-$\pi$) & -4.64$\pm$0.08 & -4.70$\pm$0.02 & -0.07$\pm$0.09  \\
53 & Benzene$\cdots$AcNH$_2$ (NH-$\pi$) & -4.25$\pm$0.08 & -4.38$\pm$0.01 & -0.14$\pm$0.08  \\
54 & Benzene$\cdots$Water (OH-$\pi$) & -3.20$\pm$0.07 & -3.27$\pm$0.01 & -0.07$\pm$0.07  \\
55 & Benzene$\cdots$MeOH (OH-$\pi$) & -3.93$\pm$0.07 & -4.15$\pm$0.02 & -0.22$\pm$0.07  \\
56 & Benzene$\cdots$MeNH$_2$ (NH-$\pi$) & -3.03$\pm$0.07 & -3.18$\pm$0.01 & -0.15$\pm$0.07  \\
57 & Benzene$\cdots$Peptide (NH-$\pi$) & -5.10$\pm$0.08 & -5.23$\pm$0.02 & -0.13$\pm$0.08  \\
58 & Pyridine$\cdots$Pyridine (CH-N) & -4.22$\pm$0.12 & -4.21$\pm$0.02 & 0.01$\pm$0.13  \\
59 & Ethyne$\cdots$Water (CH-O) & -3.04$\pm$0.04 & -2.92$\pm$0.01 & 0.12$\pm$0.04  \\
60 & Ethyne$\cdots$AcOH (OH-$\pi$) & -4.98$\pm$0.06 & -4.94$\pm$0.02 & 0.04$\pm$0.07  \\
61 & Pentane$\cdots$AcOH & -2.63$\pm$0.08 & -2.89$\pm$0.02 & -0.26$\pm$0.08  \\
62 & Pentane$\cdots$AcNH$_2$ & -3.08$\pm$0.07 & -3.51$\pm$0.02 & -0.42$\pm$0.07  \\
63 & Benzene$\cdots$AcOH & -3.51$\pm$0.08 & -3.72$\pm$0.02 & -0.21$\pm$0.09  \\
64 & Peptide$\cdots$Ethene & -2.78$\pm$0.07 & -2.98$\pm$0.01 & -0.20$\pm$0.07  \\
65 & Pyridine$\cdots$Ethyne & -4.26$\pm$0.06 & -4.08$\pm$0.02 & 0.18$\pm$0.06  \\
66 & MeNH$_2$$\cdots$Pyridine & -3.81$\pm$0.06 & -3.94$\pm$0.02 & -0.13$\pm$0.06  \\
\end{longtable}\normalsize

\clearpage
\section{Comparison between DMC, CCSD(T) and CCSD(cT)-fit}\label{sec:relative_difference}
\begin{figure}[p!]
    \centering
    \includegraphics[width=1\linewidth]{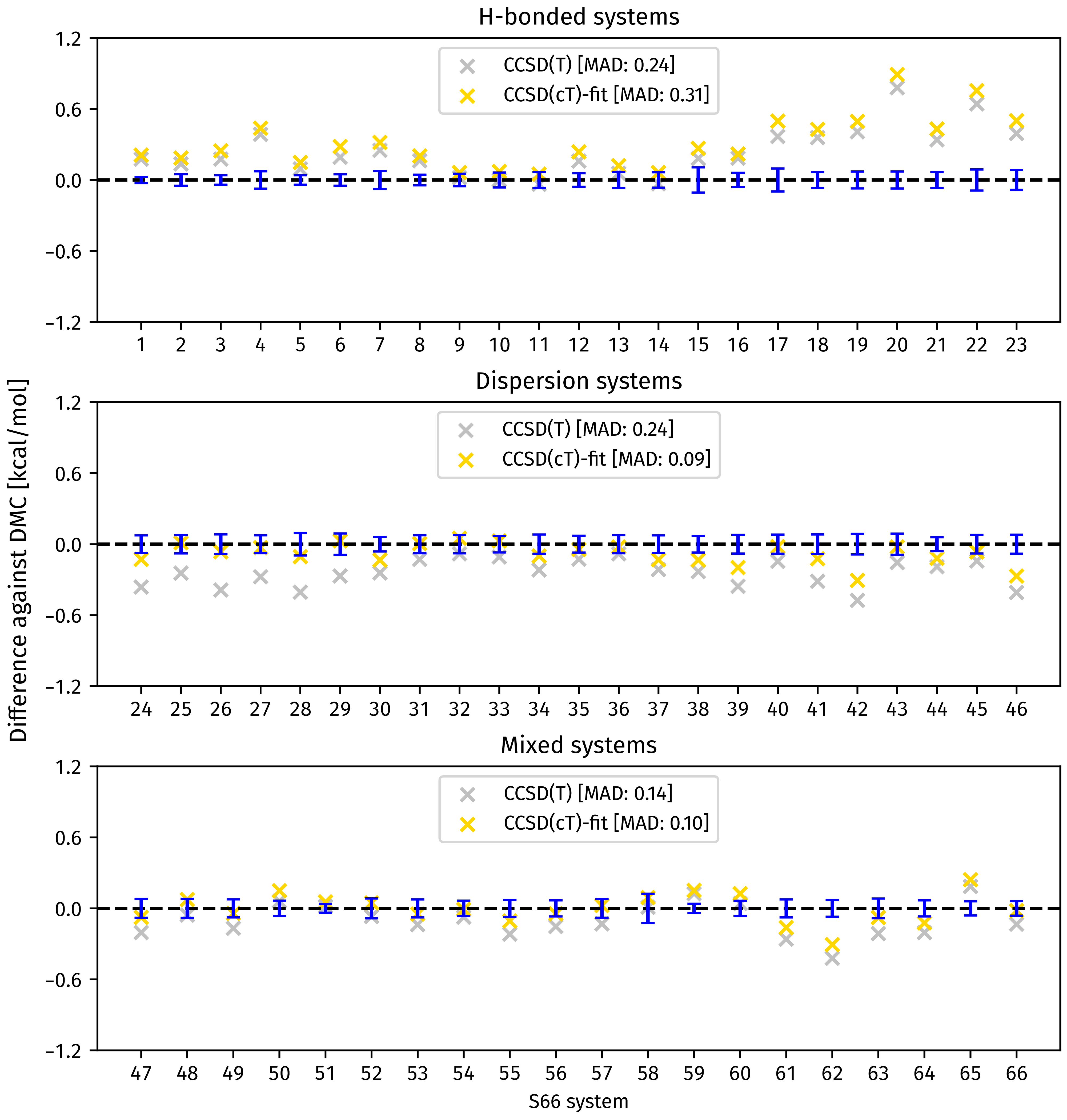}
    \caption{Difference in kcal/mol between DMC and CCSD(T) binding energies of the S66 dataset. We report the difference between previously computed CCSD(T)\cite{S66_revJM_2018,S66x8,nagy_gold} values and our DMC estimates of the interaction energies for each system in S66. The DMC statistical error bar is reported in blue. CCSD(T) estimates from from Sec.~\ref{sec:final_cc_estimates} are reported with grey stars. CCSD(cT) values estimated in this work according to the approach described in Ref.~\citenum{CCSDcT_Gruneis_2} are reported with golden stars. The interaction energies are split in three different panels according to the prevalent interaction in the molecular complex: hydrogen bond (top), dispersion (centre), and mixed (bottom).}
    \label{fig:si-differences}
\end{figure}

\begin{figure}[p!]
    \centering
    \includegraphics[width=1\linewidth]{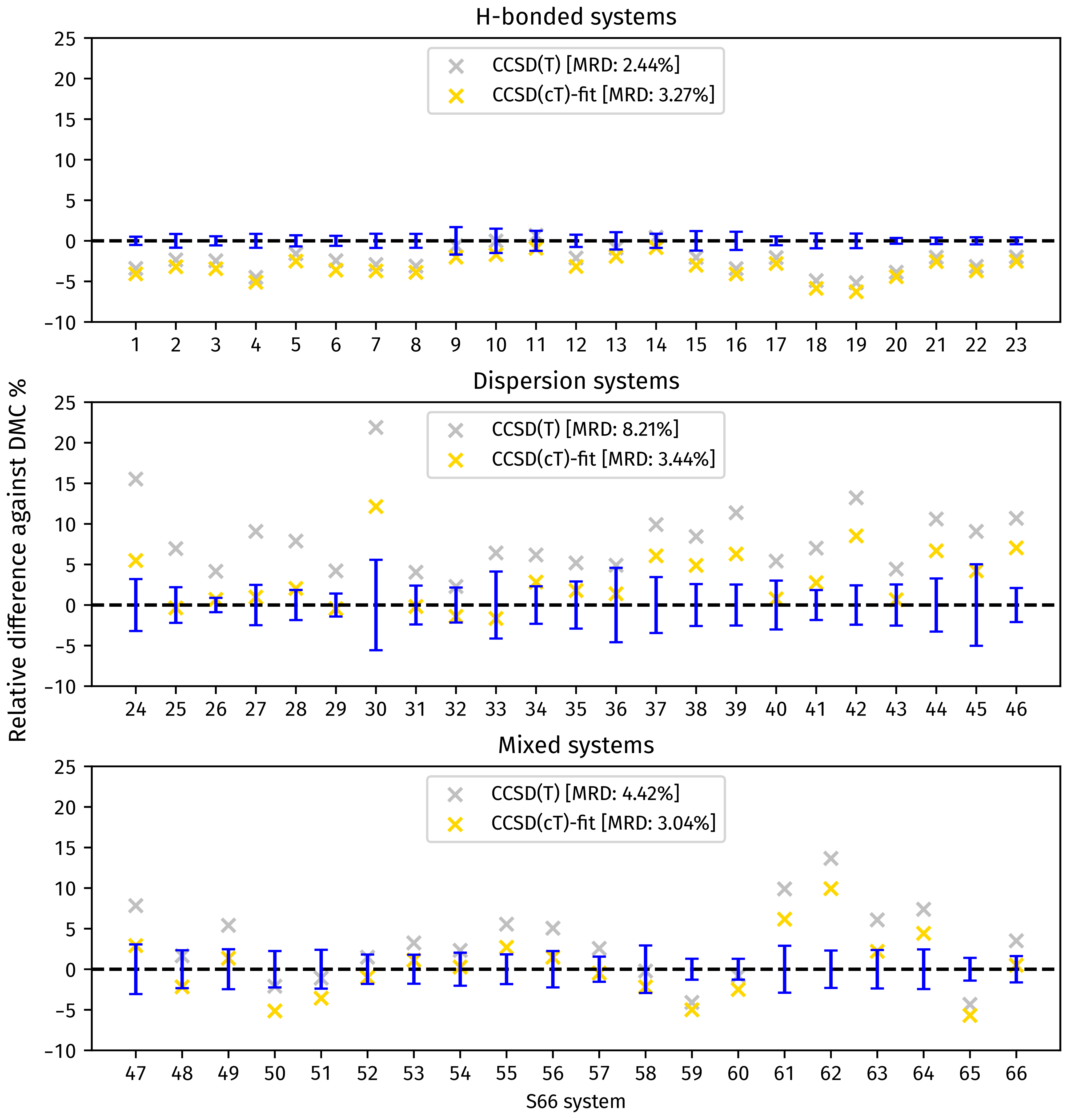}
    \caption{Relative difference between DMC and CCSD(T) binding energies of the S66 dataset. We report the difference between previously computed CCSD(T)\cite{S66_revJM_2018,S66x8,nagy_gold} values and our DMC estimates of the interaction energies for each system in S66. The DMC statistical error bar is reported in blue. CCSD(T) estimates from Sec.~\ref{sec:final_cc_estimates} are reported with grey stars. CCSD(cT) values estimated in this work according to the approach described in Ref.~\citenum{CCSDcT_Gruneis_2} are reported with
golden stars. The interaction energies are split in three different panels according to the prevalent interaction
in the molecular complex: hydrogen bond (top), dispersion (centre), and mixed (bottom).}
    \label{fig:si-relative-differences}
\end{figure}

In the main manuscript, we report the  difference between the DMC binding energy computed in this work and final estimate based on the previous CCSD(T) estimates from Refs.~\citenum{S66x8,S66_revJM_2018} and \citenum{nagy_gold}. 
In Fig.~\ref{fig:si-differences} and~\ref{fig:si-relative-differences}, we show the numerical difference (in kcal/mol) between DMC, CCSD(T) and CCSD(cT)-fit for each system in S66.
The main effect of CCSD(cT)-fit is to weaken the interaction energy $\Delta E_\text{int}$ relative to CCSD(T) in all cases.
The result is increased agreement for dispersion-dominated complexes, as discussed in the main text.
On the other hand, this causes additional errors for the H-bonded (or electrostatics-dominated) systems, with an MAD that increases from $0.24\,$kcal/mol for CCSD(T) to $0.31\,$kcal/mol for CCSD(cT)-fit.
This observation for electrostatics-dominated systems is in agreement to direct CCSD(cT) calculations computed for a subset of the S22 dataset in Ref.~\citenum{CCSDcT_Gruneis_2}.
As expected, the improvement for systems of mixed character is less clear, where CCSD(cT)-fit gets closer to DMC for some systems while not for others.
We must end with the caveat that CCSD(cT)-fit has only been parametrized to dispersion-dominated complexes, and its ability to match direct CCSD(cT) calculations is yet to be confirmed for electrostatics-dominated systems and those of mixed character.

In Fig.~\ref{fig:si-relative-differences}, we have computed the relative difference (in \%) between DMC, CCSD(T) and CCSD(cT)-fit for all systems in the S66 dataset.
It confirms the conclusion of the main manuscript, i.e., the overall good agreement between CCSD(T) and DMC.
The Mean Relative Deviation (MRD) is in fact ${\sim} 2.4 \%$ for H-bonded systems, ${\sim}8.2 \%$ for dispersion dominated systems, and ${\sim}4.4 \%$ for mixed  systems.
It shows here that while the discrepancies between DMC and CCSD(T)/CCSD(cT)-fit can be larger (in terms of absolute magnitude) than those of the dispersion-dominated systems, the overall relative difference is much smaller than the dispersion-dominated systems.
The improvement of the CCSD(cT) approach\cite{CCSDcT_Gruneis_2} for dispersion-dominated systems is also apparent, lowering MRD from ${\sim}8.2 \%$ to ${\sim} 3.4 \%$.

\clearpage

\clearpage
\section{\label{sec:localization_acetic}Validation tests for the AcOH dimer}

\begin{figure}[h!]
    \centering
    \includegraphics[width=\textwidth]{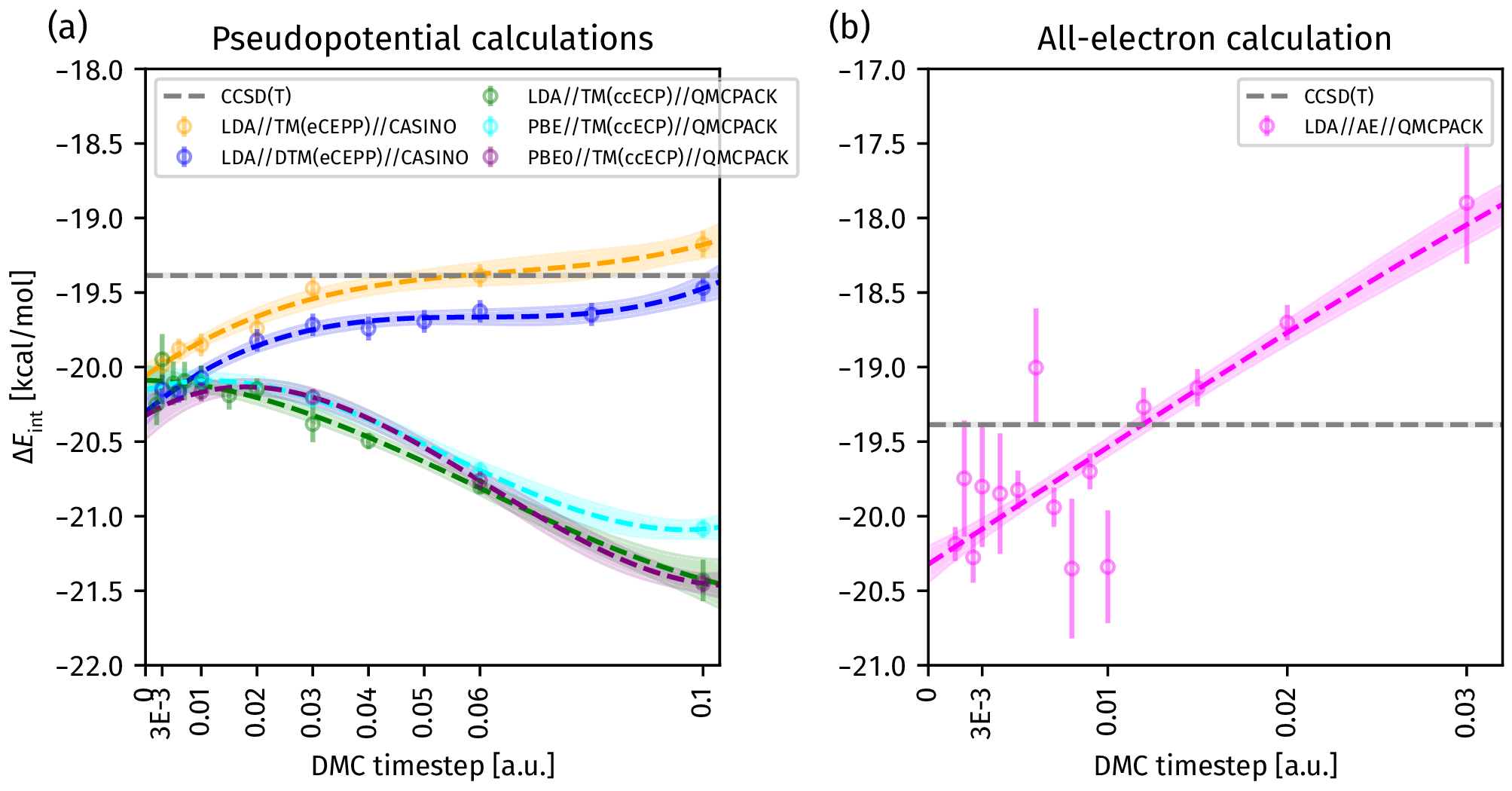}
    \caption{Analysis of the localization error for acetic acid. In (a), we report the dimer binding energy as a function of the DMC simulation time step, computed with the CASINO code and three different localization schemes (TM, DLA, and DTM) using the eCEPP\cite{CASINO_PSEUDO_eCEPP} pseudopotentials, respectively in orange, red, and blue. In addition, we show the results with the TM algorithm for the the ccECP pseudopotentials with the LDA (green), PBE (cyan) and PBE0 (purple) trial wave-functions computed with QMCPACK. In (b), we report the estimates made with an all-electron calculation, i.e. with no pseudopotentials (magenta), computed with the QMCPACK code. The cubic fits of the DMC data are plotted with dashed lines. We plot the final CCSD(T) estimate from Table~S5.}
    \label{fig:si-acetic_acid}
\end{figure}

The acetic acid (AcOH) dimer (entry 20 of the S66 dataset) is found to have the largest deviation ($\sim 1 \mathrm{kcal/mol}$) between CCSD(T) and DMC.
We thus perform additional validation tests to validate the accuracy of our estimate, utilising the DLA localization scheme~\cite{dla_zen} with an LDA trial wave-function and the eCEPP pseudopotential~\cite{CASINO_PSEUDO_eCEPP}.
In Fig.~\ref{fig:si-acetic_acid} we report the binding energy of acetic acid as a function of the simulation time step computed with following set-ups: (i) the DLA localization scheme and the eCEPP pseudopotentials with the CASINO code using an LDA nodal surface (i.e., our original setup); (ii) the TM localization scheme and the ccECP pseudopotentials with the QMCPACK code using the LDA, PBE and PBE0 nodal surface; (iii) the DTM localization scheme and the eCEPP pseudopotentials with the CASINO code for an LDA nodal surface; (iv) the all electron (AE) calculation, i.e. with no pseudopotentials, with the QMCPACK code using the LDA nodal surface.
We summarize the (cubic) extrapolated zero time step estimates in Table~\ref{tab:acetic_acid_validation} for each of the above methods.
For the AE calculations, while several data points had large stochastic error bars, the nature of the curve fitting procedure (described in Sec.~S2.4) means that these points only have a small contribution towards the fitting.
It can be seen that there is only a small range of only $0.24\,$kcal/mol between all of these procedures, suggesting that the results and conclusions reported in the main manuscript are not influenced by the simulation set up (code, localization scheme and choice of pseudopotentials) when fully converged.   


\begin{table}[h!]
\caption{\label{tab:acetic_acid_validation}Validation of the DLA localization scheme with an LDA trial wave-function for the AcOH$\cdots$AcOH dimer (ID 20). We report estimates using various trial wave-functions, localization schemes as well as with all-electron LDA. All estimates have been extrapolated to the zero time step limit for $\tau{\leq}0.1\,$au, except for the all-electron calculations, where we used data-points with $\tau{\leq}0.05\,$au.}
\begin{adjustbox}{center}
\begin{tabular}{ll}
\toprule
 & $\Delta E_\text{int}$ \\ 
\midrule
LDA//DLA(eCEPP)//CASINO & -20.17$\pm$0.07 \\
LDA//TM(eCEPP)//CASINO & -20.06$\pm$0.08 \\
LDA//DTM(eCEPP)//CASINO & -20.30$\pm$0.08 \\
LDA//TM(ccECP)//QMCPACK & -20.09$\pm$0.10 \\
PBE//TM(ccECP)//QMCPACK & -20.15$\pm$0.15 \\
PBE0//TM(ccECP)//QMCPACK & -20.33$\pm$0.16 \\
LDA//AE//QMCPACK & -20.32$\pm$0.12 \\
\bottomrule
\end{tabular}
\end{adjustbox}
\end{table}

\clearpage
\section{Convergence of the total and binding energy with respect to the DMC simulation time step}\label{sec:timestep}
In this section, we show one figure for each dimer in S66, reporting the binding energy (left panel) and the total energy (right panel) as a function of the simulation time step.
In each binding energy plot we also report the final CCSD(T) estimate obtained as described in Sec.~\ref{sec:CCSDT}.
In each plot of the binding and total energy, we also show a cubic fit (red) over the range $\tau \sim [0,0.2] \mathrm{ au} $ and a linear fit (blue) over the range $\tau \sim [0,0.01] \mathrm{ au} $.
The cubic fit is always used by the linear fit can be used to gauge the expected level of error for systems where the time step behavior changes at small time steps, as discussed in Sec.~\ref{sec:timestep_error}.
We observe that the $0.003\,$au estimates are converged to within $0.15\,$ kcal/mol w.r.t. the zero time step limit estimates across the entire S66 dataset, with an MAD of $0.03\,$kcal/mol.

\newpage

\begin{figure}[!h]
    \includegraphics[width=6.69in]{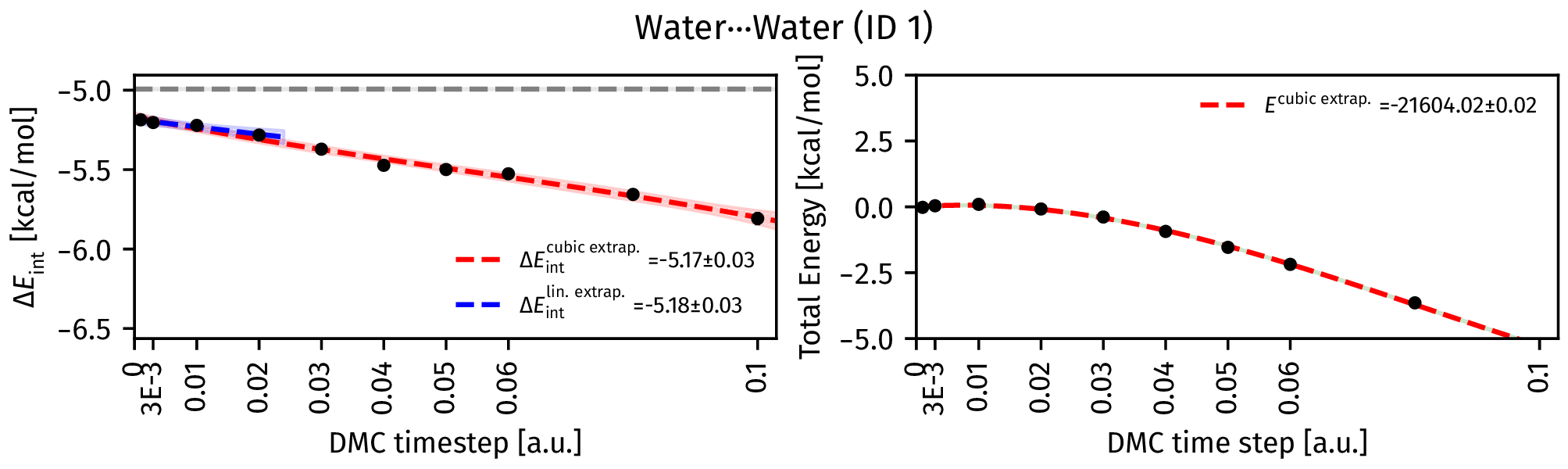}
    \caption{\label{fig:dimer_01} The time step dependence of $\Delta E_\text{int}$ and the total energy of the dimer complex for the Water$\cdots$Water (ID 1) dimer.The dotted gray line represents the CCSD(T) reference in Table~\ref{sec:final_cc_estimates} and the black markers with stochastic 1$\sigma$ error bars represent the DMC estimate for each time step.}
\end{figure}
    
\begin{figure}[!h]
    \includegraphics[width=6.69in]{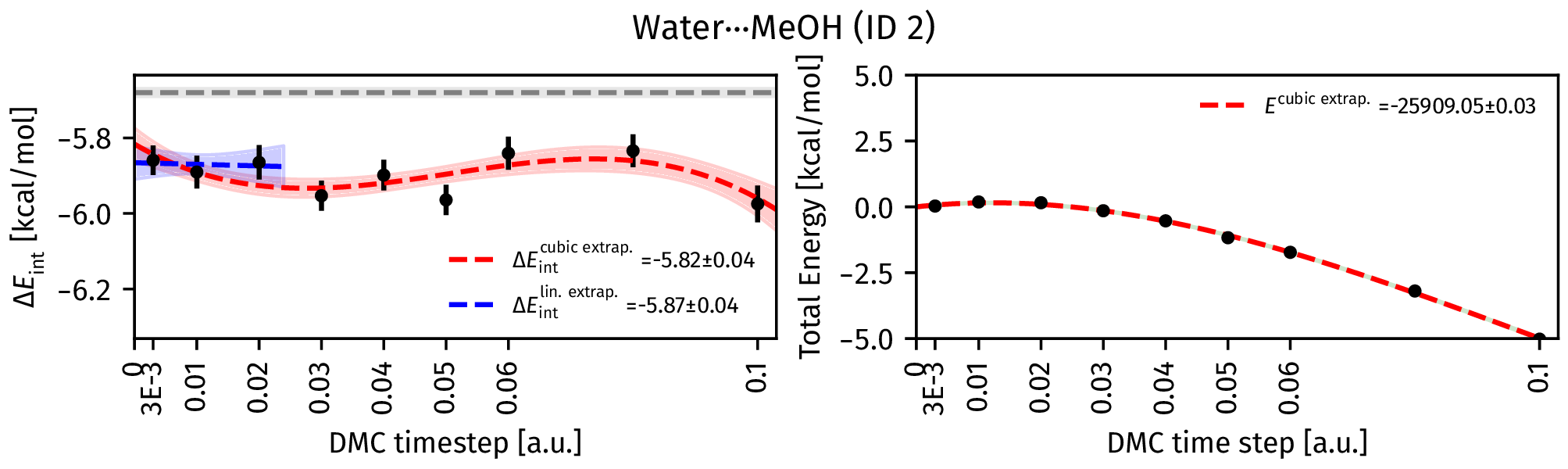}
    \caption{\label{fig:dimer_02} The time step dependence of $\Delta E_\text{int}$ and the total energy of the dimer complex for the Water$\cdots$MeOH (ID 2) dimer.The dotted gray line represents the CCSD(T) reference in Table~\ref{sec:final_cc_estimates} and the black markers with stochastic 1$\sigma$ error bars represent the DMC estimate for each time step.}
\end{figure}
    
\begin{figure}[!h]
    \includegraphics[width=6.69in]{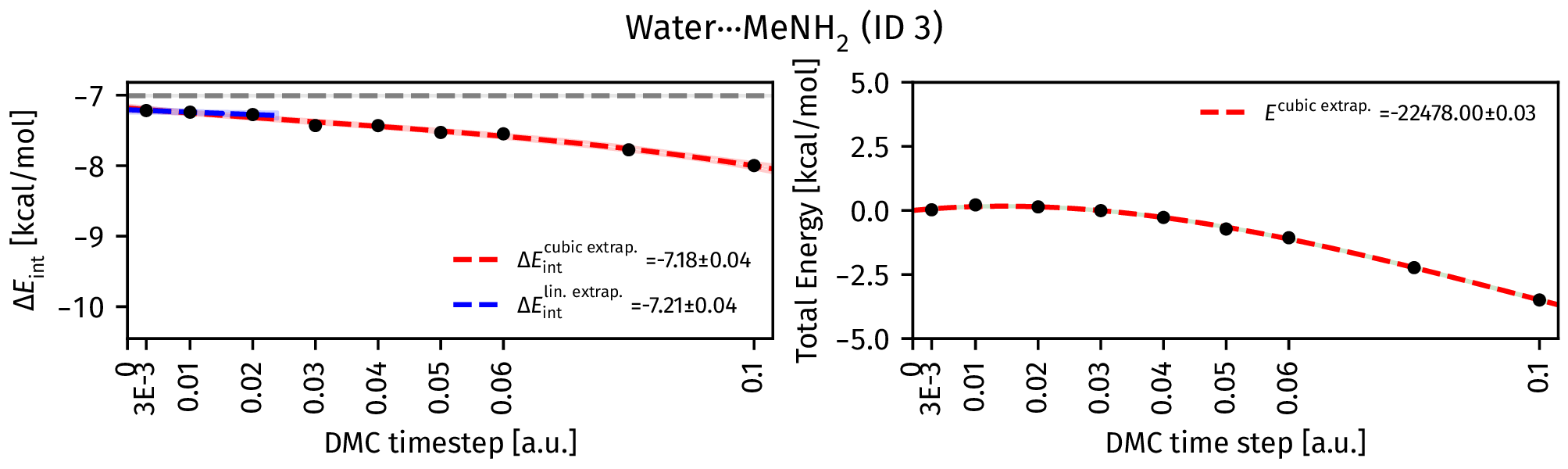}
    \caption{\label{fig:dimer_03} The time step dependence of $\Delta E_\text{int}$ and the total energy of the dimer complex for the Water$\cdots$MeNH$_2$ (ID 3) dimer.The dotted gray line represents the CCSD(T) reference in Table~\ref{sec:final_cc_estimates} and the black markers with stochastic 1$\sigma$ error bars represent the DMC estimate for each time step.}
\end{figure}
    
\begin{figure}[!h]
    \includegraphics[width=6.69in]{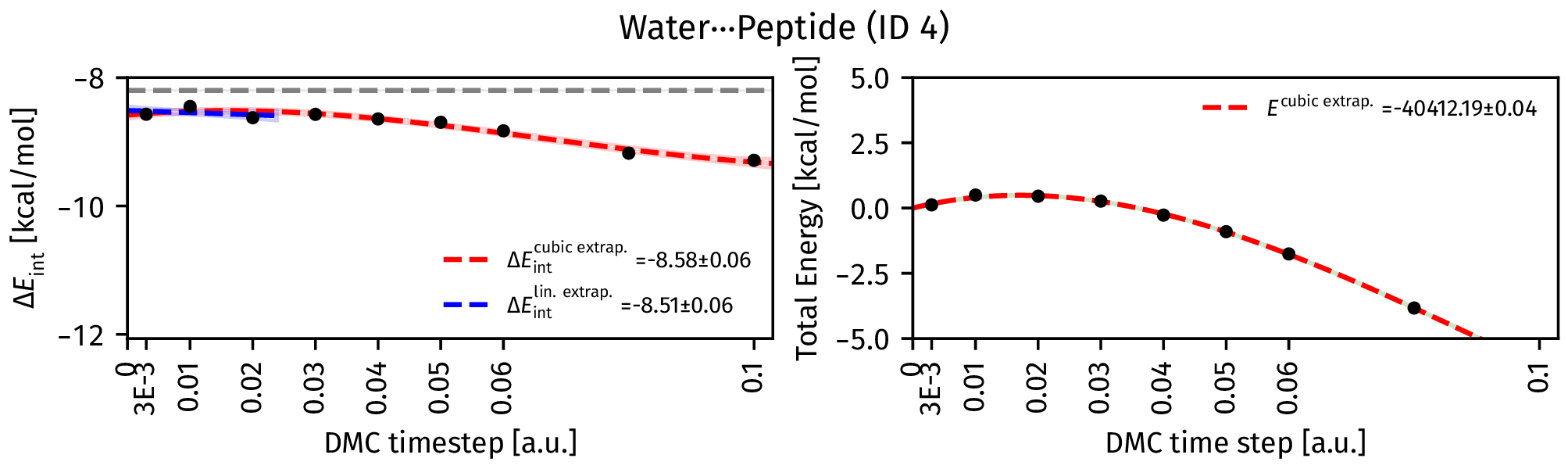}
    \caption{\label{fig:dimer_04} The time step dependence of $\Delta E_\text{int}$ and the total energy of the dimer complex for the Water$\cdots$Peptide (ID 4) dimer.The dotted gray line represents the CCSD(T) reference in Table~\ref{sec:final_cc_estimates} and the black markers with stochastic 1$\sigma$ error bars represent the DMC estimate for each time step.}
\end{figure}
    
\begin{figure}[!h]
    \includegraphics[width=6.69in]{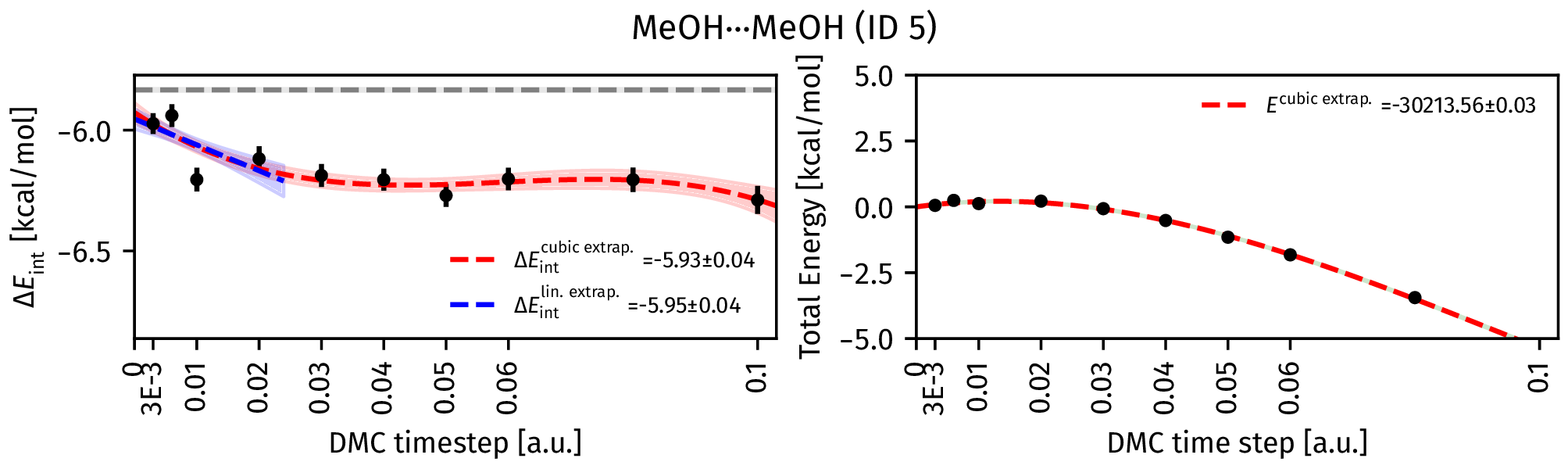}
    \caption{\label{fig:dimer_05} The time step dependence of $\Delta E_\text{int}$ and the total energy of the dimer complex for the MeOH$\cdots$MeOH (ID 5) dimer.The dotted gray line represents the CCSD(T) reference in Table~\ref{sec:final_cc_estimates} and the black markers with stochastic 1$\sigma$ error bars represent the DMC estimate for each time step.}
\end{figure}
    
\begin{figure}[!h]
    \includegraphics[width=6.69in]{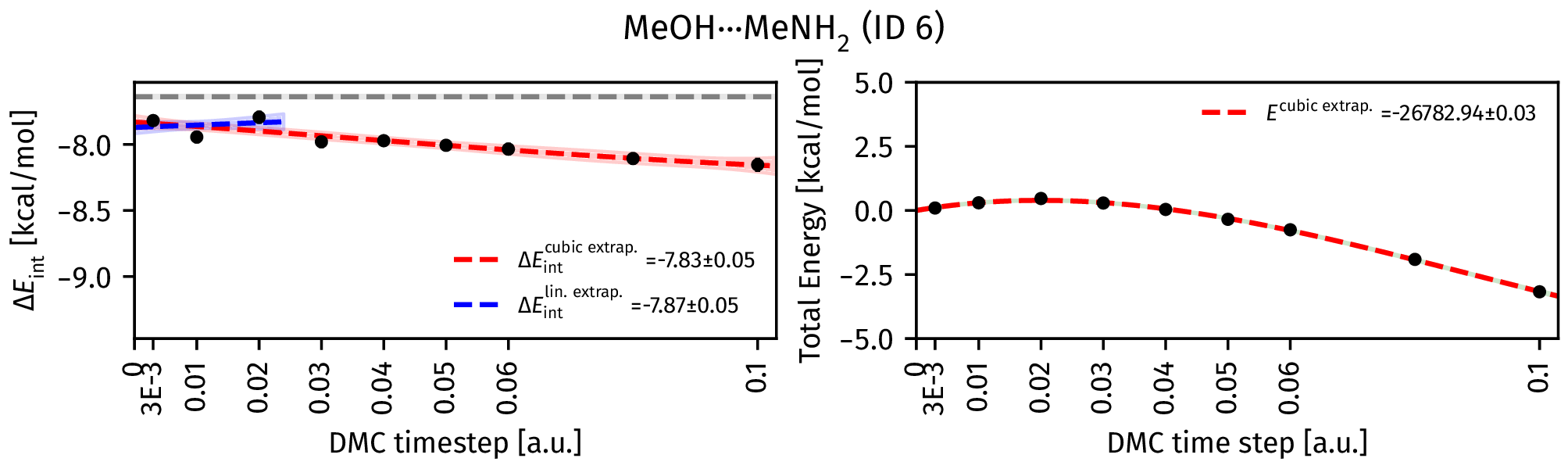}
    \caption{\label{fig:dimer_06} The time step dependence of $\Delta E_\text{int}$ and the total energy of the dimer complex for the MeOH$\cdots$MeNH$_2$ (ID 6) dimer.The dotted gray line represents the CCSD(T) reference in Table~\ref{sec:final_cc_estimates} and the black markers with stochastic 1$\sigma$ error bars represent the DMC estimate for each time step.}
\end{figure}
    
\begin{figure}[!h]
    \includegraphics[width=6.69in]{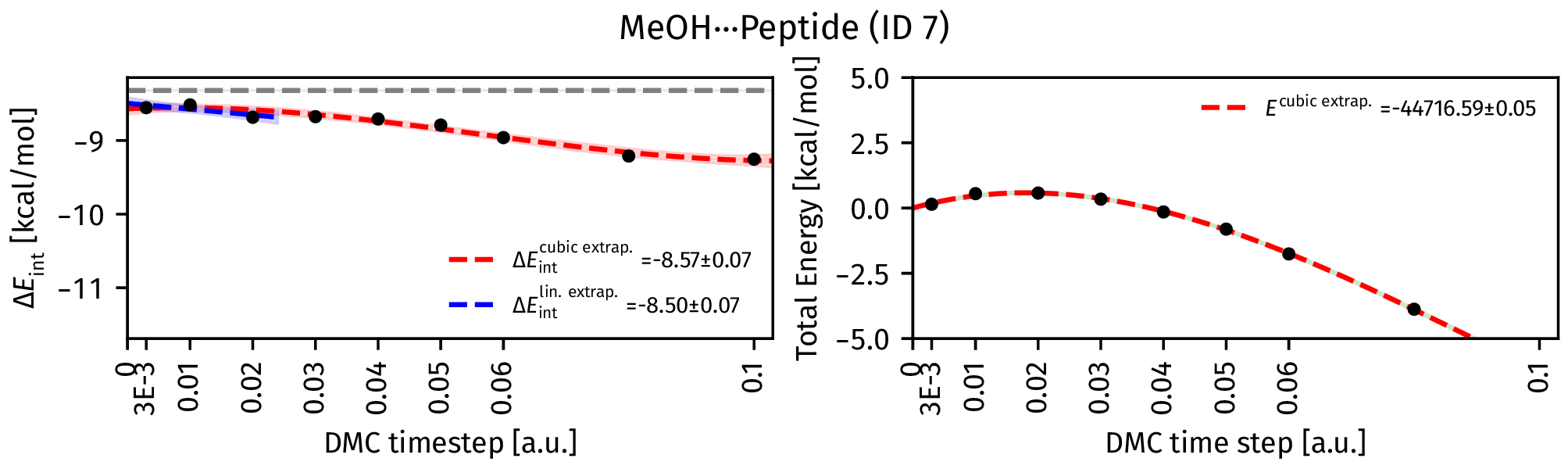}
    \caption{\label{fig:dimer_07} The time step dependence of $\Delta E_\text{int}$ and the total energy of the dimer complex for the MeOH$\cdots$Peptide (ID 7) dimer.The dotted gray line represents the CCSD(T) reference in Table~\ref{sec:final_cc_estimates} and the black markers with stochastic 1$\sigma$ error bars represent the DMC estimate for each time step.}
\end{figure}
    
\begin{figure}[!h]
    \includegraphics[width=6.69in]{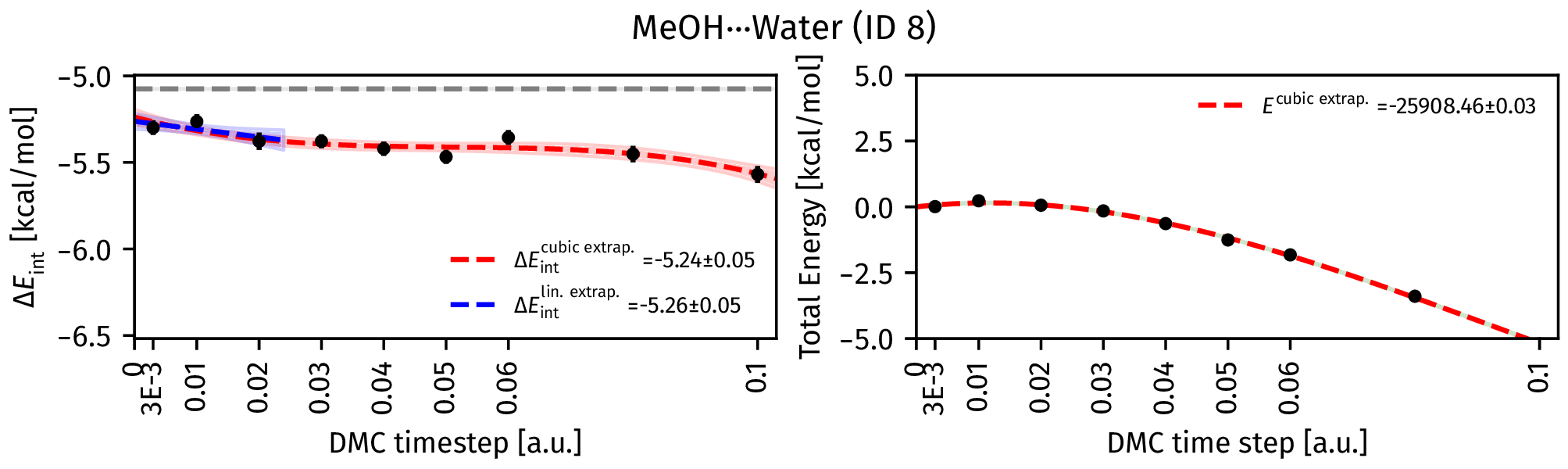}
    \caption{\label{fig:dimer_08} The time step dependence of $\Delta E_\text{int}$ and the total energy of the dimer complex for the MeOH$\cdots$Water (ID 8) dimer.The dotted gray line represents the CCSD(T) reference in Table~\ref{sec:final_cc_estimates} and the black markers with stochastic 1$\sigma$ error bars represent the DMC estimate for each time step.}
\end{figure}
    
\begin{figure}[!h]
    \includegraphics[width=6.69in]{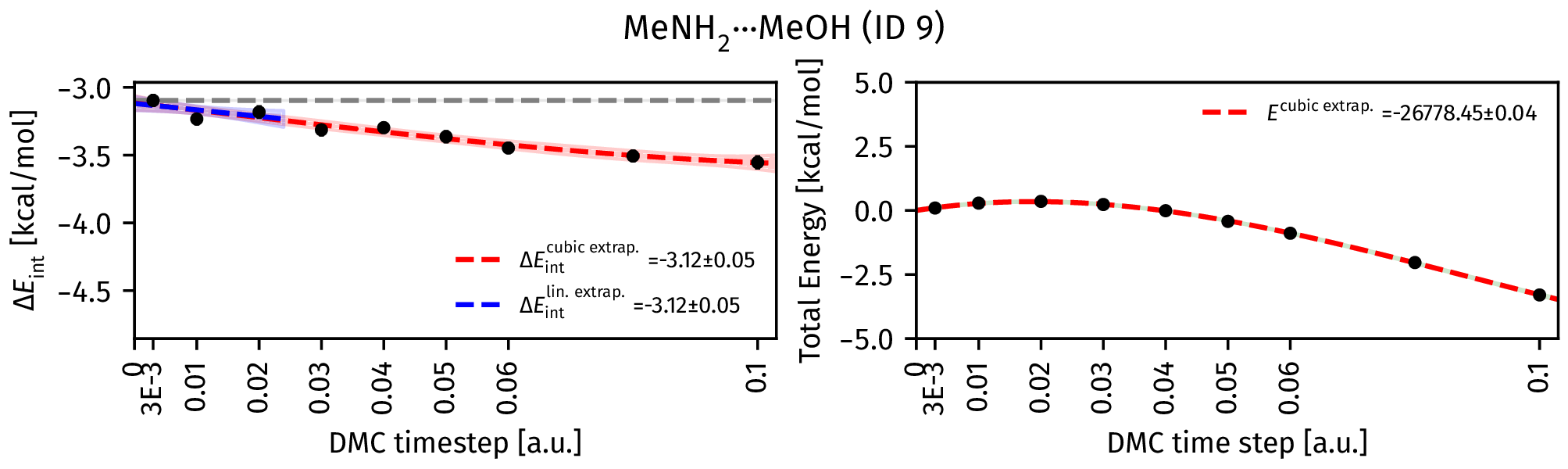}
    \caption{\label{fig:dimer_09} The time step dependence of $\Delta E_\text{int}$ and the total energy of the dimer complex for the MeNH$_2$$\cdots$MeOH (ID 9) dimer.The dotted gray line represents the CCSD(T) reference in Table~\ref{sec:final_cc_estimates} and the black markers with stochastic 1$\sigma$ error bars represent the DMC estimate for each time step.}
\end{figure}
    
\begin{figure}[!h]
    \includegraphics[width=6.69in]{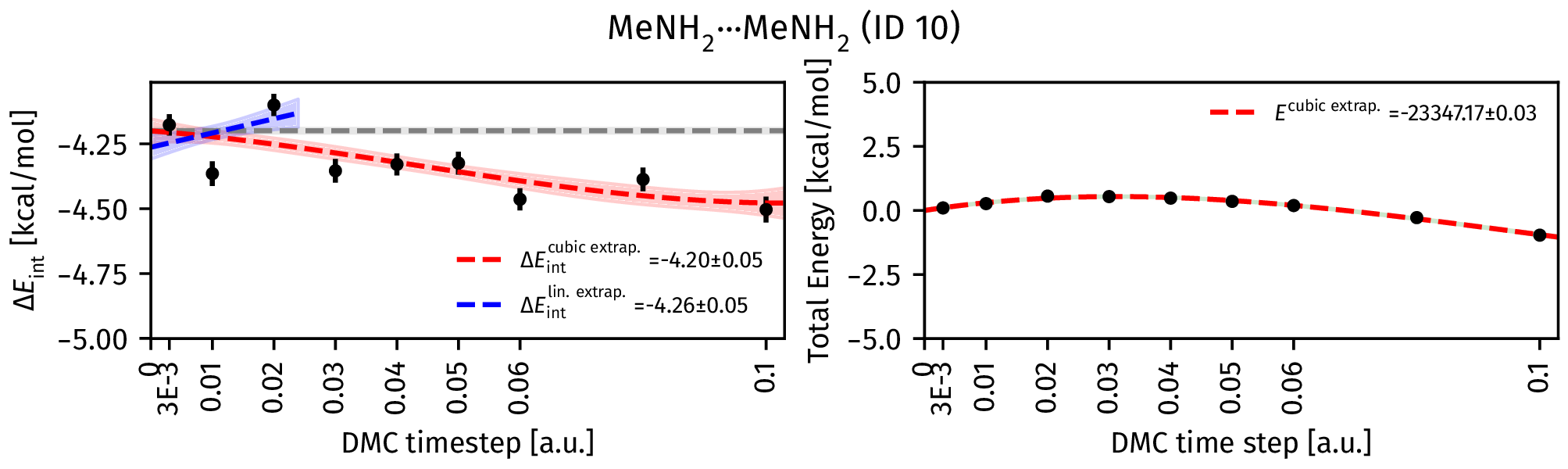}
    \caption{\label{fig:dimer_10} The time step dependence of $\Delta E_\text{int}$ and the total energy of the dimer complex for the MeNH$_2$$\cdots$MeNH$_2$ (ID 10) dimer.The dotted gray line represents the CCSD(T) reference in Table~\ref{sec:final_cc_estimates} and the black markers with stochastic 1$\sigma$ error bars represent the DMC estimate for each time step.}
\end{figure}
    
\begin{figure}[!h]
    \includegraphics[width=6.69in]{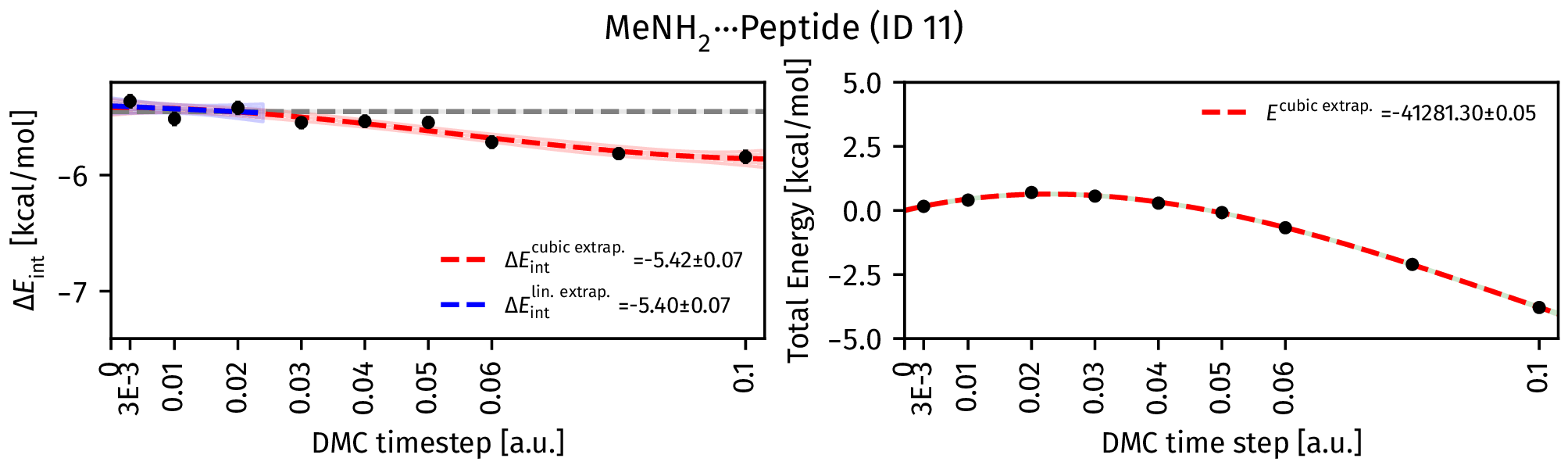}
    \caption{\label{fig:dimer_11} The time step dependence of $\Delta E_\text{int}$ and the total energy of the dimer complex for the MeNH$_2$$\cdots$Peptide (ID 11) dimer.The dotted gray line represents the CCSD(T) reference in Table~\ref{sec:final_cc_estimates} and the black markers with stochastic 1$\sigma$ error bars represent the DMC estimate for each time step.}
\end{figure}
    
\begin{figure}[!h]
    \includegraphics[width=6.69in]{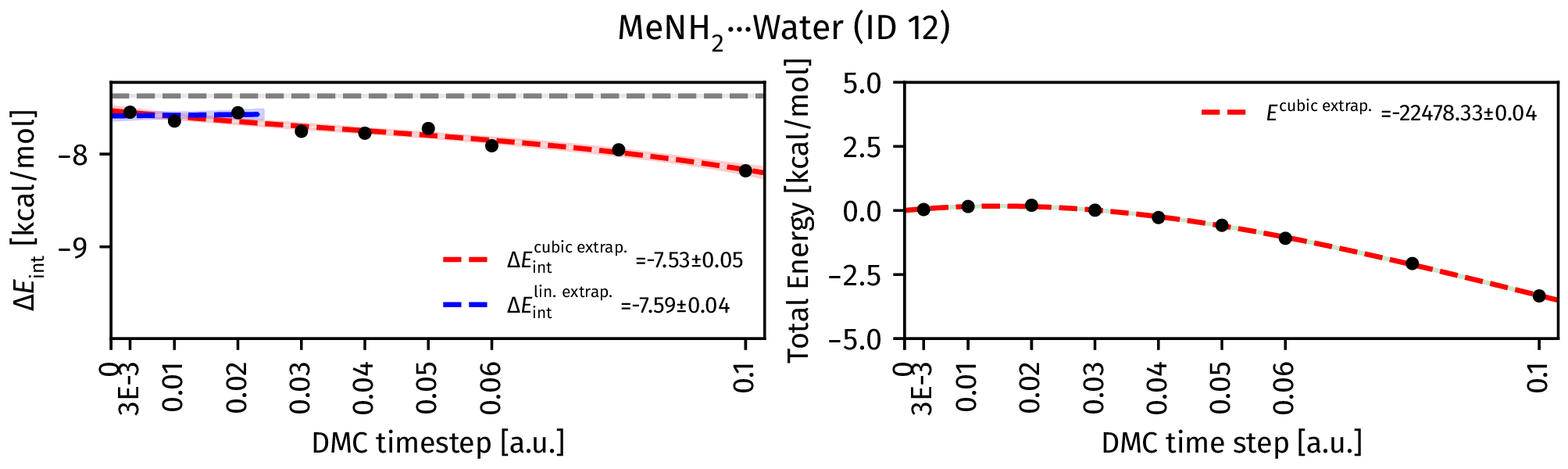}
    \caption{\label{fig:dimer_12} The time step dependence of $\Delta E_\text{int}$ and the total energy of the dimer complex for the MeNH$_2$$\cdots$Water (ID 12) dimer.The dotted gray line represents the CCSD(T) reference in Table~\ref{sec:final_cc_estimates} and the black markers with stochastic 1$\sigma$ error bars represent the DMC estimate for each time step.}
\end{figure}
    
\begin{figure}[!h]
    \includegraphics[width=6.69in]{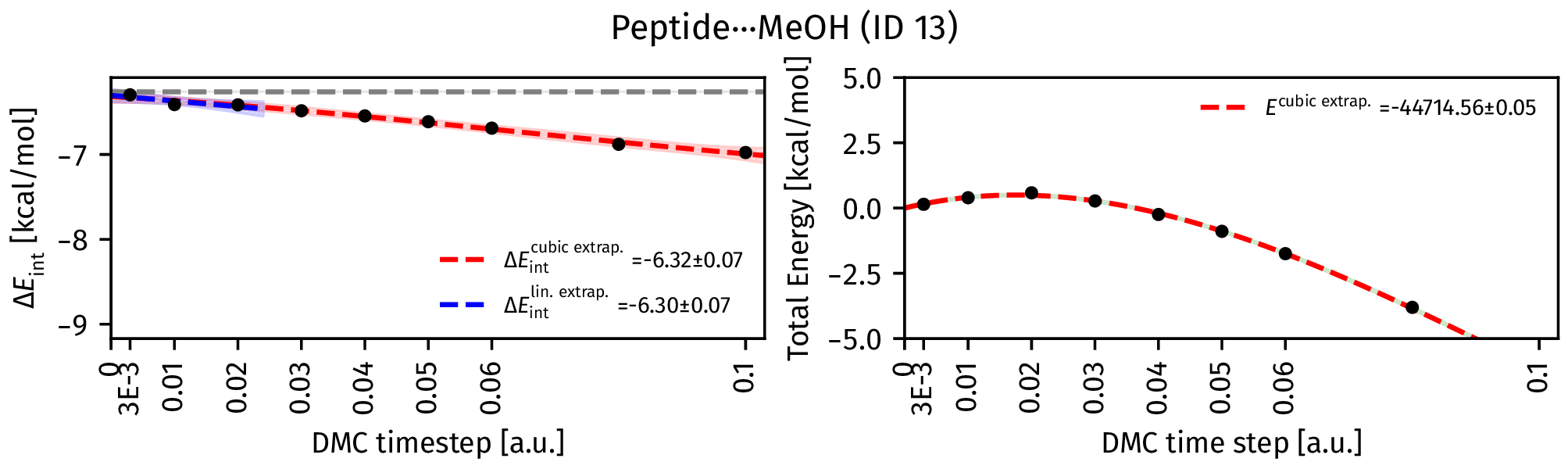}
    \caption{\label{fig:dimer_13} The time step dependence of $\Delta E_\text{int}$ and the total energy of the dimer complex for the Peptide$\cdots$MeOH (ID 13) dimer.The dotted gray line represents the CCSD(T) reference in Table~\ref{sec:final_cc_estimates} and the black markers with stochastic 1$\sigma$ error bars represent the DMC estimate for each time step.}
\end{figure}
    
\begin{figure}[!h]
    \includegraphics[width=6.69in]{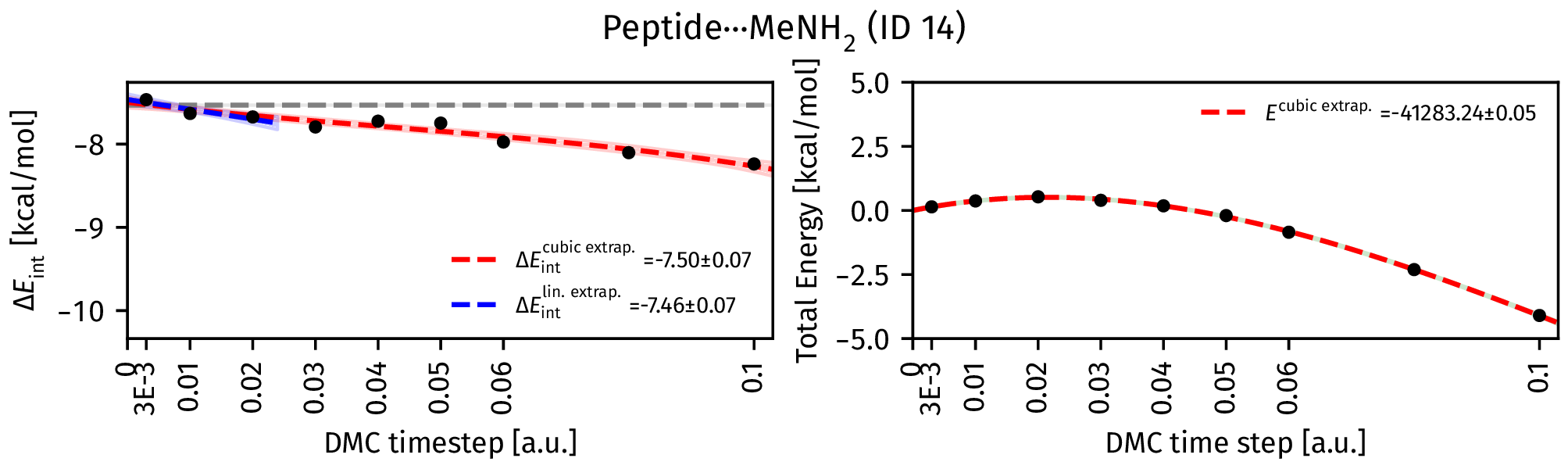}
    \caption{\label{fig:dimer_14} The time step dependence of $\Delta E_\text{int}$ and the total energy of the dimer complex for the Peptide$\cdots$MeNH$_2$ (ID 14) dimer.The dotted gray line represents the CCSD(T) reference in Table~\ref{sec:final_cc_estimates} and the black markers with stochastic 1$\sigma$ error bars represent the DMC estimate for each time step.}
\end{figure}
    
\begin{figure}[!h]
    \includegraphics[width=6.69in]{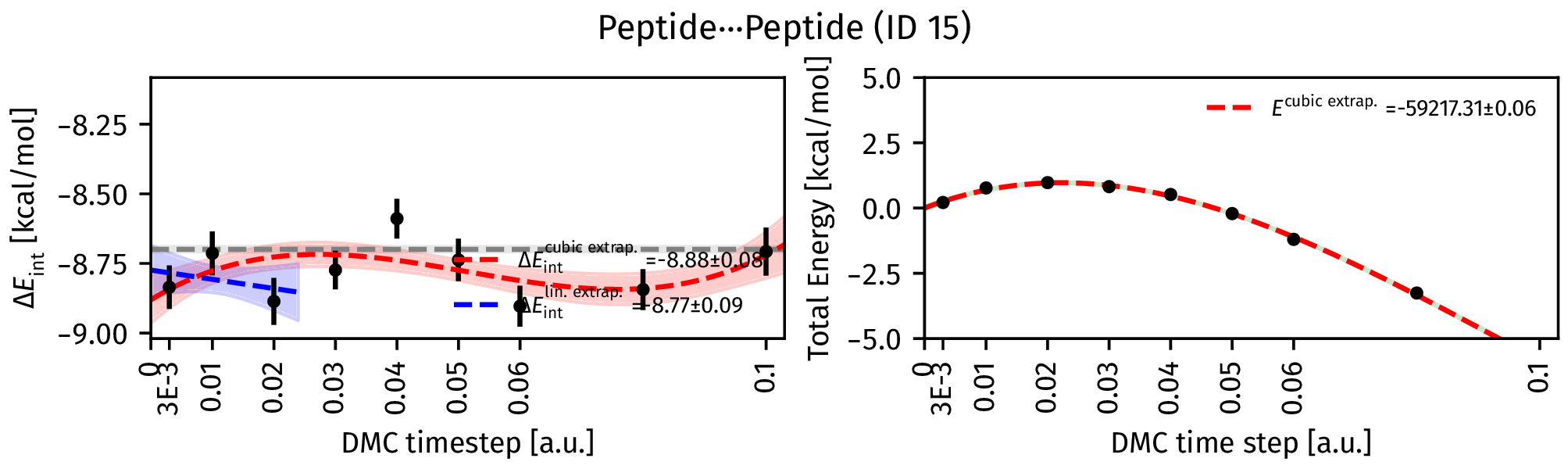}
    \caption{\label{fig:dimer_15} The time step dependence of $\Delta E_\text{int}$ and the total energy of the dimer complex for the Peptide$\cdots$Peptide (ID 15) dimer.The dotted gray line represents the CCSD(T) reference in Table~\ref{sec:final_cc_estimates} and the black markers with stochastic 1$\sigma$ error bars represent the DMC estimate for each time step.}
\end{figure}
    
\begin{figure}[!h]
    \includegraphics[width=6.69in]{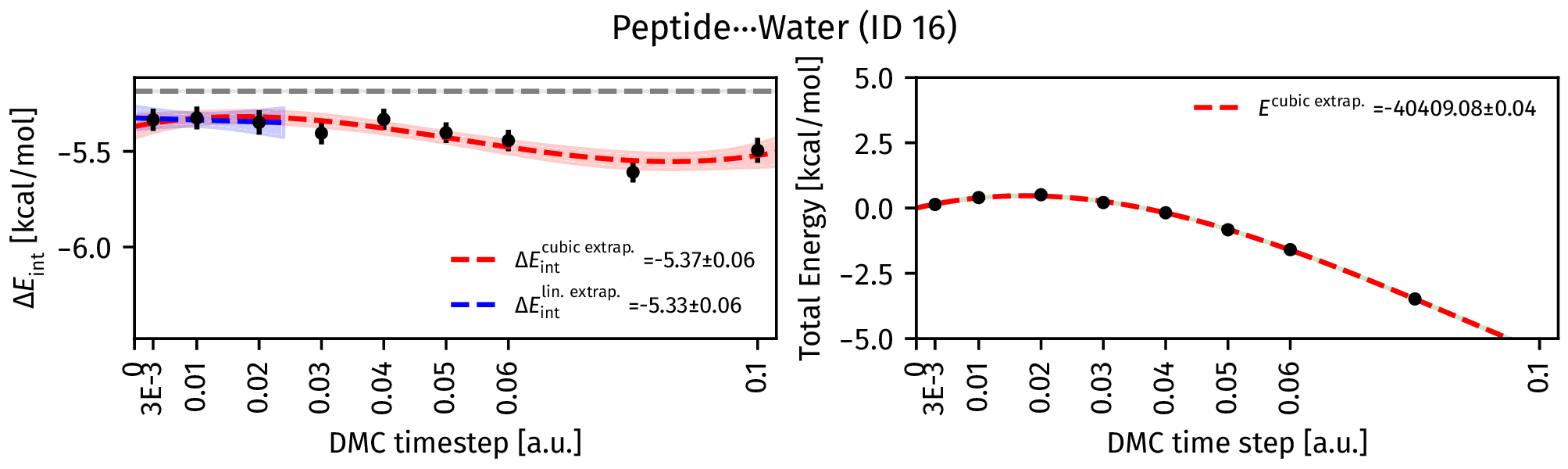}
    \caption{\label{fig:dimer_16} The time step dependence of $\Delta E_\text{int}$ and the total energy of the dimer complex for the Peptide$\cdots$Water (ID 16) dimer.The dotted gray line represents the CCSD(T) reference in Table~\ref{sec:final_cc_estimates} and the black markers with stochastic 1$\sigma$ error bars represent the DMC estimate for each time step.}
\end{figure}
    
\begin{figure}[!h]
    \includegraphics[width=6.69in]{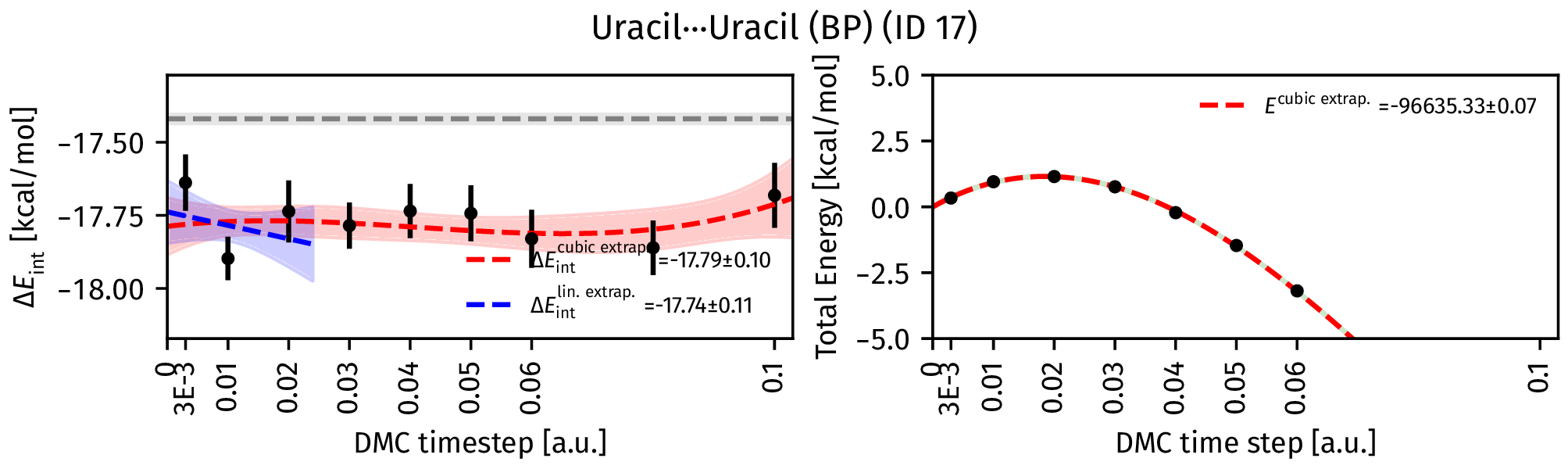}
    \caption{\label{fig:dimer_17} The time step dependence of $\Delta E_\text{int}$ and the total energy of the dimer complex for the Uracil$\cdots$Uracil (BP) (ID 17) dimer.The dotted gray line represents the CCSD(T) reference in Table~\ref{sec:final_cc_estimates} and the black markers with stochastic 1$\sigma$ error bars represent the DMC estimate for each time step.}
\end{figure}
    
\begin{figure}[!h]
    \includegraphics[width=6.69in]{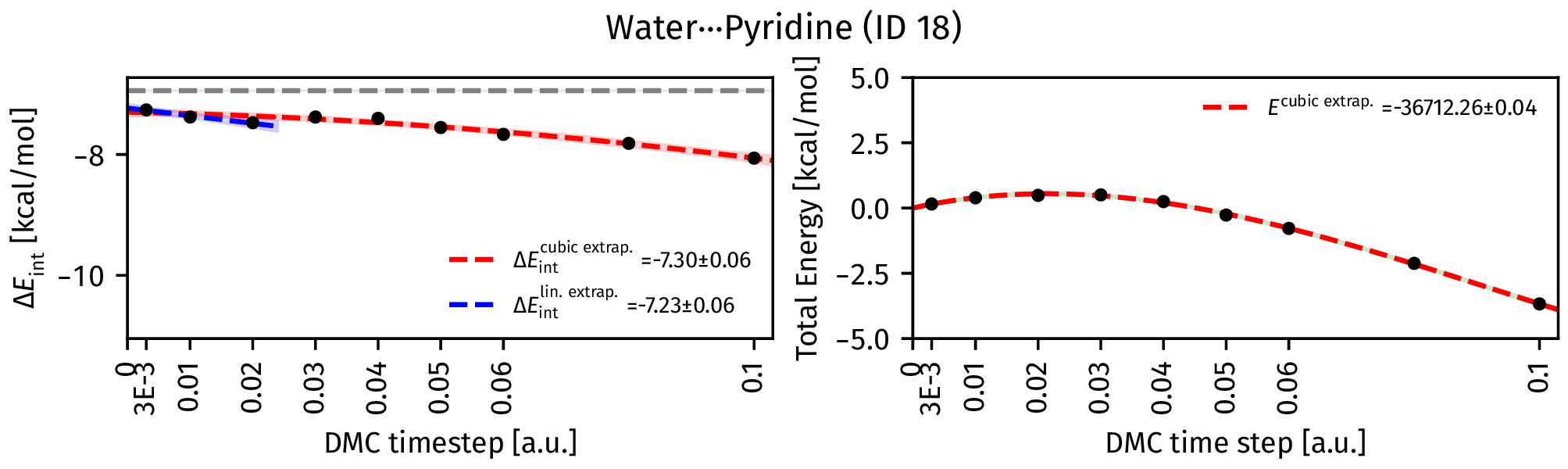}
    \caption{\label{fig:dimer_18} The time step dependence of $\Delta E_\text{int}$ and the total energy of the dimer complex for the Water$\cdots$Pyridine (ID 18) dimer.The dotted gray line represents the CCSD(T) reference in Table~\ref{sec:final_cc_estimates} and the black markers with stochastic 1$\sigma$ error bars represent the DMC estimate for each time step.}
\end{figure}
    
\begin{figure}[!h]
    \includegraphics[width=6.69in]{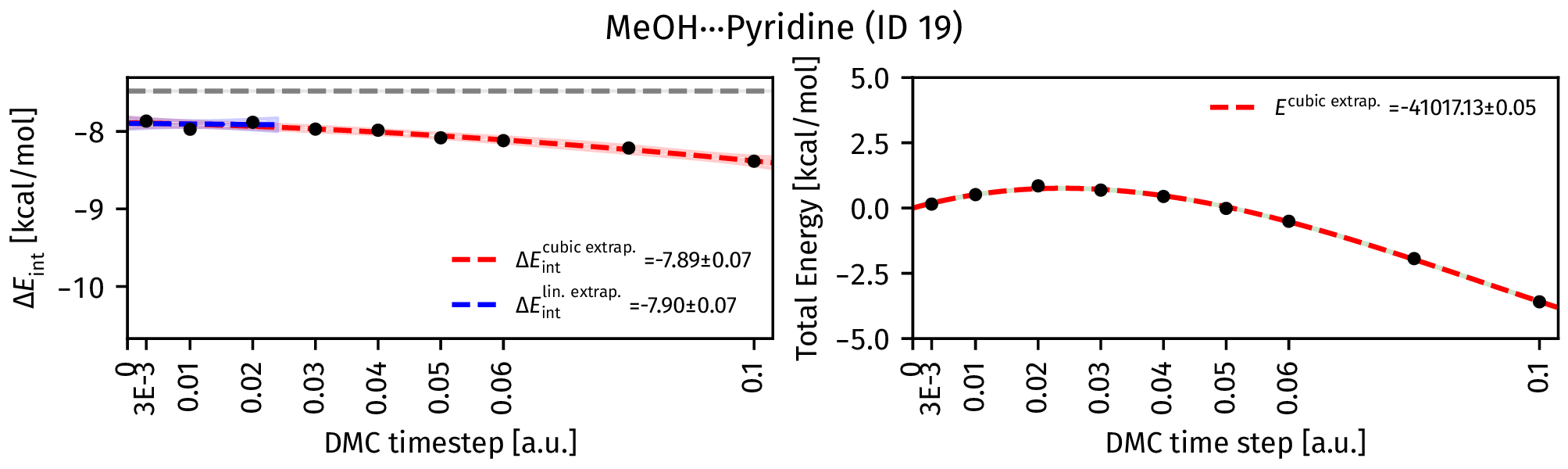}
    \caption{\label{fig:dimer_19} The time step dependence of $\Delta E_\text{int}$ and the total energy of the dimer complex for the MeOH$\cdots$Pyridine (ID 19) dimer.The dotted gray line represents the CCSD(T) reference in Table~\ref{sec:final_cc_estimates} and the black markers with stochastic 1$\sigma$ error bars represent the DMC estimate for each time step.}
\end{figure}
    
\begin{figure}[!h]
    \includegraphics[width=6.69in]{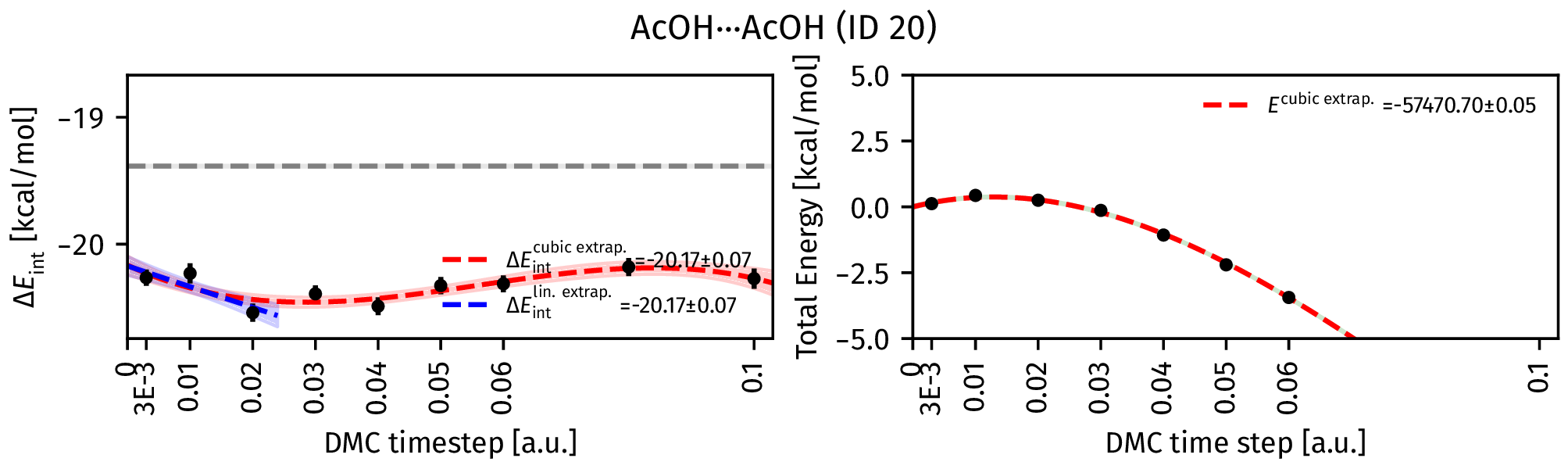}
    \caption{\label{fig:dimer_20} The time step dependence of $\Delta E_\text{int}$ and the total energy of the dimer complex for the AcOH$\cdots$AcOH (ID 20) dimer.The dotted gray line represents the CCSD(T) reference in Table~\ref{sec:final_cc_estimates} and the black markers with stochastic 1$\sigma$ error bars represent the DMC estimate for each time step.}
\end{figure}
    
\begin{figure}[!h]
    \includegraphics[width=6.69in]{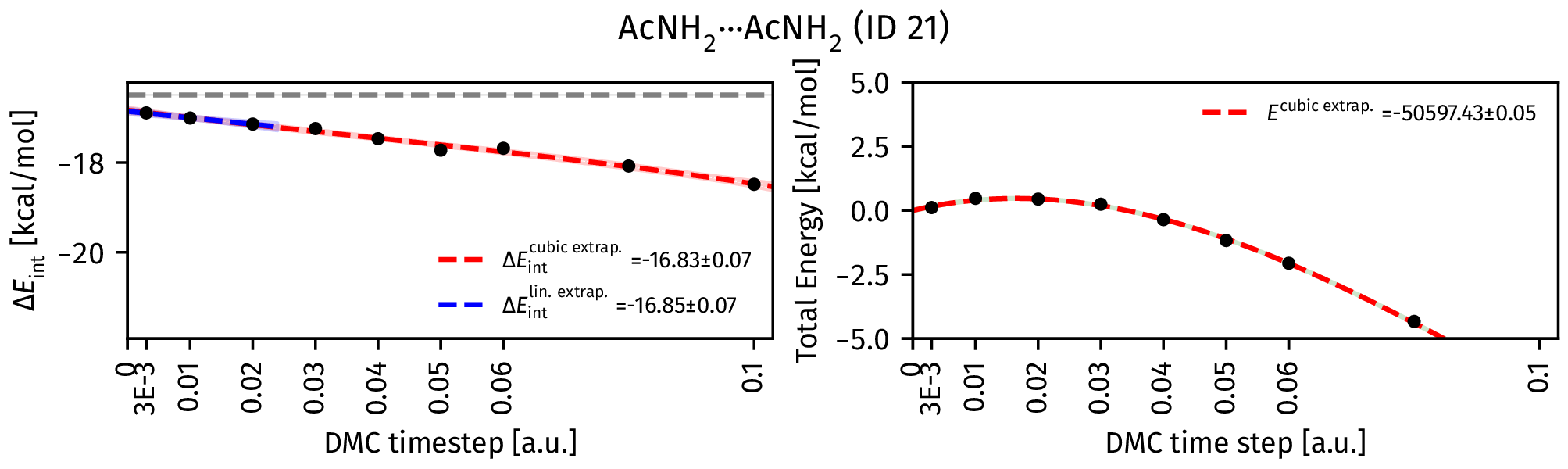}
    \caption{\label{fig:dimer_21} The time step dependence of $\Delta E_\text{int}$ and the total energy of the dimer complex for the AcNH$_2$$\cdots$AcNH$_2$ (ID 21) dimer.The dotted gray line represents the CCSD(T) reference in Table~\ref{sec:final_cc_estimates} and the black markers with stochastic 1$\sigma$ error bars represent the DMC estimate for each time step.}
\end{figure}
    
\begin{figure}[!h]
    \includegraphics[width=6.69in]{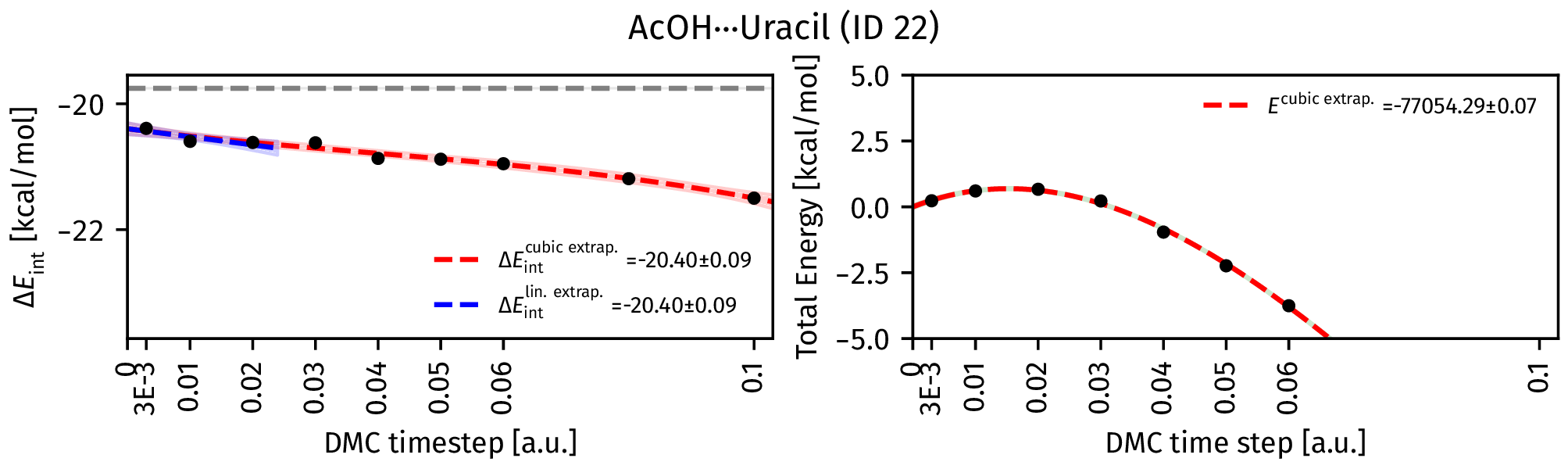}
    \caption{\label{fig:dimer_22} The time step dependence of $\Delta E_\text{int}$ and the total energy of the dimer complex for the AcOH$\cdots$Uracil (ID 22) dimer.The dotted gray line represents the CCSD(T) reference in Table~\ref{sec:final_cc_estimates} and the black markers with stochastic 1$\sigma$ error bars represent the DMC estimate for each time step.}
\end{figure}
    
\begin{figure}[!h]
    \includegraphics[width=6.69in]{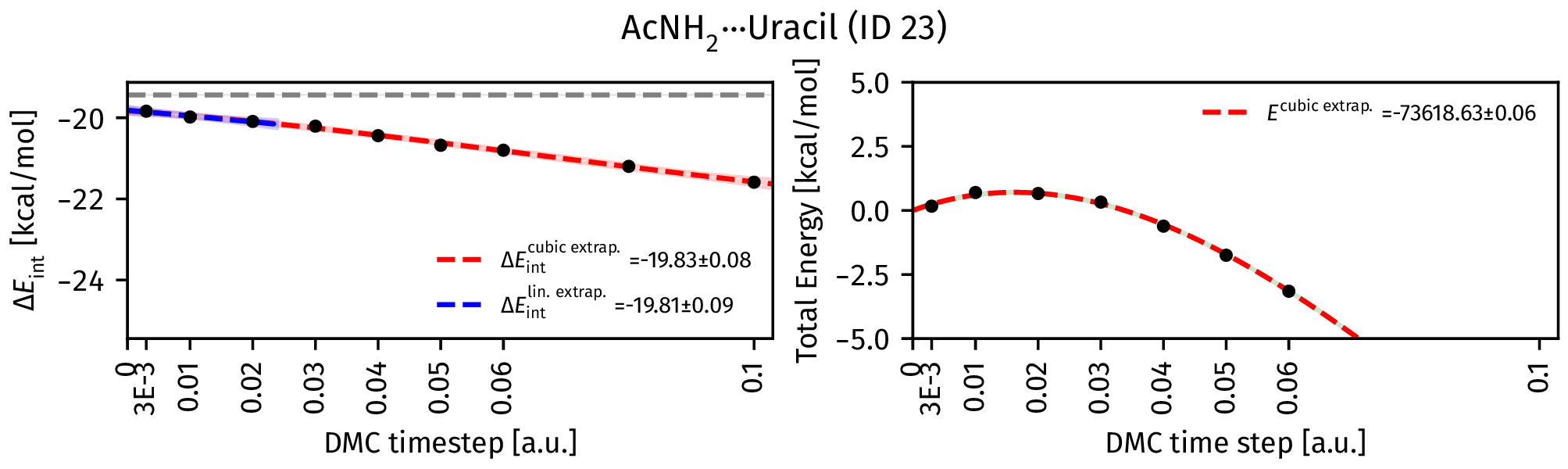}
    \caption{\label{fig:dimer_23} The time step dependence of $\Delta E_\text{int}$ and the total energy of the dimer complex for the AcNH$_2$$\cdots$Uracil (ID 23) dimer.The dotted gray line represents the CCSD(T) reference in Table~\ref{sec:final_cc_estimates} and the black markers with stochastic 1$\sigma$ error bars represent the DMC estimate for each time step.}
\end{figure}

\clearpage
    
\begin{figure}[!h]
    \includegraphics[width=6.69in]{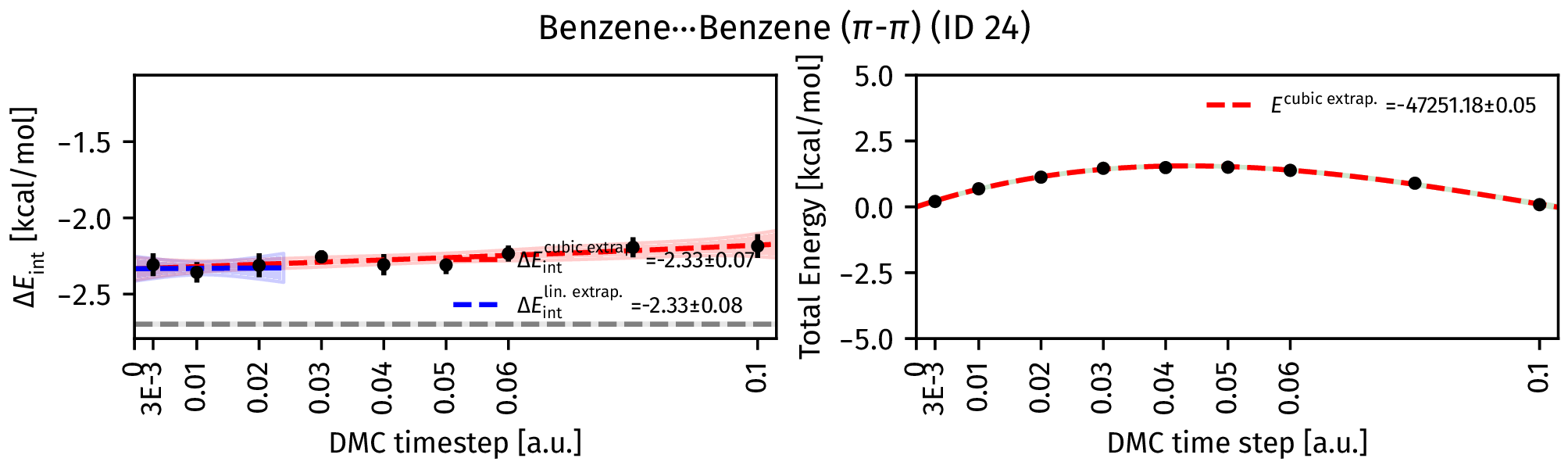}
    \caption{\label{fig:dimer_24} The time step dependence of $\Delta E_\text{int}$ and the total energy of the dimer complex for the Benzene$\cdots$Benzene ($\pi$-$\pi$) (ID 24) dimer.The dotted gray line represents the CCSD(T) reference in Table~\ref{sec:final_cc_estimates} and the black markers with stochastic 1$\sigma$ error bars represent the DMC estimate for each time step.}
\end{figure}
    
\begin{figure}[!h]
    \includegraphics[width=6.69in]{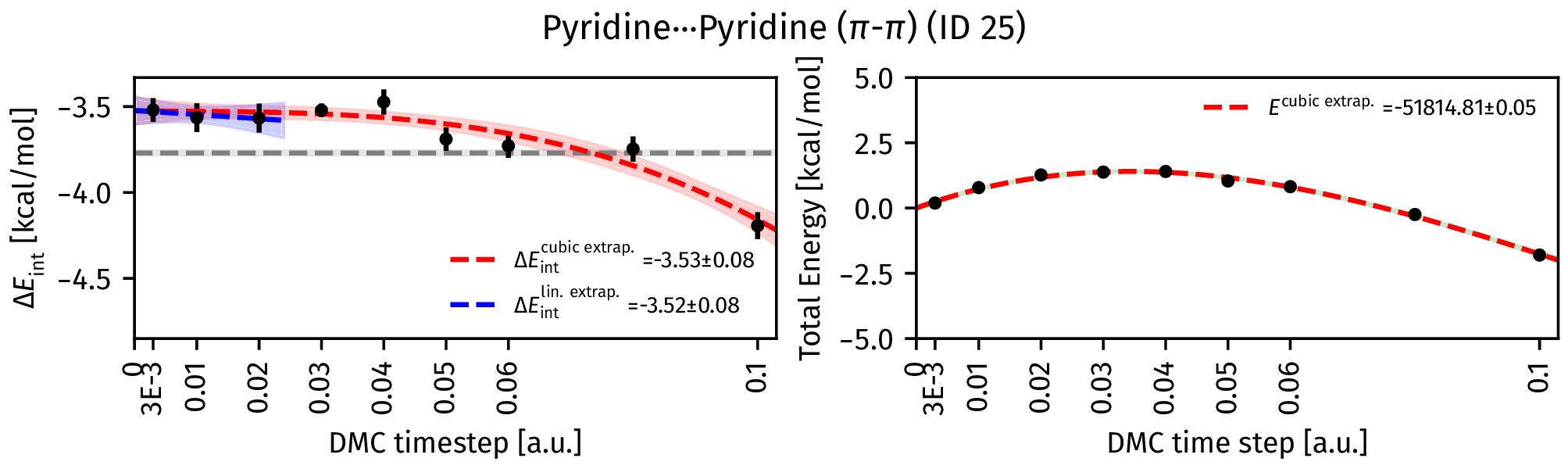}
    \caption{\label{fig:dimer_25} The time step dependence of $\Delta E_\text{int}$ and the total energy of the dimer complex for the Pyridine$\cdots$Pyridine ($\pi$-$\pi$) (ID 25) dimer.The dotted gray line represents the CCSD(T) reference in Table~\ref{sec:final_cc_estimates} and the black markers with stochastic 1$\sigma$ error bars represent the DMC estimate for each time step.}
\end{figure}
    
\begin{figure}[!h]
    \includegraphics[width=6.69in]{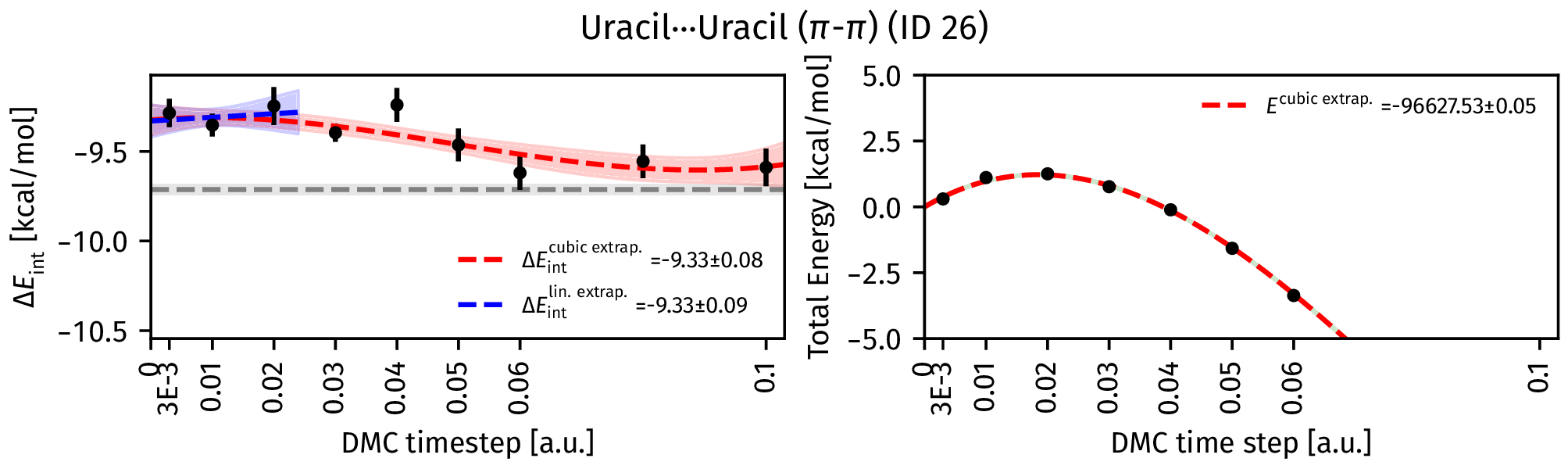}
    \caption{\label{fig:dimer_26} The time step dependence of $\Delta E_\text{int}$ and the total energy of the dimer complex for the Uracil$\cdots$Uracil ($\pi$-$\pi$) (ID 26) dimer.The dotted gray line represents the CCSD(T) reference in Table~\ref{sec:final_cc_estimates} and the black markers with stochastic 1$\sigma$ error bars represent the DMC estimate for each time step.}
\end{figure}
    
\begin{figure}[!h]
    \includegraphics[width=6.69in]{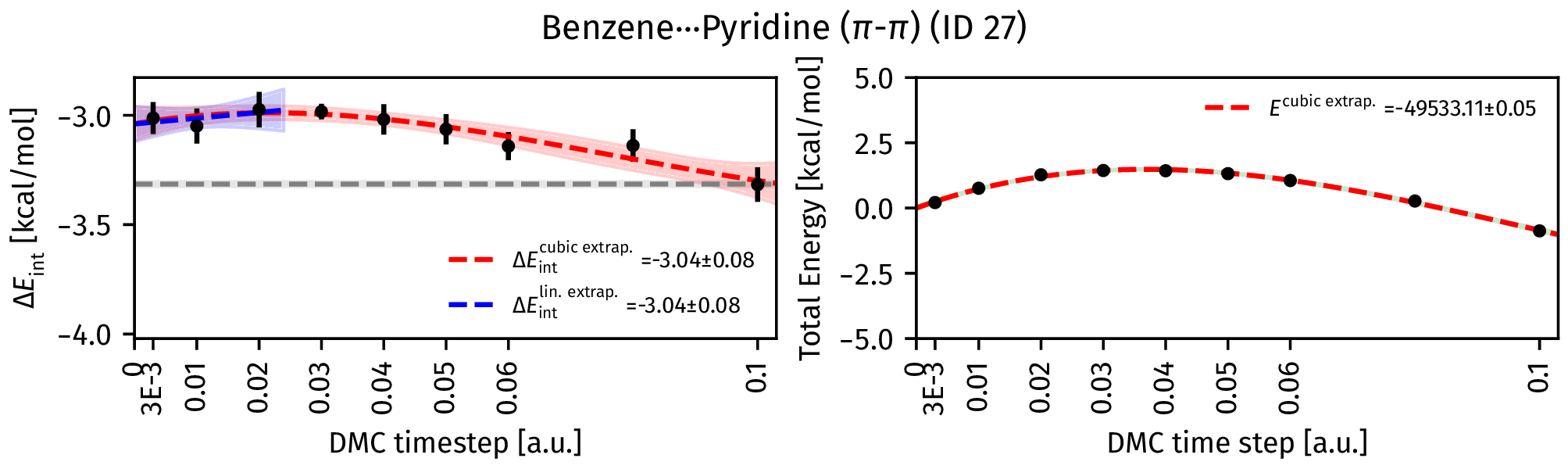}
    \caption{\label{fig:dimer_27} The time step dependence of $\Delta E_\text{int}$ and the total energy of the dimer complex for the Benzene$\cdots$Pyridine ($\pi$-$\pi$) (ID 27) dimer.The dotted gray line represents the CCSD(T) reference in Table~\ref{sec:final_cc_estimates} and the black markers with stochastic 1$\sigma$ error bars represent the DMC estimate for each time step.}
\end{figure}
    
\begin{figure}[!h]
    \includegraphics[width=6.69in]{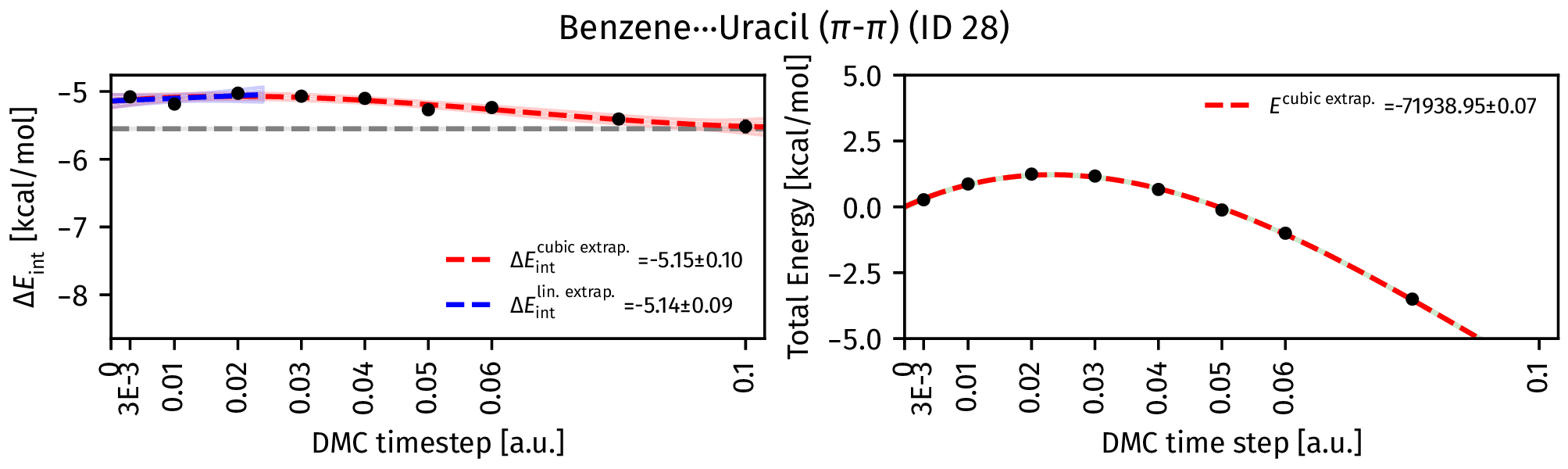}
    \caption{\label{fig:dimer_28} The time step dependence of $\Delta E_\text{int}$ and the total energy of the dimer complex for the Benzene$\cdots$Uracil ($\pi$-$\pi$) (ID 28) dimer.The dotted gray line represents the CCSD(T) reference in Table~\ref{sec:final_cc_estimates} and the black markers with stochastic 1$\sigma$ error bars represent the DMC estimate for each time step.}
\end{figure}
    
\begin{figure}[!h]
    \includegraphics[width=6.69in]{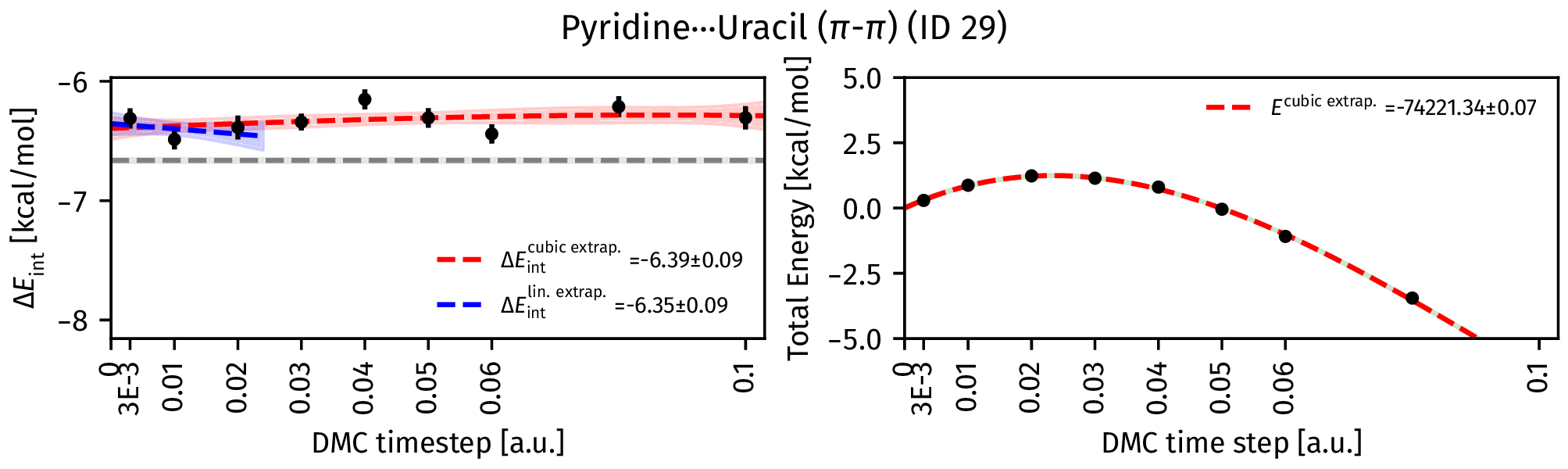}
    \caption{\label{fig:dimer_29} The time step dependence of $\Delta E_\text{int}$ and the total energy of the dimer complex for the Pyridine$\cdots$Uracil ($\pi$-$\pi$) (ID 29) dimer.The dotted gray line represents the CCSD(T) reference in Table~\ref{sec:final_cc_estimates} and the black markers with stochastic 1$\sigma$ error bars represent the DMC estimate for each time step.}
\end{figure}
    
\begin{figure}[!h]
    \includegraphics[width=6.69in]{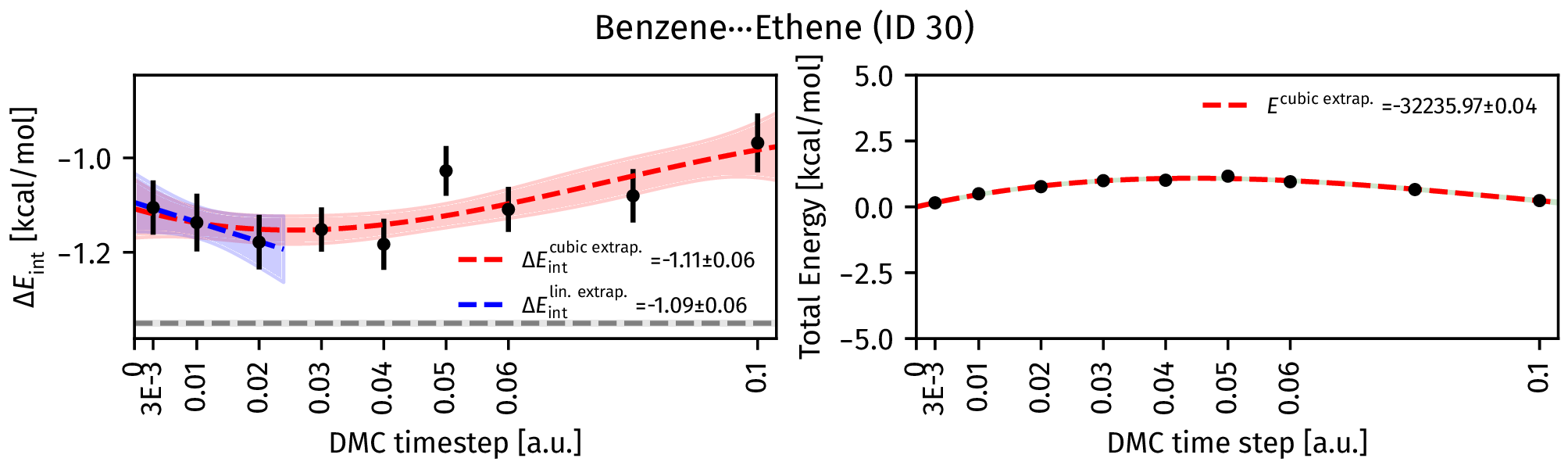}
    \caption{\label{fig:dimer_30} The time step dependence of $\Delta E_\text{int}$ and the total energy of the dimer complex for the Benzene$\cdots$Ethene (ID 30) dimer.The dotted gray line represents the CCSD(T) reference in Table~\ref{sec:final_cc_estimates} and the black markers with stochastic 1$\sigma$ error bars represent the DMC estimate for each time step.}
\end{figure}
    
\begin{figure}[!h]
    \includegraphics[width=6.69in]{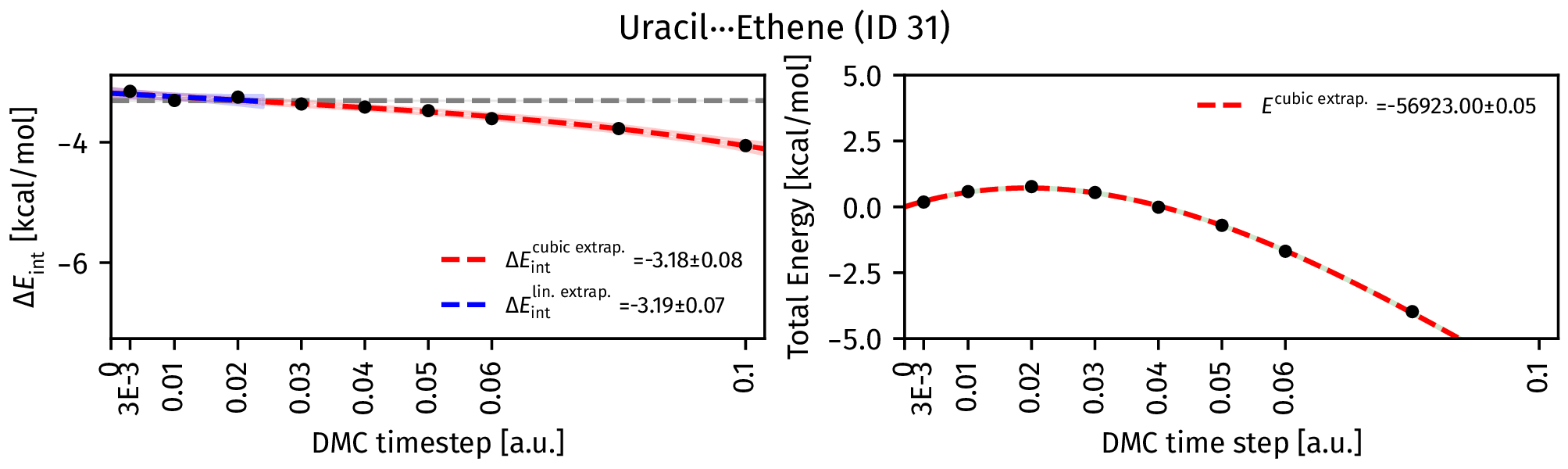}
    \caption{\label{fig:dimer_31} The time step dependence of $\Delta E_\text{int}$ and the total energy of the dimer complex for the Uracil$\cdots$Ethene (ID 31) dimer.The dotted gray line represents the CCSD(T) reference in Table~\ref{sec:final_cc_estimates} and the black markers with stochastic 1$\sigma$ error bars represent the DMC estimate for each time step.}
\end{figure}
    
\begin{figure}[!h]
    \includegraphics[width=6.69in]{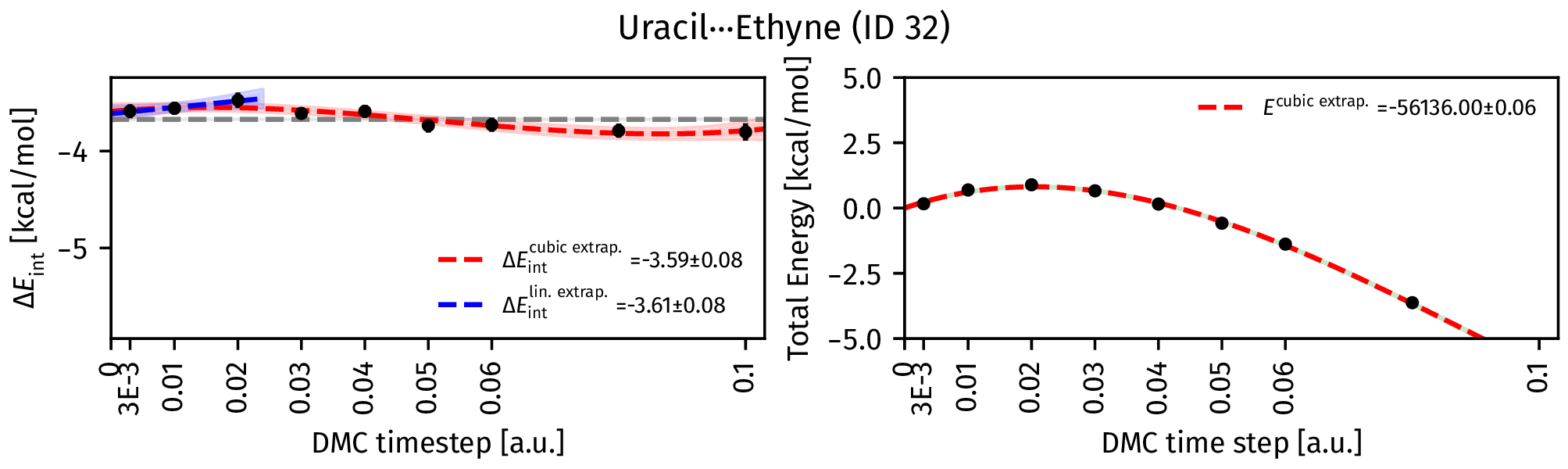}
    \caption{\label{fig:dimer_32} The time step dependence of $\Delta E_\text{int}$ and the total energy of the dimer complex for the Uracil$\cdots$Ethyne (ID 32) dimer.The dotted gray line represents the CCSD(T) reference in Table~\ref{sec:final_cc_estimates} and the black markers with stochastic 1$\sigma$ error bars represent the DMC estimate for each time step.}
\end{figure}
    
\begin{figure}[!h]
    \includegraphics[width=6.69in]{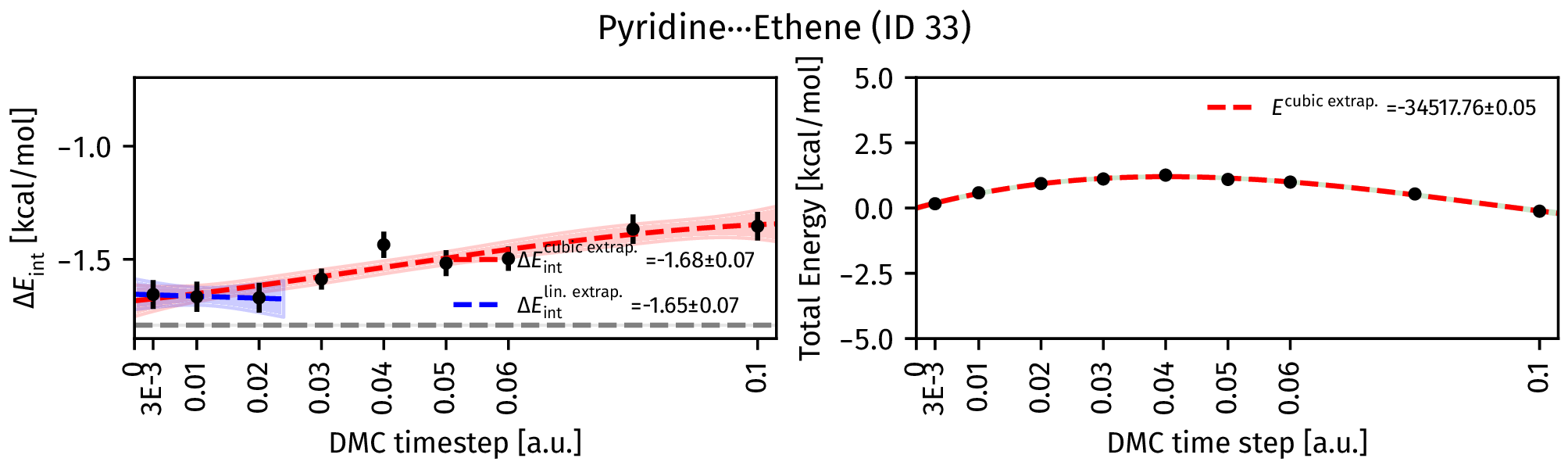}
    \caption{\label{fig:dimer_33} The time step dependence of $\Delta E_\text{int}$ and the total energy of the dimer complex for the Pyridine$\cdots$Ethene (ID 33) dimer.The dotted gray line represents the CCSD(T) reference in Table~\ref{sec:final_cc_estimates} and the black markers with stochastic 1$\sigma$ error bars represent the DMC estimate for each time step.}
\end{figure}
    
\begin{figure}[!h]
    \includegraphics[width=6.69in]{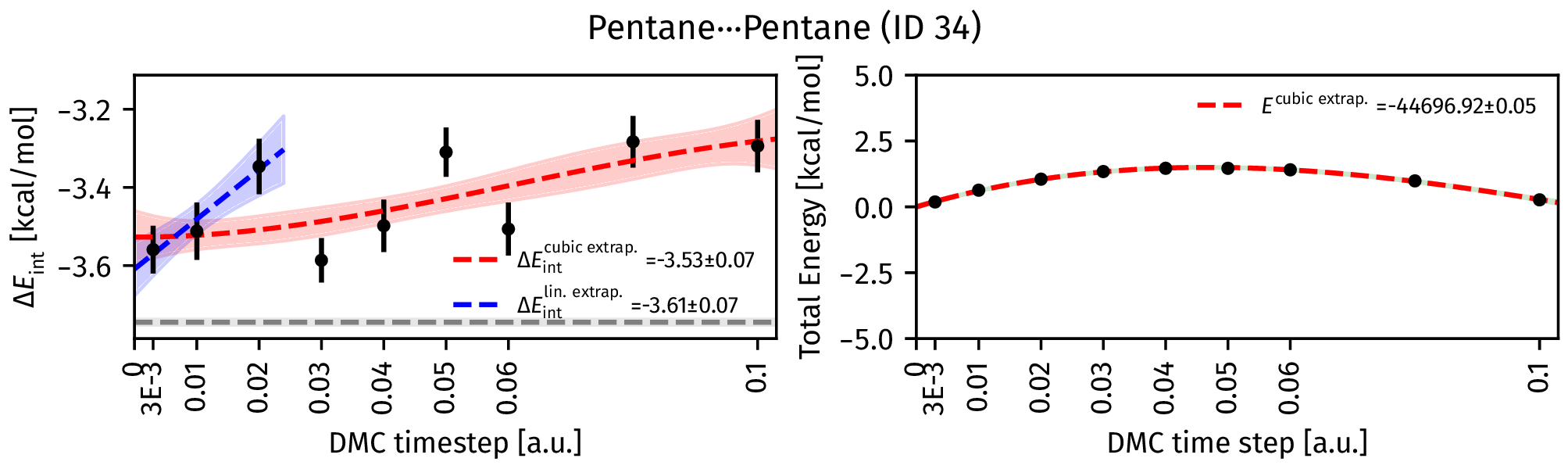}
    \caption{\label{fig:dimer_34} The time step dependence of $\Delta E_\text{int}$ and the total energy of the dimer complex for the Pentane$\cdots$Pentane (ID 34) dimer.The dotted gray line represents the CCSD(T) reference in Table~\ref{sec:final_cc_estimates} and the black markers with stochastic 1$\sigma$ error bars represent the DMC estimate for each time step.}
\end{figure}
    
\begin{figure}[!h]
    \includegraphics[width=6.69in]{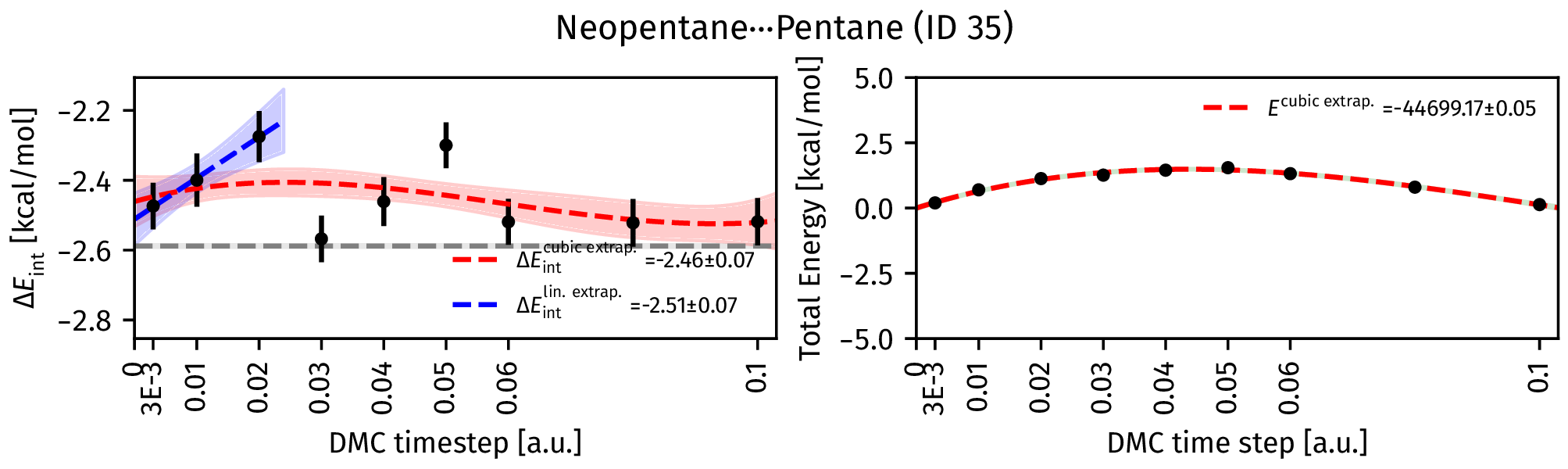}
    \caption{\label{fig:dimer_35} The time step dependence of $\Delta E_\text{int}$ and the total energy of the dimer complex for the Neopentane$\cdots$Pentane (ID 35) dimer.The dotted gray line represents the CCSD(T) reference in Table~\ref{sec:final_cc_estimates} and the black markers with stochastic 1$\sigma$ error bars represent the DMC estimate for each time step.}
\end{figure}
    
\begin{figure}[!h]
    \includegraphics[width=6.69in]{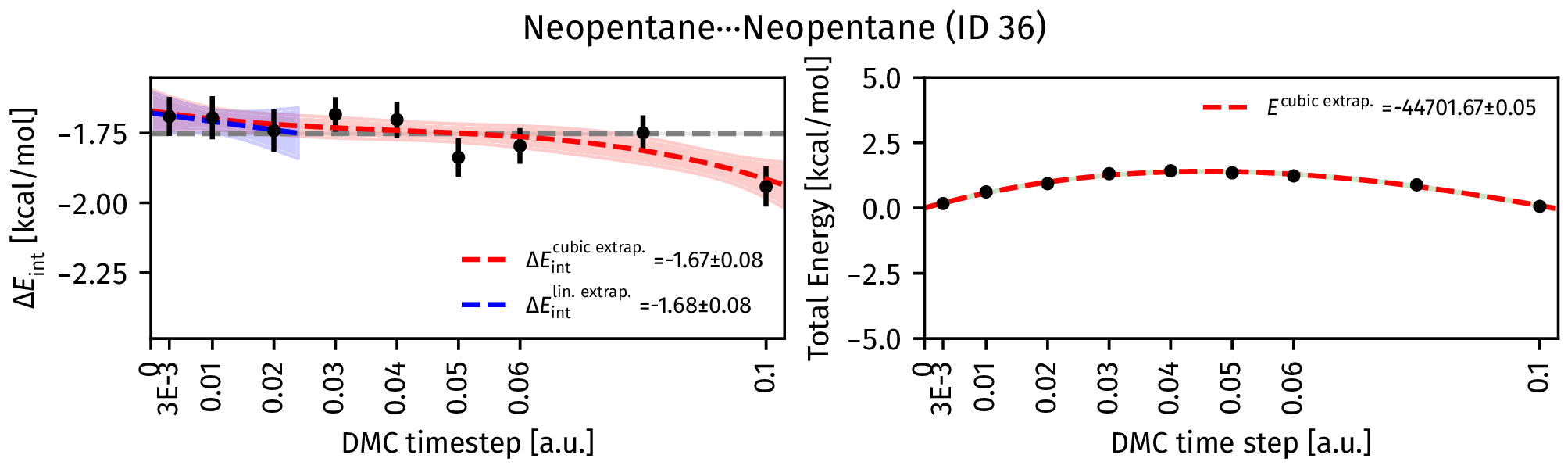}
    \caption{\label{fig:dimer_36} The time step dependence of $\Delta E_\text{int}$ and the total energy of the dimer complex for the Neopentane$\cdots$Neopentane (ID 36) dimer.The dotted gray line represents the CCSD(T) reference in Table~\ref{sec:final_cc_estimates} and the black markers with stochastic 1$\sigma$ error bars represent the DMC estimate for each time step.}
\end{figure}
    
\begin{figure}[!h]
    \includegraphics[width=6.69in]{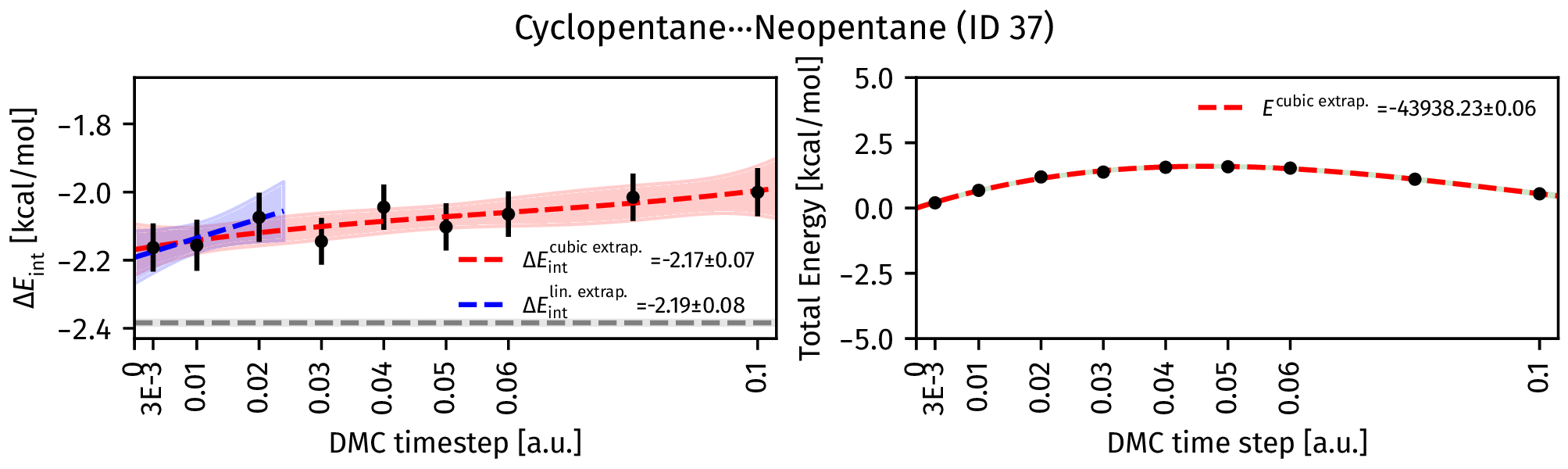}
    \caption{\label{fig:dimer_37} The time step dependence of $\Delta E_\text{int}$ and the total energy of the dimer complex for the Cyclopentane$\cdots$Neopentane (ID 37) dimer.The dotted gray line represents the CCSD(T) reference in Table~\ref{sec:final_cc_estimates} and the black markers with stochastic 1$\sigma$ error bars represent the DMC estimate for each time step.}
\end{figure}
    
\begin{figure}[!h]
    \includegraphics[width=6.69in]{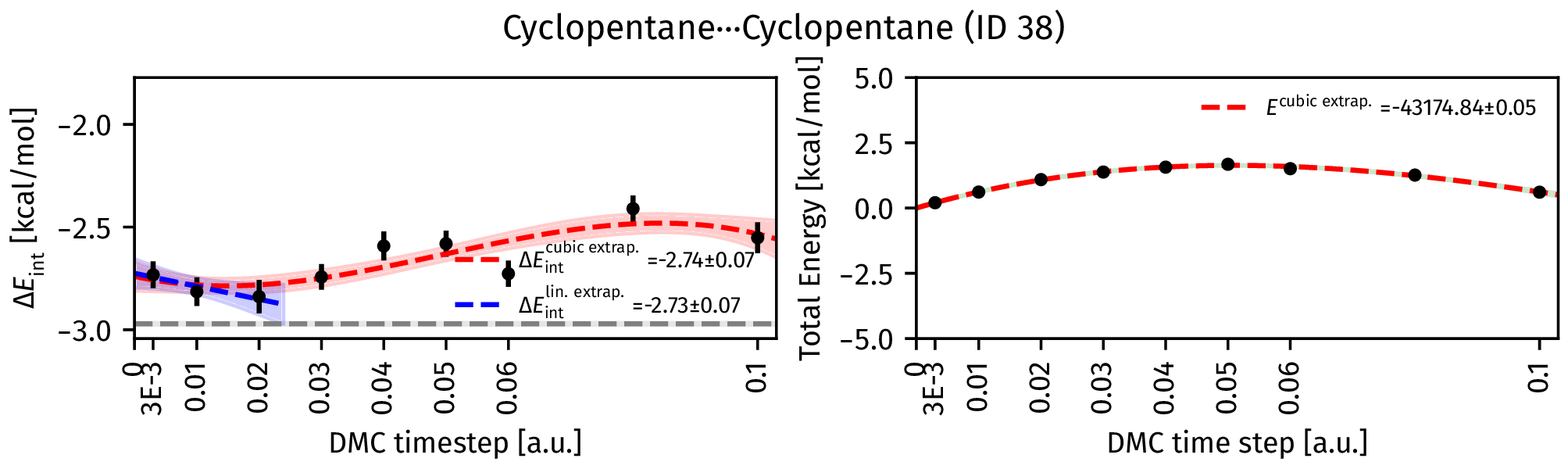}
    \caption{\label{fig:dimer_38} The time step dependence of $\Delta E_\text{int}$ and the total energy of the dimer complex for the Cyclopentane$\cdots$Cyclopentane (ID 38) dimer.The dotted gray line represents the CCSD(T) reference in Table~\ref{sec:final_cc_estimates} and the black markers with stochastic 1$\sigma$ error bars represent the DMC estimate for each time step.}
\end{figure}
    
\begin{figure}[!h]
    \includegraphics[width=6.69in]{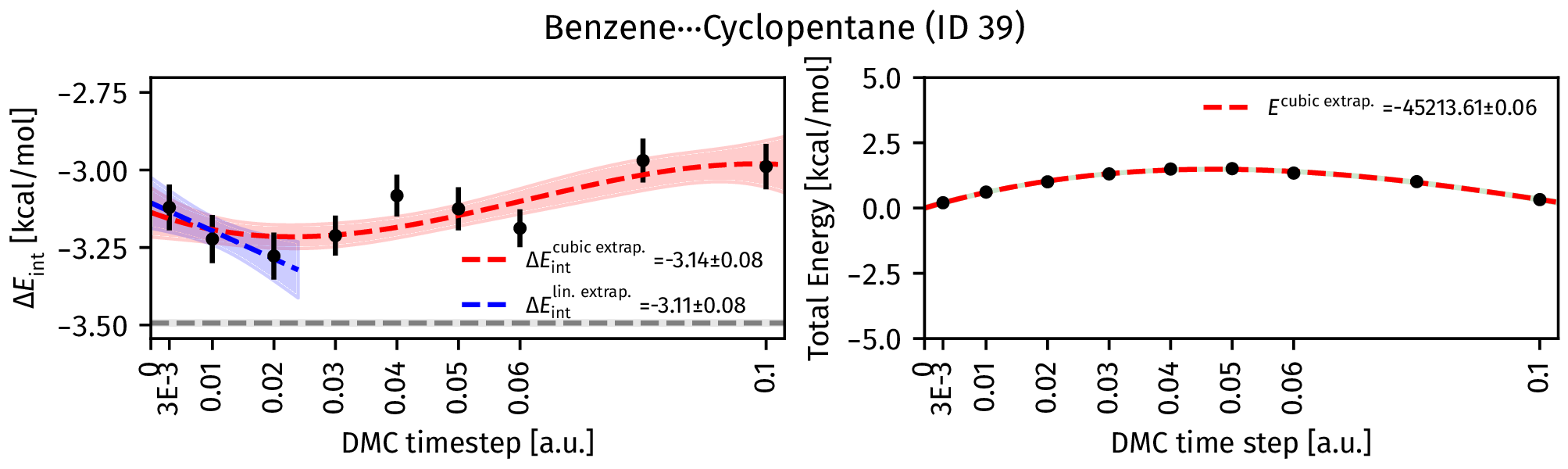}
    \caption{\label{fig:dimer_39} The time step dependence of $\Delta E_\text{int}$ and the total energy of the dimer complex for the Benzene$\cdots$Cyclopentane (ID 39) dimer.The dotted gray line represents the CCSD(T) reference in Table~\ref{sec:final_cc_estimates} and the black markers with stochastic 1$\sigma$ error bars represent the DMC estimate for each time step.}
\end{figure}
    
\begin{figure}[!h]
    \includegraphics[width=6.69in]{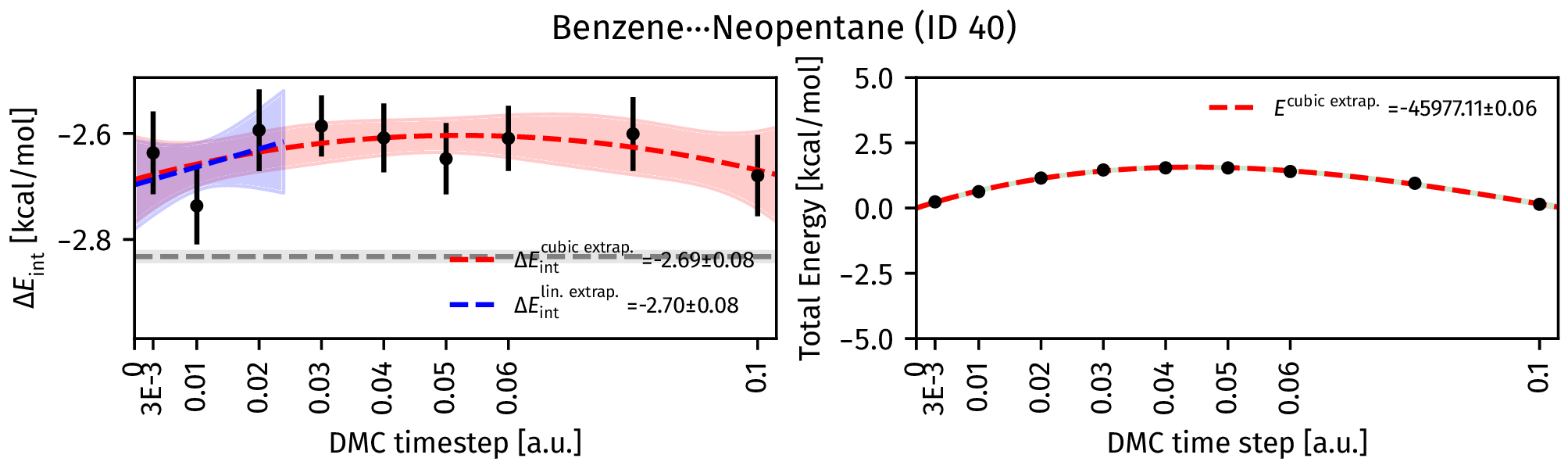}
    \caption{\label{fig:dimer_40} The time step dependence of $\Delta E_\text{int}$ and the total energy of the dimer complex for the Benzene$\cdots$Neopentane (ID 40) dimer.The dotted gray line represents the CCSD(T) reference in Table~\ref{sec:final_cc_estimates} and the black markers with stochastic 1$\sigma$ error bars represent the DMC estimate for each time step.}
\end{figure}
    
\begin{figure}[!h]
    \includegraphics[width=6.69in]{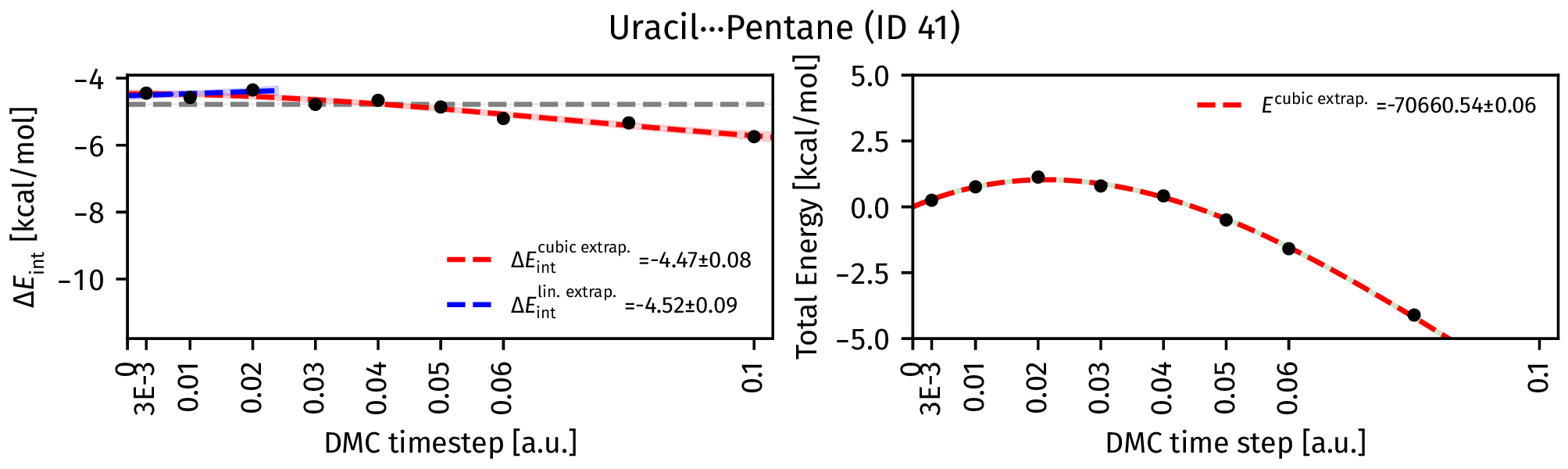}
    \caption{\label{fig:dimer_41} The time step dependence of $\Delta E_\text{int}$ and the total energy of the dimer complex for the Uracil$\cdots$Pentane (ID 41) dimer.The dotted gray line represents the CCSD(T) reference in Table~\ref{sec:final_cc_estimates} and the black markers with stochastic 1$\sigma$ error bars represent the DMC estimate for each time step.}
\end{figure}
    
\begin{figure}[!h]
    \includegraphics[width=6.69in]{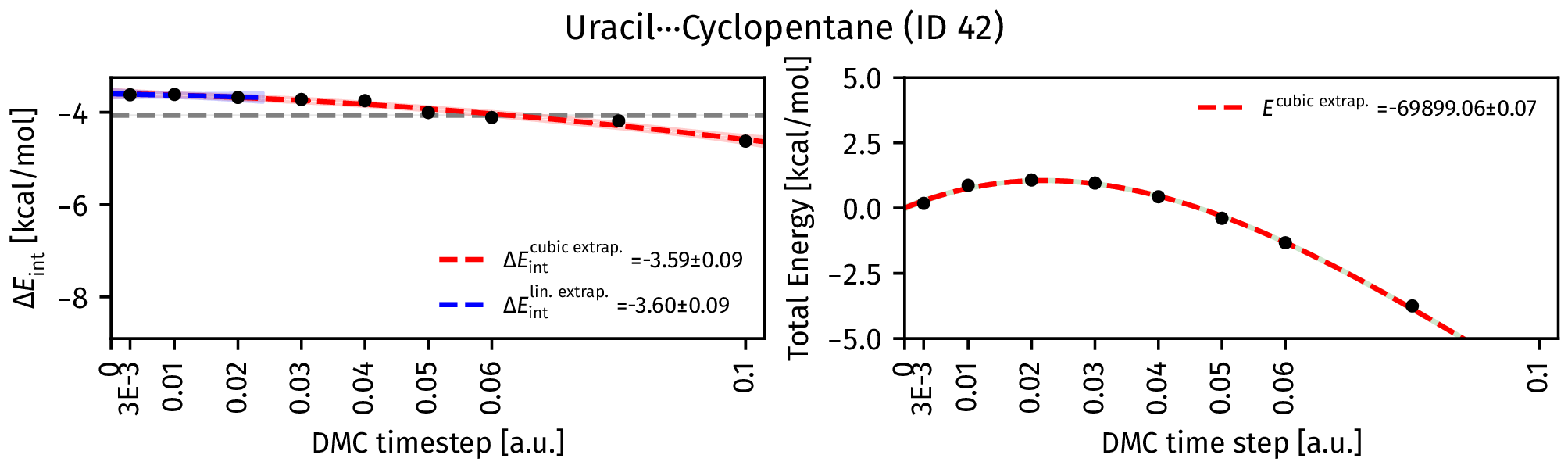}
    \caption{\label{fig:dimer_42} The time step dependence of $\Delta E_\text{int}$ and the total energy of the dimer complex for the Uracil$\cdots$Cyclopentane (ID 42) dimer.The dotted gray line represents the CCSD(T) reference in Table~\ref{sec:final_cc_estimates} and the black markers with stochastic 1$\sigma$ error bars represent the DMC estimate for each time step.}
\end{figure}
    
\begin{figure}[!h]
    \includegraphics[width=6.69in]{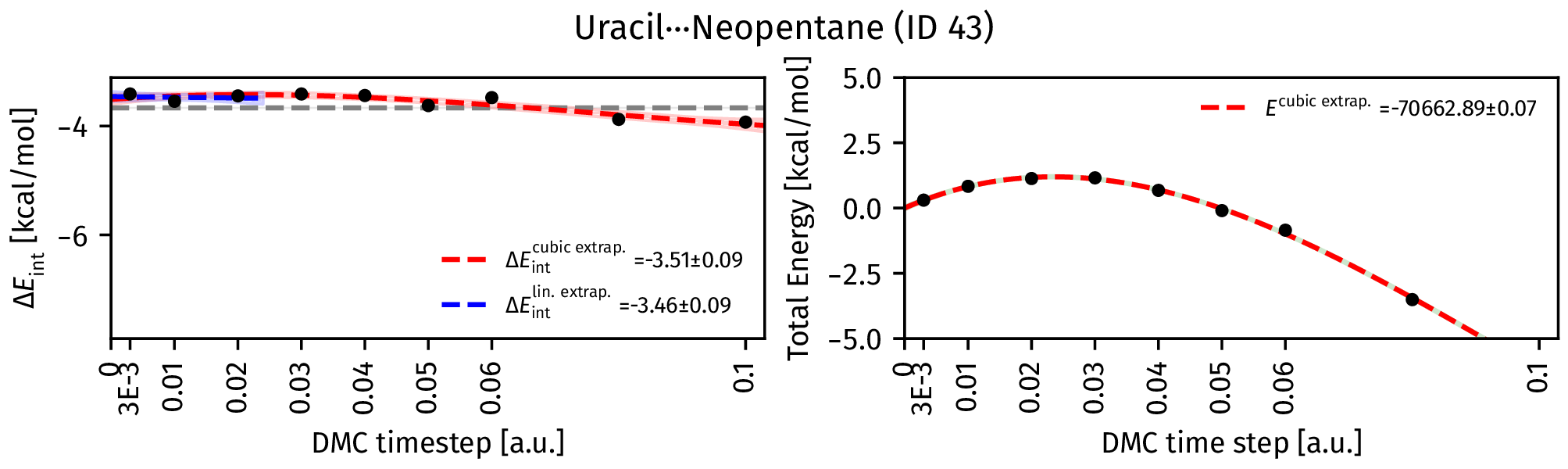}
    \caption{\label{fig:dimer_43} The time step dependence of $\Delta E_\text{int}$ and the total energy of the dimer complex for the Uracil$\cdots$Neopentane (ID 43) dimer.The dotted gray line represents the CCSD(T) reference in Table~\ref{sec:final_cc_estimates} and the black markers with stochastic 1$\sigma$ error bars represent the DMC estimate for each time step.}
\end{figure}
    
\begin{figure}[!h]
    \includegraphics[width=6.69in]{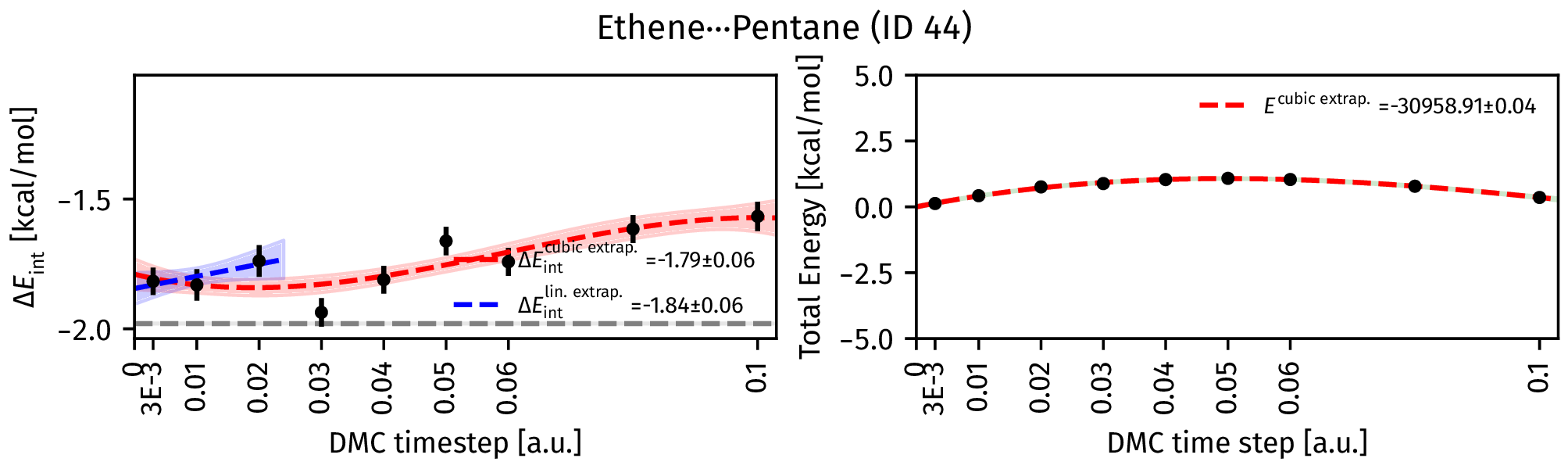}
    \caption{\label{fig:dimer_44} The time step dependence of $\Delta E_\text{int}$ and the total energy of the dimer complex for the Ethene$\cdots$Pentane (ID 44) dimer.The dotted gray line represents the CCSD(T) reference in Table~\ref{sec:final_cc_estimates} and the black markers with stochastic 1$\sigma$ error bars represent the DMC estimate for each time step.}
\end{figure}
    
\begin{figure}[!h]
    \includegraphics[width=6.69in]{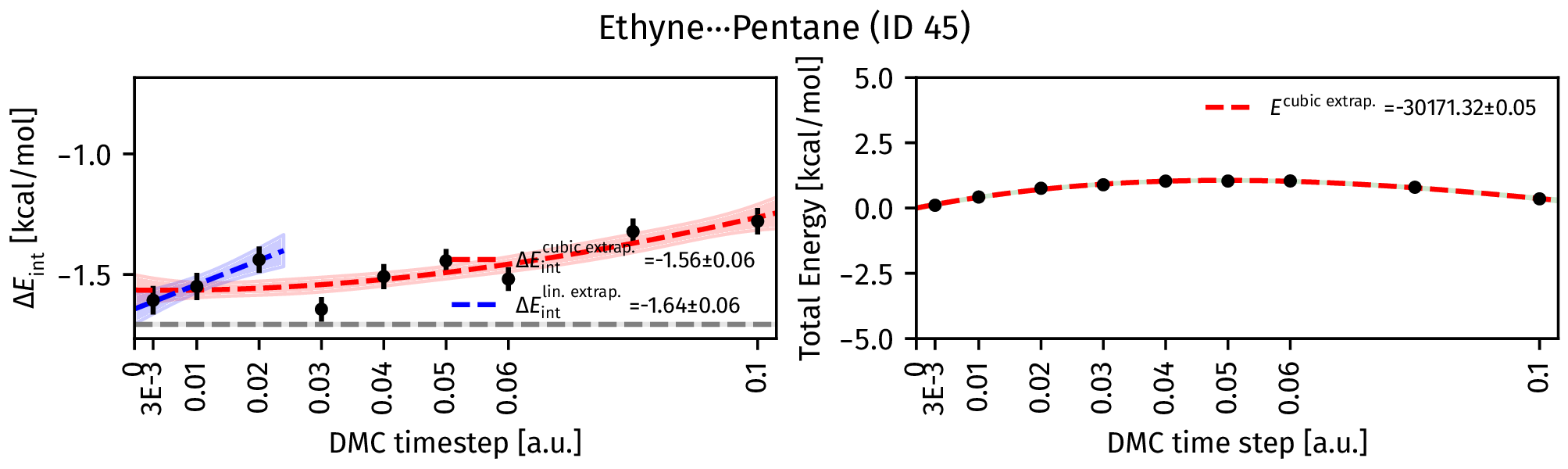}
    \caption{\label{fig:dimer_45} The time step dependence of $\Delta E_\text{int}$ and the total energy of the dimer complex for the Ethyne$\cdots$Pentane (ID 45) dimer.The dotted gray line represents the CCSD(T) reference in Table~\ref{sec:final_cc_estimates} and the black markers with stochastic 1$\sigma$ error bars represent the DMC estimate for each time step.}
\end{figure}
    
\begin{figure}[!h]
    \includegraphics[width=6.69in]{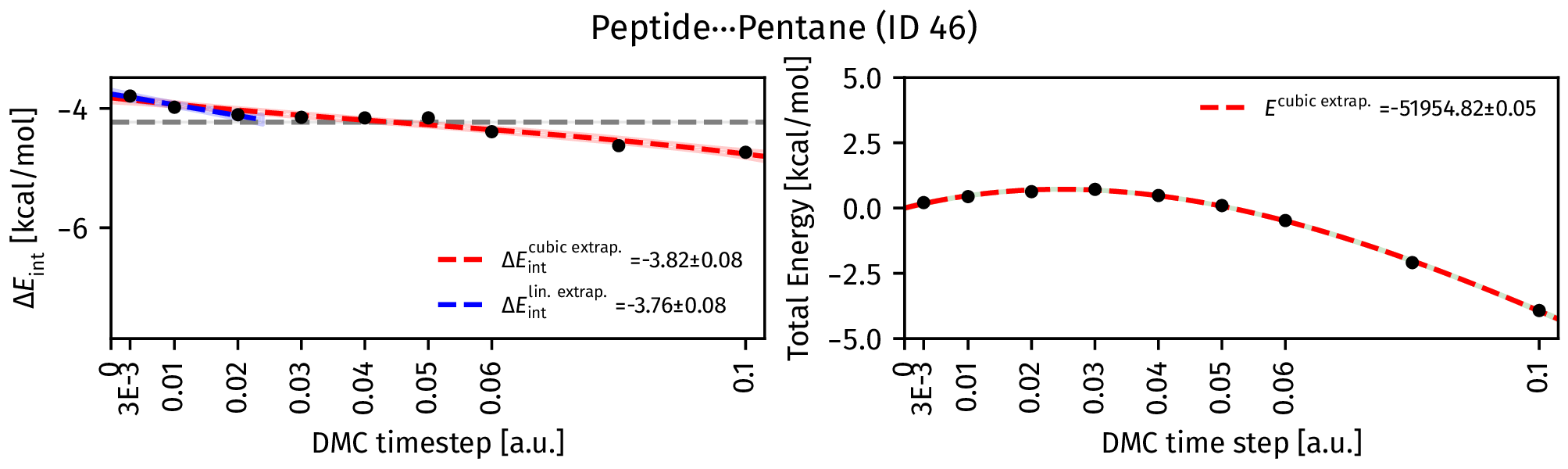}
    \caption{\label{fig:dimer_46} The time step dependence of $\Delta E_\text{int}$ and the total energy of the dimer complex for the Peptide$\cdots$Pentane (ID 46) dimer.The dotted gray line represents the CCSD(T) reference in Table~\ref{sec:final_cc_estimates} and the black markers with stochastic 1$\sigma$ error bars represent the DMC estimate for each time step.}
\end{figure}
    
\clearpage

\begin{figure}[!h]
    \includegraphics[width=6.69in]{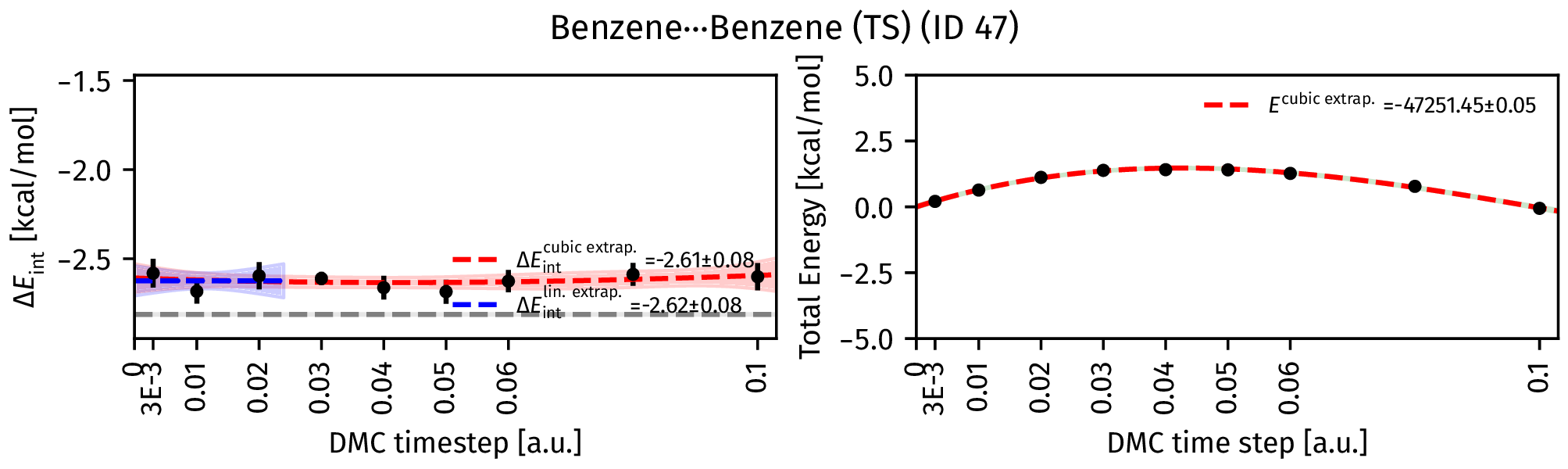}
    \caption{\label{fig:dimer_47} The time step dependence of $\Delta E_\text{int}$ and the total energy of the dimer complex for the Benzene$\cdots$Benzene (TS) (ID 47) dimer.The dotted gray line represents the CCSD(T) reference in Table~\ref{sec:final_cc_estimates} and the black markers with stochastic 1$\sigma$ error bars represent the DMC estimate for each time step.}
\end{figure}
    
\begin{figure}[!h]
    \includegraphics[width=6.69in]{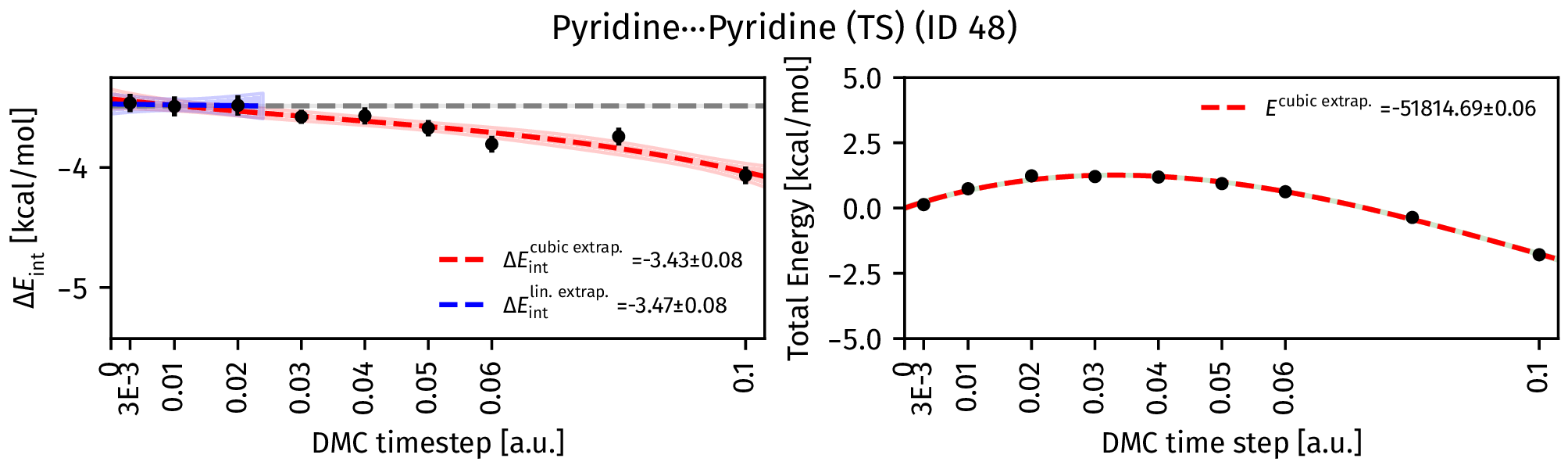}
    \caption{\label{fig:dimer_48} The time step dependence of $\Delta E_\text{int}$ and the total energy of the dimer complex for the Pyridine$\cdots$Pyridine (TS) (ID 48) dimer.The dotted gray line represents the CCSD(T) reference in Table~\ref{sec:final_cc_estimates} and the black markers with stochastic 1$\sigma$ error bars represent the DMC estimate for each time step.}
\end{figure}
    
\begin{figure}[!h]
    \includegraphics[width=6.69in]{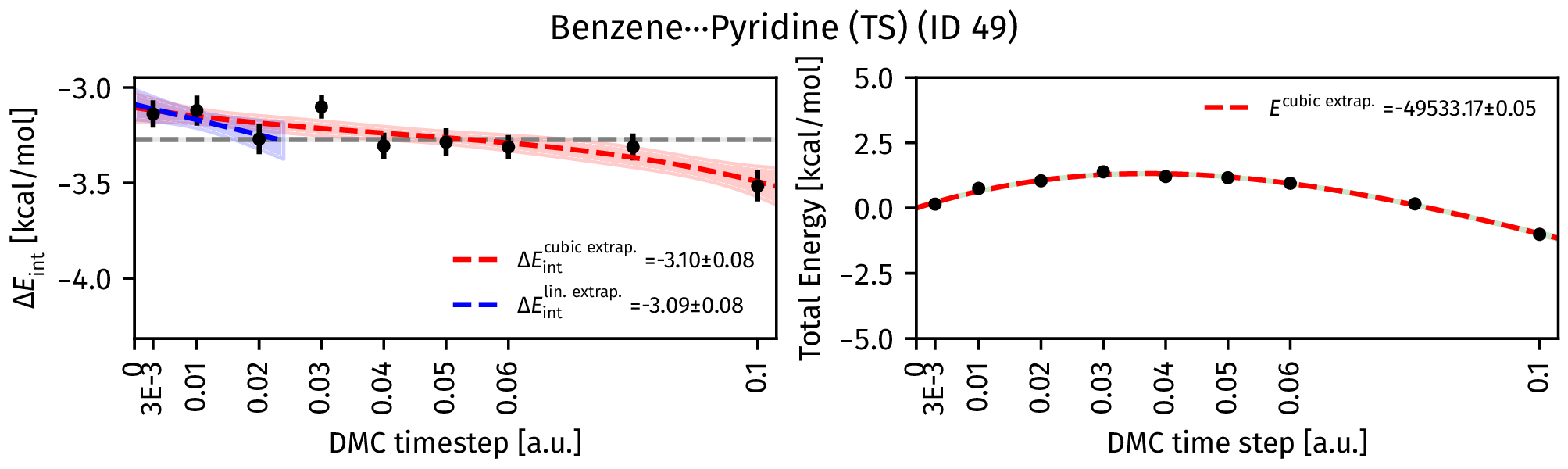}
    \caption{\label{fig:dimer_49} The time step dependence of $\Delta E_\text{int}$ and the total energy of the dimer complex for the Benzene$\cdots$Pyridine (TS) (ID 49) dimer.The dotted gray line represents the CCSD(T) reference in Table~\ref{sec:final_cc_estimates} and the black markers with stochastic 1$\sigma$ error bars represent the DMC estimate for each time step.}
\end{figure}
    
\begin{figure}[!h]
    \includegraphics[width=6.69in]{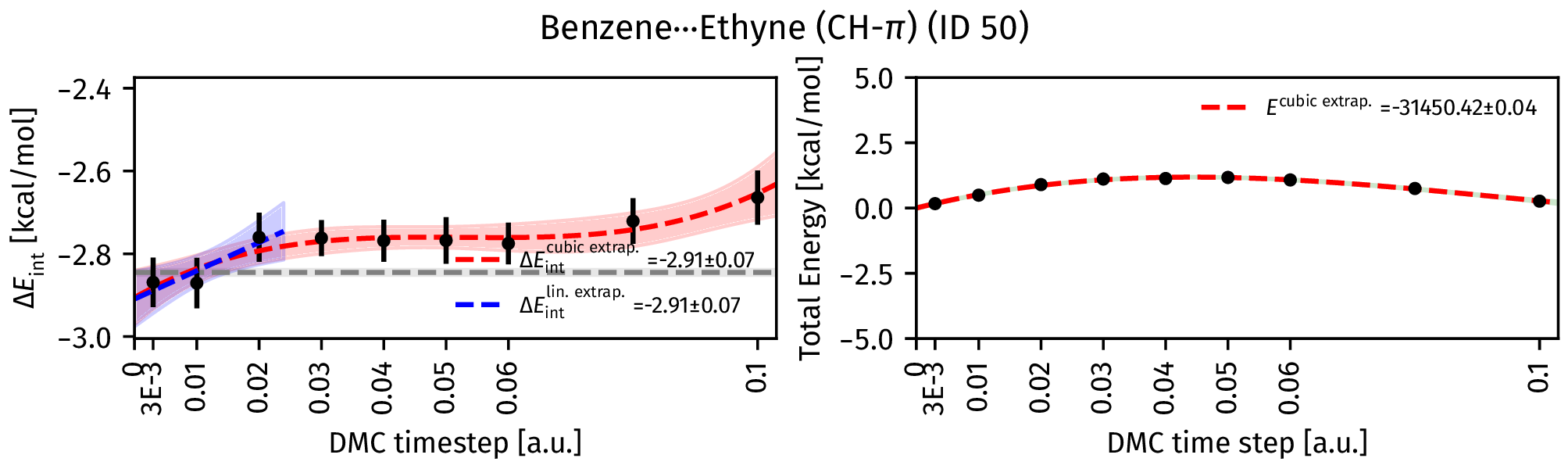}
    \caption{\label{fig:dimer_50} The time step dependence of $\Delta E_\text{int}$ and the total energy of the dimer complex for the Benzene$\cdots$Ethyne (CH-$\pi$) (ID 50) dimer.The dotted gray line represents the CCSD(T) reference in Table~\ref{sec:final_cc_estimates} and the black markers with stochastic 1$\sigma$ error bars represent the DMC estimate for each time step.}
\end{figure}
    
\begin{figure}[!h]
    \includegraphics[width=6.69in]{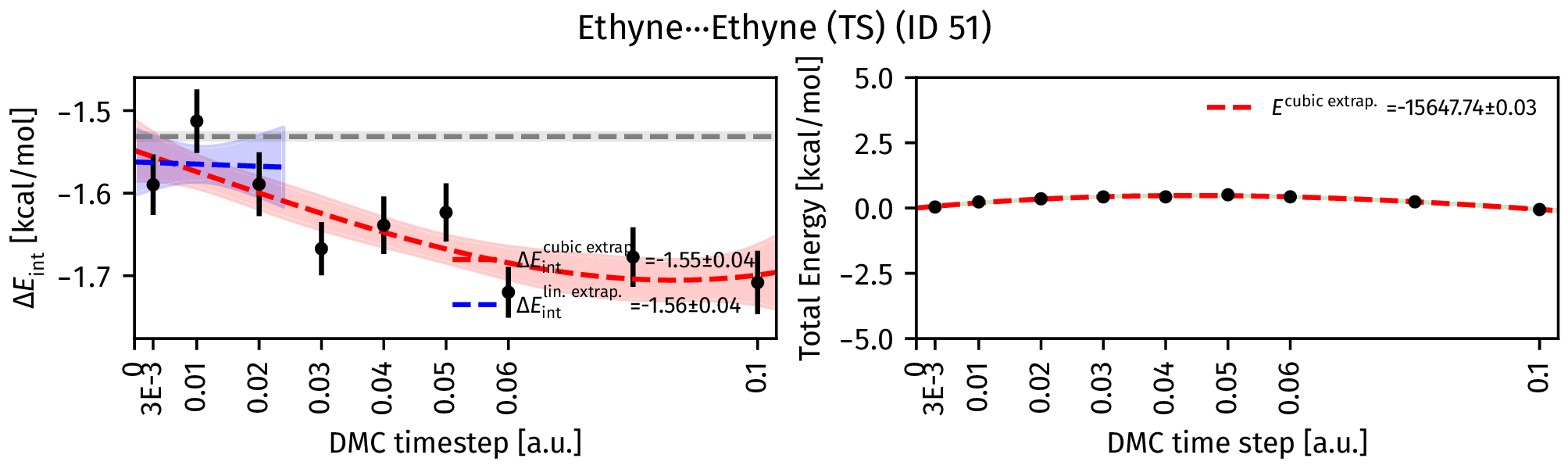}
    \caption{\label{fig:dimer_51} The time step dependence of $\Delta E_\text{int}$ and the total energy of the dimer complex for the Ethyne$\cdots$Ethyne (TS) (ID 51) dimer.The dotted gray line represents the CCSD(T) reference in Table~\ref{sec:final_cc_estimates} and the black markers with stochastic 1$\sigma$ error bars represent the DMC estimate for each time step.}
\end{figure}
    
\begin{figure}[!h]
    \includegraphics[width=6.69in]{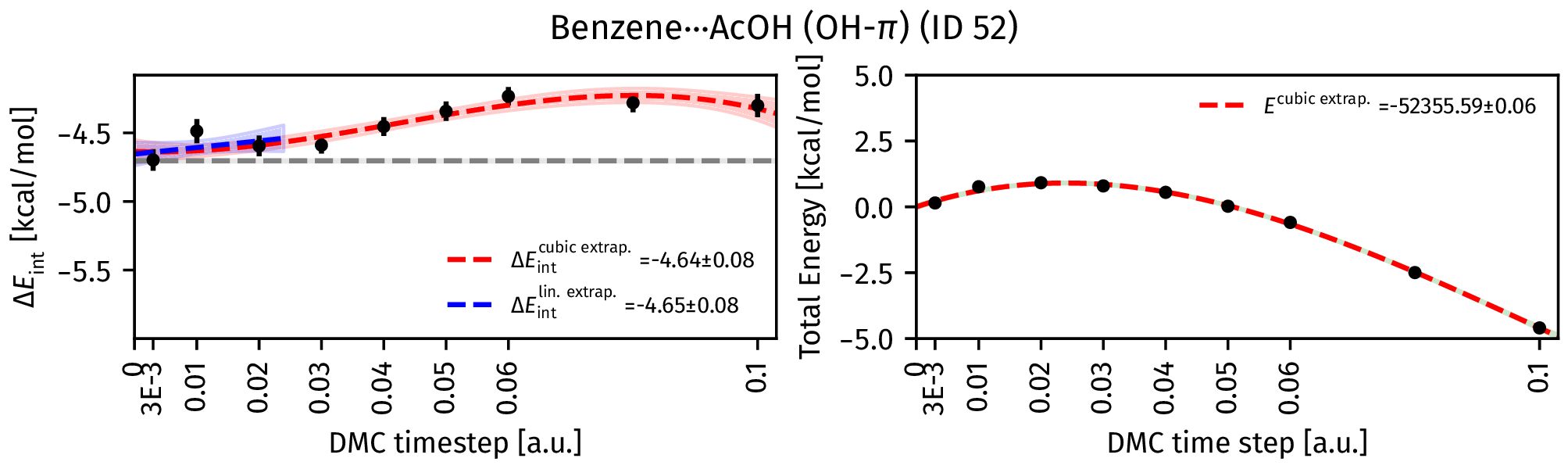}
    \caption{\label{fig:dimer_52} The time step dependence of $\Delta E_\text{int}$ and the total energy of the dimer complex for the Benzene$\cdots$AcOH (OH-$\pi$) (ID 52) dimer.The dotted gray line represents the CCSD(T) reference in Table~\ref{sec:final_cc_estimates} and the black markers with stochastic 1$\sigma$ error bars represent the DMC estimate for each time step.}
\end{figure}

\clearpage
    
\begin{figure}[!h]
    \includegraphics[width=6.69in]{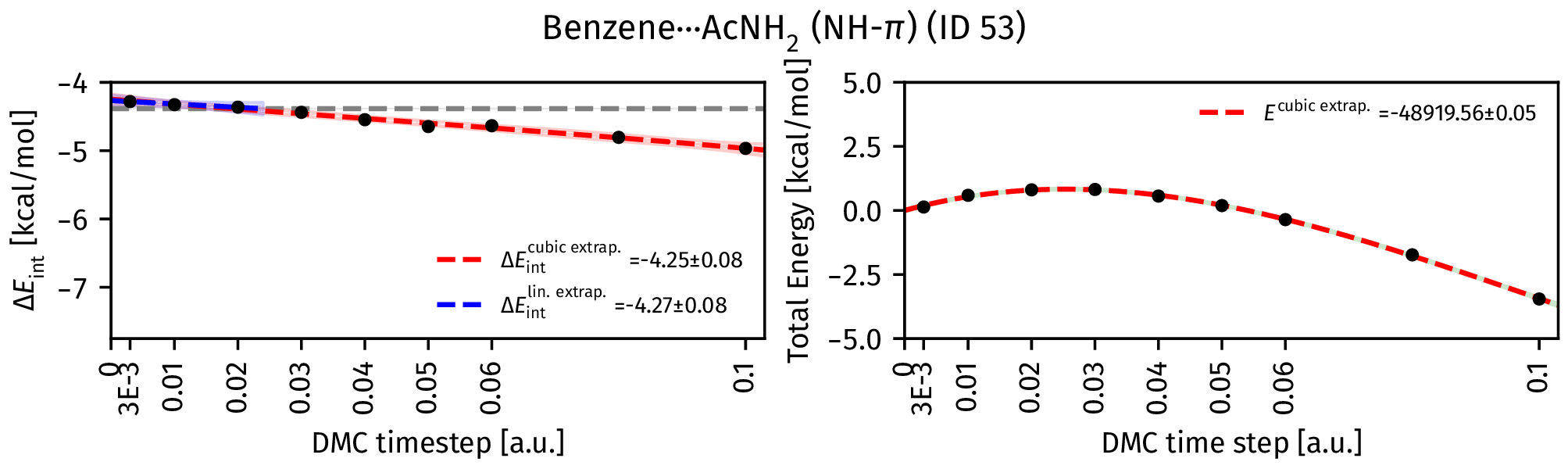}
    \caption{\label{fig:dimer_53} The time step dependence of $\Delta E_\text{int}$ and the total energy of the dimer complex for the Benzene$\cdots$AcNH$_2$ (NH-$\pi$) (ID 53) dimer.The dotted gray line represents the CCSD(T) reference in Table~\ref{sec:final_cc_estimates} and the black markers with stochastic 1$\sigma$ error bars represent the DMC estimate for each time step.}
\end{figure}
    
\begin{figure}[!h]
    \includegraphics[width=6.69in]{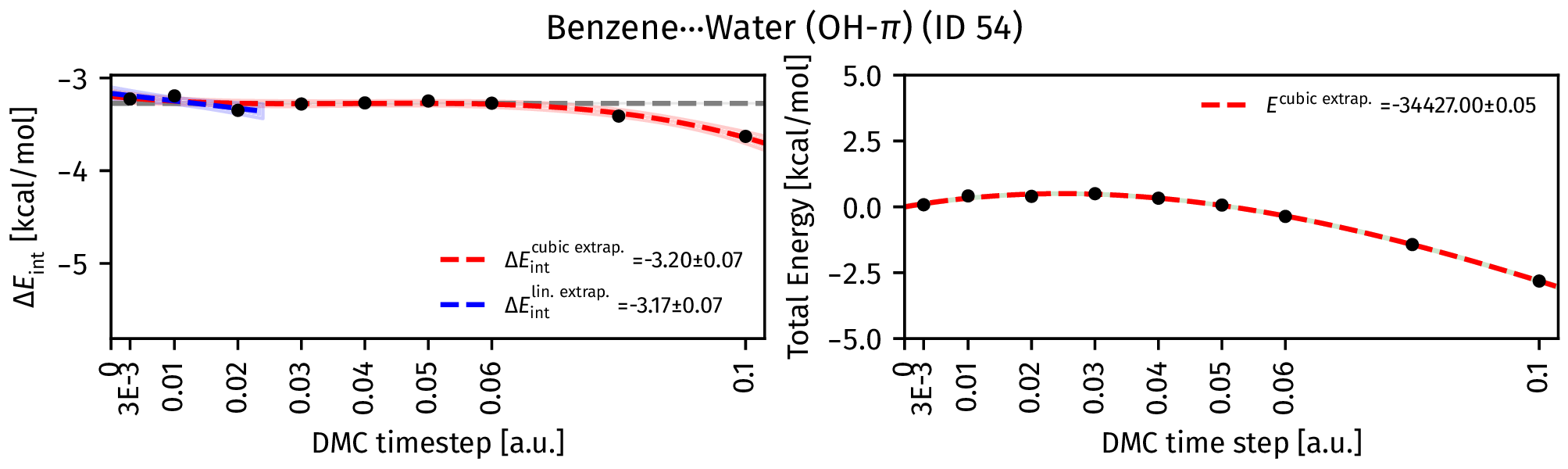}
    \caption{\label{fig:dimer_54} The time step dependence of $\Delta E_\text{int}$ and the total energy of the dimer complex for the Benzene$\cdots$Water (OH-$\pi$) (ID 54) dimer.The dotted gray line represents the CCSD(T) reference in Table~\ref{sec:final_cc_estimates} and the black markers with stochastic 1$\sigma$ error bars represent the DMC estimate for each time step.}
\end{figure}
    
\begin{figure}[!h]
    \includegraphics[width=6.69in]{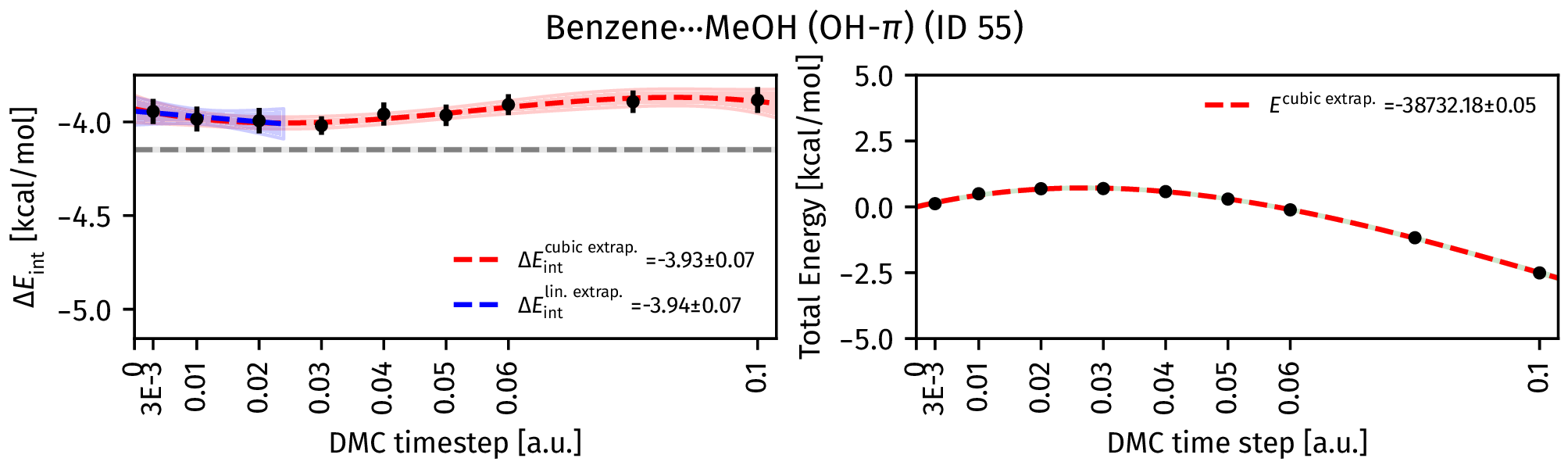}
    \caption{\label{fig:dimer_55} The time step dependence of $\Delta E_\text{int}$ and the total energy of the dimer complex for the Benzene$\cdots$MeOH (OH-$\pi$) (ID 55) dimer.The dotted gray line represents the CCSD(T) reference in Table~\ref{sec:final_cc_estimates} and the black markers with stochastic 1$\sigma$ error bars represent the DMC estimate for each time step.}
\end{figure}
    
\begin{figure}[!h]
    \includegraphics[width=6.69in]{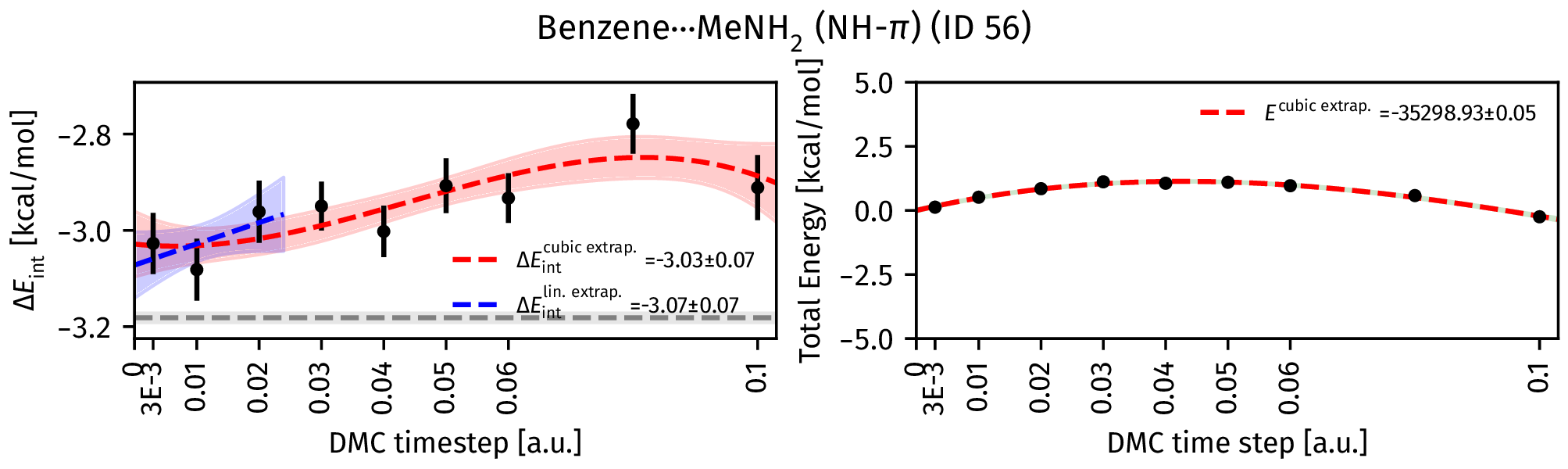}
    \caption{\label{fig:dimer_56} The time step dependence of $\Delta E_\text{int}$ and the total energy of the dimer complex for the Benzene$\cdots$MeNH$_2$ (NH-$\pi$) (ID 56) dimer.The dotted gray line represents the CCSD(T) reference in Table~\ref{sec:final_cc_estimates} and the black markers with stochastic 1$\sigma$ error bars represent the DMC estimate for each time step.}
\end{figure}
    
\begin{figure}[!h]
    \includegraphics[width=6.69in]{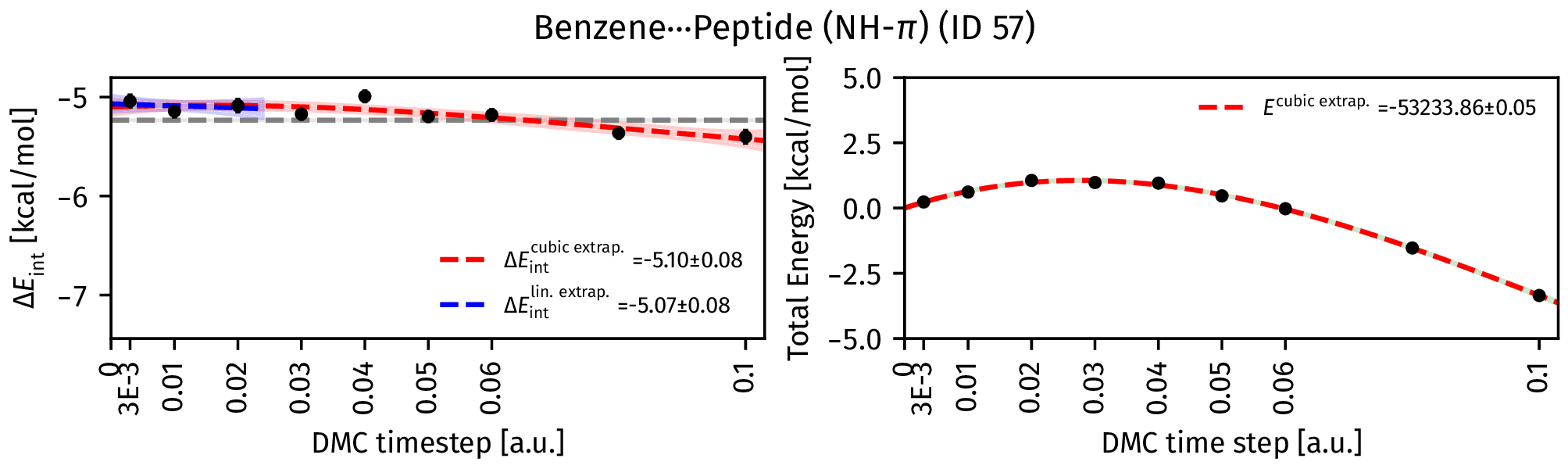}
    \caption{\label{fig:dimer_57} The time step dependence of $\Delta E_\text{int}$ and the total energy of the dimer complex for the Benzene$\cdots$Peptide (NH-$\pi$) (ID 57) dimer.The dotted gray line represents the CCSD(T) reference in Table~\ref{sec:final_cc_estimates} and the black markers with stochastic 1$\sigma$ error bars represent the DMC estimate for each time step.}
\end{figure}
    
\begin{figure}[!h]
    \includegraphics[width=6.69in]{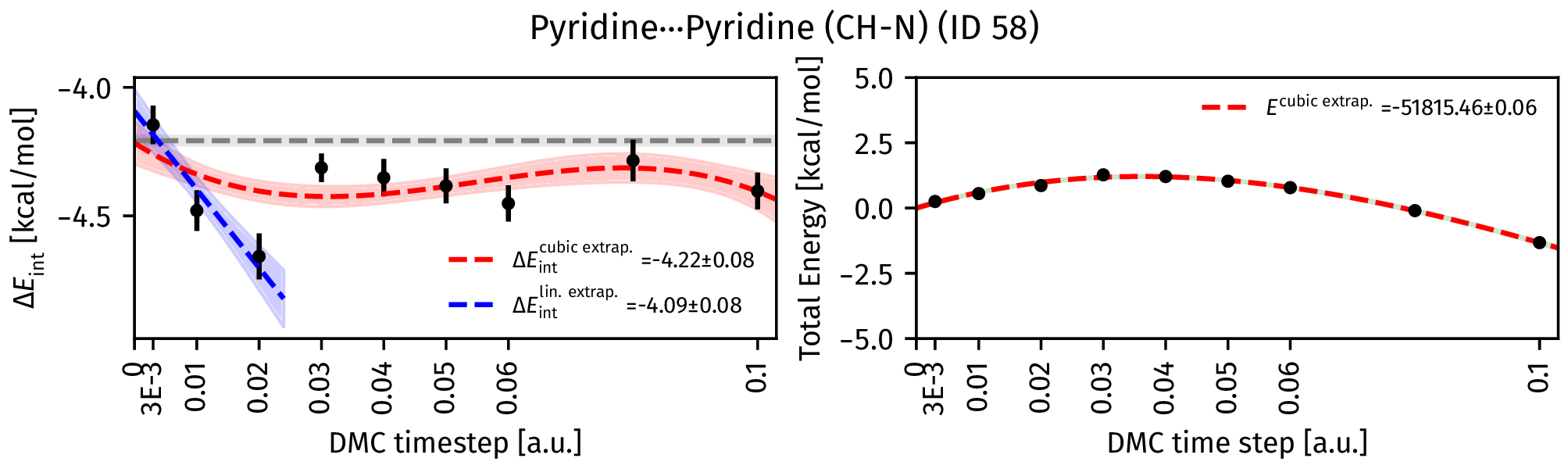}
    \caption{\label{fig:dimer_58} The time step dependence of $\Delta E_\text{int}$ and the total energy of the dimer complex for the Pyridine$\cdots$Pyridine (CH-N) (ID 58) dimer.The dotted gray line represents the CCSD(T) reference in Table~\ref{sec:final_cc_estimates} and the black markers with stochastic 1$\sigma$ error bars represent the DMC estimate for each time step.}
\end{figure}
    
\begin{figure}[!h]
    \includegraphics[width=6.69in]{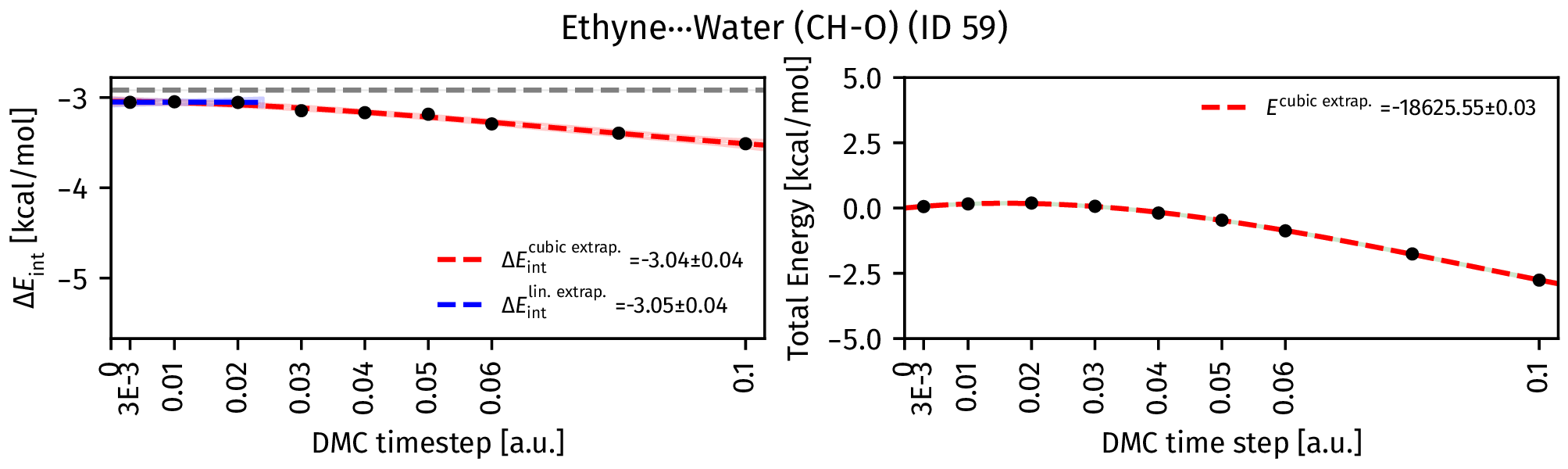}
    \caption{\label{fig:dimer_59} The time step dependence of $\Delta E_\text{int}$ and the total energy of the dimer complex for the Ethyne$\cdots$Water (CH-O) (ID 59) dimer.The dotted gray line represents the CCSD(T) reference in Table~\ref{sec:final_cc_estimates} and the black markers with stochastic 1$\sigma$ error bars represent the DMC estimate for each time step.}
\end{figure}
    
\begin{figure}[!h]
    \includegraphics[width=6.69in]{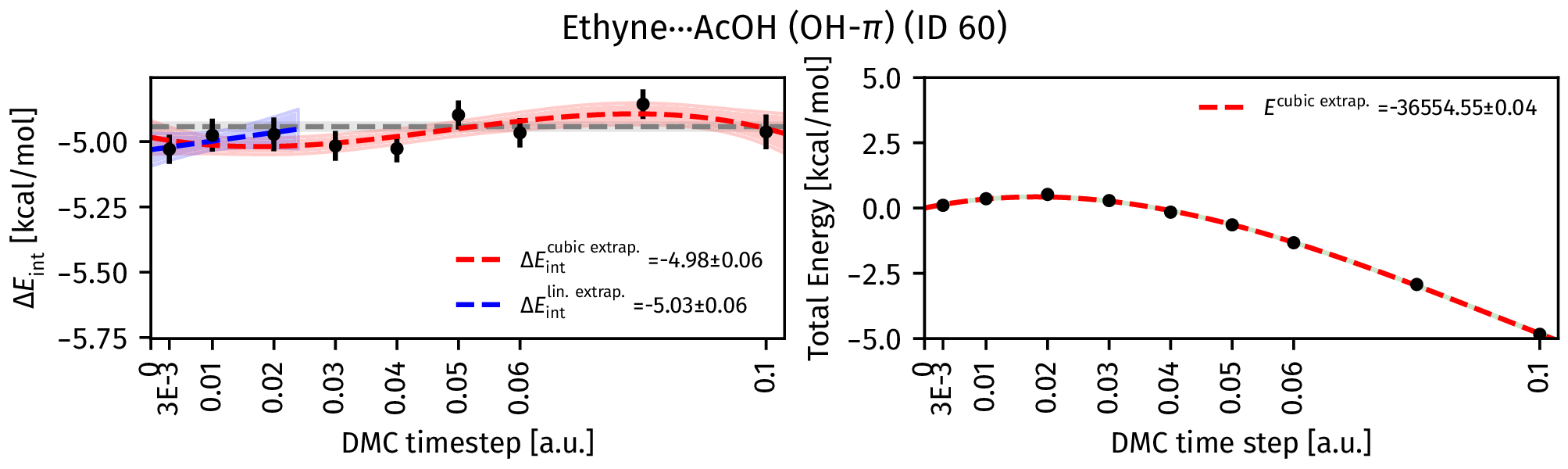}
    \caption{\label{fig:dimer_60} The time step dependence of $\Delta E_\text{int}$ and the total energy of the dimer complex for the Ethyne$\cdots$AcOH (OH-$\pi$) (ID 60) dimer.The dotted gray line represents the CCSD(T) reference in Table~\ref{sec:final_cc_estimates} and the black markers with stochastic 1$\sigma$ error bars represent the DMC estimate for each time step.}
\end{figure}
    
\begin{figure}[!h]
    \includegraphics[width=6.69in]{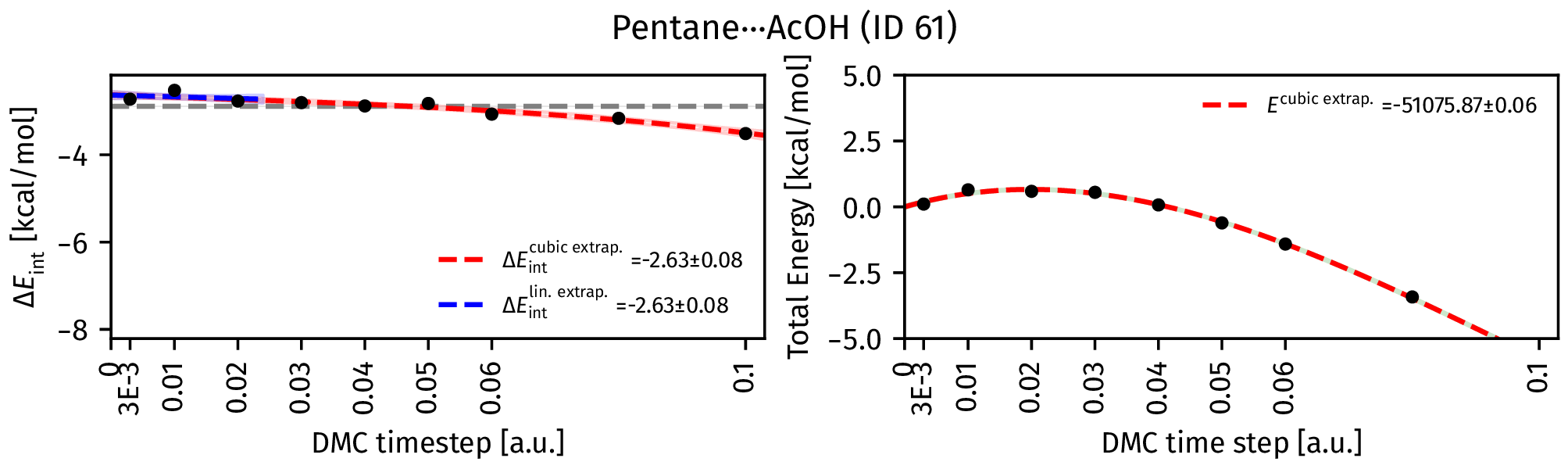}
    \caption{\label{fig:dimer_61} The time step dependence of $\Delta E_\text{int}$ and the total energy of the dimer complex for the Pentane$\cdots$AcOH (ID 61) dimer.The dotted gray line represents the CCSD(T) reference in Table~\ref{sec:final_cc_estimates} and the black markers with stochastic 1$\sigma$ error bars represent the DMC estimate for each time step.}
\end{figure}
    
\begin{figure}[!h]
    \includegraphics[width=6.69in]{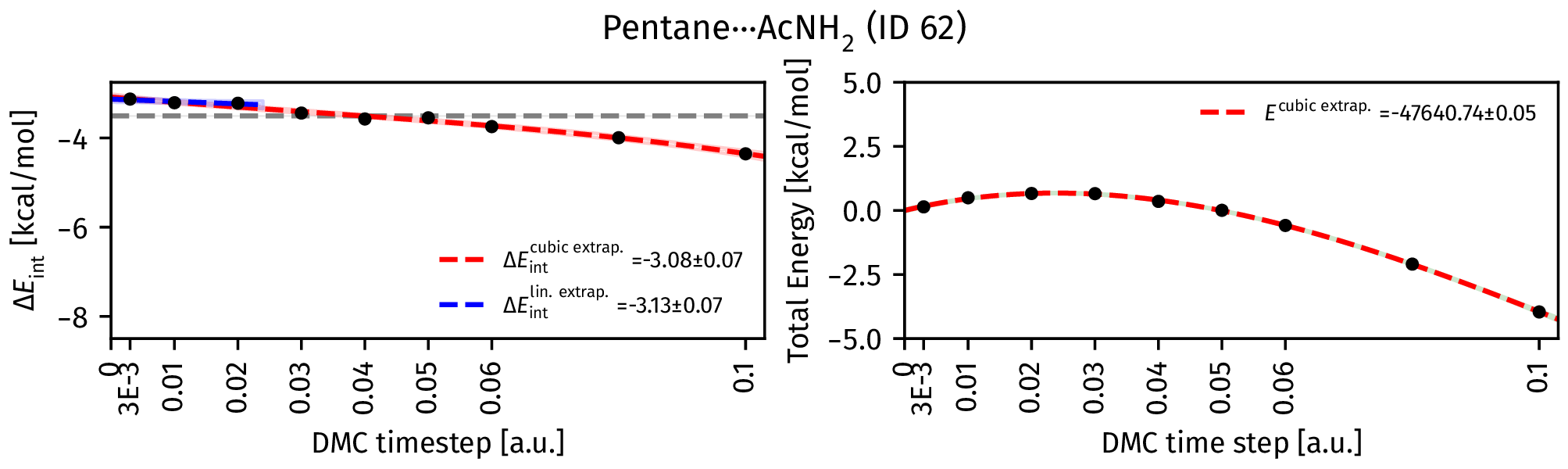}
    \caption{\label{fig:dimer_62} The time step dependence of $\Delta E_\text{int}$ and the total energy of the dimer complex for the Pentane$\cdots$AcNH$_2$ (ID 62) dimer.The dotted gray line represents the CCSD(T) reference in Table~\ref{sec:final_cc_estimates} and the black markers with stochastic 1$\sigma$ error bars represent the DMC estimate for each time step.}
\end{figure}
    
\begin{figure}[!h]
    \includegraphics[width=6.69in]{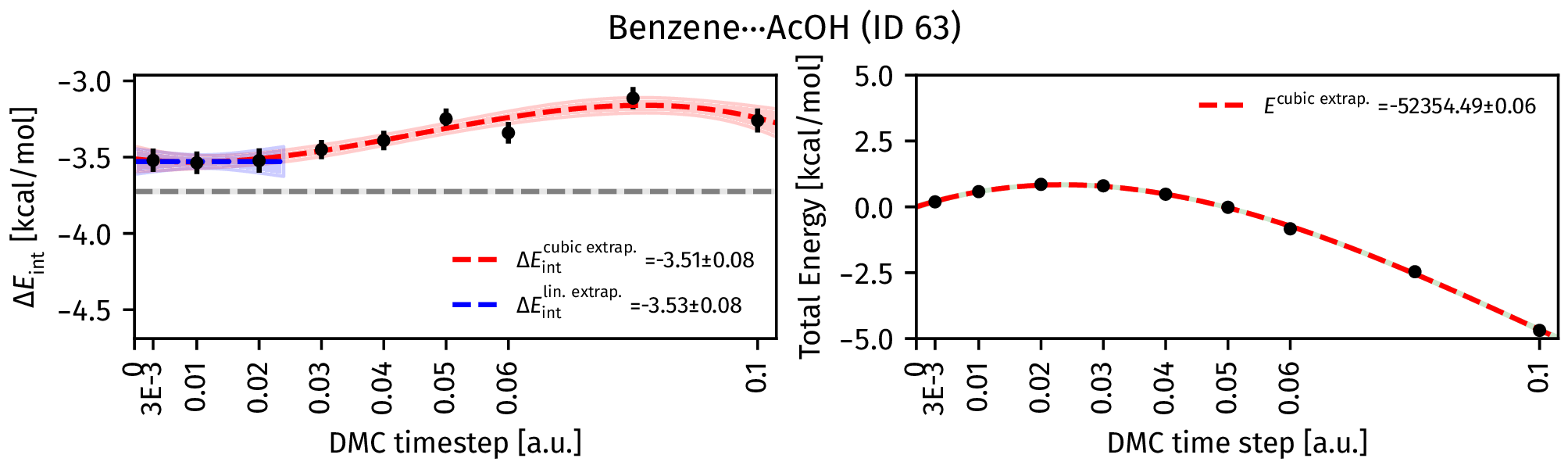}
    \caption{\label{fig:dimer_63} The time step dependence of $\Delta E_\text{int}$ and the total energy of the dimer complex for the Benzene$\cdots$AcOH (ID 63) dimer.The dotted gray line represents the CCSD(T) reference in Table~\ref{sec:final_cc_estimates} and the black markers with stochastic 1$\sigma$ error bars represent the DMC estimate for each time step.}
\end{figure}
    
\begin{figure}[!h]
    \includegraphics[width=6.69in]{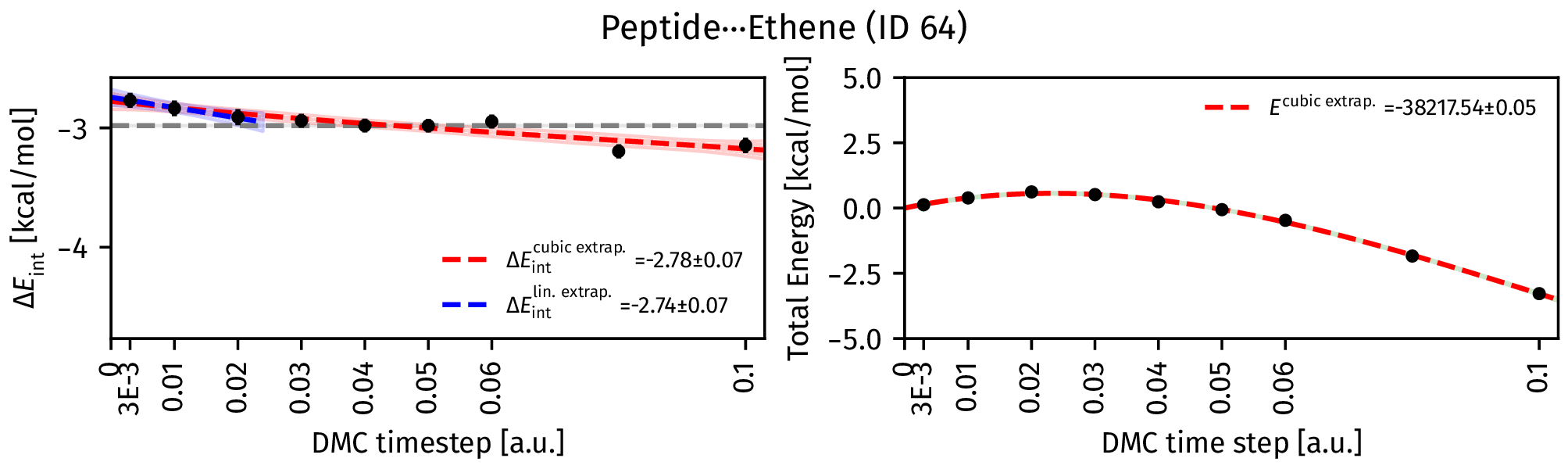}
    \caption{\label{fig:dimer_64} The time step dependence of $\Delta E_\text{int}$ and the total energy of the dimer complex for the Peptide$\cdots$Ethene (ID 64) dimer.The dotted gray line represents the CCSD(T) reference in Table~\ref{sec:final_cc_estimates} and the black markers with stochastic 1$\sigma$ error bars represent the DMC estimate for each time step.}
\end{figure}
    
\begin{figure}[!h]
    \includegraphics[width=6.69in]{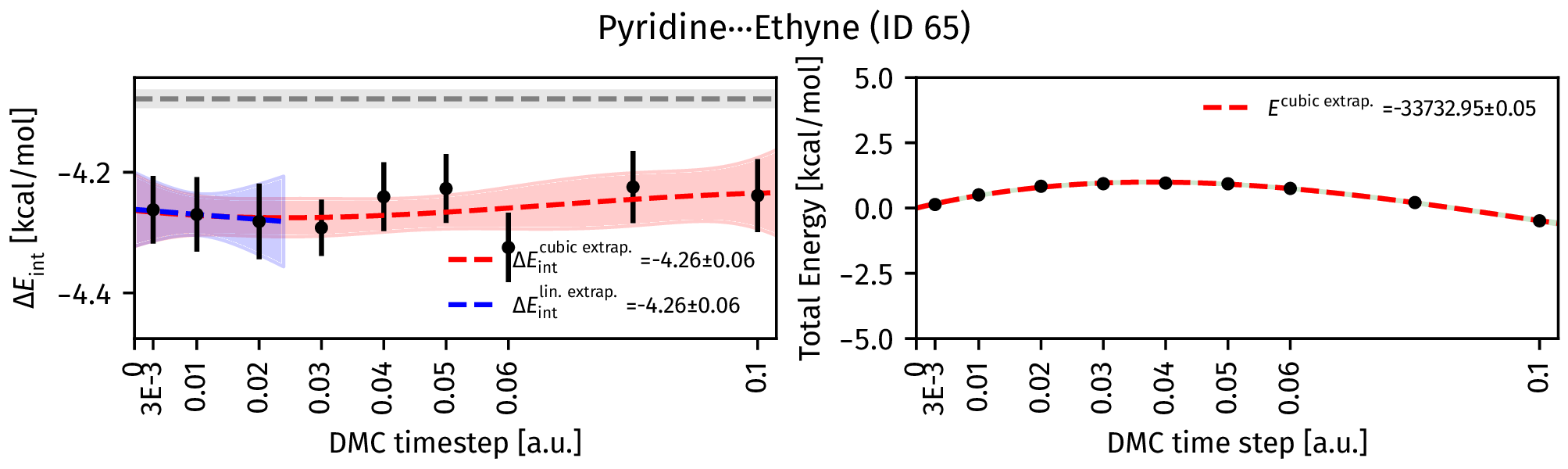}
    \caption{\label{fig:dimer_65} The time step dependence of $\Delta E_\text{int}$ and the total energy of the dimer complex for the Pyridine$\cdots$Ethyne (ID 65) dimer.The dotted gray line represents the CCSD(T) reference in Table~\ref{sec:final_cc_estimates} and the black markers with stochastic 1$\sigma$ error bars represent the DMC estimate for each time step.}
\end{figure}
    
\begin{figure}[!h]
    \includegraphics[width=6.69in]{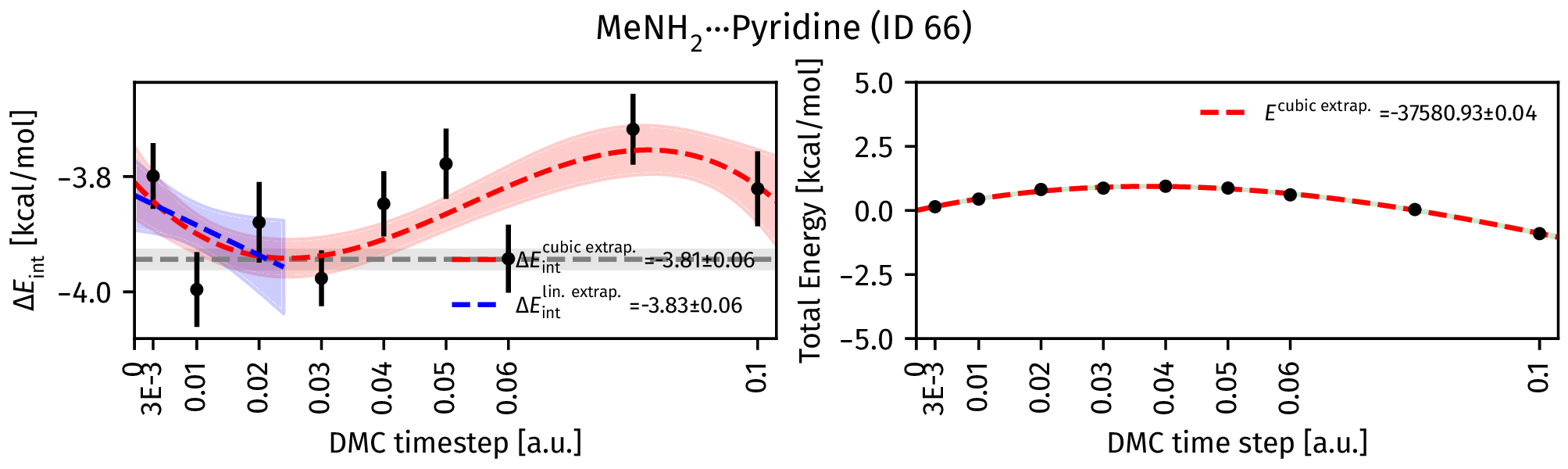}
    \caption{\label{fig:dimer_66} The time step dependence of $\Delta E_\text{int}$ and the total energy of the dimer complex for the MeNH$_2$$\cdots$Pyridine (ID 66) dimer.The dotted gray line represents the CCSD(T) reference in Table~\ref{sec:final_cc_estimates} and the black markers with stochastic 1$\sigma$ error bars represent the DMC estimate for each time step.}
\end{figure}

\clearpage
\section{\label{sec:int_ene_eda}Interaction energy decomposition analysis}

\begin{figure}[h!]
    \centering
    \includegraphics[width=.6\linewidth]{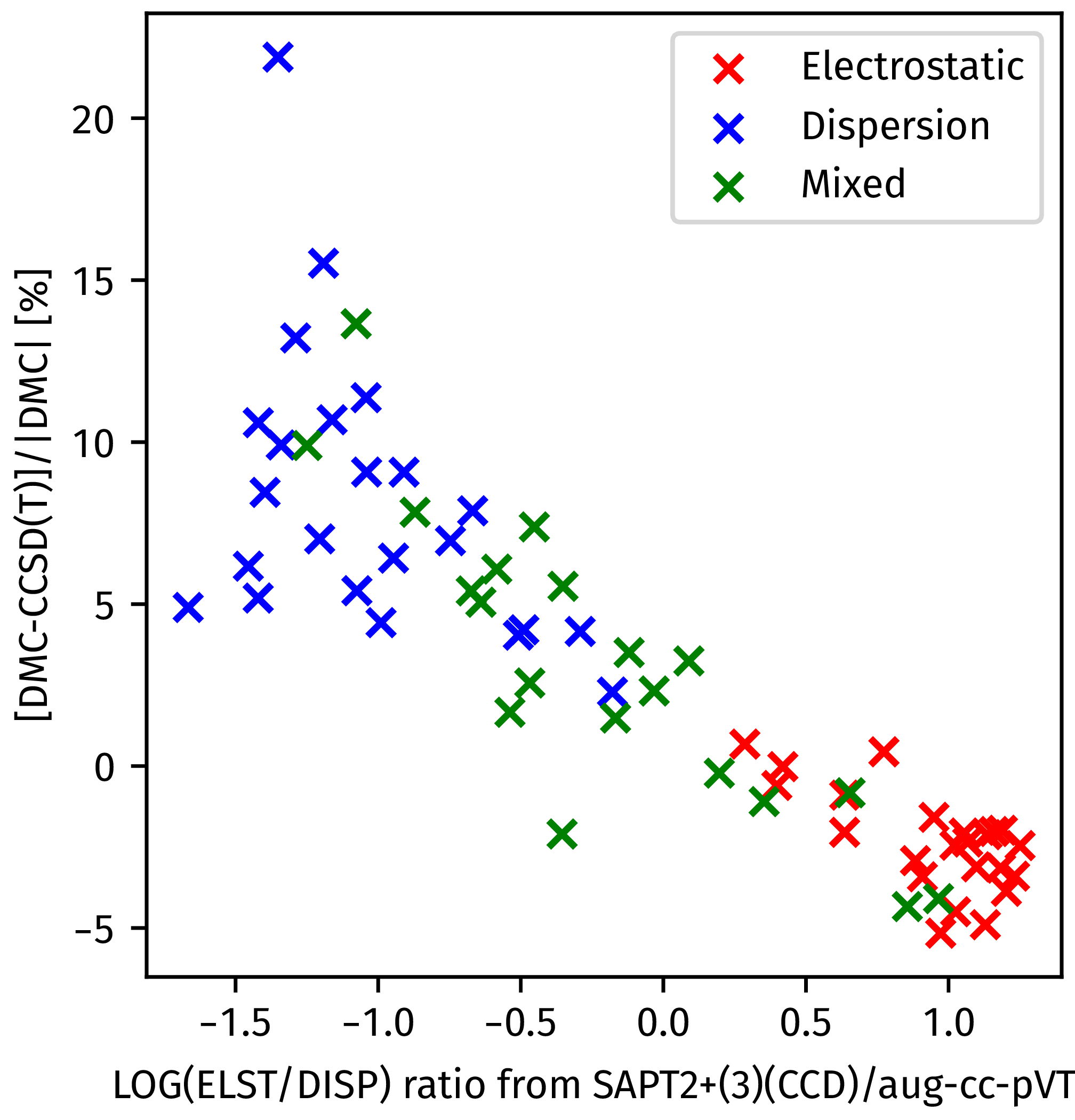}
    \caption{Error decomposition analysis. We report the absolute difference between DMC and CCSD(T) relative to the DMC values, i.e. $\left(E_{\mathrm{DMC}} - E_{\mathrm{CCSD(T)}}\right)/\left|E_{\mathrm{DMC}}\right|$, as a function of the natural logarithm of the electrostatic (ELST) to dispersion (DISP) ratio contribution to the binding energy. The ELST to DISP ratio is determined from the SAPT analysis from Ref.~\citenum{si-villotInitioDispersionPotentials2024}. The color code is red for H-bonded systems (ID from 1 to 23), blue for dispersion dominated systems (ID from 24 to 46), and green for mixed systems (ID from 47 to 66). There is a strong linear trend, with $R^2{=}0.77$. }
    \label{fig:sapt_cc23}
\end{figure}

In this section, we provide the electrostatic (ELST), exchange (EXCH), induction (IND) and dispersion (DISP) contributions to $\Delta E_\text{int.}$ at the $s$SAPT0/jun-cc-pVDZ level in Table~S9.
These values were taken from Burns \etal{}~\cite{SAPT_Sherrill} and we also provide the natural logarithm of ELST and DISP [$\log(\frac{\text{ELST}}{\text{DISP}})$].
In addition, we have collated the same terms for another SAPT level -- SAPT2+(3)(CCD)/aug-cc-pVTZ -- and compared $\log(\frac{\text{ELST}}{\text{DISP}})$ against the relative difference (in \%) between CCSD(T) and DMC in Fig.~\ref{fig:sapt_cc23}. 
These estimates were taken from Ref.~\citenum{si-villotInitioDispersionPotentials2024} for the S66x8 dataset~\cite{si_rezacExtensionsS66Data2011}, where we have used the equilibrium geometries (corresponding closely to the original S66 dataset).
There is a strong linear trend, comparable to the trend observed for $s$SAPT0/jun-cc-pVDZ in Fig.~4 of the main text, with $R^2{=}0.77$.

\LTcapwidth=\textwidth
    \begin{longtable}{lrrrrr}
\caption{\label{tab:sapt_s66_decomposition}Symmetry-adapted perturbation theory (SAPT) energy decomposition for the S66 dataset taken from Ref.~\citenum{SAPT_Sherrill} using the $s$SAPT0 level with the jun-cc-pVDZ basis set. The electrostatic (ELST), exchange (EXCH), induction (IND) and dispersion (DISP) energy components to the interaction energy are reported. The natural logarithm of the ratio between the electrostatic and dispersion energy is also reported.} \\

\toprule
System & ELST & EXCH & IND & DISP & LOG(ELST/DISP) \\ 
\midrule
\endfirsthead

\caption[]{(continued)}\\
\toprule
System & ELST & EXCH & IND & DISP & LOG(ELST/DISP) \\ 
\midrule
\endhead

\multicolumn{6}{r}{(Continued on next page)}\\
\endfoot

\bottomrule
\endlastfoot

Water$\cdots$Water & -8.569 & 6.651 & -1.992 & -1.222 & 1.947 \\
Water$\cdots$MeOH & -9.517 & 8.040 & -2.456 & -1.747 & 1.695 \\
Water$\cdots$MeNH$_2$ & -12.719 & 11.830 & -3.785 & -2.120 & 1.792 \\
Water$\cdots$Peptide & -13.376 & 11.329 & -3.791 & -2.639 & 1.623 \\
MeOH$\cdots$MeOH & -9.547 & 8.413 & -2.586 & -2.059 & 1.534 \\
MeOH$\cdots$MeNH$_2$ & -13.210 & 13.167 & -4.194 & -2.930 & 1.506 \\
MeOH$\cdots$Peptide & -13.224 & 12.114 & -3.984 & -3.199 & 1.419 \\
MeOH$\cdots$Water & -8.454 & 6.853 & -2.092 & -1.447 & 1.765 \\
MeNH$_2$$\cdots$MeOH & -4.350 & 4.261 & -0.991 & -1.622 & 0.986 \\
MeNH$_2$$\cdots$MeNH$_2$ & -5.970 & 6.435 & -1.489 & -2.472 & 0.882 \\
MeNH$_2$$\cdots$Peptide & -7.334 & 7.561 & -1.766 & -3.416 & 0.764 \\
MeNH$_2$$\cdots$Water & -12.935 & 12.208 & -3.900 & -2.469 & 1.656 \\
Peptide$\cdots$MeOH & -8.503 & 7.199 & -2.122 & -2.687 & 1.152 \\
Peptide$\cdots$MeNH$_2$ & -11.186 & 10.961 & -3.310 & -3.422 & 1.184 \\
Peptide$\cdots$Peptide & -11.743 & 10.602 & -3.379 & -4.002 & 1.076 \\
Peptide$\cdots$Water & -7.371 & 5.350 & -1.655 & -1.611 & 1.521 \\
Uracil$\cdots$Uracil (BP) & -27.486 & 26.002 & -10.821 & -6.076 & 1.509 \\
Water$\cdots$Pyridine & -11.574 & 10.775 & -3.610 & -2.482 & 1.540 \\
MeOH$\cdots$Pyridine & -12.117 & 11.904 & -3.965 & -3.196 & 1.333 \\
AcOH$\cdots$AcOH & -33.623 & 34.750 & -15.334 & -6.168 & 1.696 \\
AcNH$_2$$\cdots$AcNH$_2$ & -26.449 & 24.247 & -9.552 & -5.031 & 1.660 \\
AcOH$\cdots$Uracil & -32.158 & 30.809 & -13.446 & -6.119 & 1.659 \\
AcNH$_2$$\cdots$Uracil & -30.530 & 27.251 & -11.514 & -5.770 & 1.666 \\
Benzene$\cdots$Benzene ($\pi$-$\pi$) & -1.674 & 6.316 & -0.686 & -6.794 & -1.401 \\
Pyridine$\cdots$Pyridine ($\pi$-$\pi$) & -3.379 & 7.458 & -0.809 & -7.290 & -0.769 \\
Uracil$\cdots$Uracil ($\pi$-$\pi$) & -9.511 & 11.714 & -1.645 & -9.648 & -0.014 \\
Benzene$\cdots$Pyridine ($\pi$-$\pi$) & -2.693 & 7.030 & -0.762 & -7.116 & -0.972 \\
Benzene$\cdots$Uracil ($\pi$-$\pi$) & -5.521 & 10.078 & -1.126 & -8.824 & -0.469 \\
Pyridine$\cdots$Uracil ($\pi$-$\pi$) & -6.619 & 9.909 & -1.215 & -8.736 & -0.278 \\
Benzene$\cdots$Ethene & -0.892 & 4.271 & -0.481 & -3.707 & -1.425 \\
Uracil$\cdots$Ethene & -3.788 & 5.816 & -0.545 & -4.334 & -0.135 \\
Uracil$\cdots$Ethyne & -4.676 & 5.709 & -0.584 & -3.999 & 0.156 \\
Pyridine$\cdots$Ethene & -1.712 & 4.865 & -0.521 & -3.954 & -0.837 \\
Pentane$\cdots$Pentane & -1.649 & 5.536 & -0.500 & -5.765 & -1.252 \\
Neopentane$\cdots$Pentane & -1.164 & 3.911 & -0.378 & -4.014 & -1.238 \\
Neopentane$\cdots$Neopentane & -0.659 & 2.637 & -0.284 & -2.850 & -1.465 \\
Cyclopentane$\cdots$Neopentane & -1.144 & 3.890 & -0.393 & -3.894 & -1.225 \\
Cyclopentane$\cdots$Cyclopentane & -1.324 & 4.371 & -0.438 & -4.554 & -1.235 \\
Benzene$\cdots$Cyclopentane & -2.394 & 5.864 & -0.646 & -5.972 & -0.914 \\
Benzene$\cdots$Neopentane & -1.814 & 4.464 & -0.521 & -4.635 & -0.938 \\
Uracil$\cdots$Pentane & -2.468 & 6.869 & -0.813 & -6.973 & -1.039 \\
Uracil$\cdots$Cyclopentane & -1.990 & 5.843 & -0.608 & -6.154 & -1.129 \\
Uracil$\cdots$Neopentane & -2.350 & 5.004 & -0.490 & -4.913 & -0.738 \\
Ethene$\cdots$Pentane & -1.018 & 3.218 & -0.335 & -2.956 & -1.065 \\
Ethyne$\cdots$Pentane & -1.241 & 3.069 & -0.342 & -2.542 & -0.717 \\
Peptide$\cdots$Pentane & -2.332 & 6.161 & -0.931 & -5.673 & -0.889 \\
Benzene$\cdots$Benzene (TS) & -2.025 & 4.016 & -0.574 & -4.277 & -0.748 \\
Pyridine$\cdots$Pyridine (TS) & -3.033 & 4.592 & -0.723 & -4.350 & -0.361 \\
Benzene$\cdots$Pyridine (TS) & -2.626 & 4.241 & -0.707 & -4.330 & -0.500 \\
Benzene$\cdots$Ethyne (CH-$\pi$) & -2.490 & 3.106 & -0.805 & -2.798 & -0.117 \\
Ethyne$\cdots$Ethyne (TS) & -2.111 & 2.203 & -0.521 & -0.981 & 0.767 \\
Benzene$\cdots$AcOH (OH-$\pi$) & -4.395 & 5.735 & -1.792 & -3.786 & 0.149 \\
Benzene$\cdots$AcNH$_2$ (NH-$\pi$) & -5.326 & 6.017 & -1.515 & -3.321 & 0.473 \\
Benzene$\cdots$Water (OH-$\pi$) & -3.360 & 3.520 & -0.976 & -2.203 & 0.422 \\
Benzene$\cdots$MeOH (OH-$\pi$) & -3.680 & 5.051 & -1.168 & -3.761 & -0.022 \\
Benzene$\cdots$MeNH$_2$ (NH-$\pi$) & -2.649 & 4.407 & -0.677 & -3.780 & -0.356 \\
Benzene$\cdots$Peptide (NH-$\pi$) & -4.265 & 6.178 & -1.329 & -5.458 & -0.247 \\
Pyridine$\cdots$Pyridine (CH-N) & -4.964 & 5.017 & -1.336 & -3.047 & 0.488 \\
Ethyne$\cdots$Water (CH-O) & -4.482 & 3.051 & -0.950 & -0.878 & 1.630 \\
Ethyne$\cdots$AcOH (OH-$\pi$) & -7.787 & 7.906 & -2.401 & -2.516 & 1.130 \\
Pentane$\cdots$AcOH & -1.574 & 4.290 & -0.533 & -3.843 & -0.893 \\
Pentane$\cdots$AcNH$_2$ & -2.141 & 5.398 & -0.992 & -4.457 & -0.733 \\
Benzene$\cdots$AcOH & -3.688 & 5.718 & -0.821 & -4.734 & -0.250 \\
Peptide$\cdots$Ethene & -2.991 & 4.378 & -0.804 & -2.948 & 0.015 \\
Pyridine$\cdots$Ethyne & -6.249 & 5.446 & -1.749 & -1.926 & 1.177 \\
MeNH$_2$$\cdots$Pyridine & -4.622 & 6.071 & -1.035 & -3.863 & 0.179 \\
\end{longtable}

\clearpage

\bibliography{si-refs}